\documentclass[twocolumn]{aastex62}
\usepackage{amsmath, amsfonts, amssymb, cancel, gensymb, graphicx, float, parskip}
\usepackage{natbib}
\usepackage{hyperref}

\newcommand{\um}{ \text{ } \mu \text{m}}

\shorttitle{AO Survey of Stellar Variability at the Galactic Center}
\shortauthors{Gautam et al.}

\begin{document}
\title{An Adaptive Optics Survey of Stellar Variability at the Galactic Center}

\author[0000-0002-2836-117X]{Abhimat Krishna Gautam}
\affiliation{Department of Physics and Astronomy, University of California, Los Angeles}

\author{Tuan Do}
\affiliation{Department of Physics and Astronomy, University of California, Los Angeles}

\author[0000-0003-3230-5055]{Andrea M. Ghez}
\affiliation{Department of Physics and Astronomy, University of California, Los Angeles}

\author[0000-0002-6753-2066]{Mark R. Morris}
\affiliation{Department of Physics and Astronomy, University of California, Los Angeles}

\author{Gregory D. Martinez}
\affiliation{Department of Physics and Astronomy, University of California, Los Angeles}

\author[0000-0003-2874-1196]{Matthew W. Hosek Jr.}
\affiliation{Institute for Astronomy, University of Hawaii}

\author[0000-0001-9611-0009]{Jessica R. Lu}
\affiliation{Department of Astronomy, University of California, Berkeley}

\author[0000-0001-5972-663X]{Shoko Sakai}
\affiliation{Department of Physics and Astronomy, University of California, Los Angeles}

\author[0000-0003-2618-797X]{Gunther Witzel}
\affiliation{Department of Physics and Astronomy, University of California, Los Angeles}

\author[0000-0001-5341-0765]{Siyao Jia}
\affiliation{Department of Astronomy, University of California, Berkeley}

\author{Eric E. Becklin}
\affiliation{Department of Physics and Astronomy, University of California, Los Angeles}

\author{Keith Matthews}
\affiliation{Division of Physics, Mathematics, and Astronomy, California Institute of Technology}

\correspondingauthor{Abhimat Gautam}
\email{abhimat@astro.ucla.edu}

\begin{abstract}
    \raggedright
    We present a $\approx 11.5$ year adaptive optics (AO) study of stellar variability and search for eclipsing binaries in the central $\sim 0.4$ pc ($\sim 10''$) of the Milky Way nuclear star cluster. We measure the photometry of 563 stars using the Keck II NIRC2 imager ($K'$-band, $\lambda_0 = 2.124 \um$). We achieve a photometric uncertainty floor of $\Delta m_{K'} \sim 0.03$ ($\approx 3\%$), comparable to the highest precision achieved in other AO studies. Approximately half of our sample ($50 \pm 2 \%$) shows variability. $52 \pm 5\%$ of known early-type young stars and $43 \pm 4 \%$ of known late-type giants are variable. These variability fractions are higher than those of other young, massive star populations or late-type giants in globular clusters, and can be largely explained by two factors. First, our experiment time baseline is sensitive to long-term intrinsic stellar variability. Second, the proper motion of stars behind spatial inhomogeneities in the foreground extinction screen can lead to variability. We recover the two known Galactic center eclipsing binary systems: IRS 16SW and S4-258 (E60). We constrain the Galactic center eclipsing binary fraction of known early-type stars to be at least $2.4 \pm 1.7\%$. We find no evidence of an eclipsing binary among the young S-stars nor among the young stellar disk members. These results are consistent with the local OB eclipsing binary fraction. We identify a new periodic variable, S2-36, with a 39.43 day period. Further observations are necessary to determine the nature of this source.
\end{abstract}


\section{Introduction} 
\label{sec:Introduction}
At a distance of $\approx 8 \text{ kpc}$, the Milky Way Galactic center contains the closest nuclear star cluster and a supermassive black hole (SMBH) with a mass of $\approx 4 \times 10^6 \text{ } M_{\odot}$ at the location of the radio source Sgr A* \citep{2016ApJ...830...17B, 2008ApJ...689.1044G,2009ApJ...692.1075G}. Adaptive optics (AO) on near-infrared (NIR) 8--10 m class telescopes has allowed diffraction-limited, resolved imaging and spectroscopic studies of the stellar population in the crowded central regions of the Galactic center. The NIR spectroscopic observations have revealed a population of more than 100 young, massive stars \citep{2007ApJ...669.1024M, 2010ApJ...708..834B, 2011ApJ...741..108P, 2013ApJ...764..154D} within the central 0.5 pc \citep{2015ApJ...808..106S, 2013ApJ...764..154D} of age $\approx 4$--8 Myr \citep{2013ApJ...764..155L}. This young star cluster is among the most massive in the Milky Way.
Most members of the nuclear star cluster are old stars with ages $> 1$ Gyr \citep{2013ApJ...764..154D}. NIR observations sample the bright end of the old population, primarily composed of late-type M and K giant stars.

While AO observations have improved knowledge of the Galactic center stellar population, no general stellar photometric variability study has yet been conducted of this population with NIR AO.
\citet{2014ApJ...782..101P} used NIR AO photometry from the Very Large Telescope (VLT) to search for periodic variability, indicative of eclipsing binary systems. However the study only searched for binary variability and the sample was limited to the spectroscopically confirmed young star population at the Galactic center. Other AO photometric studies, such as that of \citet{2010A&A...511A..18S}, only reported single-epoch photometry. Previous studies without AO observations have largely focused on wider fields of view centered at the Galactic center \citep[e.g.][]{2007AcA....57..173P, 2017MNRAS.470.3427D}. These experiments suffered from confusion in the central regions of the nuclear star cluster, where rising stellar population density leads to crowding.

\citet{2007ApJ...659.1241R} studied photometric variability in the resolved stellar populations of the central $5'' \times 5''$ of the Galactic center. This study used Keck Observatory speckle data over a time baseline of 10 years. However, the speckle data and the ``shift-and-add" image combination technique implemented in the study faced limitations with sensitivity and photometric precision, especially for stars fainter than $m_{K} \sim 14$, along with a smaller field of view. Using LGSAO data, our study is able to achieve greater depth with much higher precision at fainter magnitudes. Additionally, the NIRC2 imager used in our study affords a larger stellar sample with a wider, $10'' \times 10''$ field of view.

In this work, we performed a general variability study of the stellar populations in the Galactic center using 10 years of Keck AO imaging data. These long-term Galactic center monitoring data have previously been used primarily to derive astrometric measurements of the stars in the nuclear star cluster \citep[e.g.][]{2008ApJ...689.1044G, 2014ApJ...783..131Y, 2016ApJ...830...17B}. Using these data, we investigated the following scientific questions:

\begin{itemize}
    \item \emph{The long-term variability of a young star cluster:} While various sources contribute to the variability of young, massive stars, NIR observations are especially sensitive to phenomena such as dust extinction and accretion activity common in very young stars. While several recent studies have been conducted of NIR variability in other young star clusters \citep[e.g.][]{1999MNRAS.304L..10G, 2012ApJ...755...65R, 2015AJ....150..132R, 2016MNRAS.456.2505L}, the time baselines of such studies only span a few months to a few years. Our experiment's $\approx 11.5$ year time baseline offers a unique opportunity to study the long-term variability of stars in a young star cluster.
    \item \emph{A search for binaries:} Binary systems are especially useful to learn about the Galactic center environment. Stellar multiplicity is typically a direct result of fragmentation during star formation \citep[see e.g.][]{2013ARA&A..51..269D}. Dynamical interactions with the dense Galactic center stellar environment and its central SMBH can further affect the observed binary fraction \citep[e.g.][]{1988Natur.331..687H, 2014ApJ...780..148A, 2016MNRAS.460.3494S}. The observed binary fraction can therefore constrain Galactic center star formation and dynamical evolution models. Photometry offers a method to search for binary systems, allowing for the detection of eclipsing binaries or tidally distorted systems. Our experiment offers the largest photometric sample of stars in the central half parsec of the Galactic center to search for binary systems.
    \item \emph{Stars on the instability strip:} Precision photometry can reveal interesting classes of variable stars undergoing pulsations during periods of instabilities. Such stars (e.g. Classical Cepheids and Type II Cepheids, AGB stars, and Miras) often have characteristic periods, luminosities, and variability amplitudes that can reveal specific populations having associated ages or metallicities to which they belong \citep[see e.g.][]{2006MNRAS.370.1979M, 2010ApJ...723.1195R, 2017MNRAS.464.1119C}.
    \item \emph{Search for microlensing events:} The high stellar density at the Galactic center makes microlensing events likely. Such events can be revealed through photometric monitoring, with brightening events associated with the passing of a foreground massive object in front of a background star.
    \item \emph{Constraints on dust column size and identification of stars whose variability can be ascribed to extinction.:} Wide-field studies of the Galactic center have found that the extinguishing material in the environment is clumpy and has structure on approximately arcsecond spatial scales \citep[e.g.][]{2004A&A...426...81P, 2010A&A...511A..18S, 2018A&A...610A..83N}. Stars can display variability while passing behind such variations in the extinction screen due to the stellar proper motions. Examples of non-periodic variability on long time-scales therefore can probe fluctuations in the extinction screen towards the Galactic center and constrain the dust column size of possible extinguishing dust structures.
    \item \emph{Investigate properties of AO photometry and anisoplanatism:} AO data faces challenges for obtaining precision photometry. In a crowded field, flux is estimated by point spread function (PSF) fitting to isolate flux contributions of individual stars \citep[see e.g.][]{2010A&A...511A..18S}. However due to anisoplanatic effects, atmospheric conditions, and performance of the AO system during observations, the PSF shape varies over time and across a field of view such as that used in this work. In this work, we investigated the properties of such effects and developed a method to perform corrections to single PSF AO photometry estimates.
\end{itemize}

Section~\ref{sec:observations_calibration_and_stellar_sample} describes our observations, data reduction methods, and our photometric calibration process.
Section~\ref{sec:observations_calibration_and_stellar_sample} also details the selection of the stellar sample used in this work.
In Section~\ref{sec:stellar_variability}, we describe our methods to identify variable stars and to constrain the variability fraction.
Section~\ref{sec:periodic_variability} details our methods to identify periodically variable stars.
Our results are detailed in Section~\ref{sec:results}.
In Section~\ref{sec:Discussion}, we review what our results reveal about the Galactic center stellar population and environment.
We summarize our findings in Section~\ref{sec:Conclusion}.


\startlongtable
\begin{deluxetable*}{llrrrccccc}
    \tablecolumns{9}
    \tablecaption{Observations used in this work\label{tab:Observations}}
    \tablehead{
        \colhead{Date}                      &
        \colhead{MJD}                       &
        \colhead{Frames}                    &
        \colhead{Total Stars}               &
        \colhead{Stars in}                  &
        \colhead{Absolute Phot.}            &
        \colhead{Relative}                  &
        \colhead{Astrometric}               &
        \colhead{Med.}                      &
        \colhead{Med.}                      \\
        \colhead{(UTC)}                     &
        \colhead{}                          &
        \colhead{}                          &
        \colhead{Detected}                  &
        \colhead{Sample}                    &
        \colhead{Zeropoint}                 &
        \colhead{Phot. Med.}                &
        \colhead{Med.}                      &
        \colhead{FWHM}                      &
        \colhead{Strehl}                    \\
        \colhead{}                          &
        \colhead{}                          &
        \colhead{}                          &
        \colhead{}                          &
        \colhead{}                          &
        \colhead{Error ($K'$ mag)}          &
        \colhead{Error ($K'$ mag)}          &
        \colhead{Error (mas)}               &
        \colhead{(mas)}                     &
        \colhead{Ratio}
    }
    \startdata
    2006-05-03                  & 53858.512 & 107   & 1768  & 500   & 0.179 & 0.035 & 0.332 & 57.61     & 0.35  \\
    2006-06-20\tablenotemark{S} & 53906.392 & 50    & 1456  & 493   & 0.197 & 0.049 & 0.347 & 60.10     & 0.31  \\
    2006-06-21\tablenotemark{S} & 53907.411 & 119   & 1759  & 508   & 0.181 & 0.041 & 0.320 & 56.59     & 0.38  \\
    2006-07-17                  & 53933.344 & 64    & 2179  & 501   & 0.172 & 0.031 & 0.320 & 57.73     & 0.37  \\
    2007-05-17                  & 54237.551 & 76    & 2514  & 511   & 0.202 & 0.066 & 0.334 & 58.02     & 0.36  \\
    2007-08-10\tablenotemark{S} & 54322.315 & 35    & 1246  & 479   & 0.189 & 0.045 & 0.385 & 63.57     & 0.24  \\
    2007-08-12\tablenotemark{S} & 54324.304 & 54    & 1539  & 503   & 0.185 & 0.054 & 0.352 & 55.66     & 0.34  \\
    2008-05-15                  & 54601.492 & 134   & 2089  & 524   & 0.193 & 0.039 & 0.298 & 53.47     & 0.30  \\
    2008-07-24                  & 54671.323 & 104   & 2189  & 515   & 0.165 & 0.022 & 0.297 & 58.95     & 0.33  \\
    2009-05-01\tablenotemark{S} & 54952.543 & 127   & 1650  & 506   & 0.181 & 0.019 & 0.341 & 63.82     & 0.32  \\
    2009-05-02\tablenotemark{S} & 54953.517 & 49    & 1302  & 507   & 0.179 & 0.021 & 0.361 & 58.26     & 0.36  \\
    2009-05-04\tablenotemark{S} & 54955.552 & 56    & 1788  & 519   & 0.182 & 0.020 & 0.339 & 53.49     & 0.43  \\
    2009-07-24                  & 55036.333 & 75    & 1701  & 501   & 0.185 & 0.026 & 0.332 & 61.82     & 0.27  \\
    2009-09-09                  & 55083.249 & 43    & 1921  & 517   & 0.174 & 0.031 & 0.357 & 58.20     & 0.36  \\
    2010-05-04\tablenotemark{S} & 55320.546 & 105   & 1235  & 490   & 0.178 & 0.043 & 0.389 & 63.24     & 0.31  \\
    2010-05-05\tablenotemark{S} & 55321.583 & 60    & 1631  & 522   & 0.177 & 0.038 & 0.325 & 60.37     & 0.34  \\
    2010-07-06                  & 55383.351 & 117   & 1956  & 502   & 0.184 & 0.036 & 0.326 & 61.11     & 0.32  \\
    2010-08-15                  & 55423.284 & 127   & 1826  & 515   & 0.176 & 0.037 & 0.314 & 58.16     & 0.30  \\
    2011-05-27                  & 55708.505 & 114   & 1563  & 494   & 0.200 & 0.027 & 0.402 & 64.00     & 0.29  \\
    2011-07-18                  & 55760.346 & 167   & 2031  & 506   & 0.210 & 0.033 & 0.331 & 58.14     & 0.28  \\
    2011-08-23\tablenotemark{S} & 55796.280 & 102   & 2052  & 516   & 0.214 & 0.025 & 0.361 & 59.76     & 0.36  \\
    2011-08-24\tablenotemark{S} & 55797.274 & 102   & 1640  & 492   & 0.212 & 0.028 & 0.371 & 62.13     & 0.31  \\
    2012-05-15\tablenotemark{S} & 56062.518 & 178   & 1778  & 522   & 0.209 & 0.030 & 0.339 & 59.69     & 0.31  \\
    2012-05-18\tablenotemark{S} & 56065.494 & 68    & 1252  & 494   & 0.208 & 0.020 & 0.389 & 68.25     & 0.26  \\
    2012-07-24                  & 56132.310 & 162   & 2344  & 517   & 0.206 & 0.020 & 0.319 & 58.41     & 0.35  \\
    2013-04-26\tablenotemark{S} & 56408.564 & 75    & 1418  & 475   & 0.162 & 0.075 & 0.368 & 65.63     & 0.25  \\
    2013-04-27\tablenotemark{S} & 56409.566 & 79    & 1313  & 478   & 0.168 & 0.042 & 0.376 & 70.80     & 0.25  \\
    2013-07-20                  & 56493.325 & 193   & 1805  & 509   & 0.161 & 0.035 & 0.347 & 58.63     & 0.36  \\
    2014-05-19                  & 56796.524 & 147   & 1483  & 497   & 0.159 & 0.033 & 0.384 & 64.20     & 0.30  \\
    2014-08-06                  & 56875.290 & 127   & 1778  & 508   & 0.156 & 0.034 & 0.347 & 56.89     & 0.36  \\
	2015-08-09\tablenotemark{S} & 57243.298 & 43    & 1435  & 490   & 0.163 & 0.041 & 0.553 & 62.63     & 0.32  \\
	2015-08-10\tablenotemark{S} & 57244.291 & 98    & 1884  & 497   & 0.161 & 0.026 & 0.499 & 57.02	    & 0.38  \\
	2015-08-11\tablenotemark{S} & 57245.303 & 74    & 1662  & 499   & 0.162 & 0.032 & 0.573 & 56.72     & 0.38  \\
	2016-05-03                  & 57511.515 & 166   & 1661  & 490   & 0.197 & 0.022 & 0.552 & 61.10     & 0.34  \\
	2016-07-13                  & 57582.363 & 144   & 1389  & 476   & 0.170 & 0.034 & 0.658 & 60.00     & 0.30  \\
    2017-05-04\tablenotemark{S} & 57877.536 & 112   & 1307  & 471   & 0.168 & 0.036 & 0.721 & 70.77     & 0.26  \\
    2017-05-05\tablenotemark{S} & 57878.531 & 177   & 1705  & 489   & 0.160 & 0.023 & 0.588 & 58.06     & 0.35  \\
    2017-07-18                  & 57952.402 & 9     & 1125  & 469   & 0.168 & 0.033 & 0.693 & 65.10     & 0.27  \\
    2017-07-27                  & 57961.274 & 23    & 652   & 361   & 0.151 & 0.077 & 1.348 & 88.22     & 0.15  \\
    2017-08-09\tablenotemark{S} & 57974.321 & 23    & 1168  & 472   & 0.164 & 0.028 & 0.828 & 62.73     & 0.30  \\
    2017-08-10\tablenotemark{S} & 57975.285 & 29    & 1264  & 472   & 0.173 & 0.026 & 0.799 & 59.12     & 0.32  \\
    2017-08-11\tablenotemark{S} & 57976.283 & 87    & 1495  & 483   & 0.176 & 0.026 & 0.770 & 53.19     & 0.37  \\
    2017-08-23\tablenotemark{S} & 57988.268 & 59    & 1311  & 477   & 0.192 & 0.027 & 0.802 & 65.07     & 0.29  \\
    2017-08-24\tablenotemark{S} & 57989.268 & 41    & 1016  & 469   & 0.200 & 0.029 & 0.825 & 61.48     & 0.33  \\
    2017-08-26\tablenotemark{S} & 57991.255 & 33    & 1377  & 475   & 0.183 & 0.027 & 0.757 & 59.67     & 0.33  \\
    \enddata
    \tablecomments{Median astrometric and photometric errors were computed for stars in our study's sample detected in the corresponding observation. Absolute photometric zeropoint errors were calculated after conducting initial calibration, using bandpass corrected reference fluxes for non-variable stars from \citet{1996ApJ...470..864B} in our experiment's field of view. Relative photometric errors were determined after our calibration and local correction method were applied. The median FWHM and Strehl quantities were calculated for IRS 33N across all frames used to construct the final image for the corresponding observation.}
    \tablenotetext{S}{Denotes consecutive nights of observations that were combined into single epochs in previous publications from our group for astrometric study. In this work, we split multiple night combined epochs into single night epochs for greater time precision.}
\end{deluxetable*}

\section{Observations, Photometric Calibration, and Stellar Sample} 
\label{sec:observations_calibration_and_stellar_sample}

\subsection{Observations and Data Reduction} 
\label{sub:observations_and_data_reduction}
We used laser guide star adaptive optics (LGSAO) high-resolution imaging of the Galactic center obtained at the 10-m W. M. Keck II telescope with the NIRC2 near-infrared facility imager (PI: K. Matthews) through the $K'$ bandpass ($\lambda_0 = 2.124 \text{ } \mu \text{m}$, $\Delta \lambda = 0.351 \text{ } \mu \text{m}$). Observations were centered near the location of Sgr A* in the nuclear star cluster, with a field of view of the NIRC2 images extending about $10'' \times 10''$ ($10'' \approx 0.4$ pc at Sgr A*'s distance of $R_0 \approx 8 \text{ kpc}$ \citep{2016ApJ...830...17B}) and a plate scale of 9.952 mas/pix \citep[up to 2014 data]{2010ApJ...725..331Y} or 9.971 mas/pix \citep[post 2014 data]{2016PASP..128i5004S}. Observations used in this work were obtained over 45 nights spanning May 2006 -- August 2017. We list details about individual observations in Table~\ref{tab:Observations}, and the observational setup is further detailed by \citet{2008ApJ...689.1044G} and \citet{2014ApJ...783..131Y}. Observations taken until 2013 have been reported in previous studies by our group \citep{2008ApJ...689.1044G, 2014ApJ...783..131Y, 2016ApJ...830...17B}. Observations used in this work taken during 2014--2017 have not been previously reported.

Final images for each night were created following the same methods as reported by \citet{2008ApJ...689.1044G,2008ApJ...675.1278S}. We combined frames to construct final images separately for each night to achieve higher time precision, whereas in previous studies by our group, frames separated by a few days were combined into single final images \citep{2008ApJ...689.1044G, 2014ApJ...783..131Y, 2016ApJ...830...17B}. Each frame was sky-subtracted, flat-fielded, bad-pixel-corrected, and corrected for the effects of optical distortion \citep{2010ApJ...725..331Y, 2016PASP..128i5004S}. The bright, isolated star IRS 33N ($m_{K'} \sim 11.3$) was used to measure a Strehl ratio and full width at half maximum of the AO-corrected stellar image (FWHM) to evaluate the quality of each frame. We constructed the final image for each observation by averaging the individual frames (weighted by Strehl ratio) collected over that night. We selected frames to create the final nightly image by a cut in the FWHM: frames used for the final nightly image passed the condition $\text{FWHM}_{\text{33N}} \leq 1.25 \times \text{min}(\text{FWHM}_{\text{33N}})$. This cut was implemented to reduce the impact of lower quality frames in making the nightly images. The Strehl ratio weights used to average the individual frames were additionally used to calculate a weighted Modified Julian Date (MJD) time for the final image from the observation times of the individual frames used. This weighted MJD was adopted as the observation time for each data point used in this work.

The frames used to construct final images for each observation night were further divided into three independent subsets. Each subset received frames of similar Strehl and FWHM statistics, and the frames in each of the three subsets were averaged (weighted by Strehl ratio) to create three submaps. The standard deviation of the measured astrometric and photometric values in the three submaps were used for initial estimates of the astrometric and photometric uncertainties before additional sources of error were included during the astrometric transformation and photometric calibration processes.

We used the PSF-fitting software \textsc{StarFinder} \citep{2000A&AS..147..335D} to identify point sources in the observation epoch and submap images \citep[detailed further by][]{2008ApJ...689.1044G}. The identifications yield measurements of flux and position on the image for each source. Importantly for this work, this step also involved computing the photometric uncertainty originating from our stellar flux measurements, $F$, during the point source identification. We use the variance in the three submaps as our estimate of the instrumental flux uncertainty ($\sigma_F^2$). The instrumental flux uncertainty was converted to an instrumental magnitude uncertainty, $\sigma_m$, with the following equation:

\begin{eqnarray}
	\sigma_{m} = 1.0857 \frac{\sigma_F}{F}.
\end{eqnarray}

Observations from individual epochs were matched and placed in a common reference frame (Jia et al. in prep.). The process provides astrometric positions for detected sources in each observation and an estimate of the proper motion of each source. The reference frame is constructed using the same method outlined in \citet{2010ApJ...725..331Y} and further improved by Sakai et al. (in prep). 


\subsection{Systematics from Stellar Confusion and Resolved~Sources} 
\label{sub:stellar_confusion_and_resolved_sources}

Stellar confusion and proximity to resolved sources introduces biases in our photometric flux measurements. Stellar confusion originates from the individual proper motions of stars causing multiple stars to be positioned so that they can be confused during the PSF-fitting and cross-matching stage. In photometry, confusion results in misestimation of the stellar flux by biasing it when the PSFs of confused stars are blended together. During the cross-matching step, the proper motion of each star was fitted to an acceleration model. With the acceleration model, if the expected positions of two or more stars intersected with each other during an observation within $0.1''$ and had brightnesses within 5 magnitudes, the stars were identified as confused (Jia et al. in prep.). If all intersecting stars were not each identified as separate detections, the photometric and astrometric measurements obtained for each confused star in that epoch were then removed from our dataset.

A similar problem can arise for resolved sources, leading to biases in photometry. During PSF-fitting, the flux from resolved sources was not modeled accurately with a single PSF, and therefore led to residual flux in an extended halo not captured by the fitted PSF. The flux in the extended halo could subsequently bias flux measurements derived for any sources lying in that halo. In this experiment, we identified resolved sources by visual inspection of the residual image for an observation night. This residual image was constructed by subtracting the PSF fits to each source from the observation's final image. Resolved sources appeared as those with extended flux still remaining in the residual image. We found that the extended flux for all resolved sources could be captured within $\approx 5 \times \text{median FWHM}_{\text{33N}} \text{ of observations}$ or typically $0.3''$. In each observation night, we therefore removed photometric and astrometric measurements for a star from our dataset if it passed within $0.3''$ of a resolved source. Sources identified as resolved are shown with their respective $0.3''$ boundaries on our experiment's field of view in Figure~\ref{fig:stellar_sample}.


\subsection{Artifact Sources from Elongated PSFs} 
\label{sub:artifact_sources_from_elongated_psfs}
Due to anisoplanatic effects, the PSF shape near the edges of our experiment's field of view was often elongated. During some observations, the elongated PSFs could lead our PSF fitting routine to report artifactual sources alongside stars located near the edges of our field of view.
As a consequence of assigning some flux to the artifact, the PSF-fitting routine would report fluxes for the actual associated star that are too low. We therefore dismissed observations of any stars where they were affected by such an artifact.

We identified possible artifact sources by performing fits to their proper motion during the cross-matching stage, building on the methods outlined by Jia et al. (in prep.). The presence of artifact sources in our images was greatly dependent on the performance of the AO correction during a given observation, and therefore these sources were not present in every observation. We found, however, that artifact sources, when present, typically had the same position offset from their respective associated stars across observations. Artifact sources and their associated stars therefore have similar fitted values for proper motion. Any two apparent stars having positional separation $\leq 0.07''$ and proper motion difference $\leq 3 \text{ mas/yr}$ were identified as a possible primary and artifact source candidate pair. The fainter object in such pairs is then added to a list of candidate artifact sources. We then removed from the list of candidate artifact sources any stars judged to be real stars by visual inspection of the images. Once the artifact sources were thereby verified, we removed any flux measurements of their associated stars from our dataset in the observations where the artifact source was present.

\begin{figure*}
    \epsscale{1.17}
    \plottwo{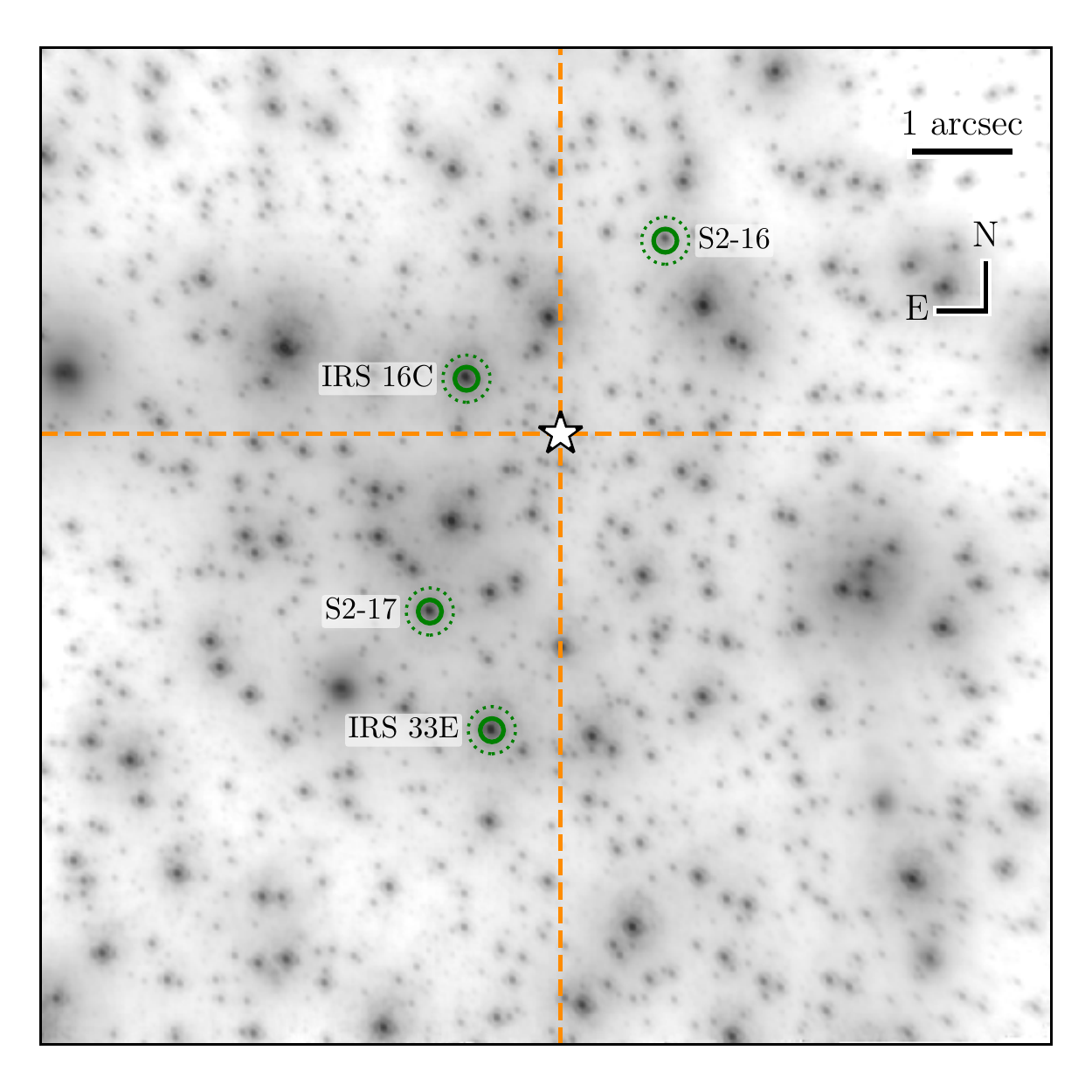}{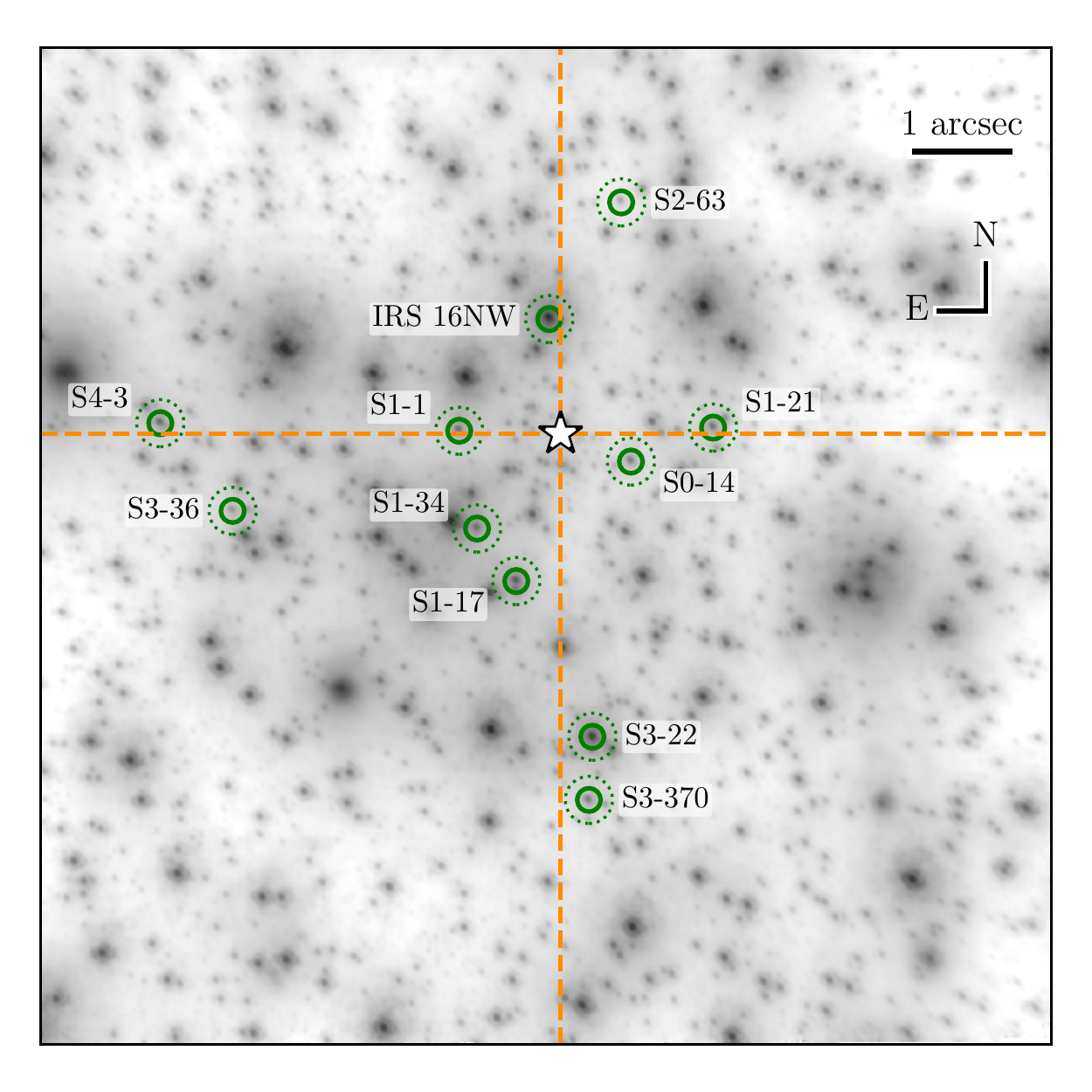}
    \caption{Calibration stars used for our initial and final calibration iterations are circled and labeled (initial calibration iteration on left and final calibration iteration on right). The white star symbol indicates the position of Sgr A*. The dashed lines indicate the boundaries of the four quadrants centered on Sgr A*. These quadrants were used to select calibration stars distributed across our field of view. The background image is from the 10 August 2015 observation. We chose the final calibration stars so that at least one and no more than 3 would lie in each of the four quadrants. Dashed circles around each calibration star indicate $0.25''$ around the position of each star. We selected the final calibration stars so that none were located within $\approx 0.25''$ of each other.}
    \label{fig:FieldStars_Cals}
\end{figure*}

\subsection{Photometric Calibration} 
\label{sub:photometric_calibration}

We performed absolute photometric calibration of the stars in our dataset using photometry reported by \citet{1996ApJ...470..864B}. In our experiment's field of view, several stars have $K$-band flux measurements from \citet{1996ApJ...470..864B}. Four of these stars (IRS~16C, IRS~33E, S2-16, and S2-17) are not identified as variable by \citet{2007ApJ...659.1241R} and do not appear as resolved sources in our images. We performed a bandpass-correction process, described in Section~\ref{sub:reference_flux_bandpass_correction}, to convert the \citet{1996ApJ...470..864B} $K$-band fluxes for these four stars to NIRC2 $K'$-band flux. We then used these four stars as calibrator stars to perform an initial photometric calibration of all stars in our image across all observation epochs. The error in zeropoint correction from this initial calibration represents our experiment's error in absolute photometry, and is listed for each of our observations in Table~\ref{tab:Observations}.

We next performed an iterative procedure to select stable, non-variable stars in our experiment's field of view as photometric calibrators. In each calibration iteration, we selected the most photometrically stable stars distributed throughout our field of view and isolated enough to not be confused during the cross-matching process. Using these stable stars as photometric calibrators helped us obtain precise, \emph{relative} photometric calibration, necessary for identifying variability. Our iterative process to select calibration stars is described in more detail in Section~\ref{sub:iterative_calibrator_selection}. Our final set of calibration stars selected by this process consists of IRS~16NW, S3-22, S1-17, S1-34, S4-3, S1-1, S1-21, S3-370, S0-14, S3-36, and S2-63. Table~\ref{tab:Calibration stars} summarizes the photometric properties of our final calibration stars and Figure~\ref{fig:FieldStars_Cals} shows our initial and final set of calibration stars on our experiment's field of view.

After photometric calibration, we implemented and performed an additional correction to our photometry on local scales within the field beyond the zeropoint photometric calibration. The local photometric correction technique's implementation in our experiment is described in more detail in Section~\ref{sub:local_photometric_correction}. This correction accounted for a variable PSF across our field of view, which caused the flux measurements of stars derived by our PSF-fitting procedure to be under- or over-estimated. Since the PSF variation was spatially correlated, the bias in the flux measurement was expected to be similar for nearby stars of similar magnitudes. The photometric flux measurements and their corresponding uncertainties used in this work incorporate our local correction technique.

\begin{figure*}
    \epsscale{1.2}
    \plotone{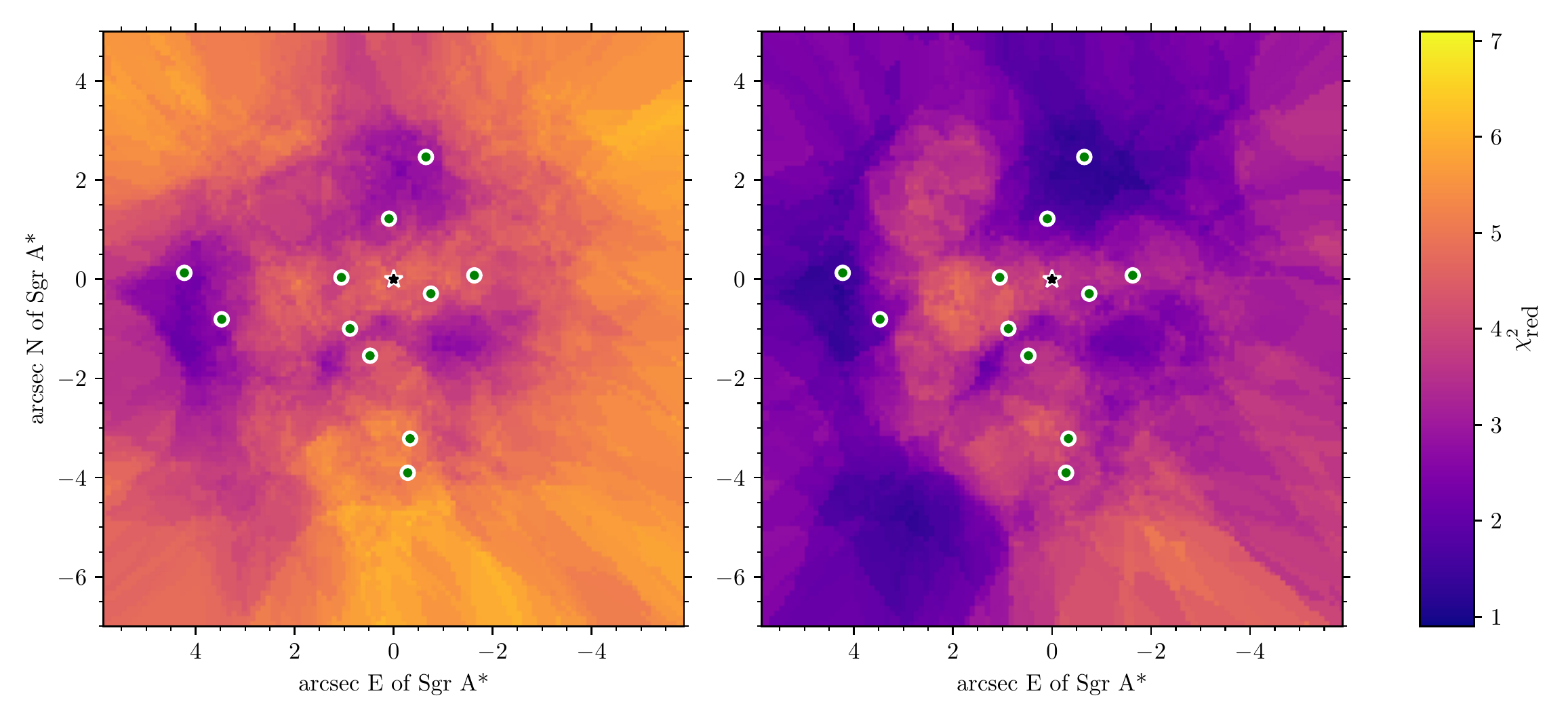}
    \caption{The variability of our stellar sample plotted on our field of view. The color of each point on the maps is determined by the mean of the $\chi^2_{\text{red}}$ of the nearest 20 stars to the point having $\chi^2_{\text{red}} < 10$ (so as to not affect the mean value with highly variable stars). The star shape indicates the position of Sgr A* while the green dots indicate the positions of our photometric calibrators. The map on the left is generated before our local correction is applied, and the map on the right is generated after our local corrections have been applied. Before the local correction is applied, the outer regions of the field demonstrate higher variability as expected from anisoplanatic effects on the PSF shape. After the local correction is applied, this spatial preference for variability is largely removed. Since this process removes systematic contributions to variability, overall $\chi^2_{\text{red}}$ values are lowered throughout the field of view.}
    \label{fig:var_field_neighbors}
\end{figure*}

To evaluate the effectiveness of our local correction method in removing photometric biases from PSF variability, we examined the distribution of variability over the field. Figure~\ref{fig:var_field_neighbors} plots overall variability of our stellar sample as a function of position on the field. Before local correction was applied, the consequences of the anisoplanatic effects on our measured photometry are evident as higher variability towards the edges of our field of view. These edge positions were located at a greater distance from the projected position of the laser guide star, towards the center of our field. After the local correction was applied, the overall variability in our sample originating from systematic effects was substantially reduced. Further, this correction reduced higher variability trends in the outer regions of the field where the influence of the anisoplanatic effects is most extreme. Section~\ref{sub:ao_photometry} further discusses the need for this correction and presents a comparison to other techniques developed for PSF variability.


\begin{figure*}
	\centering
    \epsscale{1.2}
    \plotone{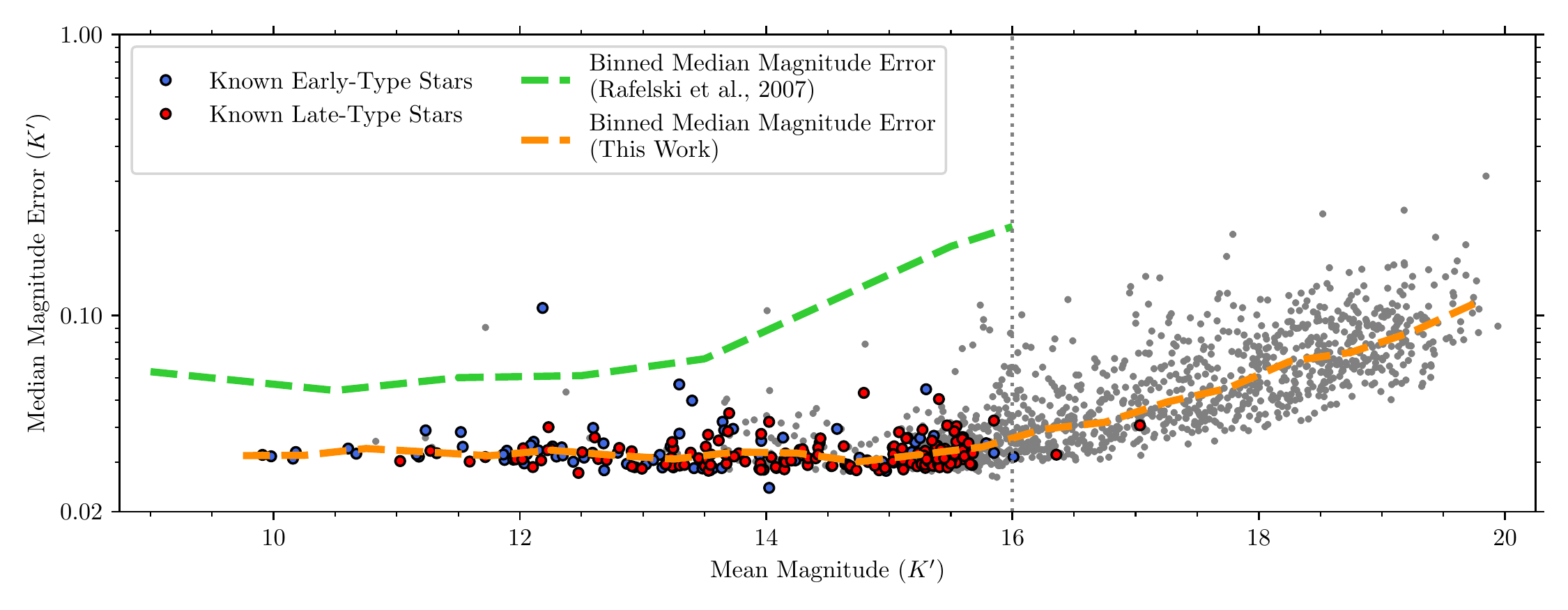}
    \caption{Median photometric uncertainty of the stellar sample in this work, identified in at least 16 nights of observation, plotted by mean magnitude. Known early-type and late-type stars are indicated as dots colored blue and red, respectively. The binned median magnitude error identifies this work's photometric precision as a function of stellar magnitude. Also shown for comparison is binned median magnitude error from previous work studying variability at the Galactic Center by \citet{2007ApJ...659.1241R} using Keck speckle photometry data. The values plotted here are calculated after conducting the local photometry correction on our dataset, detailed in Section~\ref{sub:local_photometric_correction}.\\
    The floor of the photometric uncertainty begins to rise for stars fainter than $m_{K'} \sim 16$. Based on this, we limited the sample for our variability and periodicity search to stars with $\bar{m}_{K'} \leq 16$, indicated by the vertical dashed line.
    \label{fig:Mag_MagError}
    }
\end{figure*}

\subsection{Final Photometric Quality} 
\label{sub:final_photometric_quality}
The photometric quality of our data can be quantified by analyzing the median of the photometric uncertainty ($\sigma_{m_{K'}}$) for each star across all observations, as shown in Figure~\ref{fig:Mag_MagError}. Our observation's photometric uncertainty reaches a floor of $\sigma_{m_{K'}} \sim 0.03$ to a stellar magnitude of $m_{K'} \sim 16$. This floor primarily came from the zeropoint correction error's contribution to the photometric uncertainty (see Figure~\ref{fig:Zeropoint Errors}). For fainter stars at higher mean magnitudes, the photometric uncertainty of our observations rose up to $\sigma_{m_{K'}} \sim 0.1$ at $m_{K'} \sim 19$.

\begin{figure}
    \epsscale{1.2}
    \plotone{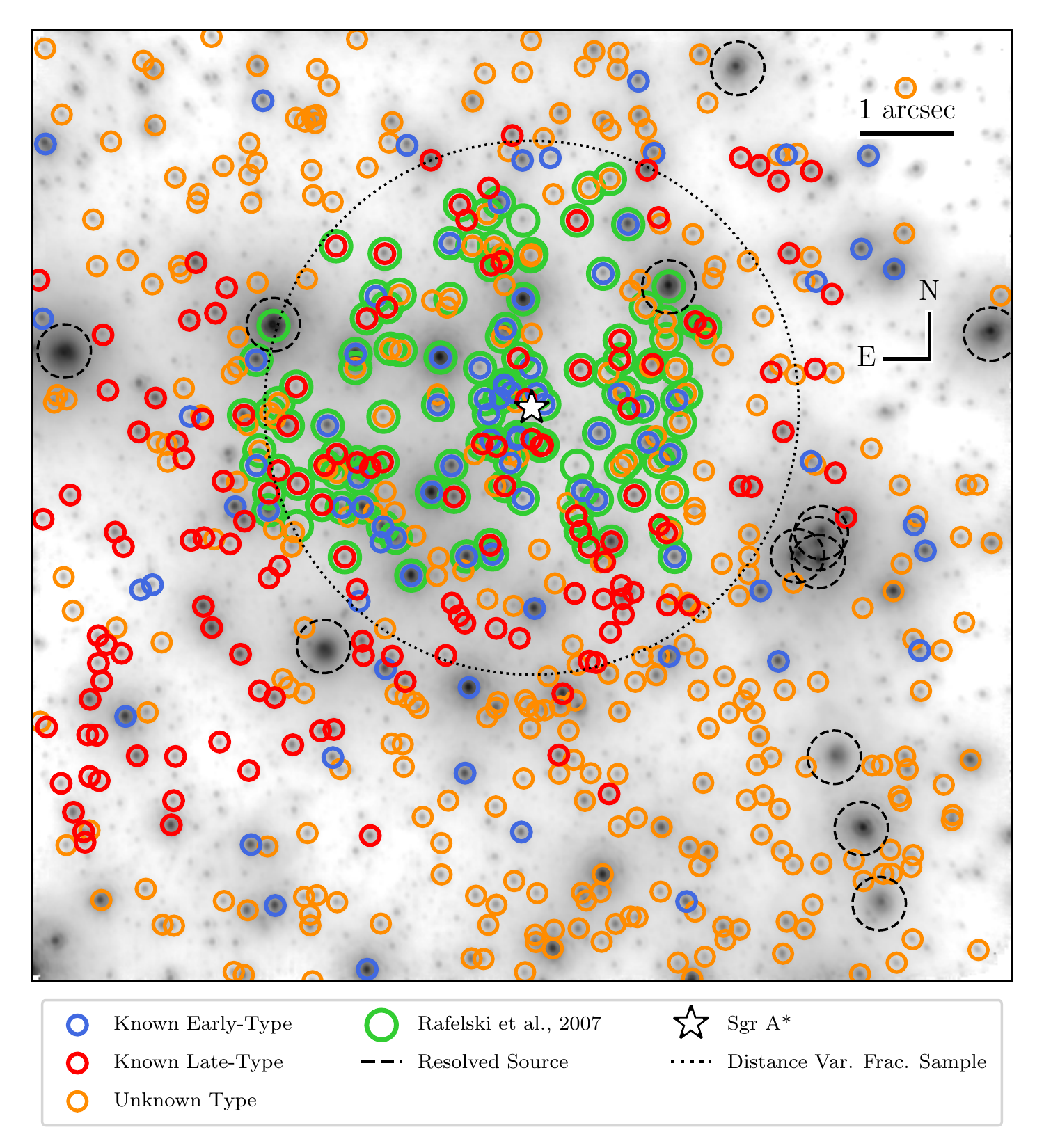}
    \caption{The stellar sample used in this work, consisting of $\bar{m}_{K'} < 16$ stars detected in at least 16 nights without confusion. The background image is from the 2012-05-15 observation. Blue, red, and orange circles indicate spectrally typed early-type, late-type, and unknown type stars respectively. Green circles indicate the stellar sample studied in the previous stellar variability analysis by \citet{2007ApJ...659.1241R} using Keck Observatory speckle data. The dashed circles indicate the region cut around resolved sources where flux measurements of sources could be biased by the presence of the resolved source. The white star symbol indicates the position of Sgr~A*, the location of the supermassive black hole. The large dotted circle indicates the region of our sample used to study the projected positional dependence of variability, out to $3''$ from Sgr~A*. $1''$ corresponds to a projected distance $\approx 0.04$ pc at the Galactic center.
    \label{fig:stellar_sample}
    }
\end{figure}

\subsection{Stellar Sample} 
\label{sub:Stellar Sample}
This experiment's stellar sample is shown in Figure~\ref{fig:stellar_sample}. The stellar sample is composed of stars passing the following conditions:
\begin{itemize}
    \item Detected in at least 16 nights out of 45 total nights, after accounting for confusion events and artifact sources, and without passing within $0.3''$ of a resolved source (see Sections~\ref{sub:stellar_confusion_and_resolved_sources} and \ref{sub:artifact_sources_from_elongated_psfs}).
    \item $\bar{m}_{K'} \leq 16$
\end{itemize}

The 16 night criterion is motivated by the contamination of our sample from artifact sources at the edges of our field of view. Our method to identify artifact sources (detailed in Section~\ref{sub:artifact_sources_from_elongated_psfs}) was not able to recover all artifact sources that appear in fewer than 16 nights. The mean magnitude cut criterion, $\bar{m}_{K'} \leq 16$, originated from our dataset's photometric quality (see Section~\ref{sub:final_photometric_quality}). At this magnitude, the floor in the photometric uncertainty, quantified by the mean magnitude error, begins to rise for stars $\bar{m}_{K'} \gtrsim 16$.

Under these criteria, 563 stars were identified and included in our photometric study. This sample of stars was further subdivided into known early- and late-type stars (identified by \citet{2006ApJ...643.1011P}, \citet{2009ApJ...703.1323D}, \citet{2009ApJ...692.1075G}, \citet{2009ApJ...697.1741B}, \citet{2012JPhCS.372a2016D}, and manually assigned a spectral type by \citet{2013ApJ...764..154D}). Under our photometric sample selection criteria, 85 stars are known early-type stars and 143 are known late-type stars. These populations were studied separately for variability, detailed in Section~\ref{sec:stellar_variability}, and are specifically indicated on our experiment's field of view in Figure~\ref{fig:stellar_sample}.



\section{Stellar Variability} 
\label{sec:stellar_variability}

\subsection{Identifying Variable Stars} 
\label{sub:identifying_variable_stars}
In order to identify variable light curves we computed the $\chi^2_{\text{red}}$ statistic for our stellar sample. We calculated the weighted mean magnitude ($\bar{m}$) of each star in our stellar sample across all observation epochs, $i$, using weights at each observation from the magnitude uncertainty:

\begin{eqnarray}
	\bar{m} = \frac{\sum \frac{1}{\sigma^2_i} m_i}{\sum \frac{1}{\sigma^2_i}}.
\end{eqnarray}

The $\chi^2_{\text{red}}$ quantity was then computed for each star to test against no variability:

\begin{eqnarray}
	\chi^2_{\text{red}} &=& \frac{1}{\nu} \sum \frac{(\bar{m} - m_i)^2}{\sigma^2_i}\\
	&=& \frac{1}{N - 1} \sum \frac{(\bar{m} - m_i)^2}{\sigma^2_i}.
\end{eqnarray}

Here, the number of degrees of freedom, $\nu$, was determined from the number of observations where a star is identified, $N$: $\nu = N - 1$. We expect a higher $\chi^2_{\text{red}}$ value for stars with photometric measurements that deviate more often and more significantly from their mean magnitude.

We use a criterion in $\chi^2_{\text{red}}$ to identify variable stars. We set the $\chi^2_{\text{red}}$ variability threshold for each star such that the probability of obtaining its $\chi^2_{\text{red}}$ value is less than $5 \sigma$ given a Gaussian distribution of deviations from the mean. The specific $\chi^2_{\text{red}}$ threshold for each star was based on its value of $\nu$, between 15 -- 44 in our sample. For $\nu = 44$ (star detected in all 45 nights), this variability threshold was at $\chi^2_{\text{red}} > 2.40$, and went up to $\chi^2_{\text{red}} > 3.87$ for $\nu = 15$ (star detected in 16 nights).

\subsection{Deriving the Variability Fraction} 
\label{sub:deriving_the_variability_fraction}
We investigated the distribution of variability in our sample as a function of projected distance from Sgr A* and observed magnitude in $K'$. These models allow us to determine whether the location or the brightness of a star is correlated with its variability. Our fit to variability as a function of distance from Sgr A* was limited to those stars within $3''$ of Sgr A*. At greater distances, near the edges of our experiment's field of view, our sample started being affected by incompleteness due to the presence of artifact sources (see Section~\ref{sub:artifact_sources_from_elongated_psfs}). We performed our fits to variability as a function of observed magnitude for all stars in our sample.

We used a mixture model analysis to model the stellar population, consisting of a variable and a non-variable population. Our models follow techniques similar to those outlined by \citet{2011ApJ...738...55M}. We assumed that the probability densities of stars in these populations at the Galactic center follow power law distributions, with $R$ as the projected distance from Sgr A*: $\Sigma_{v}(R) \propto R^{\Gamma_{v,R}}$ and $\Sigma_{n}(R) \propto R^{\Gamma_{n,R}}$ for the variable and non-variable populations, respectively. The surface density of stars at projected distances close to the central black hole ($\lesssim 2$ pc) can be well described by power law distributions \citep[see e.g.][]{2013ApJ...779L...6D, 2018A&A...609A..26G}. To fit the mixture model, we obtained the likelihood of the variability fraction as a function of distance, $\Lambda_R$, following the form of the binomial distribution:
\begin{eqnarray}
    \Lambda_{R} &\propto& \prod_i \left[(F_R \Sigma_{v})^{k_i} ((1-F_R) \Sigma_{n})^{k_i} \right].
\end{eqnarray}
Here, the parameter $F_R$ represents the variability fraction in the sub-sample used in our positional variability analysis and $i$ represents the index of the individual stars of the sub-sample. We assigned $k=1$ for variable stars and $k=0$ for non-variable stars.

Similar to the projected distance probability density distributions, we assumed that the probability density distributions of the variable and non-variable populations with respect to observed magnitude, $m$, also follow power laws: $p_{v}(m) \propto m^{\Gamma_{v,m}}$ and $p_{n}(m) \propto m^{\Gamma_{n,m}}$ for the variable and non-variable populations, respectively. The power law distribution in observed magnitude is expected to originate from the initial mass function, and has been observed previously for both early- and late-type stars at the Galactic center \citep[see e.g.][]{2010ApJ...708..834B, 2013ApJ...764..154D, 2013ApJ...764..155L}. To fit a mixture model from these distributions, we derived the likelihood of the variability fraction as a function of observed magnitude, $\Lambda_m$, again following the form of the binomial distribution:
\begin{eqnarray}
    \Lambda_{m} &\propto& \prod_i \left[(F_m p_{v})^{k_i} ((1-F_m) p_{n})^{k_i} \right].
\end{eqnarray}
Since only our fit to variability as a function of brightness used our entire stellar sample, we use its constraints on the variability fraction, $F_m$, as the overall variability fraction of our entire sample, $F$.

We used a Markov chain Monte Carlo algorithm, as implemented in the \textsc{emcee} software package \citep{2013PASP..125..306F}, to fit our model parameters. In each trial sample, we normalized the individual power law distributions for the variable and non-variable populations over our experiment's bounds: $1 = \int_{\text{bounds}} 2 \pi R \Sigma dR$ and $1 = \int_{\text{bounds}} p dm$ for our distance and brightness variability fits, respectively. We defined our variability model to have the following bounds in projected distance ($R$) and observed magnitude ($m$):
\begin{eqnarray}
    0.05'' \leq &R& \leq 3.00'',\\
    9 \leq &m& \leq 16.
\end{eqnarray}
Our final variability models fitted the overall variability fraction of our sample, $F_R$ and $F_m$, and two parameters each for the variable and non-variable population distributions with projected distance ($\Gamma_{v,R}$, $\Gamma_{n,R}$) and with magnitude ($\Gamma_{v,m}$, $\Gamma_{n,m}$). This gives each of our variability models a total of three parameters.

We can express our model as the fraction of variable stars as a function of distance from Sgr A*:
\begin{eqnarray}
    f_{v,R} &=& \frac{F_R \Sigma_{v}}{F_R \Sigma_{v} + (1-F_R) \Sigma_{n}}\\
    &=& \frac{1}{1 + \frac{1-F_R}{F_R} \frac{\Sigma_{n}}{\Sigma_{v}}}\\
    &=& \frac{1}{1 + c_R \frac{1-F_R}{F_R} R^{\alpha_R}}.
\end{eqnarray}
Here, $\alpha_R \equiv \Gamma_{n,R} - \Gamma_{v,R}$ and $c_R$ is a constant factor originating from $\Sigma_n / \Sigma_v$ used to obtain this relation.

Similarly, for observed magnitude we obtained:
\begin{eqnarray}
    f_{v,m} &=& \frac{1}{1 + c_m \frac{1-F_m}{F_m} m^{\alpha_m}}.
\end{eqnarray}

We additionally applied our brightness variability model to the known early- and late-type stars in our sample. Since the spectral typing originates from different spectroscopic surveys with incomplete spatial sampling across our experiment's field of view, we did not apply our distance variability model separately to the spectrally typed subsamples.


\section{Periodic Variability} 
\label{sec:periodic_variability}

A major focus of the variability study in our stellar sample was to identify periodically variable stars. Periodic variability in observed flux has multiple origins. We were especially interested in identifying eclipsing or ellipsoidal binary systems and periodic variables such as Cepheids, RR Lyrae, and Mira variables.

The individual observations in our data set were unevenly spaced temporally, making it difficult to search for periodic signals through several commonly implemented periodicity search techniques, such as Fourier transforms, that rely on regular sampling. For our periodicity searches, we instead employed the Lomb-Scargle periodogram method, devised by \citet{1976Ap&SS..39..447L} and \citet{1982ApJ...263..835S}. The Lomb-Scargle technique is specifically developed for uneven temporal spacing and works by fitting Fourier components to the observed measurements. This makes it particularly optimized for detecting periodic signals that have an overall sinusoidal shape in their phased light curves.

\subsection{Periodicity Search Implementation} 
\label{sub:periodicity_search_implementation}
We computed the Lomb-Scargle periodogram for all stars in our sample using the algorithm by \citet{1989ApJ...338..277P}, implemented as part of the Astropy package \citep{2013A&A...558A..33A}.

Our uneven temporal spacing makes establishing detectability limits of periods in our periodicity search difficult. With regularly sampled data, the Nyquist limit establishes that the highest detectable frequency of a periodic signal is half of the sampling frequency. However, with sampling at a cadence with no underlying regularity in observation spacing, no similar limit can be determined \citep{2017arXiv170309824V}. In practice, due to the irregular spacing of observations, periods even shorter than the smallest observational spacing can still be detected. We used a period search range between 1.11 days and 10,000 days (between frequencies of 0.9 day$^{-1}$ to $10^{-4}$ day$^{-1}$), as detailed in Appendix~\ref{sub:period_search_range}. Our trial periods for the Lomb-Scargle periodogram were derived from a uniform frequency grid. With our total observation span of $T = 4132.74$ days, our frequency spacing was dictated by the expected width of a peak in the periodogram: $\sim 1/T$ \citep{2017arXiv170309824V}. We chose an oversampling factor, $n_0 = 10$, to ensure that every peak in our periodogram is sufficiently sampled. This gave our final frequency grid spacing of $\Delta f = \frac{1}{n_0 T} = 2.420 \times 10^{-5} \text{ day}^{-1}$.

Our Lomb-Scargle periodicity searches were performed with standard normalization and a floating mean model. We additionally removed long-term linear trends from the light curve of each star before computing a periodogram. This removal of long-term linear trends is further detailed in Appendix~\ref{sub:removal_of_long_term_linear_trends}.


\subsection{Definition of Significance} 
\label{sub:definition_of_significance}
We implemented a bootstrap false alarm test to assign significance to powers in our periodograms. We derive an estimates of false alarm probability (FAP) via the bootstrap methods outlined by \citet{2014sdmm.book.....I, 2017arXiv170309824V}, using 10,000 mock light curves for each star. We define the significance of each power as $1 - \text{FAP}$. This technique estimated the likelihood of a power to appear in the periodogram given true observation cadence, typical brightnesses, and associated errors on the brightness for each star, but with no actual periodicity since measurements were shuffled when constructing each mock light curve. Importantly, this test does not give the probability that a given detection corresponds to a true periodic signal. Instead, the test estimates the likelihood that a periodogram peak does not originate from a non-periodic signal.


\subsection{Aliasing in Periodicity Searches} 
\label{sub:aliasing_in_periodicity_searches}
The temporal spacing of our observations could introduce aliasing for real periodic signals in our data set, where secondary periodogram peaks could be introduced. Any true periodic signal is sampled by a window function at our observation times, and this window function's power spectrum (discussed in more detail in Section~\ref{sub:period_search_range}) is convolved with the true signal's power spectrum to create the observed power spectrum that can have secondary peaks or aliases. Based on our photometric data set alone, distinguishing between a periodic signal at the true periodic signal's period and its alias(es) on a periodogram is difficult.

Common aliases occur from typical observing cadences of an experiment. A true periodic signal is expected to have secondary aliased peaks appearing at $|f_{\text{true}} \pm \delta f|$, where $\delta f$ is a strong feature in the observing window function \citep{2017arXiv170309824V}. In our experiment, the most common cadence was that originating from the length of a sidereal day: $\delta f = 1.0027 \text{ day}^{-1}$, leading to the strongest aliases of peaks in the periodogram. Other prominent features leading to aliases in our experiment came from our nightly observing cadence, $\delta f = 1.0 \text{ day}^{-1}$, and yearly observing cadence, $\delta f \approx 2.7 \times 10^{-3} \text{ day}^{-1}$.

When considering detections in our periodogram, we excluded those that may originate from aliasing by long-term variations ($\gtrsim 1,000$ days). On such long time scales, we could not establish periodicity without observations of multiple periods. However, these long-term variations could be aliased to appear as strong detections in our periodicity search at periods shorter than 1,000 days. An example of this behavior is the star S4-172, shown in Figure~\ref{fig:S4-172_lc_per}, the long-term variability of which led to strong detections of periodicity at $\sim 100$ and $\sim 365$ days from aliasing. In our experiment, we found that stars with power $\gtrsim 50\%$ significance at periods longer than about a quarter of our observing baseline ($\frac{1}{4} \times T = \frac{1}{4} \times 4132.74 \text{ days} = 1033.19 \text{ days}$) could lead to strong detections at shorter periods.


\section{Results} 
\label{sec:results}

\begin{figure*}[ht]
    \epsscale{1.2}
    \plotone{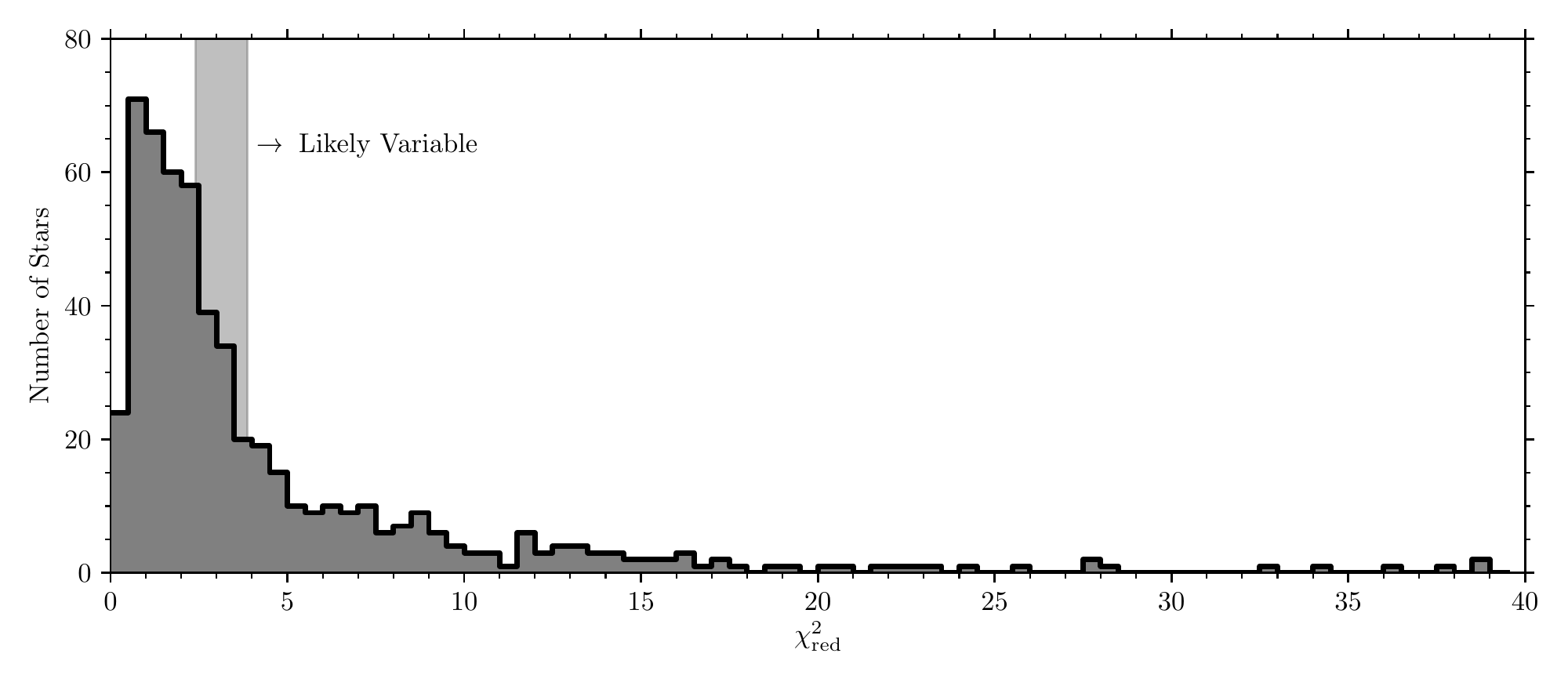}
    \caption{Binned $\chi^2_{\text{red}}$ distribution for our stellar sample identified in at least 23 observations. For variability, we drew a cut in this distribution at $5\sigma$, which for stars identified in 16 observations (with $\nu = 16 - 1 = 15$) corresponds to $\chi^2_{\text{red}} > 3.87$. Stars identified in a greater number of observations have a corresponding higher $\nu$ resulting in a $5\sigma$ cut for variability at lower $\chi^2_{\text{red}}$ values, going down to $\chi^2_{\text{red}} > 2.40$ for stars identified in all 45 nights. These $\chi^2_{\text{red}}$ cuts for variability depending on the number of nights are indicated by the vertical shaded region. In this sample with the $5\sigma$ variability cut, $50\pm2\%$ of stars are variable.
    \label{fig:RedChiSqDist}
    }
\end{figure*}

\begin{figure*}[ht]
    \epsscale{1.17}
    \plottwo{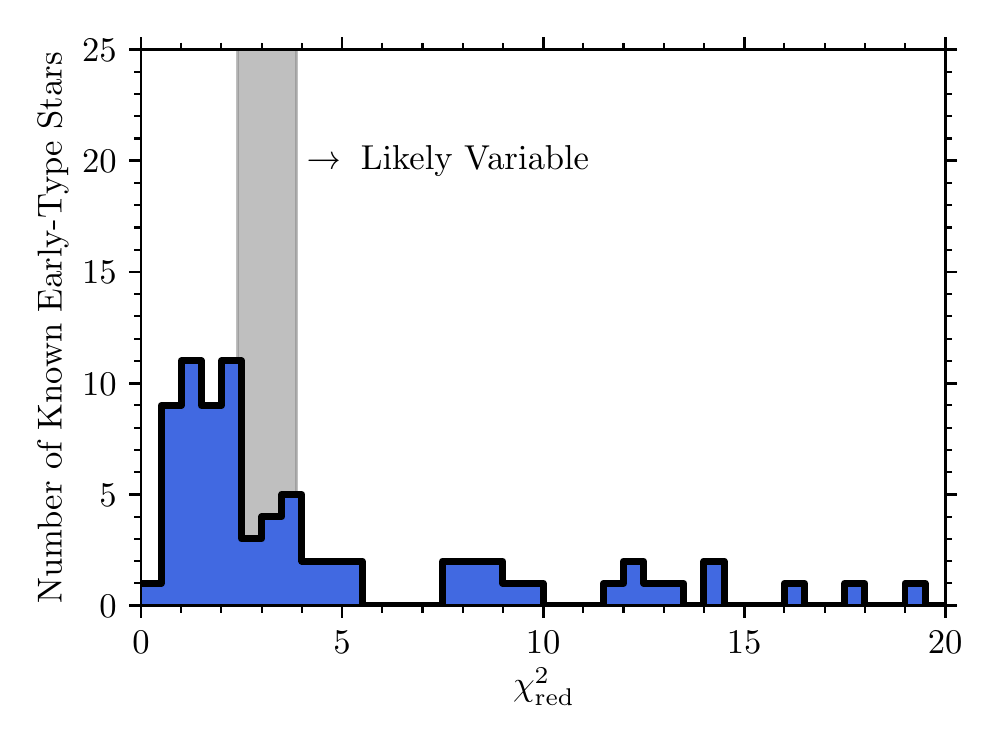}{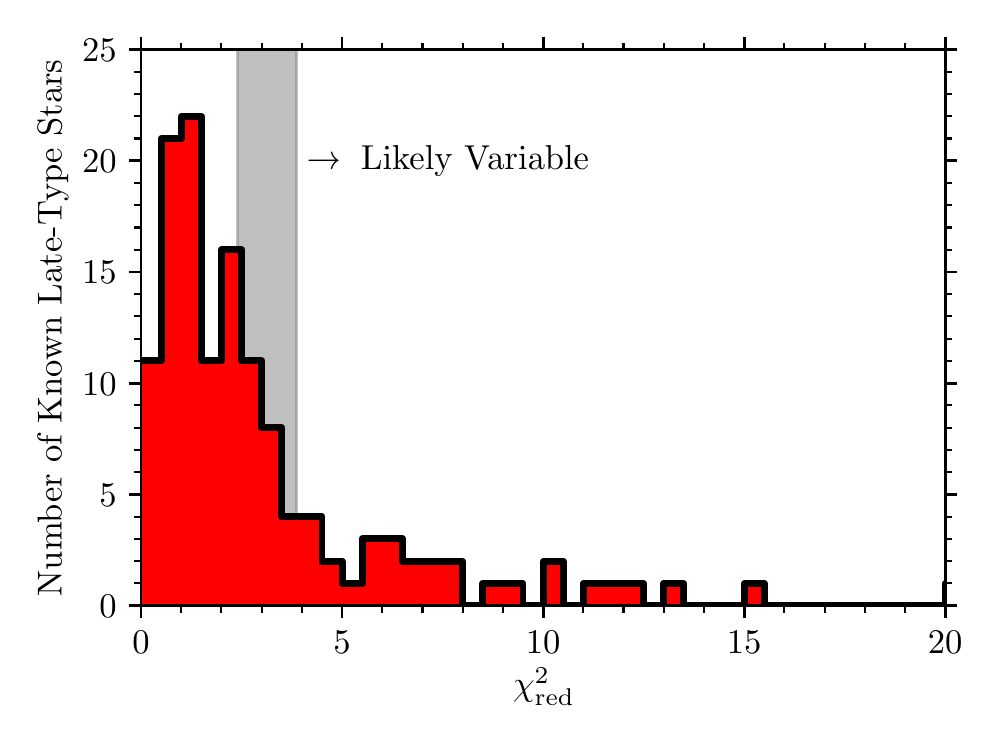}
    \caption{Same as Figure~\ref{fig:RedChiSqDist}, but for our spectroscopically confirmed early-type stellar sample (left) and late-type stellar sample (right) identified in at least 16 observations. $52\pm5\%$ of spectroscopically confirmed early-type stars are variable and $43\pm4\%$ of spectroscopically confirmed late-type stars are variable. The $\chi^2_{\text{red}}$ cuts for variability depending on the number of nights are indicated by the vertical shaded region.
    \label{fig:RedChiSqDist_spectyped}
    }
\end{figure*}

\begin{figure}[ht]
    \epsscale{1.2}
    \plotone{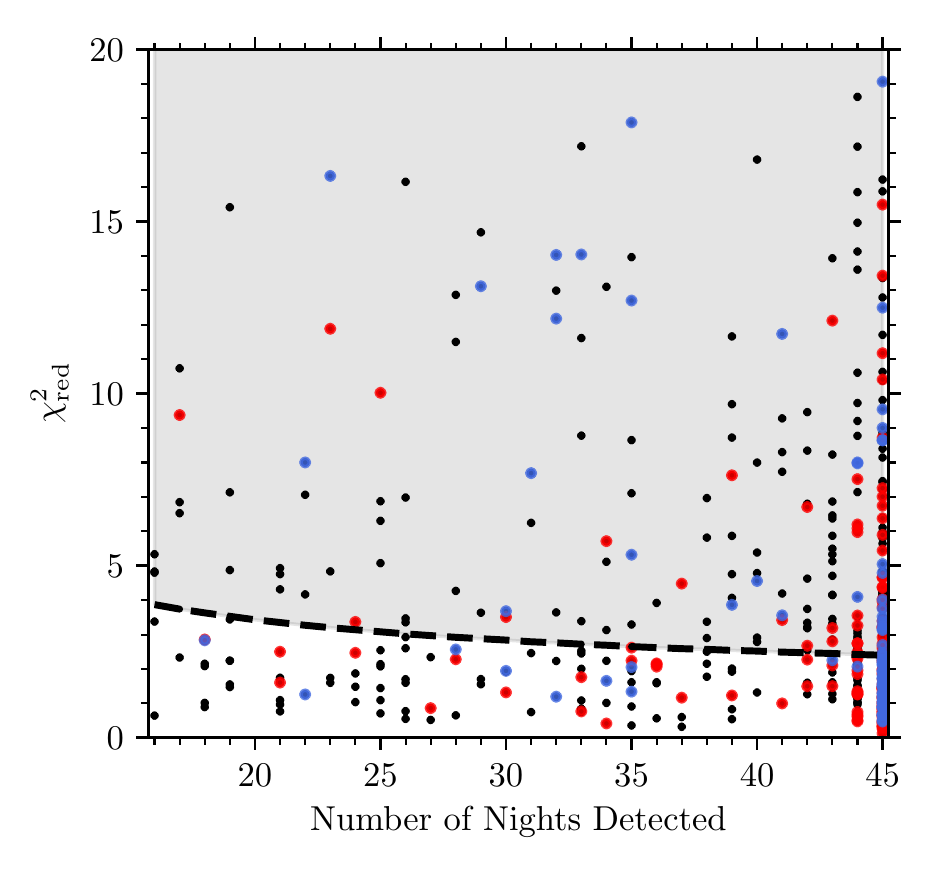}
    \caption{The dashed line indicates our $5\sigma$ $\chi^2_{\text{red}}$ cut for variability as a function of number of nights. The stars identified as variable with this cut are in the shaded gray region. Dots colored blue/red are spectroscopically confirmed early-/late-type stars, while black dots correspond to stars that have unknown type.}
    \label{fig:Nights_ChiSq}
\end{figure}

\begin{figure}[ht]
    \epsscale{1.2}
    \plotone{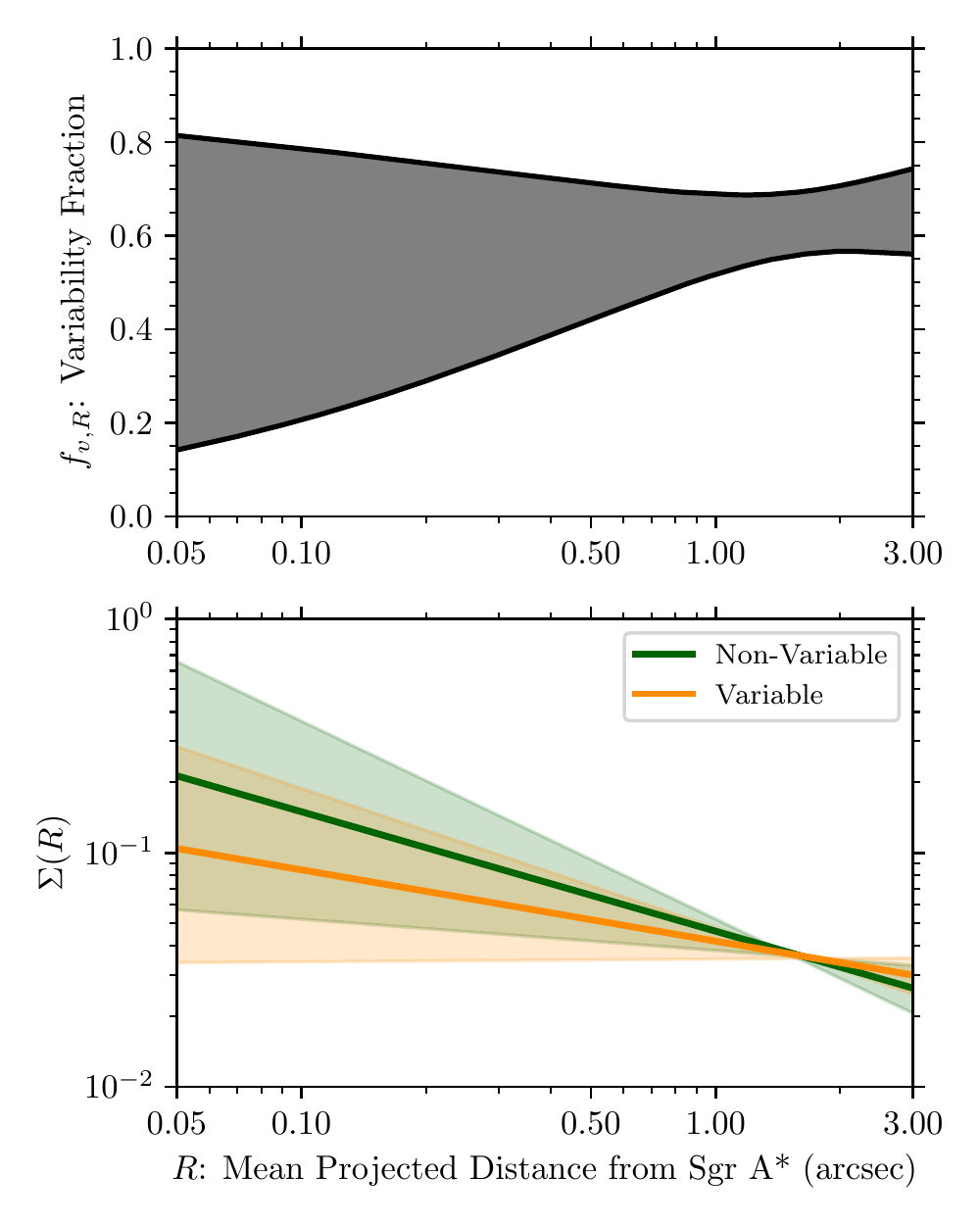}
    \caption{\emph{Top:} The variability fraction as a function of projected distance from Sgr A*, $R$. The solid black lines indicate the median $2 \sigma$ region of this relationship using stars with $R \leq 3''$ from Sgr A*. \emph{Bottom:} The surface density distribution of our non-variable and variable star populations as a function of projected distance from Sgr A*, $\Sigma_{n}(R)$ and $\Sigma_{v}(R)$. Solid lines indicate median fit across all MCMC samples and the shaded regions indicate $2 \sigma$ significance regions of this fit.}
    \label{fig:varFrac_models_dist}
\end{figure}

\begin{figure}[ht]
    \epsscale{1.2}
    \plotone{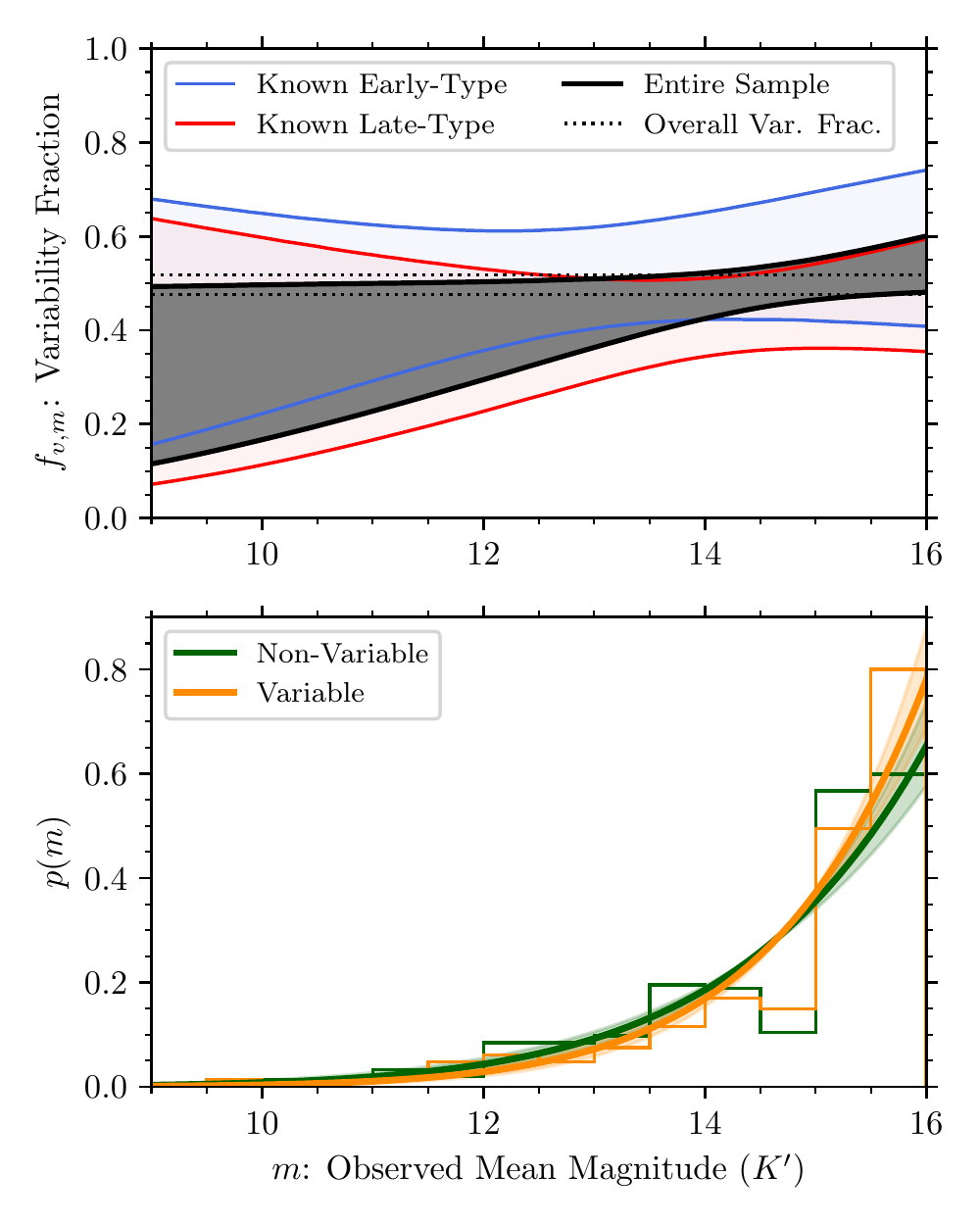}
    \caption{\emph{Top:} Variability fraction as a function of observed magnitude, $m$. The solid black lines indicate the median $2 \sigma$ region of this relationship using our entire stellar sample across all our MCMC samples. The blue and red lines indicate the same regions for the known young- and late-type stars in our stellar sample. The dotted lines indicate the $1 \sigma$ constraints on the overall variability fraction in our sample. \emph{Bottom:} Probability distribution of our non-variable and variable star populations as a function of observed magnitude, $p_{n}(m)$ and $p_{v}(m)$. Solid lines indicate median fit across all MCMC samples and the shaded regions indicate $2 \sigma$ significance regions of this fit. The non-variable and variable star populations in our data are shown as binned histograms.}
    \label{fig:varFrac_models_mag}
\end{figure}

\begin{figure*}
    \epsscale{1.15}
    \plotone{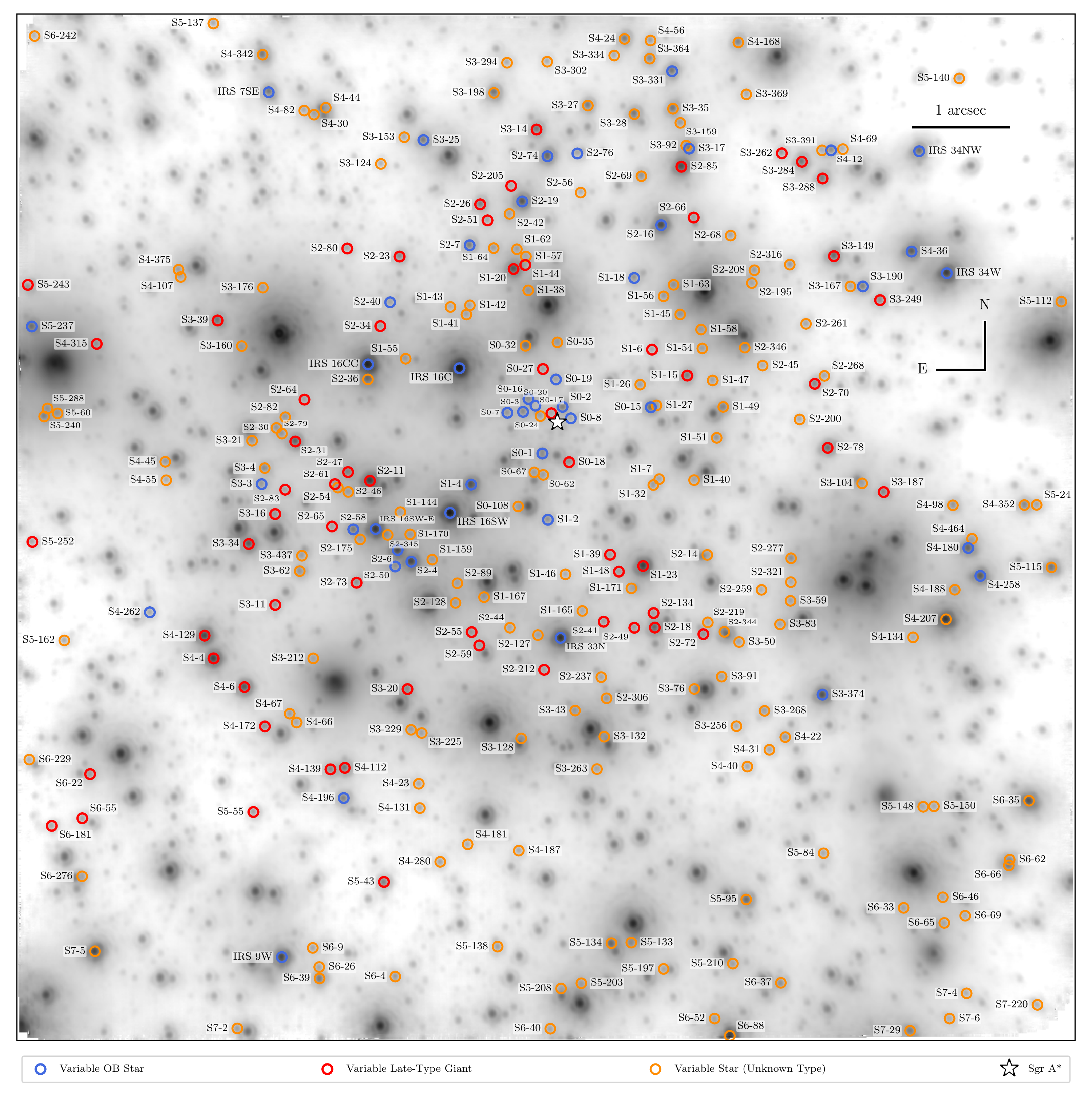}
    \caption{Stars identified as variable on our experiment's field of view. Blue, red, and orange circles indicate spectrally typed early-type, late-type, or unknown type variable stars respectively. The background image is from the 2012-05-15 observation.
    \label{fig:stellar_sample_var}
    }
\end{figure*}

\subsection{Variability Fraction} 
\label{sub:variability_fraction}

\begin{deluxetable}{lr}
    \tablecolumns{2}
    \tablecaption{Fits to parameters of variability models\label{tab:varFrac_modelParams_fits}}
    \tablehead{
        \colhead{Parameter} &
        \colhead{Fit}
    }
    \startdata
    \multicolumn{2}{l}{\emph{Variability with distance}}   \\
    $F_R$               & $0.63 \pm 0.03$  \\
    $\Gamma_{v,R}$      & $-0.30^{+0.16}_{-0.15}$ \\
    $\Gamma_{n,R}$      & $-0.51^{+0.18}_{-0.17}$ \\
    \\
    \multicolumn{2}{l}{\emph{Variability with brightness}} \\
    $F = F_m$           & $0.50 \pm 0.02$  \\
    $\Gamma_{v,m}$      & $11.5 \pm 0.8$ \\
    $\Gamma_{n,m}$      & $9.4 \pm 0.6$  \\
    \\
    \multicolumn{2}{l}{\emph{Variability with brightness,}}\\
    \multicolumn{2}{l}{\emph{known early-type stars}}   \\
    $F_m$               & $0.52 \pm 0.05$  \\
    $\Gamma_{v,m}$      & $3.5^{+1.1}_{-1.0}$  \\
    $\Gamma_{n,m}$      & $2.1 \pm 1.0$  \\
    \\
    \multicolumn{2}{l}{\emph{Variability with brightness,}}\\
    \multicolumn{2}{l}{\emph{known late-type stars}}    \\
    $F_m$               & $0.43 \pm 0.04$  \\
    $\Gamma_{v,m}$      & $8.5 \pm 1.3$  \\
    $\Gamma_{n,m}$      & $7.1 \pm 1.0$  \\
    \enddata
\end{deluxetable}

With the $\chi^2_{\text{red}}$ test for variability, we found that approximately half of the stars in our sample are variable. The $\chi^2_{\text{red}}$ distribution for the stars in our variability sample is plotted in Figure~\ref{fig:RedChiSqDist}, and the distributions for our sample's spectroscopically typed stars are shown in Figure~\ref{fig:RedChiSqDist_spectyped}. Figure~\ref{fig:Nights_ChiSq} shows our sample's $\chi^2_{\text{red}}$ distribution as a function of nights detected, overlaying our $5 \sigma$ variability cut. Using the variable population models described in Section~\ref{sub:deriving_the_variability_fraction}, we derived a variability fraction $F = 50 \pm 2 \%$ among the stars in our sample. Light curves of ``highly variable" stars (i.e.: $\chi^2_{\text{red}} \geq 10.0$) are shown in Appendix~\ref{sec:_chi_2__text_red_geq_10_variables}.

Our models also allow us to derive the variability fraction of stars as a function of projected distance from Sgr~A* (Figure~\ref{fig:varFrac_models_dist}) and the observed magnitude (Figure~\ref{fig:varFrac_models_mag}). We do not find a significant change in the variability fraction as a function of projected distance (Table~\ref{tab:varFrac_modelParams_fits}). We also find an increasing variability fraction for fainter stars in our samples, but this trend is not significant in our dataset.

When considering the spectrally-typed stars in our sample, we measured a variability fraction of $F = 52 \pm 5 \%$ for the known early-type stars and $F = 43 \pm 4 \%$ for the known late-type stars. We did not find a significant difference in the variability fractions as a function of magnitude for known early- nor known late-type star populations (Figure~\ref{fig:varFrac_models_mag}).


\subsection{Periodically Variable Stars} 
\label{sub:periodically_variable_stars}

\begin{deluxetable}{ll}
    \tablecolumns{2}
    \tablecaption{Criteria for Possible Periodic Signal\label{tab:Per_Var}}
    \tablehead{
        \colhead{Criterion} &
        \colhead{Threshold}
    }
    \startdata
    $\chi^2_{\text{red}}$ Variability   & $\geq 5 \sigma$       \\
    Period Cut (from obs. baseline)     & $\leq 4132.74 \text{ d } / 4$ \\
                                        & $\leq 1033.19 \text{ d}$   \\
    Frequency Cut (from aliasing)       & $\leq 0.9 \text{ d}^{-1}$ \\
    \\
    Amplitude of Variability            & $\geq 3 \times \bar{\sigma}_{m}$  \\
    (likely periodic threshold)         & $\geq 5 \times \bar{\sigma}_{m}$  \\
    \\
    Bootstrap False Alarm Test          & $\geq 90\%$           \\
    (likely periodic threshold)         & $\geq 99\%$           \\
    \enddata
\end{deluxetable}

\begin{deluxetable*}{lRRRRRRR}[ht]
    \tablecolumns{8}
    \tablecaption{Likely Periodic Variable Stars\label{tab:Likely_Periodic_Vars}}
    \tablehead{
        \colhead{Star}                      &
        \colhead{Period}                    &
        \colhead{Frequency}                 &
        \colhead{$K'$ Amplitude}            &
        \colhead{Amp. / $\bar{\sigma}_{m}$} &
        \colhead{$\bar{m}_{K'}$}            &
        \colhead{Normalized Lomb-}          &
        \colhead{Bootstrap False Alarm}     \\
        \colhead{}                          &
        \colhead{(d)}                       &
        \colhead{$(\text{d}^{-1})$}         &
        \colhead{(Sinusoid Fit)}            &
        \colhead{}                          &
        \colhead{(Sinusoid Fit)}            &
        \colhead{Scargle Power}             &
        \colhead{Test Significance}
    }
    \startdata
    IRS 16SW    & 9.7238    & 0.1028    & 0.4833 \pm 0.0132 & 14.21     & 9.9760 \pm 0.0046     & 0.8579    & 100.00\%  \\
    \\
    S2-36       & 39.4296   & 0.0254    & 0.3090 \pm 0.0132 & 9.16      & 13.2899 \pm 0.0049    & 0.7513    & 100.00\%  \\
    \\
    S4-258      & 1.1380    & 0.8787    & 0.3414 \pm 0.0171 & 9.21      & 12.5947 \pm 0.0055    & 0.7650    & 99.91\%   \\
    \enddata
\end{deluxetable*}

\begin{deluxetable*}{lRRRRRRR}[ht]
    \tablecolumns{8}
    \tablecaption{Possible Periodic Signals\label{tab:Possible_Periodic_Signals}}
    \tablehead{
        \colhead{Star}                      &
        \colhead{Period}                    &
        \colhead{Frequency}                 &
        \colhead{$K'$ Amplitude}            &
        \colhead{Amp. / $\bar{\sigma}_{m}$} &
        \colhead{$\bar{m}_{K'}$}            &
        \colhead{Normalized Lomb-}          &
        \colhead{Bootstrap False Alarm}     \\
        \colhead{}                          &
        \colhead{(d)}                       &
        \colhead{$(\text{d}^{-1})$}         &
        \colhead{(Sinusoid Fit)}            &
        \colhead{}                          &
        \colhead{(Sinusoid Fit)}            &
        \colhead{Scargle Power}             &
        \colhead{Test Significance}
    }
    \startdata
    IRS 16SW\tablenotemark{S}   & 1.1112    & 0.8999    & 0.4696 \pm 0.0130 & 13.81     & 9.9755 \pm 0.0046     & 0.8527    & 100.00\%  \\
    IRS 16SW\tablenotemark{D}   & 1.1146    & 0.8972    & 0.3949 \pm 0.0137 & 11.62     & 9.9963 \pm 0.0046     & 0.5450    & 97.58\%   \\
    IRS 16SW\tablenotemark{Q}   & 10.8781   & 0.0919    & 0.3650 \pm 0.0126 & 10.74     & 9.9729 \pm 0.0046     & 0.5358    & 96.50\%   \\
    \\
    S4-258\tablenotemark{S} & 8.0637    & 0.1240    & 0.3318 \pm 0.0165 & 8.96      & 12.6085 \pm 0.0056    & 0.7511    & 99.78\%   \\
    \\
    S2-72       & 12.5572   & 0.0796    & 0.0945 \pm 0.0107 & 3.15      & 14.7411 \pm 0.0039    & 0.5004    & 99.12\%   \\
    \\
    S2-14       & 12.7509   & 0.0784    & 0.1169 \pm 0.0122 & 3.76      & 15.6733 \pm 0.0045    & 0.6553    & 98.76\%   \\
    \\
    S2-58       & 84.6643   & 0.0118    & 0.1199 \pm 0.0151 & 3.99      & 13.9289 \pm 0.0050    & 0.5344    & 98.52\%   \\
    S2-58       & 90.2084   & 0.0111    & 0.1144 \pm 0.0147 & 3.81      & 13.9564 \pm 0.0045    & 0.5240    & 97.81\%   \\
    \\
    S4-139      & 24.6270   & 0.0406    & 0.1162 \pm 0.0119 & 3.60      & 14.3908 \pm 0.0042    & 0.5450    & 98.25\%   \\
    S4-139      & 228.1610  & 0.0044    & 0.1141 \pm 0.0121 & 3.54      & 14.3939 \pm 0.0043    & 0.5155    & 95.14\%   \\
    S4-139      & 12.5154   & 0.0799    & 0.1052 \pm 0.0112 & 3.26      & 14.4095 \pm 0.0042    & 0.5096    & 93.87\%   \\
    S4-139      & 15.0768   & 0.0663    & 0.0988 \pm 0.0106 & 3.06      & 14.4031 \pm 0.0042    & 0.5011    & 91.86\%   \\
    \\
    S3-27       & 26.5578   & 0.0377    & 0.1181 \pm 0.0157 & 3.27      & 13.9328 \pm 0.0047    & 0.5283    & 98.19\%   \\
    \\
    S2-4        & 36.0896   & 0.0277    & 0.1831 \pm 0.0128 & 5.72      & 11.9297 \pm 0.0044    & 0.5276    & 97.16\%   \\
    S2-4        & 23.2159   & 0.0431    & 0.1852 \pm 0.0133 & 5.78      & 11.9258 \pm 0.0045    & 0.5052    & 94.61\%   \\
    \\
    S6-69       & 101.7584  & 0.0098    & 0.2042 \pm 0.0292 & 3.18      & 15.9514 \pm 0.0112    & 0.4964    & 96.27\%   \\
    \\
    S3-4        & 315.1572  & 0.0032    & 0.0999 \pm 0.0103 & 3.16      & 14.6326 \pm 0.0041    & 0.5247    & 95.63\%   \\
    \\
    S1-6        & 3.6810    & 0.2717    & 0.2182 \pm 0.0161 & 6.11      & 15.3949 \pm 0.0062    & 0.7864    & 93.17\%   \\
    S1-6        & 1.3679    & 0.7310    & 0.2289 \pm 0.0174 & 6.40      & 15.3951 \pm 0.0061    & 0.7764    & 90.60\%   \\
    \enddata
    \tablenotetext{S}{Indicates a sidereal day alias of known periodic signal.}
    \tablenotetext{D}{Indicates a solar day alias of known periodic signal.}
    \tablenotetext{Q}{Indicates a quarter year ($\approx 91.3$ days) alias of known periodic signal.}
\end{deluxetable*}

\begin{figure}[ht]
    \epsscale{1.2}
    \plotone{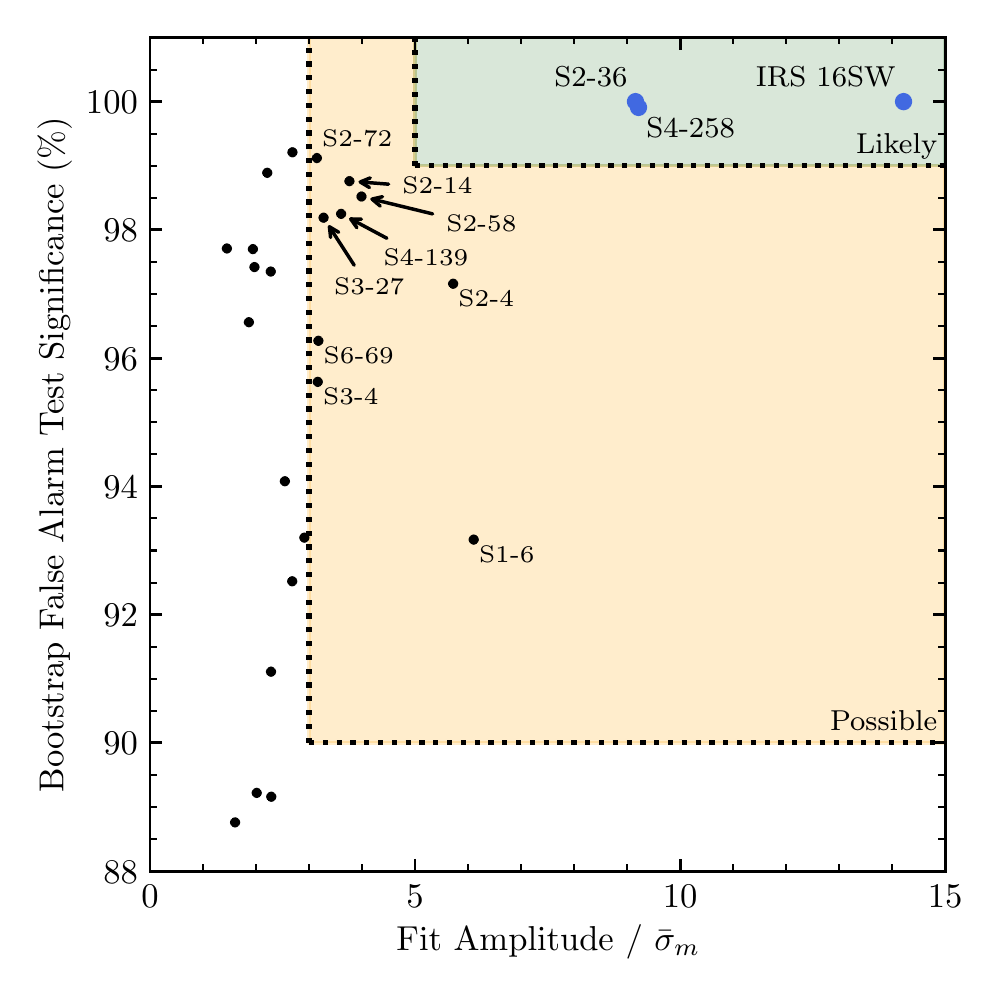}
    \caption{Periodicity detections that pass the variability, periodicity, and frequency cuts in our search, with bootstrap false alarm test significance plotted against the variability amplitude. For clarity, only the most significant periodicity search detection is plotted for stars that have multiple detections passing the variability, periodicity, and frequency cuts. The stars that we identify as \emph{likely periodic variables} (IRS 16SW, S2-36, and S4-258) stand out distinctly in significance and amplitude from other possible periodic detections identified in our experiment.}
    \label{fig:LS_fit_amp_bs_sig}
\end{figure}

We defined our possible periodic signals using a combination of criteria (summarized in Table~\ref{tab:Per_Var}) that were motivated by the characteristics of our periodicity search detailed in Section~\ref{sec:periodic_variability}. In our periodicity search, we considered stars identified as variable by our $\chi^2_{\text{red}}$ test for variability. We defined a maximum period for our periodicity search at $\frac{1}{4} \times$ our observation baseline: $\frac{1}{4} \times 4132.74 \text{ d} = 1033.19 \text{ d}$. We then removed as likely periodic any stars that had power exceeding $50\%$ significance in our bootstrap false alarm test longer than the maximum period cutoff. At such long timescales, our observation baseline was not able to sample a possible periodic signal sufficiently often enough to claim periodicity. Further, any variability leading to high power in our Lomb-Scargle test at these long periods could easily get aliased to shorter periods to falsely resemble shorter-period variability. The minimum search period in our experiment was 1.11 d (from our maximum search frequency cut of $0.9 \text{ d}^{-1}$). Higher frequencies (i.e. shorter periods) than this threshold suffered from frequently aliased peaks.

We then imposed an amplitude threshold for the remaining detections in our periodicity search. To calculate the amplitude, we constructed a sinusoidal fit to the stellar light curve phased to each periodicity detection. To pass the threshold, the amplitude of the fit must exceed $3 \times$ the mean magnitude uncertainty for the star. This threshold is imposed to remove possible peaks originating from statistical fluctuations in our photometry. We finally used our bootstrap false alarm test significance to evaluate whether a star is likely to be periodically variable. If a periodicity detection exceeded 90\% significance in the bootstrap false alarm test, the signal was then considered to be a \emph{possible periodic signal}.

\begin{figure*}[ht]
    \epsscale{1.0}
    \plotone{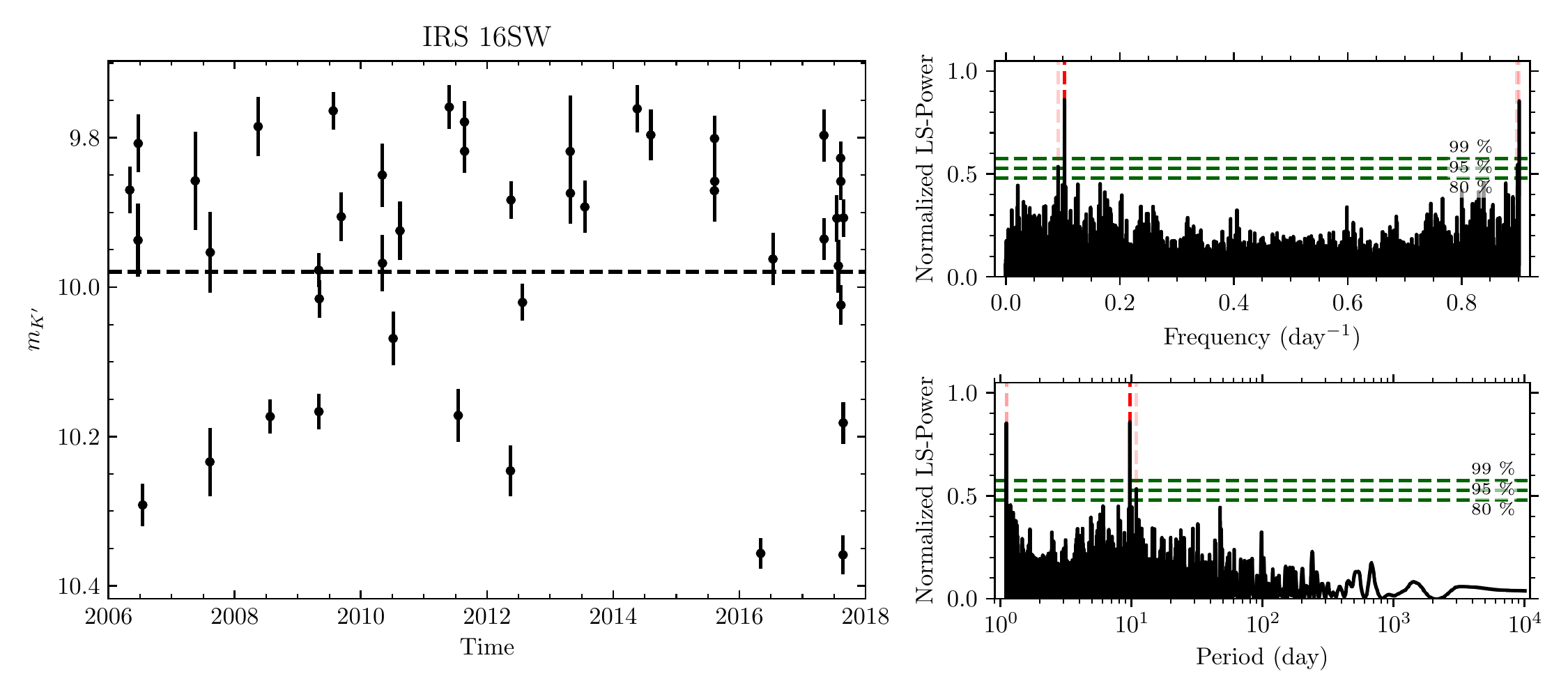}
    \plotone{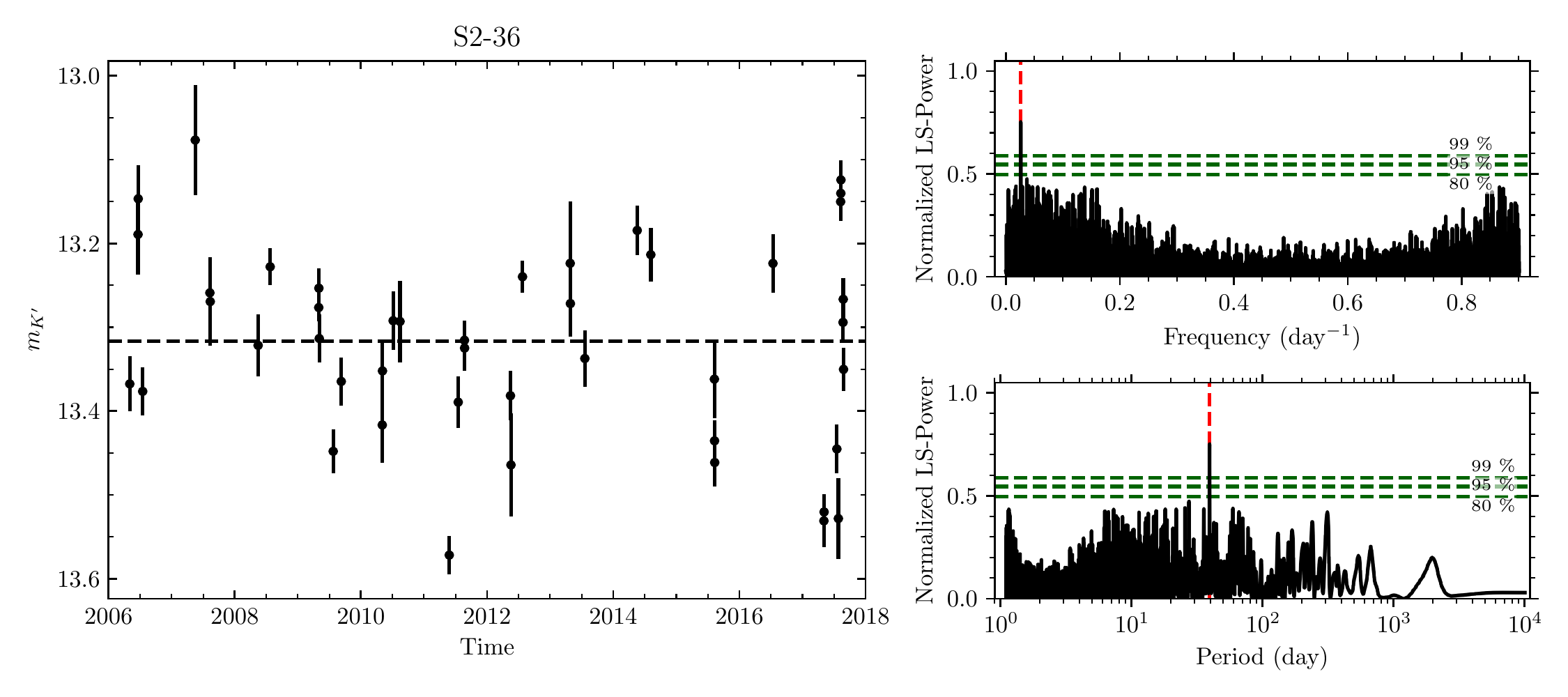}
    \plotone{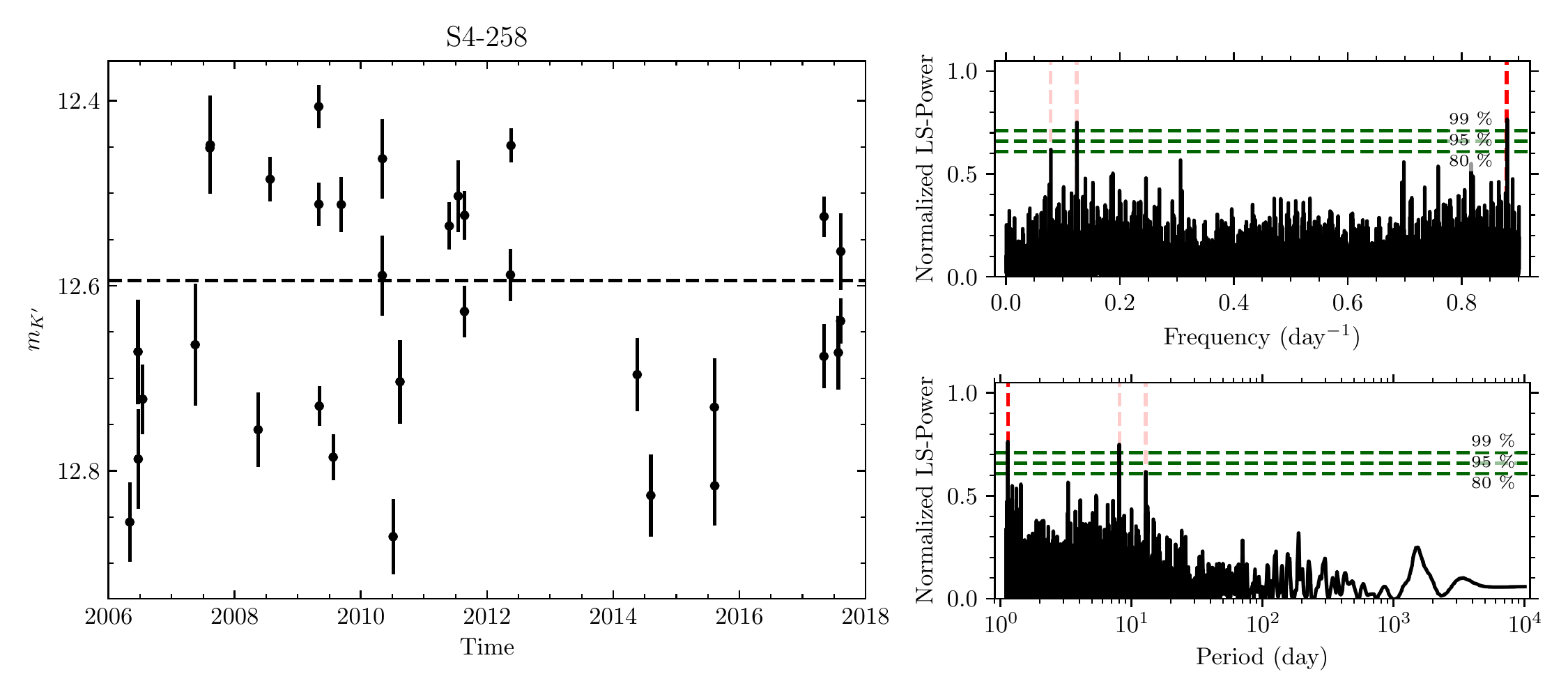}
    \caption{Light curves (\emph{left}) and periodograms (\emph{right}) for the likely periodic variable stars IRS 16SW, S2-36, and S4-258. The horizontal dashed lines in the light curves indicate the weighted mean magnitude. The horizontal dashed green lines in the periodograms indicate the bootstrap test significance levels, while the vertical dashed red lines indicate periodogram peaks above 80\% bootstrap significance.}
    \label{fig:likely_per_lc_per}
\end{figure*}

Three stars in our sample had periodic detections greatly exceeding the possible periodic signal detection amplitude and bootstrap false alarm criteria (IRS~16SW, S2-36, and S4-258; see Figures~\ref{fig:LS_fit_amp_bs_sig} and \ref{fig:likely_per_lc_per}). Based on the three stars' detections, we developed stricter thresholds for these criteria with which we identified \emph{likely periodic variables}: amplitude exceeding $5 \times$ the mean magnitude uncertainty, and detection exceeding 99\% significance in the bootstrap false alarm test. Stars identified as likely periodic variables are listed in Table~\ref{tab:Likely_Periodic_Vars} and possible periodic signal detections are listed in Table~\ref{tab:Possible_Periodic_Signals}. The significance and amplitude of these detections are plotted in Figure~\ref{fig:LS_fit_amp_bs_sig}. Phased light curves of all possible signal detections are included in Appendix~\ref{sec:_periodic_detections}.


\begin{figure*}[ht]
    \epsscale{1.1}
    \plotone{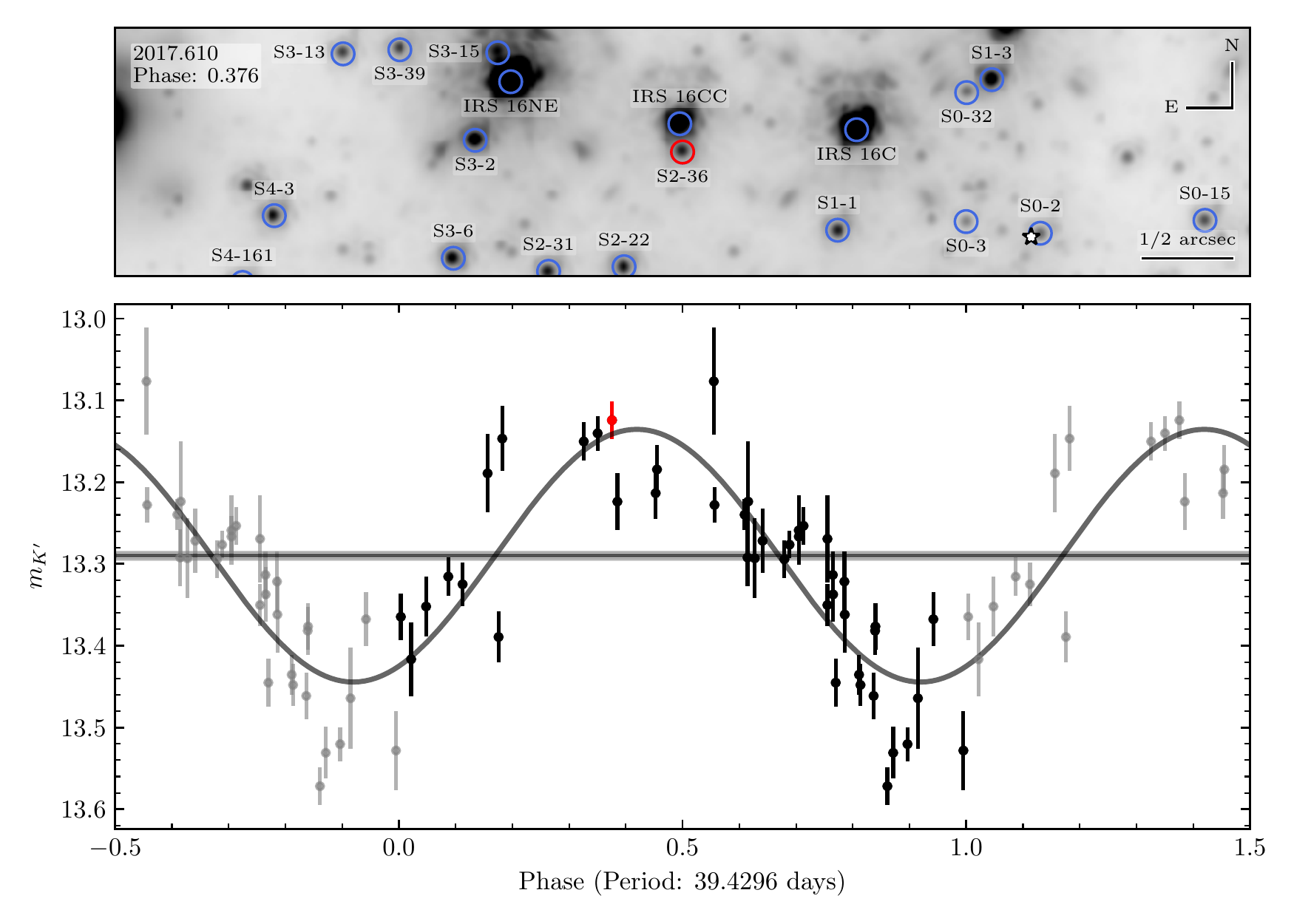}
    \caption{\emph{Top:} An image of the field near S2-36 from the 2017-08-11 observation. S2-36 is circled in red, while nearby stars brighter than $m_{K'} = 14.5$ are circled in blue. The white star symbol indicates the position of Sgr~A*, the location of the supermassive black hole. This observation is highlighted in the phased light curve as the red point. \emph{Bottom:} Phased light curve of S2-36 at the 39.43 day period found in the periodicity analysis. The best fit first order sinusoid model to the observations is overlaid. The horizontal line and surrounding shaded region indicate the fit mean magnitude and its uncertainty, respectively. The red point indicates the observation highlighted on top.}
    \label{fig:S2-36_phased}
\end{figure*}

\subsubsection{Likely periodic variable stars} 
\label{ssub:likely_periodic_variable_stars}
The stars with periodic signal detections passing our criteria for \emph{likely periodic variable stars} are listed in Table~\ref{tab:Likely_Periodic_Vars}. The likely periodic variables IRS 16SW and S4-258 are known eclipsing binary stars, which exhibit two eclipses with similar depths over their orbital period, and are therefore detected at half their binary period in the Lomb-Scargle periodicity search. Additionally, both of these stars have possible periodic signal detections at aliases originating from the length of a sidereal day (1.0027 day$^{-1}$ frequency). IRS~16SW has additional signals passing for possible periodicity which are aliases originating from the length of a solar day (1.0 day$^{-1}$ frequency) and the length of a quarter year ($1.1 \times 10^{-2}$ day$^{-1}$ frequency). These aliases are specifically indicated in Table~\ref{tab:Possible_Periodic_Signals}.

In addition to the known Galactic center eclipsing binary stars, we identified the star S2-36 as a likely periodic variable star. From our periodicity search, S2-36 has a period of 39.43 days (see Figures~\ref{fig:likely_per_lc_per} and \ref{fig:S2-36_phased}). The periodic variability in this star has not been reported previously.

\subsubsection{Possible periodic signals} 
\label{ssub:possible_periodic_signals}
The stars with periodic signal detections passing our criteria for \emph{possible periodic signals} are listed in Table~\ref{tab:Possible_Periodic_Signals}, and phased light curves are provided in Section~\ref{sub:_possible_periodic_detections}. With the limitations from our experiment's photometric precision and observational cadence, it is difficult to conclude whether these represent true periodic variability. We highlight below characteristics of the possible periodic signals in our sample, in three different period regimes.

\begin{itemize}
    \item \emph{1 -- 10 days:} Besides the aliased signals detected from the known periodic variables, IRS~16SW and S4-258, we find signals from S1-6 in this period regime. S1-6 has two signals passing for possible periodicity, at 1.37 and 3.68 days. The two periods detected correspond to sidereal day (1.0027 day$^{-1}$ frequency) aliases of each other. It is difficult to favor photometrically one period over the other as the more likely astrophysical signal if these cases are indeed detections of true periodic variability. This period regime is particularly interesting since detections could be indications of near-contact, short-period binary systems. The signals have roughly sinusoidal shaped phased light curves, but the limited significance and amplitude of these signals makes it difficult to confirm their validity as true astrophysical signals.
    \item \emph{10 -- 80 days:} In a longer period regime, we find more possible periodic signals. In this period regime, we do not expect to detect any sidereal day aliases from possible signals since aliased frequencies would be larger than our experiment's frequency search space.
    
    We found five stars with possible periodic signals in this period regime: S2-72, S2-14, S4-139, S3-27, and S2-4. As a known OB star, S2-4's possible periodic variability is difficult to explain as originating from eclipsing binary systems. The dip in its light curve is wide in phase, unexpected from eclipses at the observed period. Using NIR period-luminosity relations for these possible periodic signals at the observed periods \citep{2010ApJ...723.1195R}, the possible periodic variable signals in S2-72, S4-139, and S3-27 may be consistent with those of ellipsoidal binaries under typical Galactic center extinctions of $A_{K'} \approx 2$--3 magnitudes \citep{2010A&A...511A..18S}. However, several of the possible periodic signals in this regime are detected in stars with light curves suggesting long-term variability trends over our observation baseline (i.e. S2-72, S2-14, S3-27, S2-4). The long-term variability trends may be causing the apparent periodicities by being aliased to shorter periods. Since the long-term variability trends of these stars do not appear as significant detections at long periods, the short period detections remain as possible signals under our periodicity search criteria. Future color observations can more precisely test if the variability is indeed consistent with known periodic variable classes.
    \item \emph{$>$80 days:} In this period regime, S2-58, S4-139, S6-69, and S3-4 have possible periodic signals. While the periods and amplitudes of these stars are consistent with pulsations in evolved stars or ellipsoidal binary systems, the observed mean magnitudes are too faint to be consistent with these classes of variables. Using NIR period-luminosity relations for these possible periodic signals at the observed periods \citep{2010ApJ...723.1195R, 2009MNRAS.399.1709M}, the periodic variability detections have mean magnitudes $\sim 1$ -- $\sim 3.5$ too faint than what is expected under typical Galactic center extinctions of $A_{K'} \approx 2$--3 magnitudes \citep{2010A&A...511A..18S}. Future observations in color of these stars can more precisely test these possibilities.
\end{itemize}



\section{Discussion} 
\label{sec:Discussion}

\subsection{High stellar variability fraction at the Galactic center} 
\label{sub:high_variability_fraction}

\begin{deluxetable*}{llll}[ht]
    \tablecolumns{4}
    \tablecaption{NIR Variability Studies of Spectrally-Typed Resolved Stellar Populations\label{tab:Other_Var_Studies}}
    \tablehead{
        \colhead{Star Population}           &
        \colhead{Paper}                     &
        \colhead{Variability Fraction}      &
        \colhead{Time Baseline}
    }
    \startdata
    \sidehead{\emph{Young, Massive Stellar Populations}}
    NGC 7380    & \citet{2016MNRAS.456.2505L}   & (57 variable stars identified)     & 4 months \\
    Cygnus OB7  & \citet{2012ApJ...755...65R}   & $1.74 \pm 0.14 \%$            & 1.5 years \\
    Orion Nebula    & \citet{2015AJ....150..132R}   & $8.17 \pm 0.24 \%$        & 2.4 years \\
    Quintuplet  & \citet{1999MNRAS.304L..10G}   & $8.5 \pm 1.5 \%$              & $\approx 3$ years \\
    SMC OB Stars    & \citet{2014AnA...562A.125K}   & $40.38 \pm 0.93 \%$        & $\approx 8$ years \\
    \hline
    \sidehead{\emph{Globular Cluster Late-Type Giant Populations}}
    M71         & \citet{2014MNRAS.438.3383M}   & $0.11 \pm 0.02 \%$    & 74 days   \\
    M4          & \citet{2014MNRAS.442.2381N}   & $0.40 \pm 0.07 \%$    & 340 days  \\
    10 Galactic GCs & \citet{2016AnA...588A.128F}   & $0.49 \pm 0.06 \%$    & 1.3 years \\
    NGC 6715 & \citet{2016AnA...592A.120F}   & $5.98 \pm 0.65 \%$    & 2.3 years \\
    \enddata
    \tablecomments{We have recorded the number of variable stars identified for studies that do not report a variability fraction or total sample size.}
\end{deluxetable*}

In this study, we find that $50 \pm 2 \%$ of all stars show variability in the central 0.5 pc of Milky Way nuclear star cluster. This level of stellar variability is greater than what has been found in previous studies of both young clusters and globular clusters in the past. The long time baseline of this survey compared to previous surveys increases our sensitivity to long-term intrinsic brightness variations in stars. In addition, spatial variations in the foreground extinction and stellar confusion can cause brightness variations as the stars move.


\subsubsection{Variability from long time baseline} 
\label{ssub:long_term_variability}

\begin{deluxetable}{cccc}
    \tablecolumns{4}
    \tablecaption{Variability in Smaller Time Baseline Subsamples\label{tab:time_bl_varFrac}}
    \tablehead{
        \colhead{Data Used}             &
        \colhead{Time Baseline (yr)}    &
        \colhead{$F_{\text{young}}$}    &
        \colhead{$F_{\text{old}}$}      
    }
    \startdata
    2006 -- 2017    & 11.31 & $0.52 \pm 0.05$   & $0.43 \pm 0.04$   \\
    2006 -- 2016    & 10.20 & $0.44 \pm 0.05$   & $0.36 \pm 0.04$   \\
    2007 -- 2016    &  9.16 & $0.44 \pm 0.05$   & $0.32 \pm 0.04$   \\
    2008 -- 2016    &  8.16 & $0.42 \pm 0.05$   & $0.32 \pm 0.04$   \\
    2009 -- 2016    &  7.20 & $0.40 \pm 0.05$   & $0.32 \pm 0.04$   \\
    2010 -- 2016    &  6.19 & $0.34 \pm 0.05$   & $0.22 \pm 0.04$   \\
    2011 -- 2016    &  5.13 & $0.30 \pm 0.05$   & $0.19 \pm 0.03$   \\
    2012 -- 2016    &  4.16 & $0.24 \pm 0.05$   & $0.17 \pm 0.03$   \\
    2013 -- 2016    &  3.21 & $0.13 \pm 0.04$   & $0.07 \pm 0.02$   \\
    2014 -- 2016    &  2.15 & $0.09 \pm 0.03$   & $0.06 \pm 0.02$   \\
    2015 -- 2016    &  0.93 & $0.07 \pm 0.03$   & $0.03 \pm 0.02$   \\
    \enddata
\end{deluxetable}

\begin{figure*}[b]
    \epsscale{1.17}
    \plottwo{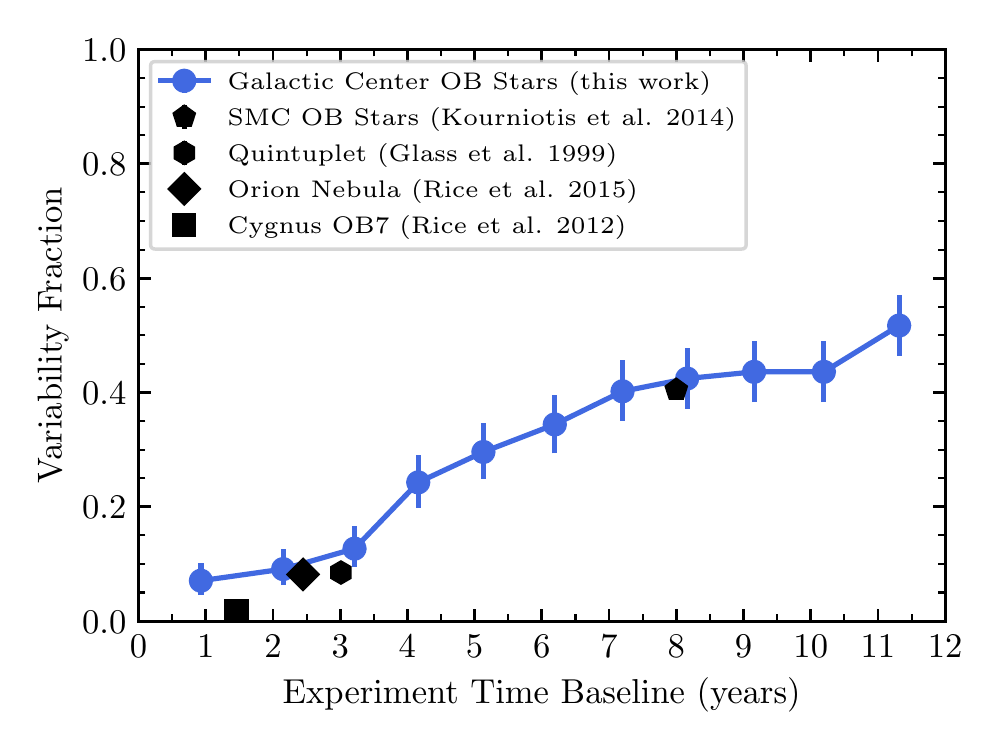}{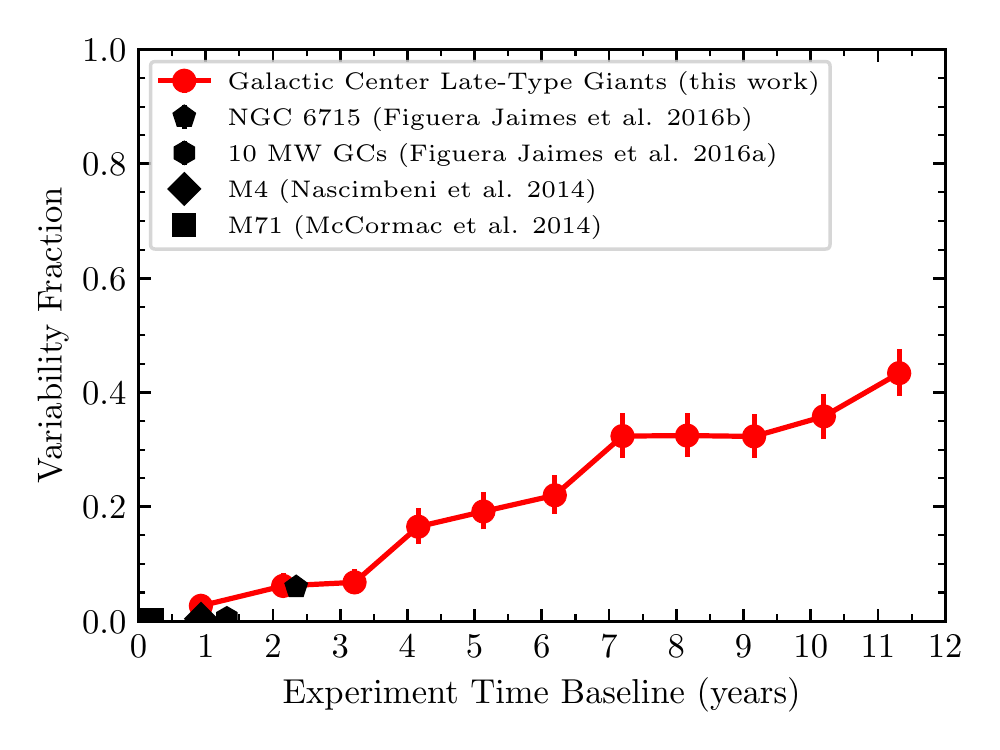}
    \caption{Variability fraction as a function of experiment time baseline for NIR studies of resolved young, massive star populations (\emph{left}) and of late-type, globular cluster (\emph{right}). Variability fractions for the Galactic center young (\emph{left}) and old (\emph{right}) stars derived in this work (entire sample and smaller time baseline subsamples) are shown as black points.}
    \label{fig:varFrac_timeBaseline}
\end{figure*}

The higher level of variability we detect at the Galactic center can be largely accounted for by our experiment's long time baseline of $\sim 11.5$ years. Most NIR stellar variability studies of other young, massive star populations or late-type giants in globular clusters have had overall time baselines on the order of several months to a few years (see Table~\ref{tab:Other_Var_Studies} and Figure~\ref{fig:varFrac_timeBaseline}). To demonstrate the increase in sensitivity to variability with long time baselines in our experiment, we ran our variability models on smaller time baseline subsamples of our data, spanning from $\approx 1$ year to $\approx 11.5$ years (see Table~\ref{tab:time_bl_varFrac}).

Our models demonstrate much lower variability fractions at shorter time baselines. As Figure~\ref{fig:varFrac_timeBaseline} and Table~\ref{tab:time_bl_varFrac} demonstrate, only $\approx 7\%$ of the known young, OB stars in our sample are variable and only $\approx 3\%$ of the known old, late-type giants are variable with an experimental time baseline of $\approx 1$ year. The variability fraction for both stellar type groups rises as the time baseline increases, reaching $\approx 52\%$ and $\approx 43\%$ in our complete time baseline for the young and old stars, respectively. When comparing to previous NIR studies of stellar variability in other resolved young or old stellar populations, the variability fractions we find in our experiment are largely consistent if we account for the time baselines of the experiments (Figure~\ref{fig:varFrac_timeBaseline}). Overall, our smaller time baseline subsamples demonstrate that the high variability fractions in our experiment are largely due to the long time baseline.


\begin{figure*}[ht]
    \epsscale{1.17}
    \plottwo{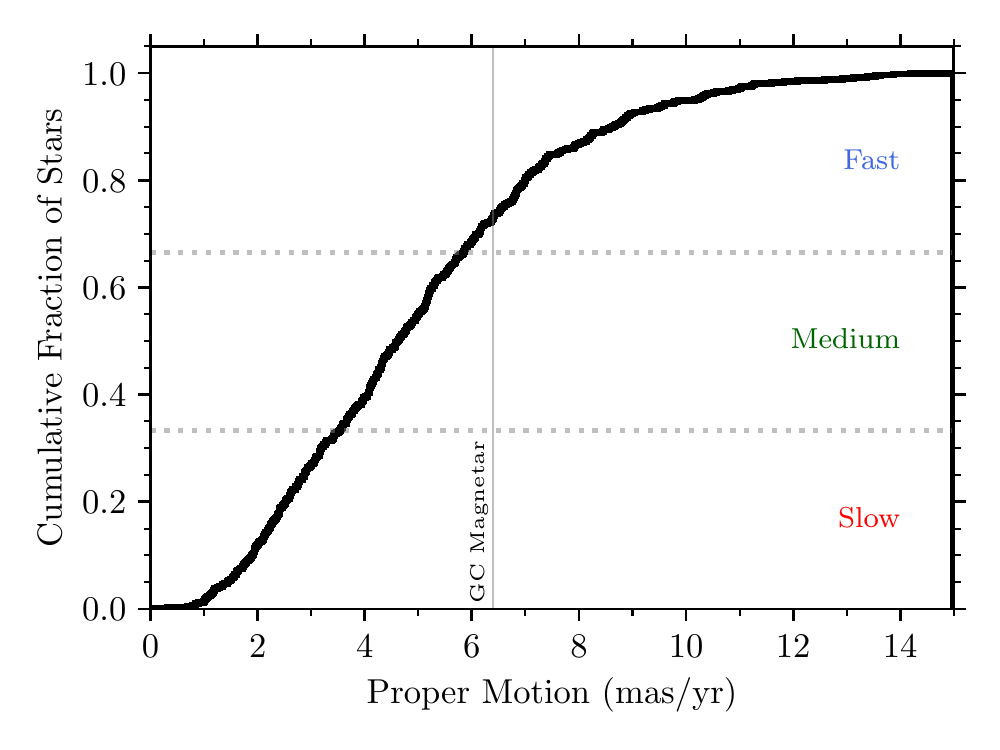}{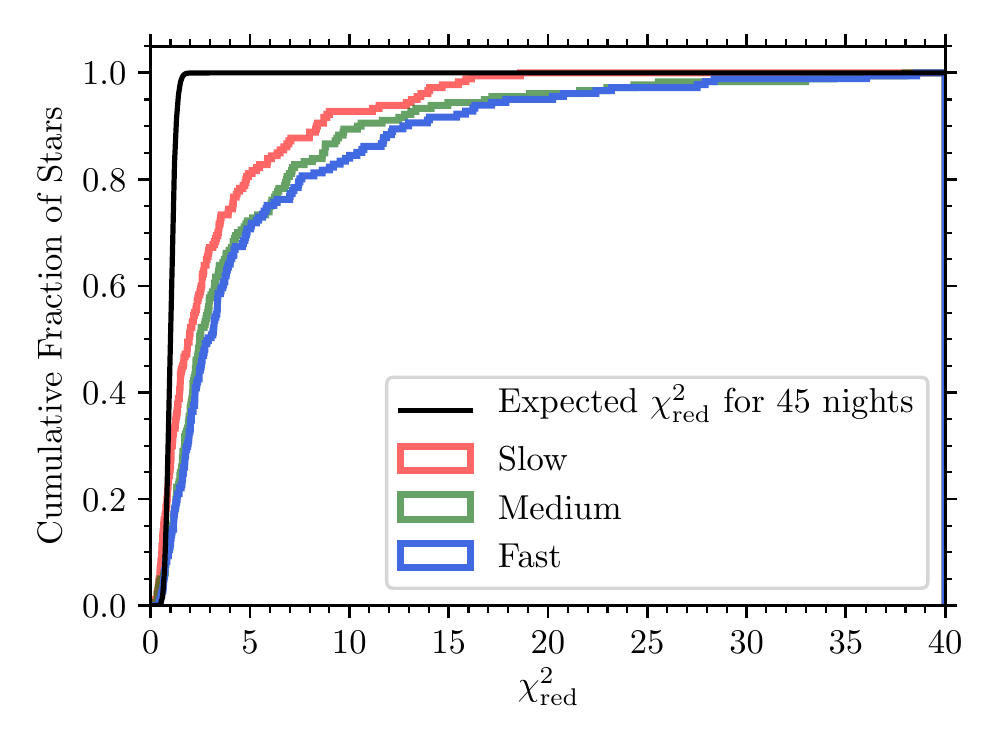}
    \caption{\emph{Left:} The cumulative distribution of proper motion velocity in our sample of stars, outside of an arcsecond of Sgr A*. We divided these stars into three groups, with the same number of stars in each group. This grouping helps select stars in the high proper motion tail of the distribution as the \emph{fast} stars. Notably, the \emph{fast} stars have proper motions comparable to or higher than that of the Galactic center magnetar, PSR J1745-2900, indicated by the vertical line. \emph{Right:} The cumulative distribution of our variability metric, $\chi^2_{\text{red}}$, amongst the three proper motion groups of stars. We found that the $\chi^2_{\text{red}}$ distribution of the \emph{slow} stars is significantly different from those of both the \emph{medium} and \emph{fast} stars ($> 2 \sigma$ and $> 3 \sigma$, respectively), while there is no significant difference amongst the distribution of \emph{medium} and \emph{fast} stars ($< 1 \sigma$).}
    \label{fig:varFrac_propMot}
\end{figure*}

\subsubsection{Variability from Extinction Screen} 
\label{ssub:variability_from_extinction_screen}
The longer 11.5 year time baseline of our experiment allowed some of the additional variability to be contributed from stellar proper motions probing the foreground extinction screen. The Galactic center has large extinction and clumpiness in the foreground extinction screen \citep[e.g.][]{2004A&A...426...81P, 2010A&A...511A..18S, 2018A&A...610A..83N}. Variability in the foreground extinction on large angular scales can result in correlated variability for several stars close together, and consequently would be lessened or removed during our local photometric correction step (Section~\ref{sub:local_photometric_correction}). The typical separation of stars in our sample is $\approx 240$ mas, with smaller separations in the central, more crowded regions of our field. Our experimental methodology would therefore not be very sensitive to features in the foreground extinction screen at much larger angular scales. However, there exist a large number of thin dust filaments identified with $L$-band observations of the Galactic center, with widths $\lesssim 100$ mas \citep{2007A&A...469..993M,2004A&A...417L..15C, 2004A&A...426...81P, 2005ApJ...635.1087G}. These filaments may be traces of gas compressed by shocks at the Galactic center and could be confined by magnetic fields in the area \citep[e.g.][]{2017ApJ...850L..23M}. Similar streamer features are also identified at other infrared and radio wavelengths \citep[e.g.][]{1998ApJ...499L.159Y, 1998ApJ...499L.163Z, 2000ASPC..195..196M, 2001A&A...366..466P, 2003ApJ...594..294S, 2017ApJ...850L..23M} and may be related. These filaments are narrow enough to extinguish light from single stars in our sample at the Galactic center, and the resulting variability would consequently not be affected by our local photometric correction.

Radio observations of the Galactic center magnetar PSR J1745-2900 provide an empirical estimate of the extinction. Rapid changes in the observed Faraday rotation measure as the magnetar's rapid proper motion allowed probing different sightlines. The observations suggest fluctuations in the Galactic center magnetic field or free electron density on size scales $\sim 2$ to $\sim 300 \text{ AU}$ \citep{2018ApJ...852L..12D}, lending evidence for the presence of a scattering screen of gas in the Galactic center environment. Previous observations have suggested that the central parsec of the Galactic center hosts well-mixed warm dust and ionized gas \citep{1991ASPC...14..214G}. If the magnetic field or free electron fluctuations implied by the Galactic center magnetar are associated with dust, they can result in NIR variability for similarly fast moving stellar sources due to varying extinction.


\begin{deluxetable}{lll}
    \tablecolumns{3}
    \tablecaption{Proper Motion Variability Groups\label{tab:propMot_var}}
    \tablehead{
        \colhead{Group}         &
        \colhead{Proper Motion ($\mu$)} &
        \colhead{Var. Frac. ($F$)}
    }
    \startdata
    Fast    &   $\mu > 5.89$ mas/yr   &   $0.55 \pm 0.04$ \\
    Medium  &   $3.56 < \mu < 5.89$ mas/yr    &   $0.51 \pm 0.04$ \\
    Slow    &   $\mu < 3.56$ mas/yr   &   $0.41 \pm 0.04$ \\
    \enddata
\end{deluxetable}

To explore the possibility that faster moving stars are more variable, we divided our stellar sample into three proper motion groups, each containing an equal number of stars: \emph{slow}, \emph{medium}, and \emph{fast}; see Table~\ref{tab:propMot_var} and Figure~\ref{fig:varFrac_propMot}. The proper motion for each star was obtained from either a velocity or acceleration model fitted to the astrometric positions, depending on which model resulted in a fit with a lower $\chi^2_{\text{red}}$ statistic. The velocity component of the chosen model's fit was then used for the proper motion analysis. To avoid stars poorly fit with the proper motion models, we excluded 8 stars from our proper motion groups that have measured orbits around Sgr A* (S0-1, S0-2, S0-3, S0-5, S0-16, S0-19, S0-20, and S0-38). The \emph{fast} proper motion group in particular consists of stars with proper motions comparable to or exceeding the proper motion observed for the Galactic center magnetar \citep[$\approx 6.4 \text{ mas yr}^{-1}$;][]{2018ApJ...852L..12D}, and we expect these stars to probe variations in the foreground extinction screen similar to those inferred for the Faraday screen of the magnetar.

We found that stars with larger proper motions in our sample are more likely to exhibit variability than stars with slower proper motions. The variability fractions of the three proper motion groups are listed in Table~\ref{tab:propMot_var}, and we find that the higher proper motion groups have significantly higher variability fractions. We further tested whether faster moving stars are more variable than slower stars by the two-sample Kolmogorov-Smirnov test (K-S test). Amongst the three proper motion groups, we derived the cumulative distribution of our photometric variability metric, $\chi^2_{\text{red}}$. We computed the two-tailed K-S test $p$-value of all pairs of distributions. The $p$-value gives the probability of the two sample distributions being drawn from the same underlying distribution. Between the \emph{medium} and \emph{fast} groups, we found $p = 60.22\%$, indicating a small difference ($< 1 \sigma$) between the groups' respective $\chi^2_{\text{red}}$ distributions. However, when comparing the \emph{slow} group with both the \emph{medium} ($p = 0.33\%$, $> 2 \sigma$) and \emph{fast} ($p = 0.04\%$, $> 3 \sigma$) groups, we found more significant differences in the $\chi^2_{\text{red}}$ distributions. Overall, our data demonstrate that slower stars have significantly lower variability in our experiment when compared to faster stars, and that variability is more likely for stars with faster proper motions. These results suggest that the foreground extinction is a contributor to our variability fraction since faster moving stars probe larger variations in the foreground extinction screen.

Furthermore, we consider in detail whether some of the most prominent long-term fluctuations in our variable star sample can be physically explained by the foreground extinction screen. Changes in the observed flux for a stellar source imply a change in optical depth, $\tau_\lambda$:

\begin{eqnarray}
    A_\lambda &=& -2.5 \log_{10} (I_\lambda / I_{\lambda, 0}) \\
    &=& -2.5 \log_{10} (e^{- \tau_\lambda}) \\
    &=& \tau_\lambda (-2.5 \log_{10} e) \approx \tau_\lambda \times 1.086
\end{eqnarray}

Assuming a constant cross section, $\sigma_\lambda$, for extinguishing dust grains, changes in optical depth, $\Delta \tau_\lambda$, correspond to changes in column density, $\Delta N_d$:

\begin{eqnarray}
    \Delta \tau_\lambda &=& \sigma_\lambda \Delta N_d
\end{eqnarray}

Amongst our highly variable stars (Section~\ref{sec:_chi_2__text_red_geq_10_variables}), stars exhibiting long-period brightening or dimming have changes in observed flux approaching $\approx 0.5$ magnitudes (e.g. S2-316, S4-12, S4-262) to $\approx 1.0$ magnitudes (e.g. S3-34). Following \citet{2004A&A...426...81P}, we assume that extinction at $K$-band is about $0.1 \times$ that in visual and that a magnitude of extinction at visual implies a column density of $\approx 2 \times 10^{21} \text{ cm}^{-2}$ H atoms. These large dips in magnitude would imply changes in column density of $\approx 10^{22} \text{ cm}^{-2}$. Since these stars exhibited only either a dimming or brightening, it is difficult to establish a physical size to inhomogeneities in the foreground material if caused by extinction. However, such scales of extinction are consistent with those observed by \citet{2004A&A...426...81P} from large gas features like the \emph{Minispiral} at the Galactic center.

Using stars that exhibit both brightening and dimming over our time baseline (e.g. S2-66, S3-249, IRS 7SE), we can estimate the density of dust in extinguishing filaments. These stars display momentary dips in flux of $\sim 1$ mag lasting $\approx 4$ years.
While there can be various physical geometries of the extinguishing material, such as dust blobs, sheets, or bow shocks, we assume here for simplicity that the dips originate from thin, filamentary structures located near the Galactic center. Under this physical assumption, the proper motion measurements of these stars in our dataset imply filament diameters of approximately $10^{-3}$ pc or 200 AU. Our diameter estimate assumes static filaments, but if the filaments themselves are also in motion near the stellar sources, the diameter estimate may increase by a factor of $\approx 2$. The typical magnitude dips then indicate number densities in the extinguishing filaments of $\approx 3 \times 10^6 \text{ cm}^{-3}$. These thin regions of high extinction could correspond to foreground high density filaments similar to those identified by \citet{2007A&A...469..993M}. The densities are consistent with models of high density bow shocks at the Galactic center \citep{2002ApJ...575..860T}. In fact, IRS 7SE's location is consistent with the X1 filament, proposed to be a bow shock source \citep{2004A&A...417L..15C, 2007A&A...469..993M}. Another highly variable star, S4-12, has a location consistent with the X4 filament \citep{2007A&A...469..993M}, a proposed bow shock source originating from IRS 3 \citep{2005A&A...433..117V, 2017ApJ...837...93Y}. The filaments could be responsible for the long-term flux dips observed in these two stars' light curves. S2-66 and S3-249, however, do not have corresponding filaments identified by \citet{2007A&A...469..993M} that would be consistent with their locations.
\citet{2007ApJ...659.1241R} highlighted the long-term variability in the light curves of three stars (particularly S2-11) using independent data as also likely originating from their passage behind thin, high-density filaments. Our experiment's observations, taken at a later time, do not reveal similar features in these stars' light curves.

Our observations suggest that variations in the extinction screen can indeed account for some of the high variability fraction found in this experiment. With our $K'$ dataset alone, however, it is difficult to assign this as the primary source of variability for any given star in our sample. Extensions of our variability study incorporating simultaneous observations at other wavebands over a long-period can add substantially to the study of extinction variations. Particularly, increased reddening during dips in flux would suggest dust extinction as the likely cause \citep[see e.g.][]{2015AJ....150..132R}.

\begin{figure}[t]
    \epsscale{1.2}
    \plotone{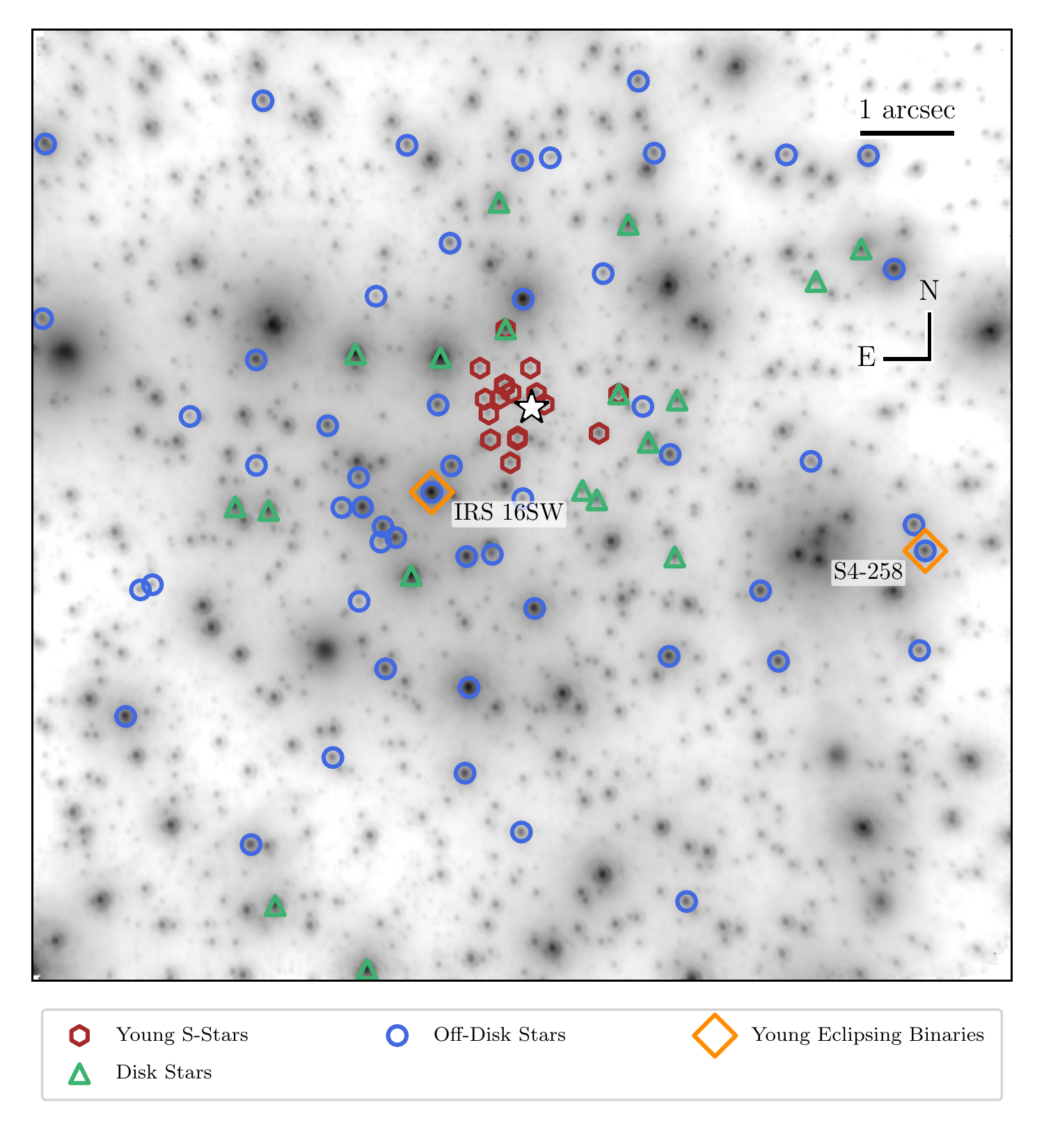}
    \caption{The young stars in our sample, indicated on the 2012-05-15 observation. We identify here the young disk and S-Star stellar populations from our sample. The young eclipsing binary systems detected in this experiment are labelled: IRS 16SW \citep{1999ApJ...523..248O, 2007AcA....57..173P, 2007ApJ...659.1241R} and S4-258 \citep[E60, discovered by][]{2014ApJ...782..101P}. We do not detect any eclipsing binary systems among the disk stars nor the young S-Star stellar populations. The star symbol indicates the location of Sgr~A*, and 1 arcsec corresponds to a projected distance of $\approx 0.04$~pc at the Galactic center distance.}
    \label{fig:young_stars_map}
\end{figure}

\subsection{Constraints on the eclipsing binary fraction of young stars} 
\label{sub:eclipsing_binary_fraction}
Our data provide the tightest constraints yet on the eclipsing binary fraction of the young stars in the nuclear star cluster by using a larger sample than previous works. In our sample of 85 stars, we recover the two previously discovered eclipsing binary systems: IRS 16SW \citep{1999ApJ...523..248O, 2007AcA....57..173P, 2007ApJ...659.1241R} and S4-258 \citep[E60, discovered by][]{2014ApJ...782..101P} (see map in Figure~\ref{fig:young_stars_map}). This places a lower limit on the eclipsing binary fraction of $2.4\% \pm 1.7\%$. Previous work using a sample of 70 young stars and detection of the same two binary systems, \citet{2014ApJ...782..101P} determined a lower limit on the young star eclipsing binary fraction of $3\% \pm 2\%$.

We do not detect any eclipsing binaries amongst the young stellar disk members. In our sample, 18 stars were identified as likely members of the young stellar disk by \citet{2014ApJ...783..131Y}. The two known eclipsing binaries are off-disk stars. While due to small number statistics this null detection is not unusual (66\% probability of a null detection in this sample from our observed eclipsing binary fraction), the lack of binaries in the disk warrants future investigation. Binaries can serve as a way to characterize the differences of formation mechanisms of stars in the disk compared to off-disk stars \citep[see e.g.][]{2008MNRAS.389.1655A, 2003ApJ...590L..33L, 2004ApJ...608..108G, 2005A&A...437..437N}. Furthermore, there may be observational biases when assigning disk membership probabilities to binaries \citep{2014ApJ...783..131Y, 2018ApJ...853L..24N}. Due to our sample size, we do not expect these biases to lead to a different conclusion about the relative eclipsing binary fraction of disk members versus non-disk members. However, these biases will be important when the sample of young stars increases.

We also do not detect any eclipsing binaries in the young S-star population (stars within a projected distance of 0.04 pc of the SMBH). Similar to the disk stars, the lack of eclipsing binaries in the young S-stars is not surprising given the small sample size (17 stars) in our experiment. However, if any S-stars are indeed binaries, we may expect to be more sensitive to eclipsing systems since they tend to be in tighter orbits \citep{2017ApJ...851..131L}. Better constraints on the binary fraction of S-stars is necessary since it can serve as an indicator of the stars' formation mechanisms. For example, if S-stars are captured components of tidally disrupted binary systems, they should no longer have a companion \citep{1988Natur.331..687H, 2003ApJ...599.1129Y}. Other recent observational constraints are consistent with this hypothesis \citep{2018ApJ...854...12C}.

The young nuclear star cluster eclipsing binary fraction is consistent with that of the local solar neighborhood.
\citet{2009A&A...507.1141L} find 40 OB binaries passing criteria similar to those of our experiment out of a sample of 2497 stars in a study of local OB variability with the HIPPARCOS satellite, giving a local OB eclipsing binary fraction of $1.60\% \pm 0.25\%$. Therefore, our estimate of the early-type eclipsing binary fraction at the Galactic center is consistent with the eclipsing binary fraction of local OB stars.

Improvements in the time sampling, sensitivity, sample size, or the addition of multiband photometry will allow tighter constraints in the eclipsing binary fraction. From our periodicity search parameters (Table~\ref{tab:Per_Var}), we are sensitive to binary periods longer than 2.22 days and amplitudes larger than $0.03 \times 5 = 0.15 \text{ mags}$. These limits to our sensitivity to binary systems can be improved by the addition of photometry in another filter to eliminate false positives during periodicity searches. Furthermore, the Lomb-Scargle periodogram and the observation cadence used in this work are particularly optimized for detecting periodic signals that have an overall sinusoidal shape in the phased light curve. Therefore, our experiment is most sensitive to those systems that have eclipses wide in phase, expected from contact or near-contact binary systems. Future work is required to infer the overall binary fraction from these detections of eclipsing binary systems at the Galactic center.


\begin{figure*}[ht]
    \epsscale{1.1}
    \plotone{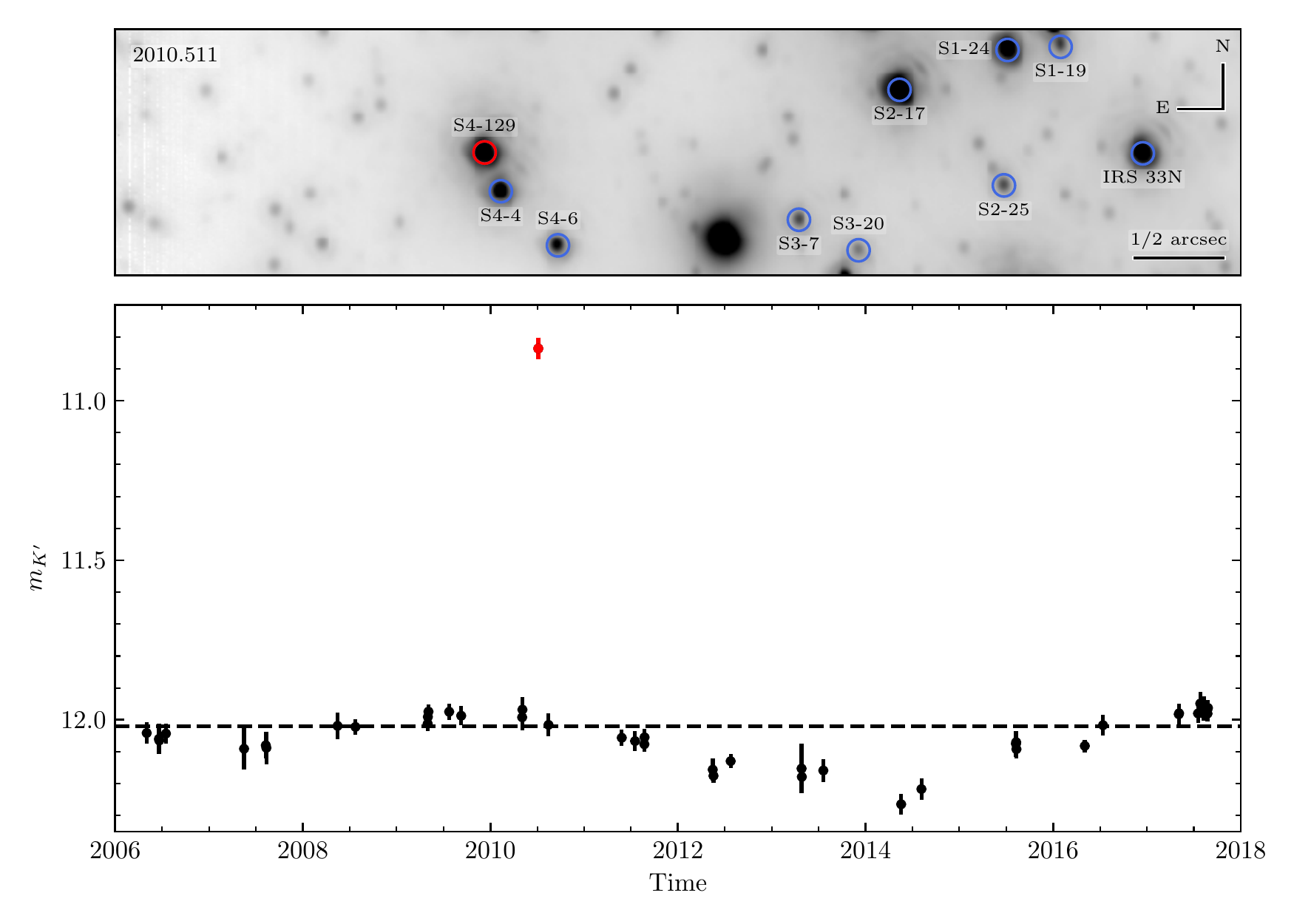}
    \caption{\emph{Top:} An image of the field near S4-129 from the 2010-07-06 observation. During this observation, S4-129 increased in brightness by $\approx 1.2$ magnitudes compared to other observations near in time. Visual inspection of the field in this and other observations near in time did not reveal any sources of potential stellar confusion. S4-129 is circled in red, while nearby stars brighter than $m_{K'} = 14.5$ are circled in blue. This observation is highlighted in the light curve as the red point. \emph{Bottom:} Light curve of S4-129 over our experiment's entire time baseline. The red point indicates the observation highlighted on top, during which we observed the brightening.}
    \label{fig:S4-129_lc_ps}
\end{figure*}

\subsection{Other periodic and variable stars} 
\label{sub:other_periodic_and_variable_stars}
This study has revealed previously unidentified periodic variability in the star S2-36, with a period of 39.43 days (Section~\ref{ssub:likely_periodic_variable_stars}). The source's period and light curve could be consistent with an ellipsoidal binary system (potentially also eclipsing) or a Type II Cepheid star. Period-luminosity relations \citep{2006MNRAS.370.1979M, 2010ApJ...723.1195R} suggest that the star's observed flux can be compatible with both classes of periodic variability under the typical range of extinctions towards the Galactic center \citep{2010A&A...511A..18S}. Determining the likely source of this star's periodic variability requires additional observations beyond just the $K'$-band dataset in this work (Gautam et al. in prep.).

Besides S2-36 and the previously discovered eclipsing binary systems at the Galactic center, we did not find evidence for other periodic variable stars. Periodic fluctuations in flux may be expected from stars during periods of instabilities, and are particularly useful in revealing membership of the corresponding stars into populations with specific ages or metallicities. Notably, our periodicity search experiment is sensitive to the period and amplitude ranges of pulsating evolved stars. The first order pulsations of these stars, known as Mira variables \citep[periods of 80 -- 1000 days, NIR amplitudes $\sim 1$ mag; see][]{2015pust.book.....C, 1997JAVSO..25...57M}, often host SiO masers and therefore can be particularly useful for Galactic center astrometric experiments \citep{2010ApJ...725..331Y}. However, we find no evidence of such stars in our experiment's field of view.

Stellar confusion is likely only a small contributor to variability in our sample. Only one star amongst our highly variable stars exhibits variability that can be clearly attributed to stellar confusion. S3-21 ($\bar{m}_{K'} = 15.28$) had a rapid brightness rise starting in 2012 due to it closely approaching the bright star S3-6 ($\bar{m}_{K'} = 12.69$).
If stellar confusion were to be a larger contributor to variability in our sample, we would expect higher variability fractions in the highly crowded central arcsecond region, where stellar crowding leads to more confusion events. However, our data did not suggest any significant increases in variability in this region (Figure~\ref{fig:RedChiSqDist}).
In general, with our implementation of checks for confused sources (see Section~\ref{sub:stellar_confusion_and_resolved_sources} and Jia et al., in prep.), we are able to largely reduce the effects of confusion.

In addition, the high density of objects at the Galactic center can lead to microlensing events, where a massive object passing in front of a star at the Galactic center can lead to a brief brightening event \citep{2001ApJ...551..223A}. Among our sample of highly variable stars, one star, S4-129, demonstrated brightening that could be the result of microlensing. S4-129 experienced a brightening of $\approx 1.2$ magnitudes ($\approx 3 \times$ increase in flux) during a single observation in our dataset (2010-07-06), and visual inspection of the star's local field in the images did not reveal any obvious sources of stellar confusion that could be the cause (see Figure~\ref{fig:S4-129_lc_ps}). With just a single point in the brightening, it is difficult to put constraints on parameters of a possible lensing system. While this is the largest short brightening event in our sample, microlensing events have been predicted in the Galactic center environment from a variety of configurations \citep[see e.g.][]{2001ApJ...551..223A, 2001ApJ...563..793C, 2005ApJ...627..790B} and may be a small contribution towards the variability fraction in our sample.

Based on the $K'$-band observations alone in our experiment, it is difficult to determine a likely physical source of variability for all of our variable stars. A future study of variability of the Galactic center stars in color space can provide additional insight into sources of difficult-to-explain variability in our sample. Variability in the $H-K'$, $H$ space can in particular reveal changes in dust extinction and accretion activity \citep{2015AJ....150..132R} or the presence of hot spots and cool spots on stars \citep{2013ApJ...773..145W}. Without this extra color variability information, it is difficult to ascribe a specific source of variability to several stars in our sample.

\subsection{The precision of single PSF AO photometry} 
\label{sub:ao_photometry}

While AO observations provide the angular resolution necessary to study the high stellar density of the Galactic center, there are two main challenges that must be overcome to achieve high-precision photometry. The extreme crowding of stellar sources in the central regions of the nuclear star cluster makes aperture photometry difficult or impossible as the point spread functions (PSFs) of the sources overlap. There is also variation in the PSF shape across the field of view and over time. Anisoplanatism results in PSF variation as a function of the position of the star with respect to the laser guide star and the tip-tilt star. Weather, atmospheric conditions, and performance of the adaptive optics system during observations further introduce fluctuations in the PSF shape. These effects cause biases when estimating the flux of stellar sources, and they therefore can be manifested in our data as a systematic variability in flux. Special efforts have to be made to account for these effects.

An approach to obtaining precise photometry from AO imaging data in a crowded field is PSF fitting and local calibration across the field. In our work, we used a single reference PSF across our entire field of view to derive initial photometric flux measurements. We expect that factors affecting PSF shape, such as anisoplanatic effects and atmospheric conditions, influence the PSF shapes and bias photometric measurements of nearby stars on the field of similar brightness in similar ways. Our local photometric correction removed these local trends in estimated flux (implementation detailed in Section~\ref{sub:local_photometric_correction}). There are two metrics with which we evaluated the precision resulting from our methodology: photometric precision per observation epoch and median photometric precision across our entire time baseline. Across several individual observations our method achieved uncertainties of $\Delta m_{K'} \sim 0.02$ ($\approx 2\%$) to $m_{K'} = 16$ (see Table~\ref{tab:Observations}). Across all our observations, our method achieved a photometric uncertainty floor of $\Delta m_{K'} \sim 0.03$ ($\approx 3\%$) out to $m_{K'} \approx 16$ (see Figure~\ref{fig:Mag_MagError}).

Another approach to precise photometry with AO imaging data is to use separate reference PSFs across the field. In their AO photometric study of Galactic center stars, \citet{2010A&A...511A..18S} partitioned their images into smaller sub-frames, where the anisoplanatic effects over the sub-frame are small. A reference PSF was separately derived in each sub-frame, accounting for a variable PSF across the field of view. With this method, they were able to obtain photometric precisions as low as $\Delta m_{K\text{s}} \sim 0.015$ ($\approx 1.5\%$) out to $m_{K\text{s}} \approx 15$. The precisions we obtain with our techniques are comparable in several individual epochs.

The most comparable previous study to this work, a study of stellar variability in the Galactic center with Keck speckle data \citep{2007ApJ...659.1241R}, achieved much lower precision than our method. With Keck speckle data, uncertainties of $\Delta m_{K} \sim 0.06$ out to $m_{K} \approx 13$ were obtained, with uncertainties reaching $\Delta m_{K} \sim 0.21$ at $m_{K} \approx 16$ (see Figure~\ref{fig:Mag_MagError}). Our method achieves much higher precision to fainter magnitudes. While much of this improvement comes from the greater depth AO imaging provides, our more robust calibration procedure and selection of stable calibrator stars also deliver more precision in relative photometry.

\section{Conclusion} 
\label{sec:Conclusion}
In this work, we presented an analysis of stellar variability and a search for eclipsing binary systems in the central $10''$ of the Galactic center with NIR laser-guide star AO data. Our photometric calibration and local correction techniques achieved photometric uncertainties reaching $\approx 3\%$ across our entire dataset and $\approx 2\%$ in several individual observations. This photometric precision is comparable to the highest precision achieved by other AO photometric studies using single-PSF fitting.

We have compiled the first catalog of photometric variables in the central half-parsec of the Galactic center with NIR AO imaging. We found that among our stellar sample of 563 stars identified across at least 16 observation nights, $50 \pm 2\%$ of stars displayed variability. Within this sample, $52 \pm 5\%$ of known early-type stars and $43 \pm 4 \%$ of known late-type stars displayed variability. The variability fractions of the typed stars in our sample are much greater than that of other young, massive star populations or late-type giants in globular clusters. The higher variability fraction relative to other studies can largely be accounted for by the longer time baseline of our experiment. Variations in the foreground extinction screen also contribute to the high variability fraction.

In a periodicity search of our photometric dataset, we recovered the two previously discovered eclipsing binary systems at the Galactic center: IRS 16SW and S4-258 (E60). We additionally identified a new periodically variable star at the Galactic center, S2-36, with a period of 39.43 days. Additional observations across other wavelengths or spectroscopic follow-up observations of this star can determine the physical source of the periodic variability.

We detected no evidence of an eclipsing binary system among the S-star population within $1''$ of the central black hole, nor among the young stellar disk. We measured a lower limit on the eclipsing binary fraction of $2.4 \pm 1.7\%$ among the young stars at the Galactic center. Our constraints on the Galactic center eclipsing binary fraction are consistent with the local OB star eclipsing binary fraction under observational limits similar to those of our experiment \citep{2009A&A...507.1141L}.

\acknowledgements
We thank the anonymous referee for helpful comments. We thank the staff of the Keck Observatory for their help in obtaining the observations presented herein. Support for this work at UCLA was provided by the Kaplan Family Student Support Fund and the W. M. Keck Foundation. The data presented herein were obtained at the W. M. Keck Observatory, which is operated as a scientific partnership among the California Institute of Technology, the University of California, and the National Aeronautics and Space Administration. The Observatory was made possible by the generous financial support of the W. M. Keck Foundation. The authors wish to recognize and acknowledge the very significant cultural role and reverence that the summit of Maunakea has always had within the indigenous Hawaiian community. We are most fortunate to have the opportunity to conduct observations from this mountain.

\facility{Keck:II (NIRC2)}

\software{NumPy \citep{vanderWalt:2011dp}, Astropy \citep{2013A&A...558A..33A}, emcee \citep{2013PASP..125..306F}, SciPy \citep{Jones01}, Matplotlib \citep{Hunter07}}

\bibliographystyle{aasjournal}
\bibliography{PaperBib,Software}

\begin{thebibliography}{}
\expandafter\ifx\csname natexlab\endcsname\relax\def\natexlab#1{#1}\fi
\providecommand{\url}[1]{\href{#1}{#1}}

\bibitem[{Alexander {et~al.}(2008)Alexander, Armitage, \&
  Cuadra}]{2008MNRAS.389.1655A}
Alexander, R.~D., Armitage, P.~J., \& Cuadra, J. 2008, MNRAS, 389, 1655

\bibitem[{Alexander \& Loeb(2001)}]{2001ApJ...551..223A}
Alexander, T., \& Loeb, A. 2001, ApJ, 551, 223

\bibitem[{Alexander \& Pfuhl(2014)}]{2014ApJ...780..148A}
Alexander, T., \& Pfuhl, O. 2014, ApJ, 780, 148

\bibitem[{{Astropy Collaboration} {et~al.}(2013){Astropy Collaboration},
  Robitaille, Tollerud, Greenfield, Droettboom, Bray, Aldcroft, Davis,
  Ginsburg, Price-Whelan, Kerzendorf, Conley, Crighton, Barbary, Muna,
  Ferguson, Grollier, Parikh, Nair, Unther, Deil, Woillez, Conseil, Kramer,
  Turner, Singer, Fox, Weaver, Zabalza, Edwards, Azalee~Bostroem, Burke, Casey,
  Crawford, Dencheva, Ely, Jenness, Labrie, Lim, Pierfederici, Pontzen, Ptak,
  Refsdal, Servillat, \& Streicher}]{2013A&A...558A..33A}
{Astropy Collaboration}, Robitaille, T.~P., Tollerud, E.~J., {et~al.} 2013,
  A{\&}A, 558, A33

\bibitem[{Bartko {et~al.}(2009)Bartko, Martins, Fritz, Genzel, Levin, Perets,
  Paumard, Nayakshin, Gerhard, Alexander, Dodds-Eden, Eisenhauer, Gillessen,
  Mascetti, Ott, Perrin, Pfuhl, Reid, Rouan, Sternberg, \&
  Trippe}]{2009ApJ...697.1741B}
Bartko, H., Martins, F., Fritz, T.~K., {et~al.} 2009, 697, 1741

\bibitem[{Bartko {et~al.}(2010)Bartko, Martins, Trippe, Fritz, Genzel, Ott,
  Eisenhauer, Gillessen, Paumard, Alexander, Dodds-Eden, Gerhard, Levin,
  Mascetti, Nayakshin, Perets, Perrin, Pfuhl, Reid, Rouan, Zilka, \&
  Sternberg}]{2010ApJ...708..834B}
Bartko, H., Martins, F., Trippe, S., {et~al.} 2010, 708, 834

\bibitem[{Blum {et~al.}(1996)Blum, Sellgren, \& Depoy}]{1996ApJ...470..864B}
Blum, R.~D., Sellgren, K., \& Depoy, D.~L. 1996, Astrophysical Journal v.470,
  470, 864

\bibitem[{Boehle {et~al.}(2016)Boehle, Ghez, Sch{\"o}del, Meyer, Yelda, Albers,
  Martinez, Becklin, Do, Lu, Matthews, Morris, Sitarski, \&
  Witzel}]{2016ApJ...830...17B}
Boehle, A., Ghez, A.~M., Sch{\"o}del, R., {et~al.} 2016, ApJ, 830, 17

\bibitem[{Bozza \& Mancini(2005)}]{2005ApJ...627..790B}
Bozza, V., \& Mancini, L. 2005, ApJ, 627, 790

\bibitem[{Castelli \& Kurucz(2004)}]{2004astro.ph..5087C}
Castelli, F., \& Kurucz, R.~L. 2004, arXiv, arXiv:astro

\bibitem[{Catelan \& Smith(2015)}]{2015pust.book.....C}
Catelan, M., \& Smith, H.~A. 2015, Pulsating Stars (Wiley-VCH)

\bibitem[{Chanam{\'e} {et~al.}(2001)Chanam{\'e}, Gould, \&
  Miralda-Escud{\'e}}]{2001ApJ...563..793C}
Chanam{\'e}, J., Gould, A., \& Miralda-Escud{\'e}, J. 2001, ApJ, 563, 793

\bibitem[{Chen {et~al.}(2017)Chen, de~Grijs, \& Deng}]{2017MNRAS.464.1119C}
Chen, X., de~Grijs, R., \& Deng, L. 2017, MNRAS, 464, 1119

\bibitem[{Chu {et~al.}(2018)Chu, Do, Hees, Ghez, Naoz, Witzel, Sakai, Chappell,
  Gautam, Lu, \& Matthews}]{2018ApJ...854...12C}
Chu, D.~S., Do, T., Hees, A., {et~al.} 2018, ApJ, 854, 12

\bibitem[{Cl{\'e}net {et~al.}(2004)Cl{\'e}net, Rouan, Gendron, Lacombe,
  Lagrange, Mouillet, Magnard, Rousset, Fusco, Montri, Genzel, Sch{\"o}del,
  Ott, Eckart, Marco, \& Tacconi-Garman}]{2004A&A...417L..15C}
Cl{\'e}net, Y., Rouan, D., Gendron, E., {et~al.} 2004, A{\&}A, 417, L15

\bibitem[{Desvignes {et~al.}(2018)Desvignes, Eatough, Pen, Lee, Mao,
  Karuppusamy, Schnitzeler, Falcke, Kramer, Wucknitz, Spitler, Torne, Liu,
  Bower, Cognard, Lyne, \& Stappers}]{2018ApJ...852L..12D}
Desvignes, G., Eatough, R.~P., Pen, U.~L., {et~al.} 2018, The Astrophysical
  Journal Letters, 852, L12

\bibitem[{Diolaiti {et~al.}(2000)Diolaiti, Bendinelli, Bonaccini, Close,
  Currie, \& Parmeggiani}]{2000A&AS..147..335D}
Diolaiti, E., Bendinelli, O., Bonaccini, D., {et~al.} 2000, Astronomy and
  Astrophysics Supplement, 147, 335

\bibitem[{Do {et~al.}(2009)Do, Ghez, Morris, Lu, Matthews, Yelda, \&
  Larkin}]{2009ApJ...703.1323D}
Do, T., Ghez, A.~M., Morris, M.~R., {et~al.} 2009, 703, 1323

\bibitem[{Do {et~al.}(2013{\natexlab{a}})Do, Lu, Ghez, Morris, Yelda, Martinez,
  Wright, \& Matthews}]{2013ApJ...764..154D}
Do, T., Lu, J.~R., Ghez, A.~M., {et~al.} 2013{\natexlab{a}}, 764, 154

\bibitem[{Do {et~al.}(2012)Do, Ghez, Lu, Morris, Yelda, Martinez, Peter,
  Wright, Bullock, Kaplinghat, \& Matthews}]{2012JPhCS.372a2016D}
Do, T., Ghez, A., Lu, J.~R., {et~al.} 2012, J. Phys.: Conf. Ser., 372, 012016

\bibitem[{Do {et~al.}(2013{\natexlab{b}})Do, Martinez, Yelda, Ghez, Bullock,
  Kaplinghat, Lu, Peter, \& Phifer}]{2013ApJ...779L...6D}
Do, T., Martinez, G.~D., Yelda, S., {et~al.} 2013{\natexlab{b}}, The
  Astrophysical Journal Letters, 779, L6

\bibitem[{Dong {et~al.}(2017)Dong, Sch{\"o}del, Williams, Nogueras-Lara,
  Gallego-Cano, Gallego-Calvente, Wang, Morris, Do, \&
  Ghez}]{2017MNRAS.470.3427D}
Dong, H., Sch{\"o}del, R., Williams, B.~F., {et~al.} 2017, MNRAS, 470, 3427

\bibitem[{Duch{\^e}ne \& Kraus(2013)}]{2013ARA&A..51..269D}
Duch{\^e}ne, G., \& Kraus, A. 2013, 51, 269

\bibitem[{Ekstr{\"o}m {et~al.}(2012)Ekstr{\"o}m, Georgy, Eggenberger, Meynet,
  Mowlavi, Wyttenbach, Granada, Decressin, Hirschi, Frischknecht, Charbonnel,
  \& Maeder}]{2012A&A...537A.146E}
Ekstr{\"o}m, S., Georgy, C., Eggenberger, P., {et~al.} 2012, A{\&}A, 537, A146

\bibitem[{Figuera~Jaimes {et~al.}(2016{\natexlab{a}})Figuera~Jaimes, Bramich,
  Skottfelt, Kains, J{\o}rgensen, Horne, Dominik, Alsubai, Bozza,
  Calchi~Novati, Ciceri, D'Ago, Galianni, Gu, Harps{\o}e, Haugb{\o}lle, Hinse,
  Hundertmark, Juncher, Korhonen, Mancini, Popovas, Rabus, Rahvar, Scarpetta,
  Schmidt, Snodgrass, Southworth, Starkey, Street, Surdej, Wang, \&
  Wertz}]{2016AnA...588A.128F}
Figuera~Jaimes, R., Bramich, D.~M., Skottfelt, J., {et~al.} 2016{\natexlab{a}},
  A{\&}A, 588, A128

\bibitem[{Figuera~Jaimes {et~al.}(2016{\natexlab{b}})Figuera~Jaimes, Bramich,
  Kains, Skottfelt, J{\o}rgensen, Horne, Dominik, Alsubai, Bozza, Burgdorf,
  Calchi~Novati, Ciceri, D'Ago, Evans, Galianni, Gu, Harps{\o}e, Haugb{\o}lle,
  Hinse, Hundertmark, Juncher, Kerins, Korhonen, Kuffmeier, Mancini, Peixinho,
  Popovas, Rabus, Rahvar, Scarpetta, Schmidt, Snodgrass, Southworth, Starkey,
  Street, Surdej, Tronsgaard, Unda-Sanzana, von Essen, Wang, Wertz, \&
  Consortium}]{2016AnA...592A.120F}
Figuera~Jaimes, R., Bramich, D.~M., Kains, N., {et~al.} 2016{\natexlab{b}},
  A{\&}A, 592, A120

\bibitem[{Foreman-Mackey {et~al.}(2013)Foreman-Mackey, Hogg, Lang, \&
  Goodman}]{2013PASP..125..306F}
Foreman-Mackey, D., Hogg, D.~W., Lang, D., \& Goodman, J. 2013, PASP, 125, 306

\bibitem[{Gallego-Cano {et~al.}(2018)Gallego-Cano, Sch{\"o}del, Dong,
  Nogueras-Lara, Gallego-Calvente, Amaro-Seoane, \&
  Baumgardt}]{2018A&A...609A..26G}
Gallego-Cano, E., Sch{\"o}del, R., Dong, H., {et~al.} 2018, A{\&}A, 609, A26

\bibitem[{Gezari \& Yusef-Zadeh(1991)}]{1991ASPC...14..214G}
Gezari, D., \& Yusef-Zadeh, F. 1991, ASP Conference Series, 14, 214

\bibitem[{Ghez {et~al.}(2005)Ghez, Hornstein, Lu, Bouchez, Le~Mignant, van Dam,
  Wizinowich, Matthews, Morris, Becklin, Campbell, Chin, Hartman, Johansson,
  Lafon, Stomski, \& Summers}]{2005ApJ...635.1087G}
Ghez, A.~M., Hornstein, S.~D., Lu, J.~R., {et~al.} 2005, ApJ, 635, 1087

\bibitem[{Ghez {et~al.}(2008)Ghez, Salim, Weinberg, Lu, Do, Dunn, Matthews,
  Morris, Yelda, Becklin, Kremenek, Milosavljevic, \&
  Naiman}]{2008ApJ...689.1044G}
Ghez, A.~M., Salim, S., Weinberg, N.~N., {et~al.} 2008, 689, 1044

\bibitem[{Gillessen {et~al.}(2009)Gillessen, Eisenhauer, Trippe, Alexander,
  Genzel, Martins, \& Ott}]{2009ApJ...692.1075G}
Gillessen, S., Eisenhauer, F., Trippe, S., {et~al.} 2009, 692, 1075

\bibitem[{Glass {et~al.}(1999)Glass, Matsumoto, Carter, \&
  Sekiguchi}]{1999MNRAS.304L..10G}
Glass, I.~S., Matsumoto, S., Carter, B.~S., \& Sekiguchi, K. 1999, MNRAS, 304,
  L10

\bibitem[{Goodman \& Tan(2004)}]{2004ApJ...608..108G}
Goodman, J., \& Tan, J.~C. 2004, 608, 108

\bibitem[{Hills(1988)}]{1988Natur.331..687H}
Hills, J.~G. 1988, 331, 687

\bibitem[{Hunter(2007)}]{Hunter07}
Hunter, J.~D. 2007, Computing In Science \& Engineering, 9, 90

\bibitem[{Ivezi{\'c} {et~al.}(2014)Ivezi{\'c}, Connelly, VanderPlas, \&
  Gray}]{2014sdmm.book.....I}
Ivezi{\'c}, {\v Z}., Connelly, A.~J., VanderPlas, J.~T., \& Gray, A. 2014,
  {Statistics, Data Mining, and Machine Learning in Astronomy}, A Practical
  Python Guide for the Analysis of Survey (Princeton University Press)

\bibitem[{Jones {et~al.}(2001--)Jones, Oliphant, Peterson, {et~al.}}]{Jones01}
Jones, E., Oliphant, T., Peterson, P., {et~al.} 2001--, {SciPy}: Open source
  scientific tools for {Python}, , .
\newblock \url{http://www.scipy.org/}

\bibitem[{Kourniotis {et~al.}(2014)Kourniotis, Bonanos, Soszynski, Poleski,
  Krikelis, Udalski, Szymanski, Kubiak, Pietrzy{\'{n}}ski, Wyrzykowski,
  Ulaczyk, Koz{\l}owski, \& Pietrukowicz}]{2014AnA...562A.125K}
Kourniotis, M., Bonanos, A.~Z., Soszynski, I., {et~al.} 2014, A{\&}A, 562, A125

\bibitem[{Lata {et~al.}(2016)Lata, Pandey, Panwar, Chen, Samal, \&
  Pandey}]{2016MNRAS.456.2505L}
Lata, S., Pandey, A.~K., Panwar, N., {et~al.} 2016, MNRAS, 456, 2505

\bibitem[{Lef{\`e}vre {et~al.}(2009)Lef{\`e}vre, Marchenko, Moffat, \&
  Acker}]{2009A&A...507.1141L}
Lef{\`e}vre, L., Marchenko, S.~V., Moffat, A. F.~J., \& Acker, A. 2009, A{\&}A,
  507, 1141

\bibitem[{Levin \& Beloborodov(2003)}]{2003ApJ...590L..33L}
Levin, Y., \& Beloborodov, A.~M. 2003, 590, L33

\bibitem[{Li {et~al.}(2017)Li, Ginsburg, Naoz, \& Loeb}]{2017ApJ...851..131L}
Li, G., Ginsburg, I., Naoz, S., \& Loeb, A. 2017, ApJ, 851, 131

\bibitem[{Lomb(1976)}]{1976Ap&SS..39..447L}
Lomb, N.~R. 1976, Astrophysics and Space Science, 39, 447

\bibitem[{Lu {et~al.}(2013)Lu, Do, Ghez, Morris, Yelda, \&
  Matthews}]{2013ApJ...764..155L}
Lu, J.~R., Do, T., Ghez, A.~M., {et~al.} 2013, ApJ, 764, 155

\bibitem[{Maness {et~al.}(2007)Maness, Martins, Trippe, Genzel, Graham, Sheehy,
  Salaris, Gillessen, Alexander, Paumard, Ott, Abuter, \&
  Eisenhauer}]{2007ApJ...669.1024M}
Maness, H., Martins, F., Trippe, S., {et~al.} 2007, 669, 1024

\bibitem[{Martinez {et~al.}(2011)Martinez, Minor, Bullock, Kaplinghat, Simon,
  \& Geha}]{2011ApJ...738...55M}
Martinez, G.~D., Minor, Q.~E., Bullock, J., {et~al.} 2011, ApJ, 738, 55

\bibitem[{Matsunaga {et~al.}(2009)Matsunaga, Kawadu, Nishiyama, Nagayama,
  Hatano, Tamura, Glass, \& Nagata}]{2009MNRAS.399.1709M}
Matsunaga, N., Kawadu, T., Nishiyama, S., {et~al.} 2009, MNRAS, 399, 1709

\bibitem[{Matsunaga {et~al.}(2006)Matsunaga, Fukushi, Nakada, Tanab{\'e},
  Feast, Menzies, Ita, Nishiyama, Baba, Naoi, Nakaya, Kawadu, Ishihara, \&
  Kato}]{2006MNRAS.370.1979M}
Matsunaga, N., Fukushi, H., Nakada, Y., {et~al.} 2006, MNRAS, 370, 1979

\bibitem[{Mattei(1997)}]{1997JAVSO..25...57M}
Mattei, J.~A. 1997, The Journal of the American Association of Variable Star
  Observers, 25, 57

\bibitem[{McCormac {et~al.}(2014)McCormac, Skillen, Pollacco, Faedi, Ramsay,
  Dhillon, Todd, \& Gonzalez}]{2014MNRAS.438.3383M}
McCormac, J., Skillen, I., Pollacco, D., {et~al.} 2014, MNRAS, 438, 3383

\bibitem[{Morris \& Maillard(2000)}]{2000ASPC..195..196M}
Morris, M.~R., \& Maillard, J.~P. 2000, in Imaging the Universe in Three
  Dimensions. Proceedings from ASP Conference Vol. 195. Edited by W. van
  Breugel and J. Bland-Hawthorn. ISBN: 1-58381-022-6 (2000), 196--

\bibitem[{Morris {et~al.}(2017)Morris, Zhao, \& Goss}]{2017ApJ...850L..23M}
Morris, M.~R., Zhao, J.-H., \& Goss, W.~M. 2017, The Astrophysical Journal
  Letters, 850, L23

\bibitem[{Muzic {et~al.}(2007)Muzic, Eckart, Sch{\"o}del, Meyer, \&
  Zensus}]{2007A&A...469..993M}
Muzic, K., Eckart, A., Sch{\"o}del, R., Meyer, L., \& Zensus, A. 2007, A{\&}A,
  469, 993

\bibitem[{Naoz {et~al.}(2018)Naoz, Ghez, Hees, Do, Witzel, \&
  Lu}]{2018ApJ...853L..24N}
Naoz, S., Ghez, A.~M., Hees, A., {et~al.} 2018, The Astrophysical Journal
  Letters, 853, L24

\bibitem[{Nascimbeni {et~al.}(2014)Nascimbeni, Bedin, Heggie, van~den Berg,
  Giersz, Piotto, Brogaard, Bellini, Milone, Rich, Pooley, Anderson, Ubeda,
  Ortolani, Malavolta, Cunial, \& Pietrinferni}]{2014MNRAS.442.2381N}
Nascimbeni, V., Bedin, L.~R., Heggie, D.~C., {et~al.} 2014, MNRAS, 442, 2381

\bibitem[{Nayakshin \& Cuadra(2005)}]{2005A&A...437..437N}
Nayakshin, S., \& Cuadra, J. 2005, A{\&}A, 437, 437

\bibitem[{Nogueras-Lara {et~al.}(2018)Nogueras-Lara, Gallego-Calvente, Dong,
  Gallego-Cano, Girard, Hilker, de~Zeeuw, Feldmeier-Krause, Nishiyama, Najarro,
  Neumayer, \& Sch{\"o}del}]{2018A&A...610A..83N}
Nogueras-Lara, F., Gallego-Calvente, A.~T., Dong, H., {et~al.} 2018, A{\&}A,
  610, A83

\bibitem[{Ott {et~al.}(1999)Ott, Eckart, \& Genzel}]{1999ApJ...523..248O}
Ott, T., Eckart, A., \& Genzel, R. 1999, 523, 248

\bibitem[{Paumard {et~al.}(2004)Paumard, Maillard, \&
  Morris}]{2004A&A...426...81P}
Paumard, T., Maillard, J.~P., \& Morris, M.~R. 2004, A{\&}A, 426, 81

\bibitem[{Paumard {et~al.}(2001)Paumard, Maillard, Morris, \&
  Rigaut}]{2001A&A...366..466P}
Paumard, T., Maillard, J.~P., Morris, M.~R., \& Rigaut, F. 2001, A{\&}A, 366,
  466

\bibitem[{Paumard {et~al.}(2006)Paumard, Genzel, Martins, Nayakshin,
  Beloborodov, Levin, Trippe, Eisenhauer, Ott, Gillessen, Abuter, Cuadra,
  Alexander, \& Sternberg}]{2006ApJ...643.1011P}
Paumard, T., Genzel, R., Martins, F., {et~al.} 2006, 643, 1011

\bibitem[{Peeples {et~al.}(2007)Peeples, Stanek, \&
  Depoy}]{2007AcA....57..173P}
Peeples, M.~S., Stanek, K.~Z., \& Depoy, D.~L. 2007, AcAau, 57, 173

\bibitem[{Pfuhl {et~al.}(2014)Pfuhl, Alexander, Gillessen, Martins, Genzel,
  Eisenhauer, Fritz, \& Ott}]{2014ApJ...782..101P}
Pfuhl, O., Alexander, T., Gillessen, S., {et~al.} 2014, 782, 101

\bibitem[{Pfuhl {et~al.}(2011)Pfuhl, Fritz, Zilka, Maness, Eisenhauer, Genzel,
  Gillessen, Ott, Dodds-Eden, \& Sternberg}]{2011ApJ...741..108P}
Pfuhl, O., Fritz, T.~K., Zilka, M., {et~al.} 2011, 741, 108

\bibitem[{Press \& Rybicki(1989)}]{1989ApJ...338..277P}
Press, W.~H., \& Rybicki, G.~B. 1989, ApJ, 338, 277

\bibitem[{Rafelski {et~al.}(2007)Rafelski, Ghez, Hornstein, Lu, \&
  Morris}]{2007ApJ...659.1241R}
Rafelski, M., Ghez, A.~M., Hornstein, S.~D., Lu, J.~R., \& Morris, M.~R. 2007,
  ApJ, 659, 1241

\bibitem[{Rice {et~al.}(2015)Rice, Reipurth, Wolk, Vaz, \&
  Cross}]{2015AJ....150..132R}
Rice, T.~S., Reipurth, B., Wolk, S.~J., Vaz, L.~P., \& Cross, N. J.~G. 2015,
  The Astronomical Journal, 150, 132

\bibitem[{Rice {et~al.}(2012)Rice, Wolk, \& Aspin}]{2012ApJ...755...65R}
Rice, T.~S., Wolk, S.~J., \& Aspin, C. 2012, 755, 65

\bibitem[{Riebel {et~al.}(2010)Riebel, Meixner, Fraser, Srinivasan, Cook, \&
  Vijh}]{2010ApJ...723.1195R}
Riebel, D., Meixner, M., Fraser, O., {et~al.} 2010, ApJ, 723, 1195

\bibitem[{Scargle(1982)}]{1982ApJ...263..835S}
Scargle, J.~D. 1982, ApJ, 263, 835

\bibitem[{Sch{\"o}del {et~al.}(2010)Sch{\"o}del, Najarro, Muzic, \&
  Eckart}]{2010A&A...511A..18S}
Sch{\"o}del, R., Najarro, F., Muzic, K., \& Eckart, A. 2010, A{\&}A, 511, A18

\bibitem[{Scoville {et~al.}(2003)Scoville, Stolovy, Rieke, Christopher, \&
  Yusef-Zadeh}]{2003ApJ...594..294S}
Scoville, N.~Z., Stolovy, S.~R., Rieke, M., Christopher, M., \& Yusef-Zadeh, F.
  2003, ApJ, 594, 294

\bibitem[{Service {et~al.}(2016)Service, Lu, Campbell, Sitarski, Ghez, \&
  Anderson}]{2016PASP..128i5004S}
Service, M., Lu, J.~R., Campbell, R., {et~al.} 2016, PASP, 128, 095004

\bibitem[{Sokolovsky {et~al.}(2017)Sokolovsky, Gavras, Karampelas, Antipin,
  Bellas-Velidis, Benni, Bonanos, Burdanov, Derlopa, Hatzidimitriou,
  Khokhryakova, Kolesnikova, Korotkiy, Lapukhin, Moretti, Popov, Pouliasis,
  Samus, Spetsieri, Veselkov, Volkov, Yang, \& Zubareva}]{2017MNRAS.464..274S}
Sokolovsky, K.~V., Gavras, P., Karampelas, A., {et~al.} 2017, MNRAS, 464, 274

\bibitem[{Stephan {et~al.}(2016)Stephan, Naoz, Ghez, Witzel, Sitarski, Do, \&
  Kocsis}]{2016MNRAS.460.3494S}
Stephan, A.~P., Naoz, S., Ghez, A.~M., {et~al.} 2016, MNRAS, 460, 3494

\bibitem[{Stolte {et~al.}(2008)Stolte, Ghez, Morris, Lu, Brandner, \&
  Matthews}]{2008ApJ...675.1278S}
Stolte, A., Ghez, A.~M., Morris, M.~R., {et~al.} 2008, ApJ, 675, 1278

\bibitem[{St{\o}stad {et~al.}(2015)St{\o}stad, Do, Murray, Lu, Yelda, \&
  Ghez}]{2015ApJ...808..106S}
St{\o}stad, M., Do, T., Murray, N., {et~al.} 2015, ApJ, 808, 106

\bibitem[{Tanner {et~al.}(2002)Tanner, Ghez, Morris, Becklin, Cotera, Ressler,
  Werner, \& Wizinowich}]{2002ApJ...575..860T}
Tanner, A., Ghez, A.~M., Morris, M.~R., {et~al.} 2002, ApJ, 575, 860

\bibitem[{van~der Walt {et~al.}(2011)van~der Walt, Colbert, \&
  Varoquaux}]{vanderWalt:2011dp}
van~der Walt, S., Colbert, S.~C., \& Varoquaux, G. 2011, Computing in Science
  {\&} Engineering, 13, 22

\bibitem[{VanderPlas(2017)}]{2017arXiv170309824V}
VanderPlas, J.~T. 2017, arXiv, arXiv:1703.09824

\bibitem[{Viehmann {et~al.}(2005)Viehmann, Eckart, Sch{\"o}del, Moultaka,
  Straubmeier, \& Pott}]{2005A&A...433..117V}
Viehmann, T., Eckart, A., Sch{\"o}del, R., {et~al.} 2005, A{\&}A, 433, 117

\bibitem[{Wolk {et~al.}(2013)Wolk, Rice, \& Aspin}]{2013ApJ...773..145W}
Wolk, S.~J., Rice, T.~S., \& Aspin, C. 2013, 773, 145

\bibitem[{Yelda {et~al.}(2014)Yelda, Ghez, Lu, Do, Meyer, Morris, \&
  Matthews}]{2014ApJ...783..131Y}
Yelda, S., Ghez, A.~M., Lu, J.~R., {et~al.} 2014, 783, 131

\bibitem[{Yelda {et~al.}(2010)Yelda, Lu, Ghez, Clarkson, Anderson, Do, \&
  Matthews}]{2010ApJ...725..331Y}
Yelda, S., Lu, J.~R., Ghez, A.~M., {et~al.} 2010, 725, 331

\bibitem[{Yu \& Tremaine(2003)}]{2003ApJ...599.1129Y}
Yu, Q., \& Tremaine, S. 2003, ApJ, 599, 1129

\bibitem[{Yusef-Zadeh {et~al.}(1998)Yusef-Zadeh, Roberts, \&
  Biretta}]{1998ApJ...499L.159Y}
Yusef-Zadeh, F., Roberts, D.~A., \& Biretta, J. 1998, ApJ, 499, L159

\bibitem[{Yusef-Zadeh {et~al.}(2017)Yusef-Zadeh, Wardle, Cotton, Sch{\"o}del,
  Royster, Roberts, \& Kunneriath}]{2017ApJ...837...93Y}
Yusef-Zadeh, F., Wardle, M., Cotton, W., {et~al.} 2017, ApJ, 837, 93

\bibitem[{Zhao \& Goss(1998)}]{1998ApJ...499L.163Z}
Zhao, J.-H., \& Goss, W.~M. 1998, ApJ, 499, L163

\end{thebibliography}

\appendix

\section{Photometric Calibration Details} 
\label{sec:photometric_calibration_details}

\begin{deluxetable}{lrcc}
    \tablewidth{0pt}
    \tablecolumns{4}
    \tablecaption{Initial Calibration Stars Bandpass Correction\label{tab:Initial_Cals}}
    \tablehead{
        \colhead{Star Name}             &
        \colhead{$K_{\text{Blum+96}}$}   &
		\colhead{$K_{\text{Blum+96}} - K'_{\text{NIRC2}}$}   &
        \colhead{$K'_{NIRC2}$}
    }
    \startdata
    IRS 16C     & $9.86 \pm 0.05$   & $-0.15 \pm 0.01$  & $10.01 \pm 0.06$  \\
    IRS 33E     & $10.02 \pm 0.05$  & $-0.16 \pm 0.01$  & $10.18 \pm 0.06$  \\
    S2-17       & $10.03 \pm 0.07$  & $-0.14 \pm 0.01$  & $10.17 \pm 0.08$  \\
    S2-16       & $11.90 \pm 0.22$  & $-0.26 \pm 0.01$  & $12.16 \pm 0.23$  \\
    \enddata
    \tablerefs{\citet{1996ApJ...470..864B}}
\end{deluxetable}

\subsection{Reference Flux Bandpass Correction} 
\label{sub:reference_flux_bandpass_correction}
Synthetic photometry was used to convert the \citet{1996ApJ...470..864B} photometry (hereafter Blum+96) of the initial calibration stars (listed in Table~\ref{tab:Initial_Cals}) into Keck NIRC2 $K'$-bandpass photometry. The calibration stars were modeled using Geneva stellar evolution models with rotation and at solar metallicity \citep{2012A&A...537A.146E} combined with ATLAS model atmospheres \citep{2004astro.ph..5087C}. Since the stars belong to the young star population, an age of 3.9 Myr was adopted \citep{2013ApJ...764..155L}. By convolving the model atmospheres with the Blum+96 and Keck NIRC2 filter functions, the photometric offset between the filters could be calculated. However, it was first necessary to calculate the extinction for each star, since the bandpass correction depends on the extinction.

The extinction of each star was calculated from the $H - K$. IRS 16C, IRS 33E, and S2-17 each have Blum+96 $H - K$ measurements that we used. S2-16 did not have a Blum+96 $H - K$, so we instead used VLT NACO $H - Ks$ measurements \citep{2010A&A...511A..18S}. The intrinsic colors of the calibrators were calculated from the model isochrones set at a distance of 10 pc and with no extinction. The intrinsic colors were constrained by the knowledge that IRS 16C, IRC 33E, and S2-16 are spectroscopically identified WR stars, and S2-17 is known to be an early-type ($M > 2 M_{\odot}$), non-WR star \citep{2013ApJ...764..154D}. We then used the \citet{2018A&A...610A..83N} extinction law to convert the color excess into a total $Ks$-band extinction (NIRC2 system), obtaining values of 2.47 mag, 2.55 mag, 4.12 mag, and 2.29 mag for IRS 16C, IRS 33E, S2-16 and S2-17, respectively. The error on the extinction values are $\pm 0.08$ mags or better, as a result of the uncertainties in the intrinsic and observed colors.

We then recalculated the synthetic photometry of the model isochrones, this time applying the extinction. From this synthetic photometry, the $\text{Blum+96 } K - \text{ NIRC2 } K'$ bandpass corrections are found to be $-0.15$ mag, $-0.16$ mag, $-0.26$ mag, and $-0.14$ mag for IRS 16C, IRS 33E, S2-16 and S2-17, respectively. The extinction uncertainty only affects the final bandpass corrections at the 0.01 mag level or lower. The S2-16 bandpass correction appears to be an outlier relative to those of the other initial calibrators, but this is due to its significantly higher extinction. The bandpass corrections and the final reference photometry we use for our initial calibrator stars are listed in Table~\ref{tab:Initial_Cals}.

\subsection{Iterative Calibrator Selection} 
\label{sub:iterative_calibrator_selection}
During each photometric calibration iteration, we used reference flux measurements and corresponding uncertainties for each of our calibrator stars. We then used the weighted mean of the calibration stars' differences from their reference values to derive a correction to the zeropoint across our observations. This zeropoint correction was used to adjust the magnitudes of every star identified in each observation. We calculated an error in the zeropoint correction for each observation and added in quadrature to the instrumental flux uncertainty measurement for each star. This gave the total measurement uncertainty in flux for each star.

Following the initial calibration, we identified a new set of secondary calibration stars that increased the precision for the relative photometry across the observations. The selection of these secondary stars was based on a set of criteria detailed below to select bright, photometrically stable stars distributed across our field of view. We updated the chosen calibration stars' reference magnitudes to the weighted mean magnitude from the previous calibration step. To select stars that are photometrically stable, we chose stars with low $\chi^2_{\text{red}}$ values (indicating low variability; further described in Section~\ref{sec:stellar_variability}), and a low mean magnitude uncertainty across all epochs (to reduce the influence of high photometric uncertainty lowering the $\chi^2_{\text{red}}$ value). These metrics used to select calibrator stars in each iteration were computed before applying local photometric correction (detailed in Section~\ref{sub:local_photometric_correction}). We included a magnitude cutoff of $\bar{m}_{K'} \leq 15.5$ to limit our calibration stars to be brighter sources. We further imposed a requirement that our calibration stars be identified in all observation epochs. Additionally, we checked a series of photometric and astrometric confusion criteria, selected to avoid choosing stars that could be confused with another nearby star during the calibration star identification process. For stars brighter than $m_{K'} = 12$, we checked if potential calibration stars that had no other neighboring stars within $0.2''$ and 1 magnitude in any observation epoch. The astrometric criterion was relaxed to $0.1''$ for stars dimmer than $m_{K'} = 12$, due to these stars having fainter PSF haloes. Finally, we imposed that at least two calibration stars and no more than three were used in each quadrant of our field of view, centered on the location of Sgr A*. We further required calibration stars to be at least $\sim 0.25''$ from each other. Imposing these final set of criteria limited the photometric calibration from biasing only small areas of the field for photometric stability with a higher density of calibration stars or resulting in other regions of the field with fewer calibrators to have more imprecise calibration. All these criteria selected bright and photometrically stable stars for each calibration iteration that were identified in all observation epochs, isolated in position and magnitude from nearby stars, and distributed across our field of view.

\begin{deluxetable}{ll}
    \tablewidth{0pt}
    \tablecolumns{2}
    \tablecaption{\small Criteria for selecting final set of photometric calibration stars
    \label{tab:Calibrator Criteria}
    }
    \tablehead{
        \colhead{Criterion}                     &
        \colhead{Calibrator Criterion Cutoff}
    }
    \startdata
    $\chi^2_{\text{red}}$       & $\leq 1.9$    \\
    Mean Mag. ($m_{K'}$)        & $\leq 15.5$   \\
    Mean Mag. Uncertainty       & $\leq 0.0295$ \\
    Number of Epochs            & $= 45$ (all)  \\
    \sidehead{\emph{Confusion criteria}}
    Nearest Star ($\bar{m}_{K'} \leq 12$)  & $> 0.2''$ \\
    Nearest Star ($\bar{m}_{K'} > 12$)     & $> 0.1''$ \\
    Nearest Star $\Delta m_{K'}$     & $> 1.0$                      \\
    \sidehead{\emph{Isolation criteria}}
    Nearest Calibrator              & $\gtrsim 0.25''$ \\
    Calibrators per FoV quadrant    & $2 \leq n \leq 3$             \\
    \enddata
    \tablecomments{These criteria were used to select photometric calibration stars from the previous calibration iteration. Therefore, these criteria are not necessarily reflected in the statistics for the photometric calibration stars in the final calibration iteration listed in Table~\ref{tab:Calibration stars}.\\
    \emph{Confusion criteria} were selected to avoid choosing calibration stars that could be confused with another star during our calibration star identification process. Calibration stars were chosen to pass both astrometric and photometric confusion criteria.\\
    \emph{Isolation criteria} were selected to avoid a high density of photometric calibrators in small regions of the field, in order to not bias only small areas of the field for photometric stability.}
\end{deluxetable}

\begin{deluxetable}{lcccc}
    \tablewidth{0pt}
    \tablecolumns{5}
    \tablecaption{Final Calibration stars\label{tab:Calibration stars}}
    \tablehead{
        \colhead{Star Name}             &
        \colhead{Mean Mag.}				&
        \colhead{Error on Mean Mag.}    &
		\colhead{$\chi^2_{\text{red}}$}	&
        \colhead{Mean of Uncertainties} \\
        \colhead{}                 		&
        \colhead{($K'$)}      		    &
        \colhead{($K'$)}      		    &
		\colhead{}						&
        \colhead{($K'$)}
    }
    \startdata
    IRS 16NW    & 10.155    & 0.019     & 1.411     & 0.029     \\
    S3-22       & 11.028    & 0.018     & 0.592     & 0.029     \\
    S1-17       & 12.171    & 0.018     & 0.517     & 0.029     \\
    S1-34       & 12.907    & 0.019     & 0.565     & 0.029     \\
    S4-3        & 12.907    & 0.019     & 1.108     & 0.029     \\
    S1-1        & 13.021    & 0.019     & 1.123     & 0.029     \\
    S1-21       & 13.214    & 0.019     & 1.644     & 0.029     \\
    S3-370      & 13.532    & 0.018     & 0.791     & 0.029     \\
    S0-14       & 13.572    & 0.018     & 1.023     & 0.029     \\
    S3-36       & 14.538    & 0.019     & 0.478     & 0.029     \\
    S2-63       & 15.341    & 0.019     & 1.880     & 0.029 
    \enddata
    \tablecomments{Metrics here are computed before application of the local photometric correction.}
    
\end{deluxetable}

\begin{figure*}[ht]
    \epsscale{0.4}
    \plotone{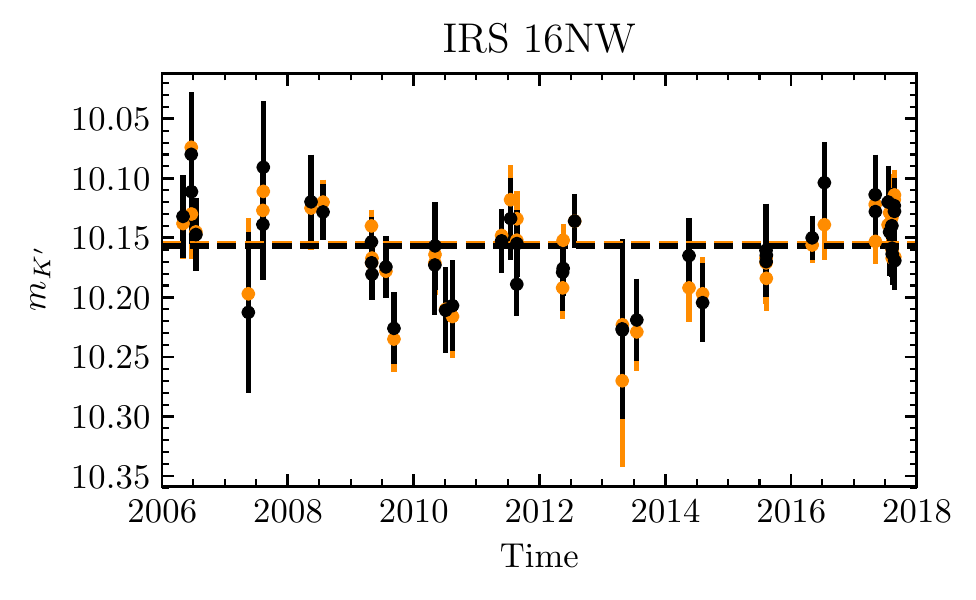}
    \plotone{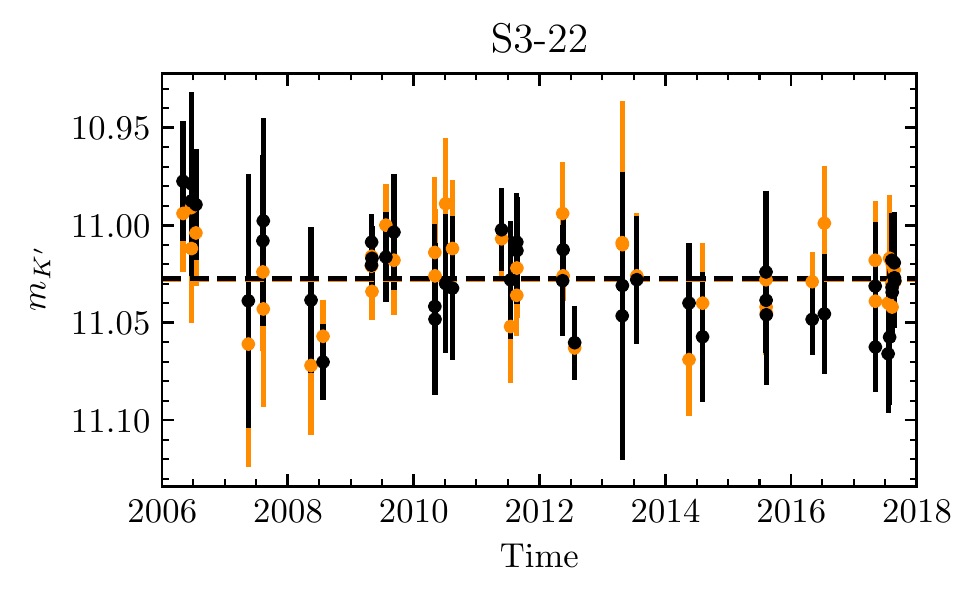}
    \plotone{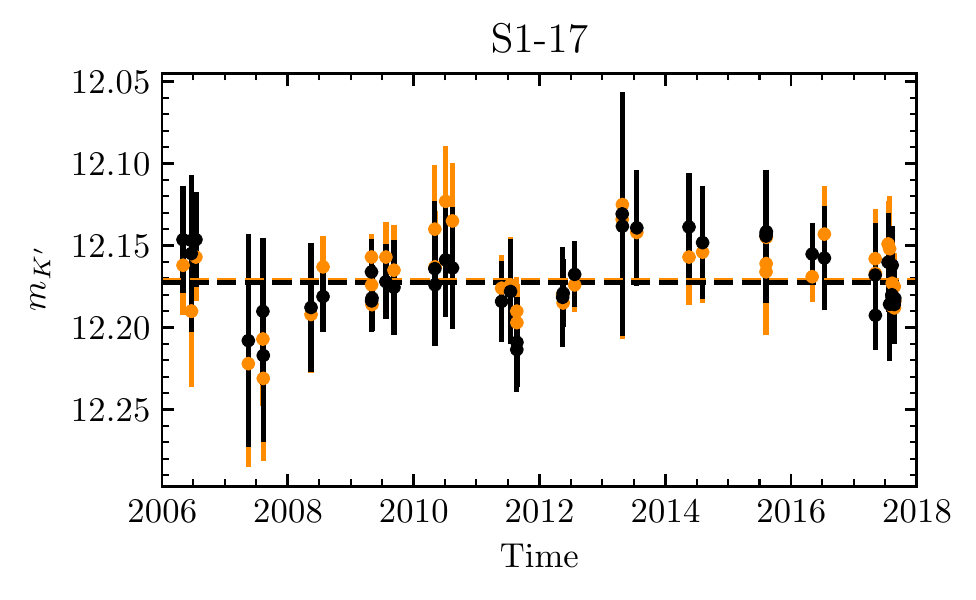}
    \plotone{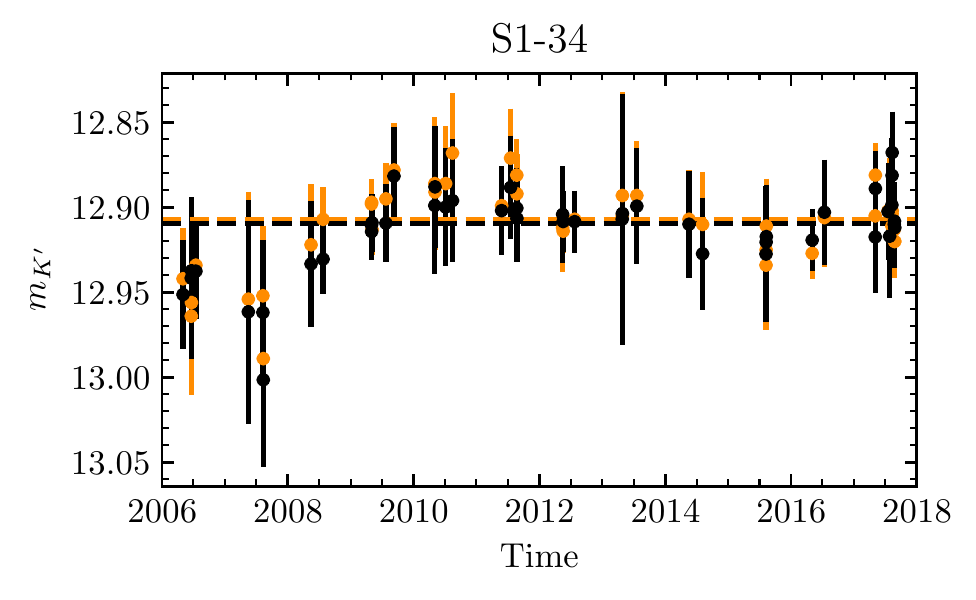}
    \plotone{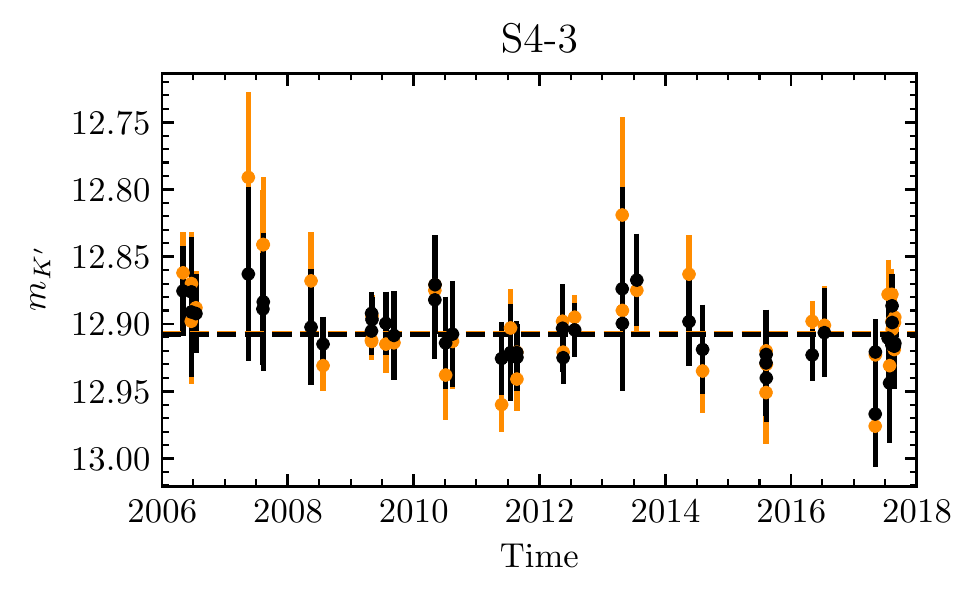}
    \plotone{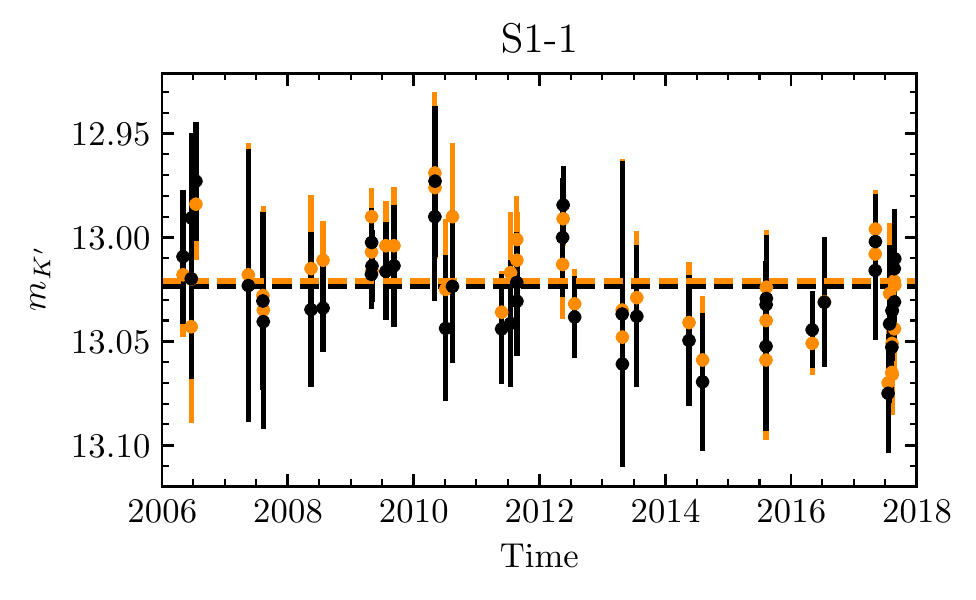}
    \plotone{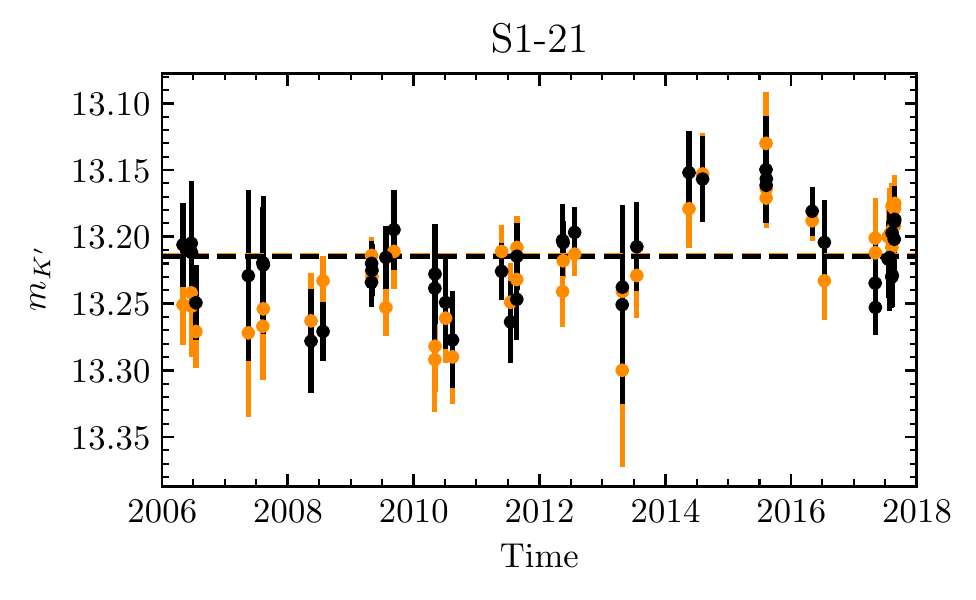}
    \plotone{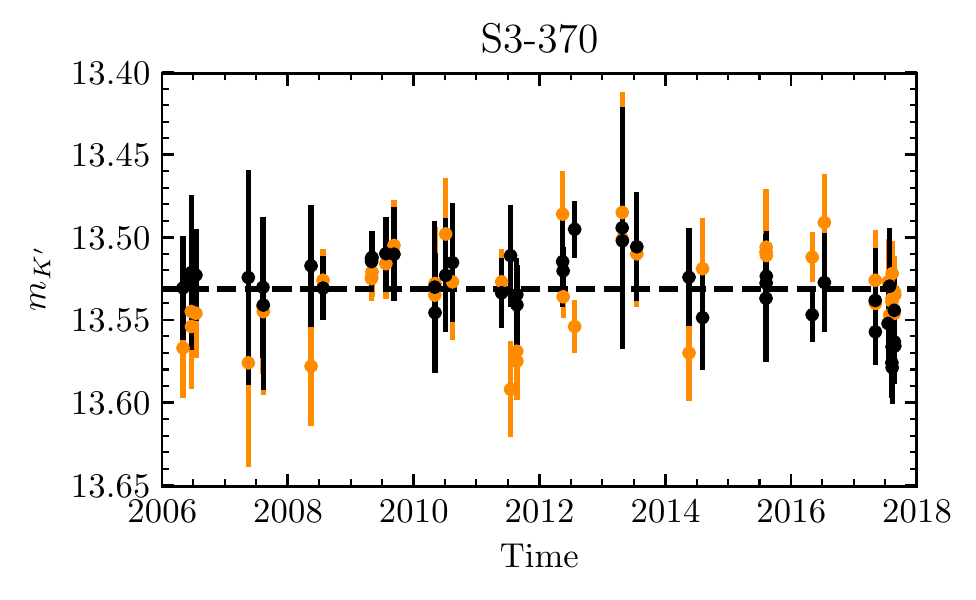}
    \plotone{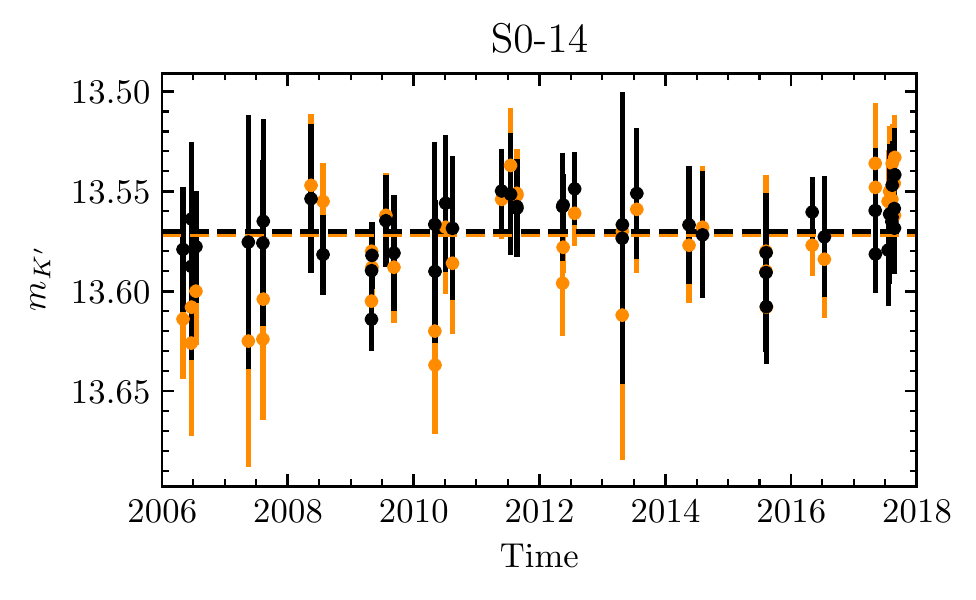}
    \plotone{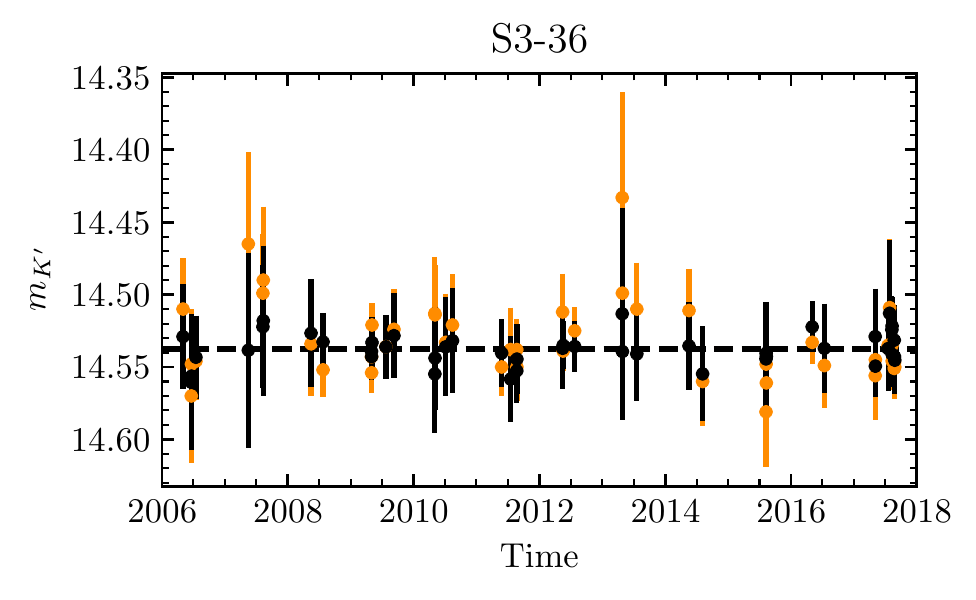}
    \plotone{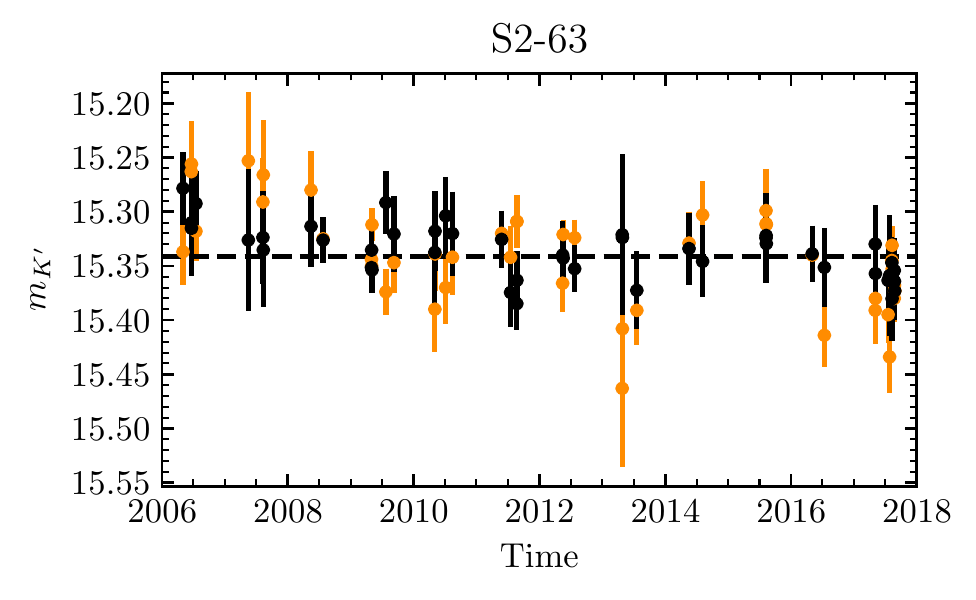}
    \caption{Light curves of the final calibration stars. Flux measurements before application of the local photometric correction are indicated in orange and flux measurements after application of the local photometric correction are indicated in black. The horizontal dashed line indicates the weighted mean magnitude.
    \label{fig:cal_light_curves}
    }
\end{figure*}

The above process to select new stable secondary calibration stars was repeated 3 times until it converged onto the same set of calibrators. Before each iteration, we refined our calibration star selection criteria ($\chi^2_{\text{red}}$ and mean magnitude uncertainty) to better isolate stable stars. We used the mean magnitude and uncertainty on the mean magnitude for each of the calibrator stars from the previous iteration as their respective reference fluxes and uncertainties. Our iterative process converged to our final calibration star selection criteria detailed in Table~\ref{tab:Calibrator Criteria}, and our final set of calibration stars are listed in Table~\ref{tab:Calibration stars} and displayed on our field of view in Figure~\ref{fig:FieldStars_Cals}. Light curves of all final calibration stars, after the local photometric correction is applied (correction detailed in Section~\ref{sub:local_photometric_correction}), are shown in Figure~\ref{fig:cal_light_curves}. By identifying stable secondary calibration stars, the iterative process effectively reduced the contribution to the photometric uncertainty originating from uncertainty in the zeropoint correction and achieve greater precision for relative photometry.

\begin{figure*}[ht]
    \epsscale{1.0}
    \plottwo{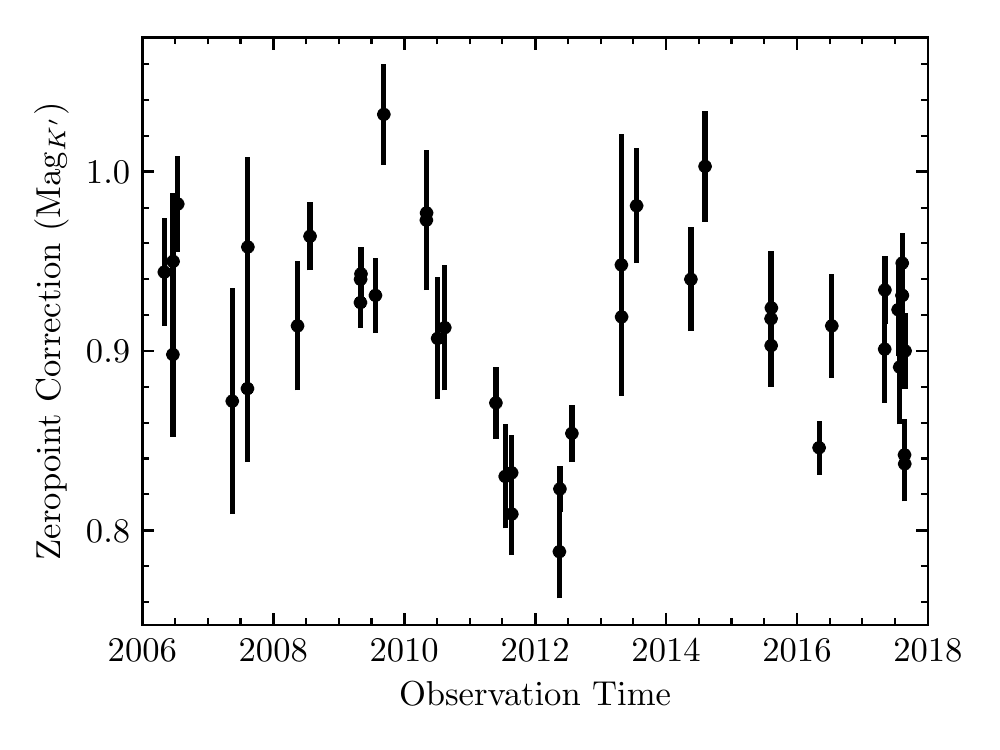}{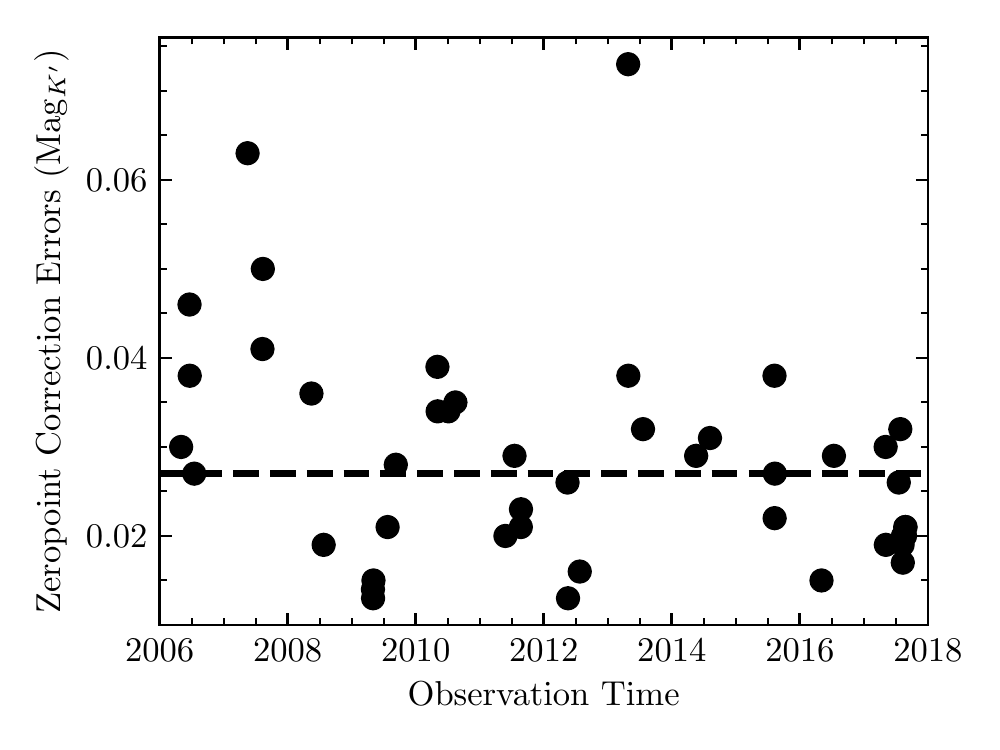}
    \caption{\emph{Left:} $K'$ magnitude zeropoint corrections from the photometric calibration used in this work. The zeropoint correction is calculated as the mean of the difference between the measured photometric flux of the calibration stars and their recorded value.\\
    \emph{Right:} The errors in the $K'$ magnitude zeropoint corrections used in this work, calculated as the variance of the zeropoint magnitude adjustment in each observation. This zeropoint correction error in each observation dominates the photometric uncertainty in our measurements, and the median zeropoint correction error (dashed line) across our observations is $\sigma_{m_{K'}} \sim 0.025$.
    \label{fig:Zeropoint Errors}
    }
\end{figure*}

\subsection{Local Photometric Correction} 
\label{sub:local_photometric_correction}

\begin{figure*}[ht]
	\centering
    \epsscale{1.2}
    \plotone{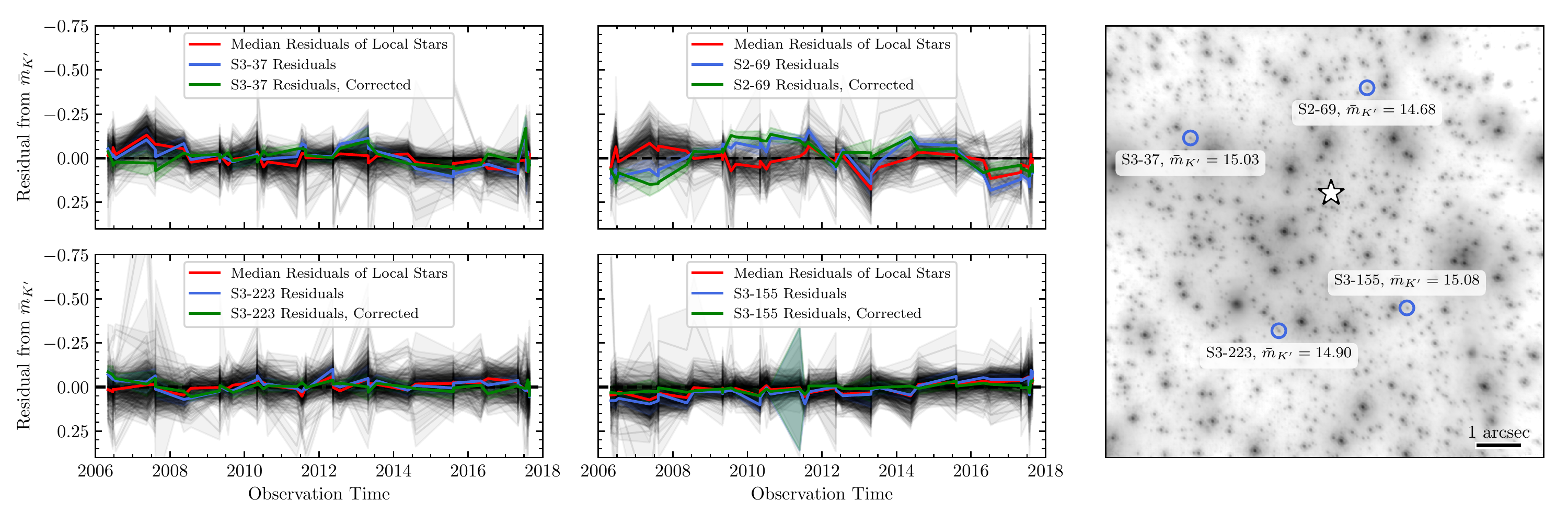}
    \caption{Four example target stars of similar brightness from different areas of our field selected to highlight our local photometry correction. Each curve displays the residual of a star's flux in an observation from its respective mean magnitude across all observations while shaded regions indicate uncertainty in flux. Blue curves indicate the residuals for the four example target stars, while the black curves in each plot indicate the residuals for all the local stars for the target star. Small trends in measured flux correlate across target stars and their respective local stars, suggesting a local photometric bias. The red curve indicates the median residual of the local stars, which is subtracted from the flux measurements of the target star to correct for the photometric bias. The green curves indicate the residuals for the four example target stars corrected for the photometric bias, and include the additional additive uncertainty during the local correction step.
    \label{fig:Neighborhood_Resid_Example}
    }
\end{figure*}

We performed an additional correction to our photometry on local scales of the field beyond the zeropoint photometric calibration. The need for this correction became evident when we observed similar changes in flux measurements for stars of similar brightness and position on the field in an observation epoch. This effect and our correction, described below, is illustrated for four example stars from different locations on our field in Figure~\ref{fig:Neighborhood_Resid_Example}.

A variable PSF across the field can cause our flux measurement of stars from PSF-fitting to be under- or over-estimated. Since the PSF variation is spatially correlated, this bias in the flux measurement is expected to be similar for nearby stars. We attempted to correct for this photometric variation in our dataset.

To determine the photometric bias, local stars were first determined for each star. These local stars were selected to be low variability stars in close proximity on the field and at a similar brightness to each target star. A star was identified as a local star if it was located within $2''$ and within 1.0 mag of the target star in any observation. From these, stars that were detected in fewer than 23 observations and had $\chi^2_{\text{red}} > 20.0$ were removed as local stars to determine the photometric bias. This was to reduce the influence on the measurement of the photometric bias by variable stars and those whose mean magnitude was not very well constrained due to detections in too few observations. If the total number of local stars determined under these constraints was fewer than 8, the astrometric search radius was increased in steps of $0.25''$ and the photometric search radius was increased in steps of 0.25 mag until the number of local stars reached the minimum of 8. This ensured that the measurement of photometric bias was not dominated by the variations of too few stars.

With the local stars determined, the photometric bias was measured for each star. In each observation for the target star, the residual in magnitudes for local star $i$ from its mean magnitude was measured, 
\begin{eqnarray}
    R_i \equiv m_i - \bar{m}_i.
\end{eqnarray}
The median value over all local stars of the residual in each observation epoch, $\text{med}(R_i)$, was subtracted from the target star's flux measurement in that observation. This corrected for the photometric bias measured from the local stars for every star in our sample.

With this correction, we also included an additive error to account for the uncertainty in flux introduced by this process. In each observation the error from the local correction was calculated with
\begin{eqnarray}
    \text{RMS}_R = \sqrt{\frac{\sum_{i}^{N}{R_i - \text{med}(R_i)}}{N}}\\
    \text{Local Correction Uncertainty} = \frac{\text{RMS}_R}{\sqrt{N}},
\end{eqnarray}
where $N$ represents the total number of local stars in each observation used to correct for the photometric bias. The local correction uncertainty was then added in quadrature into the flux uncertainty determined during the zeropoint correction.



\section{Variability Study Details} 
\label{sec:variability_study_details}

Table~\ref{tab:Sample_Catalog} summarizes the sample of stars studied in this work. In addition to the variability metric used in this work, $\chi^2_{\text{red}}$, we calculated additional variability metrics to aide in comparison of our stellar sample to other stellar samples.

The root mean square (RMS) calculated is that of the observed magnitude differences from the mean magnitude:
\begin{eqnarray}
    \text{RMS} = \sqrt{\frac{1}{N} * \sum{(m_i - \bar{m})^2}}
\end{eqnarray}

The interquartile range (IQR) is the difference between the median of the half brightest and half dimmest observations \citep{2017MNRAS.464..274S}. This method is more robust against outliers.

The von Neumann ratio, $\eta$, is the ratio of mean square of differences in successive observations to the variance of all observations \citep{2017MNRAS.464..274S}. Higher values of $1/\eta$ indicate higher variability, defined as:
\begin{eqnarray}
    \frac{1}{\eta} = \frac{\sigma^2}{\delta^2} = \frac{\sum^{N}_{i=1}{(m_i - \bar{m})^2/(N-1)}}{\sum^{N-1}_{i=1}{(m_{i+1} - m_i)^2/(N-1)}}
\end{eqnarray}
The $1/\eta$ method picks out stars that are less smoothly variable, i.e. with greater differences in successive observations.

\section{$\chi^2_{\text{red}} \geq 10$ Variables} 
\label{sec:_chi_2__text_red_geq_10_variables}

\begin{figure}[H]
    \epsscale{1.05}
    \plottwo{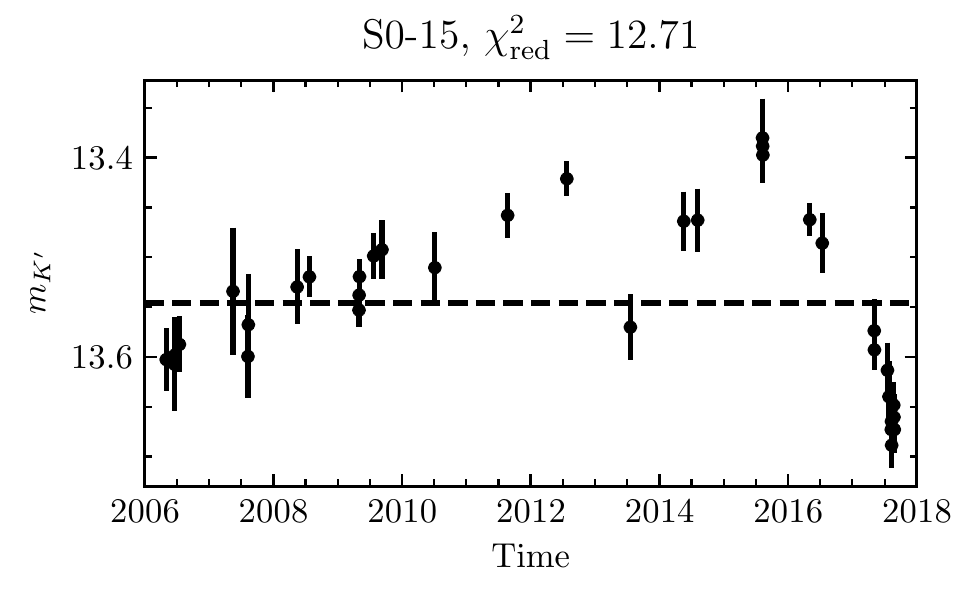}{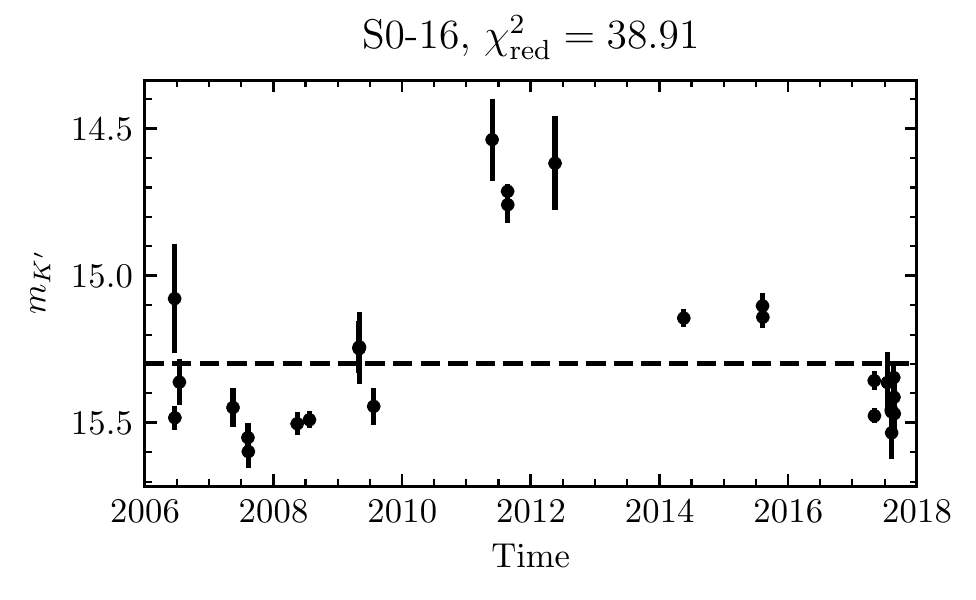}
\end{figure}

\begin{figure}[H]
    \epsscale{1.05}
    \plottwo{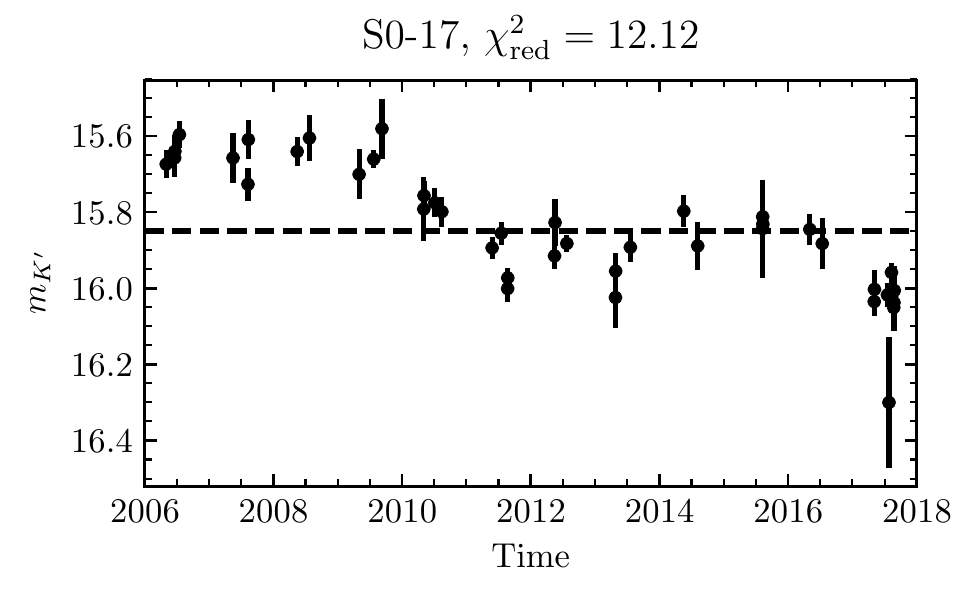}{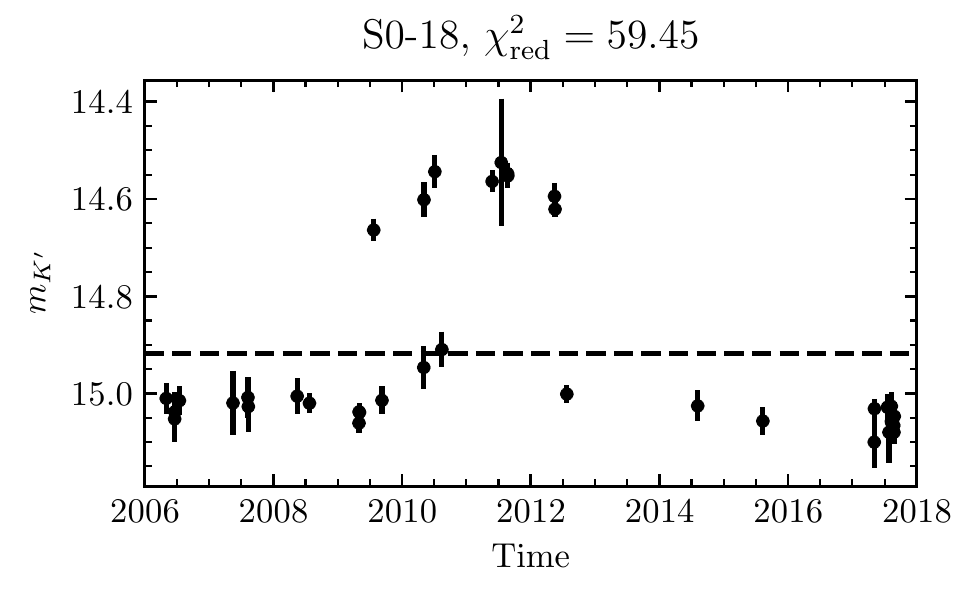}
\end{figure}

\begin{figure}[H]
    \epsscale{1.05}
    \plottwo{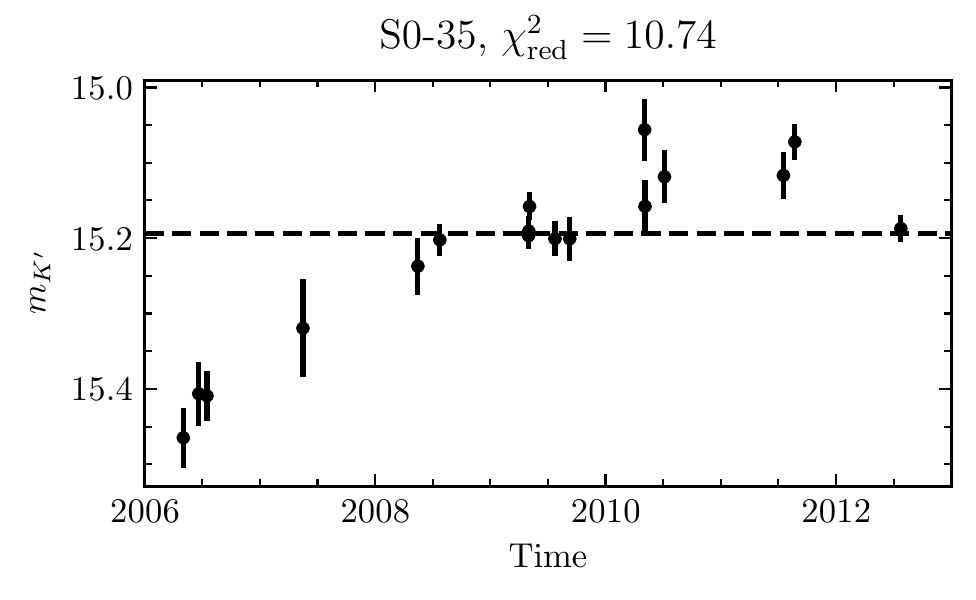}{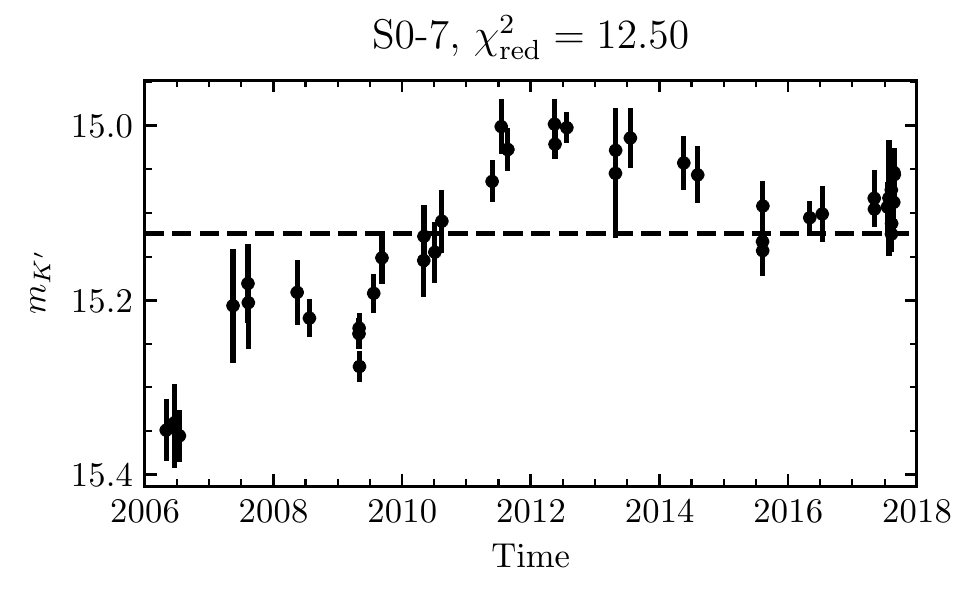}
\end{figure}

\begin{figure}[H]
    \epsscale{1.05}
    \plottwo{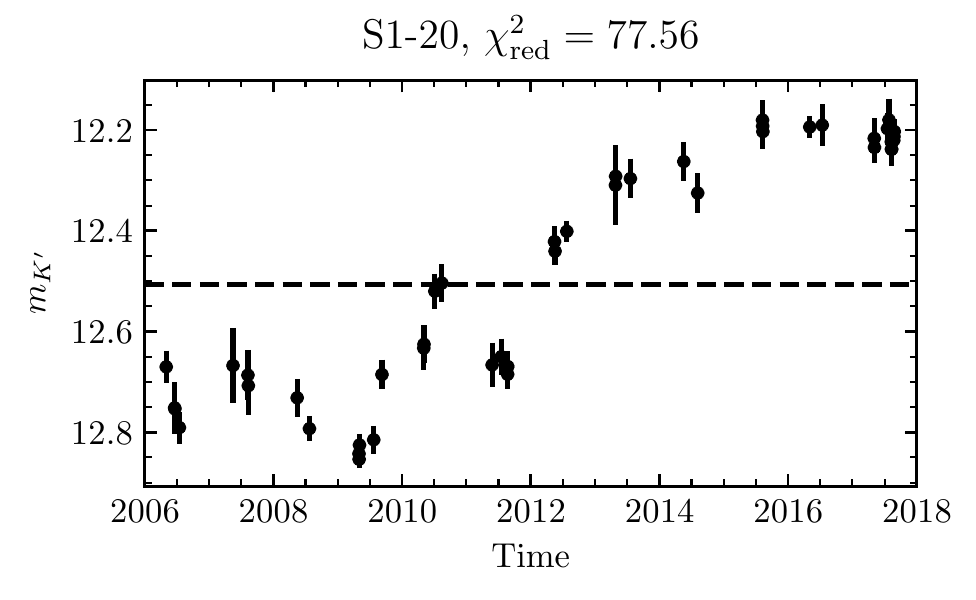}{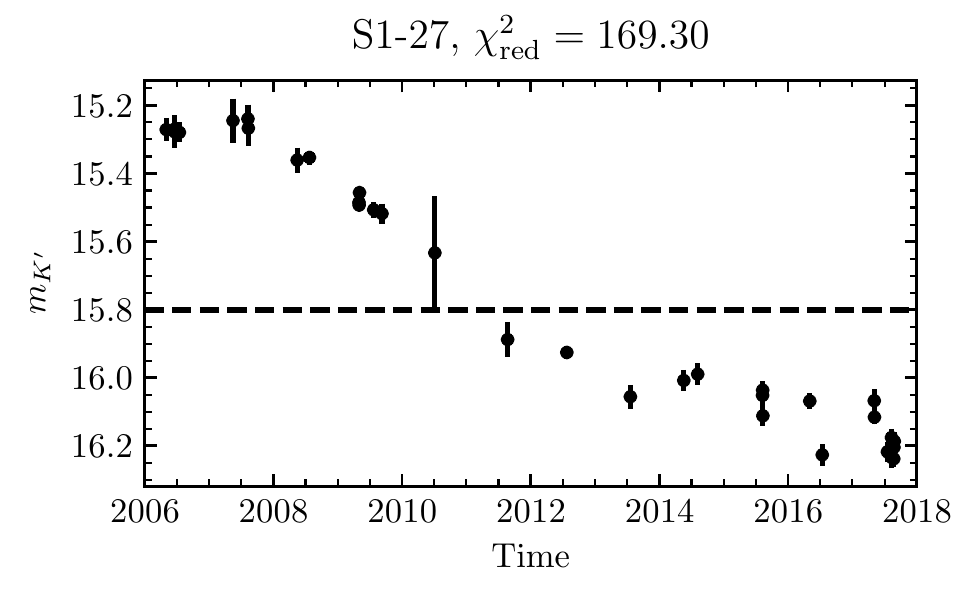}
\end{figure}

\begin{figure}[H]
    \epsscale{1.05}
    \plottwo{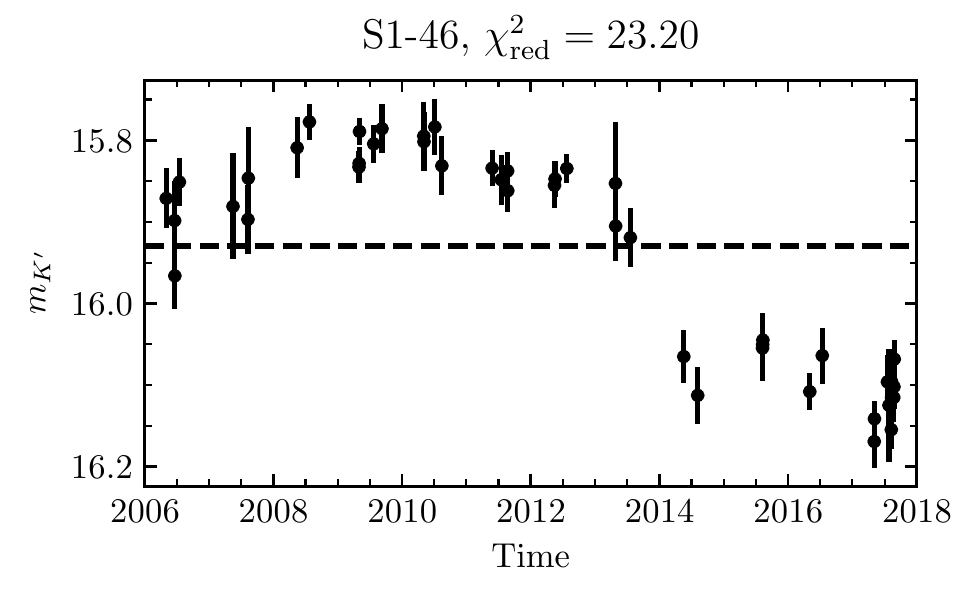}{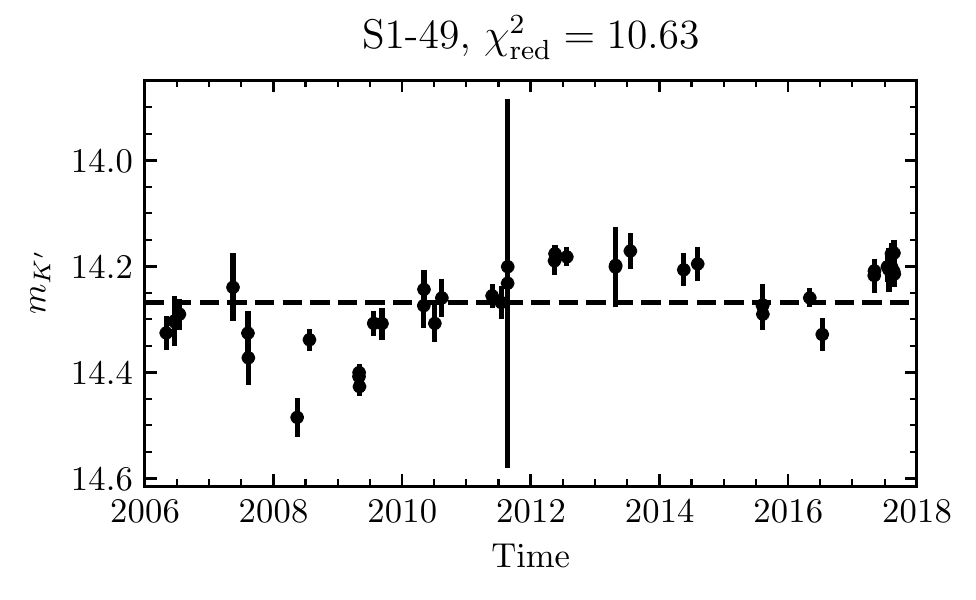}
\end{figure}

\begin{figure}[H]
    \epsscale{1.05}
    \plottwo{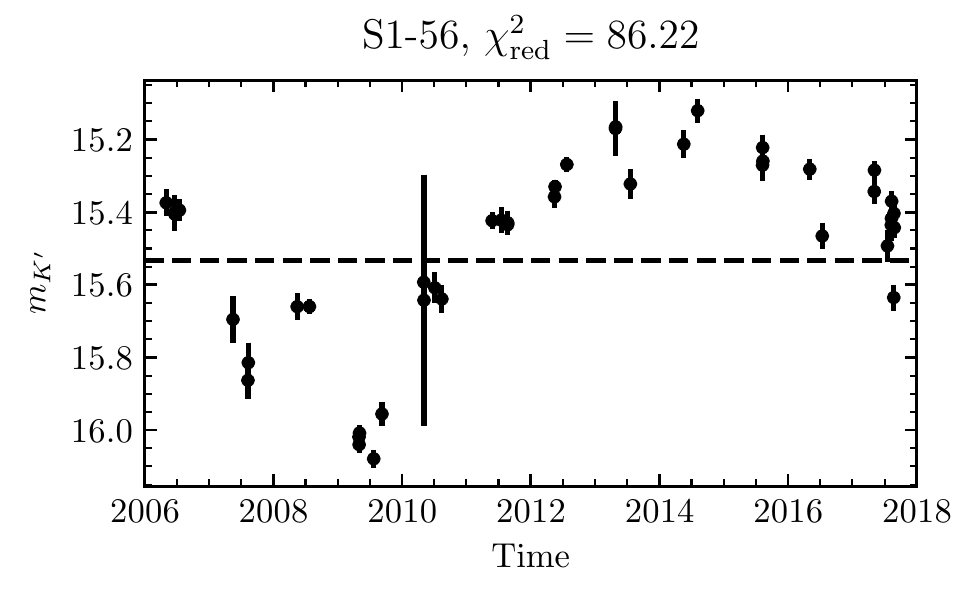}{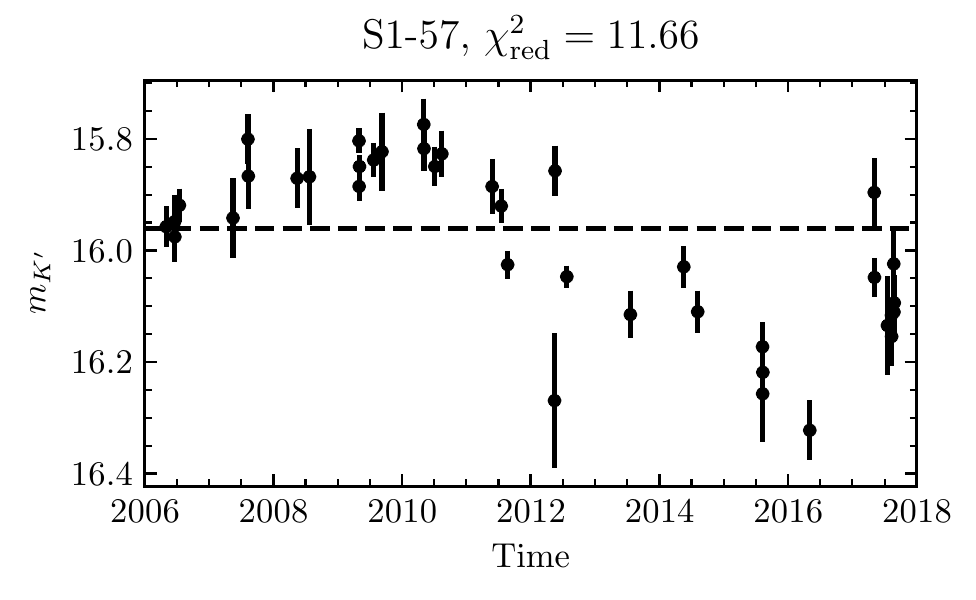}
\end{figure}

\begin{figure}[H]
    \epsscale{1.05}
    \plottwo{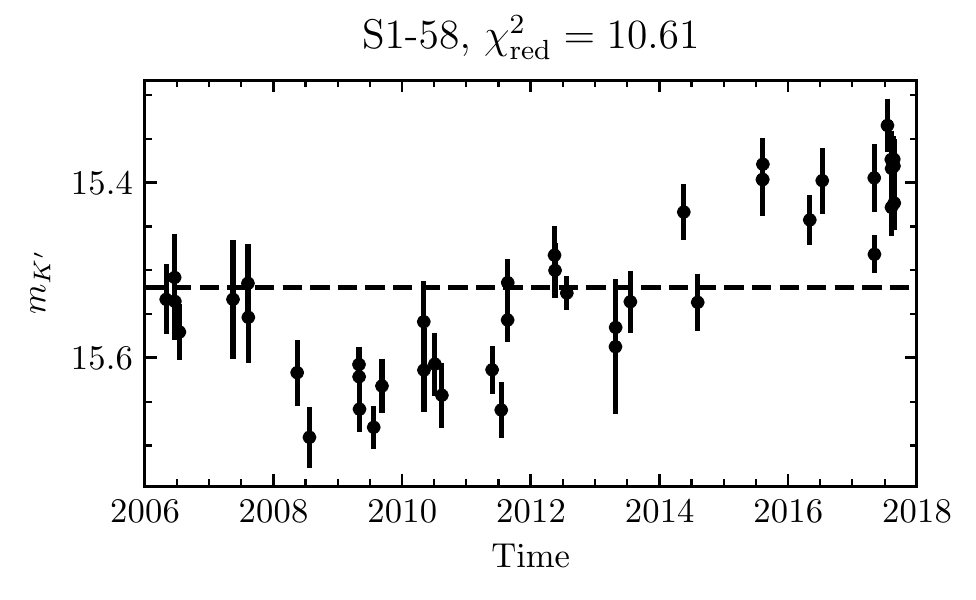}{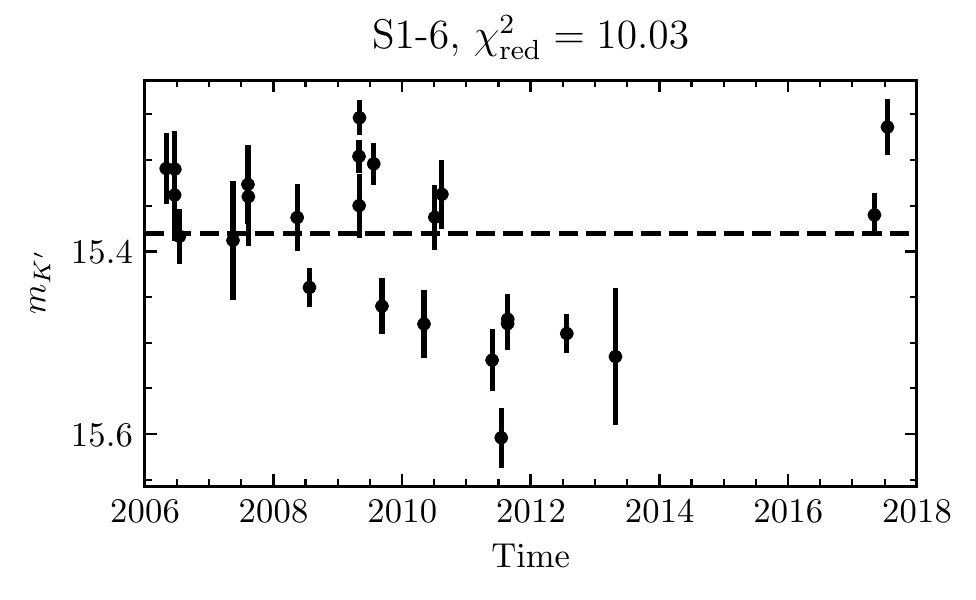}
\end{figure}

\begin{figure}[H]
    \epsscale{1.05}
    \plottwo{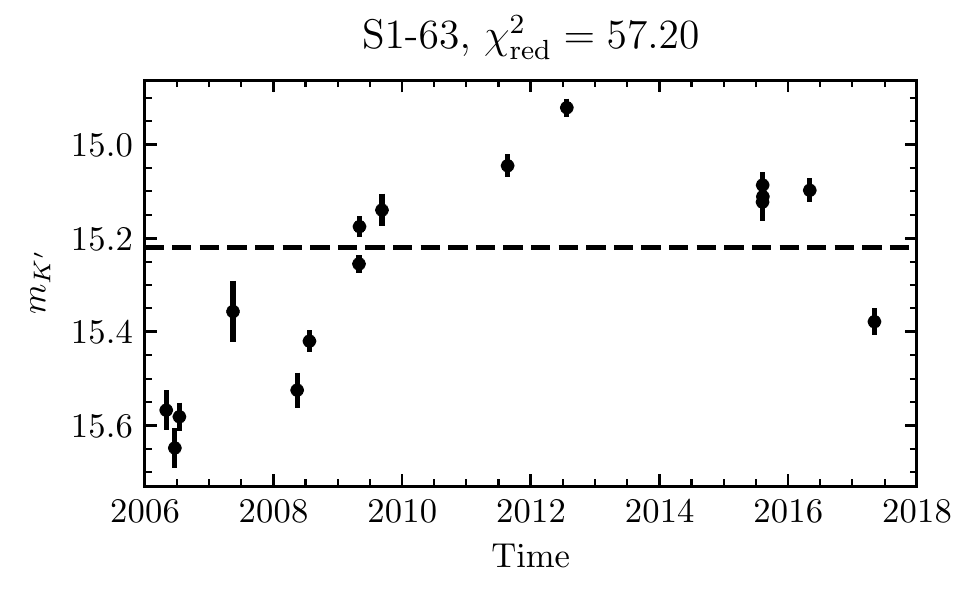}{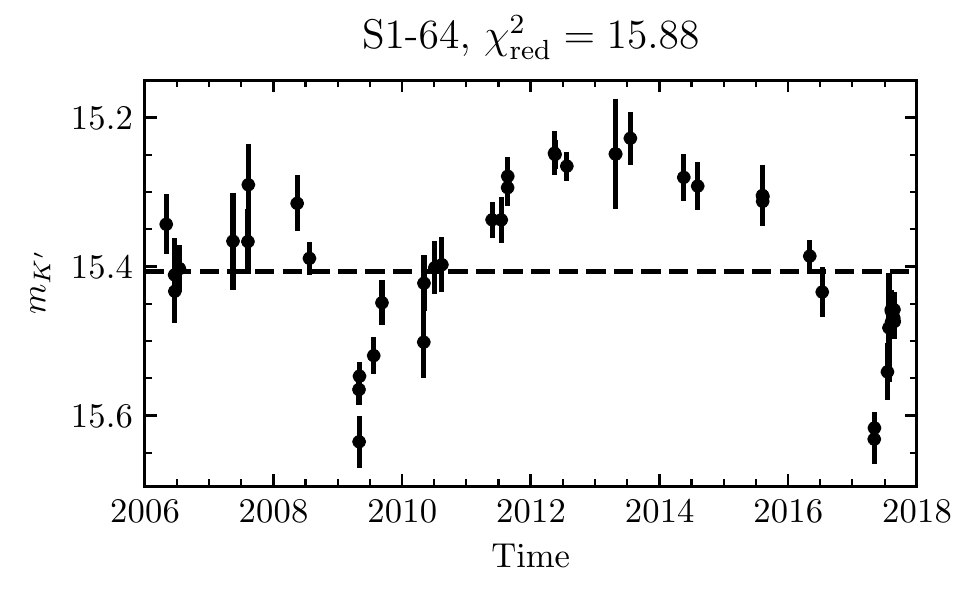}
\end{figure}

\begin{figure}[H]
    \epsscale{1.05}
    \plottwo{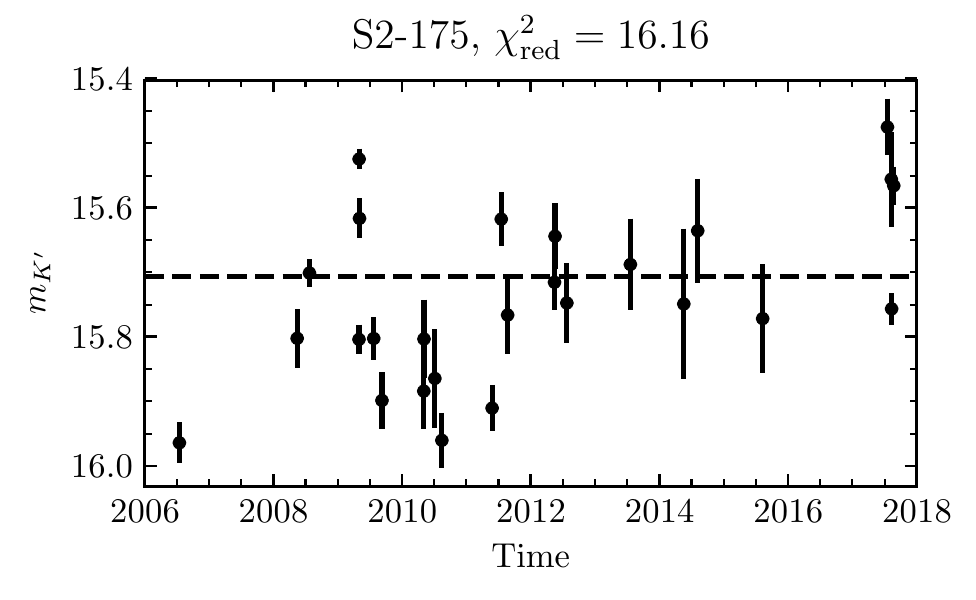}{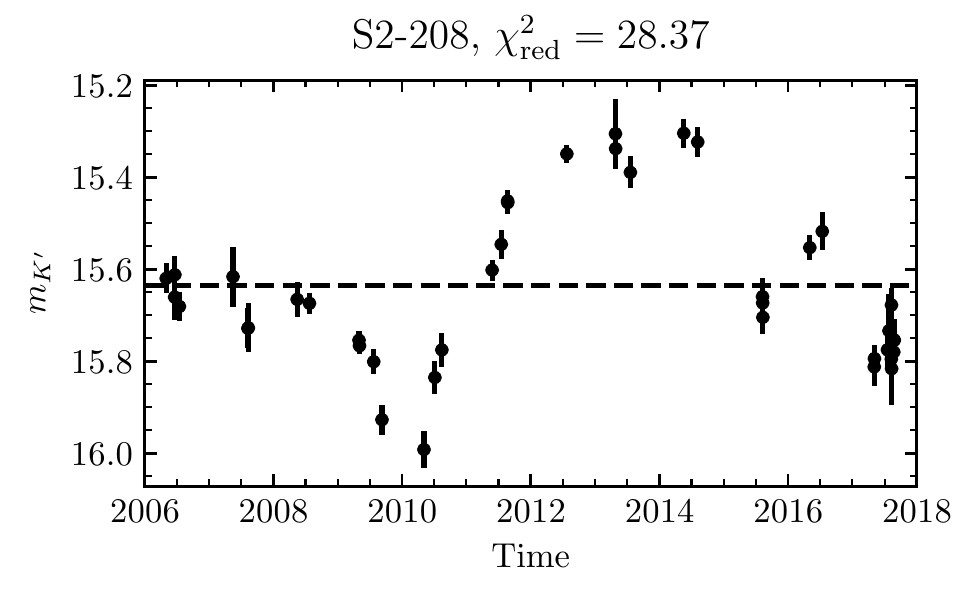}
\end{figure}

\begin{figure}[H]
    \epsscale{1.05}
    \plottwo{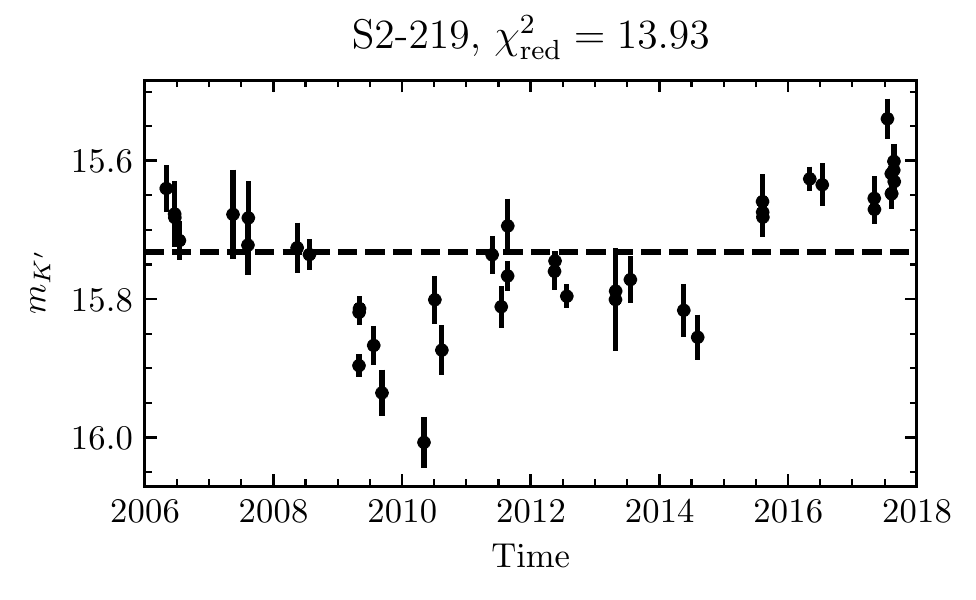}{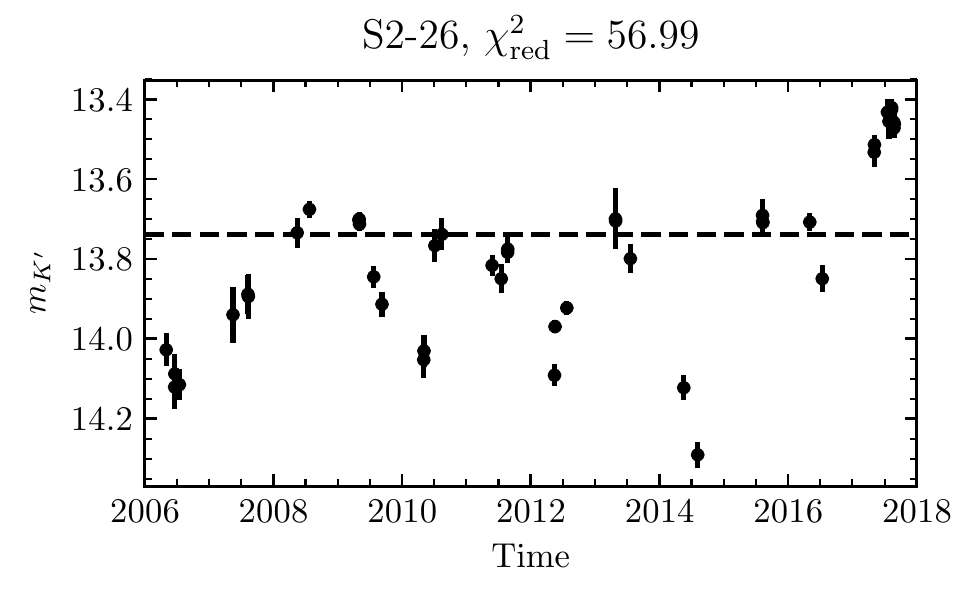}
\end{figure}

\begin{figure}[H]
    \epsscale{1.05}
    \plottwo{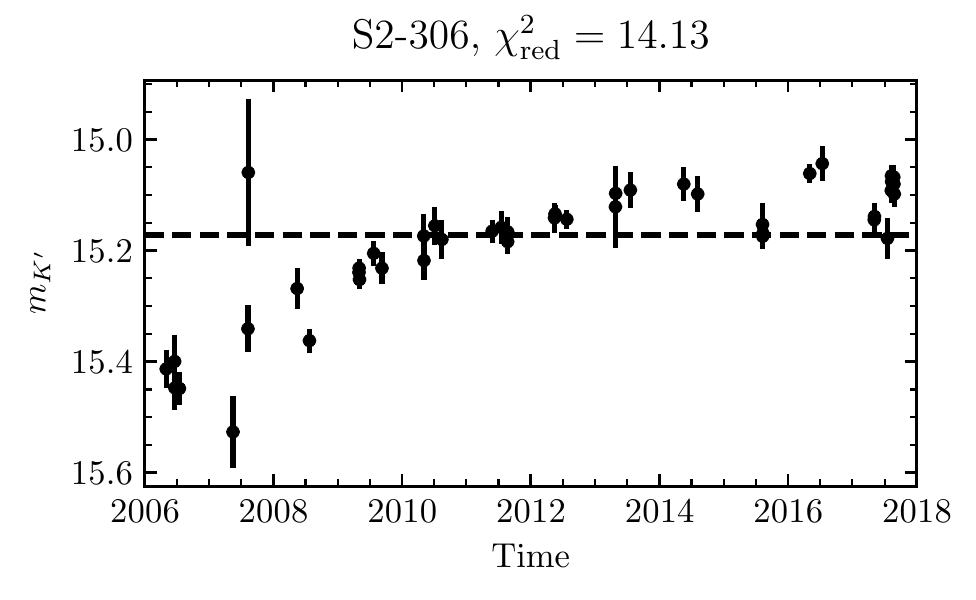}{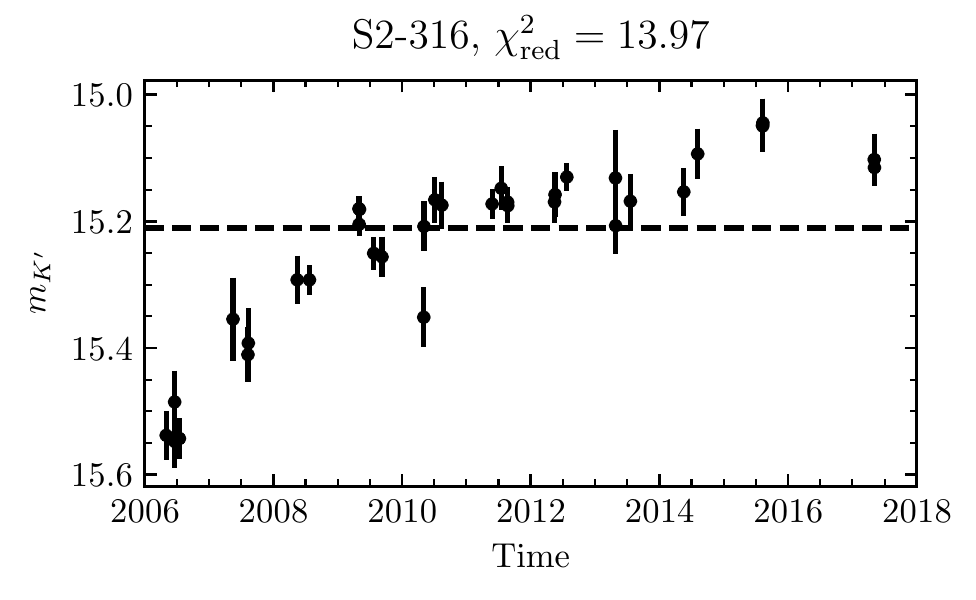}
\end{figure}

\begin{figure}[H]
    \epsscale{1.05}
    \plottwo{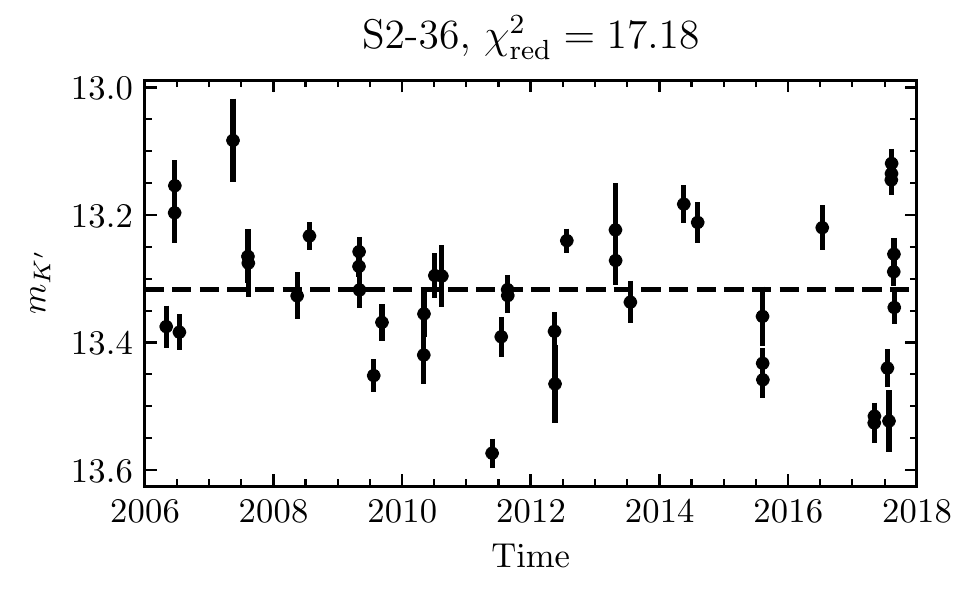}{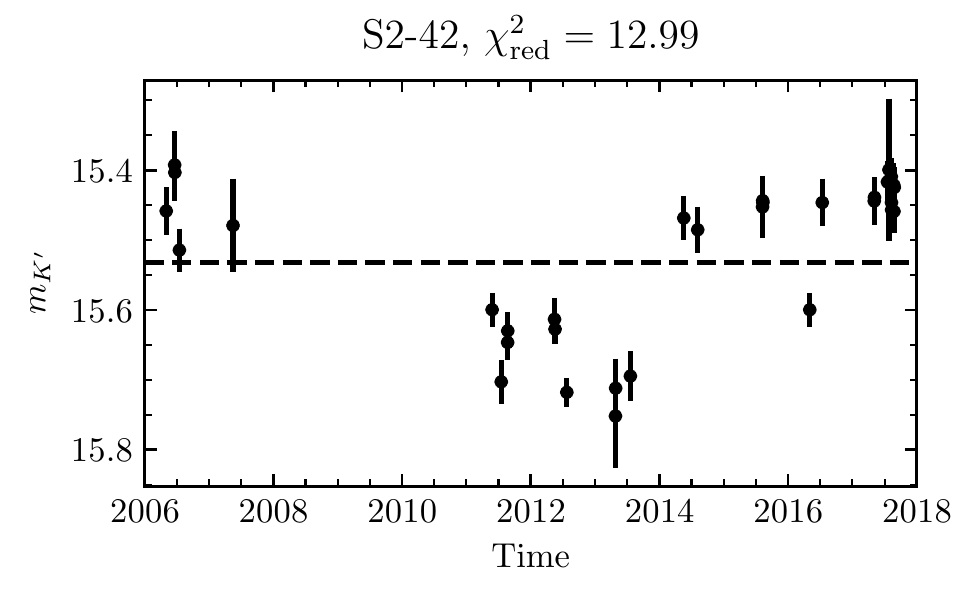}
\end{figure}

\begin{figure}[H]
    \epsscale{1.05}
    \plottwo{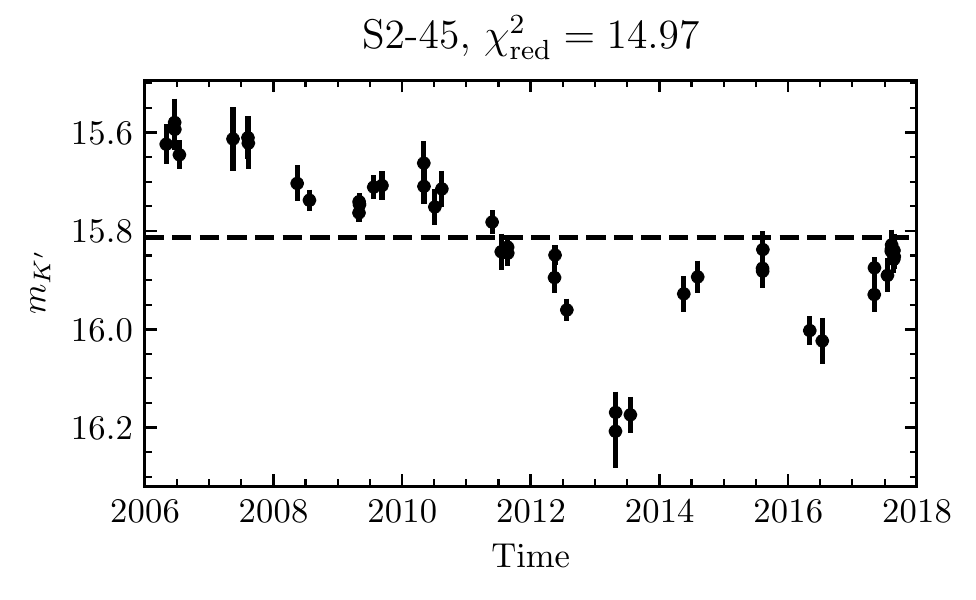}{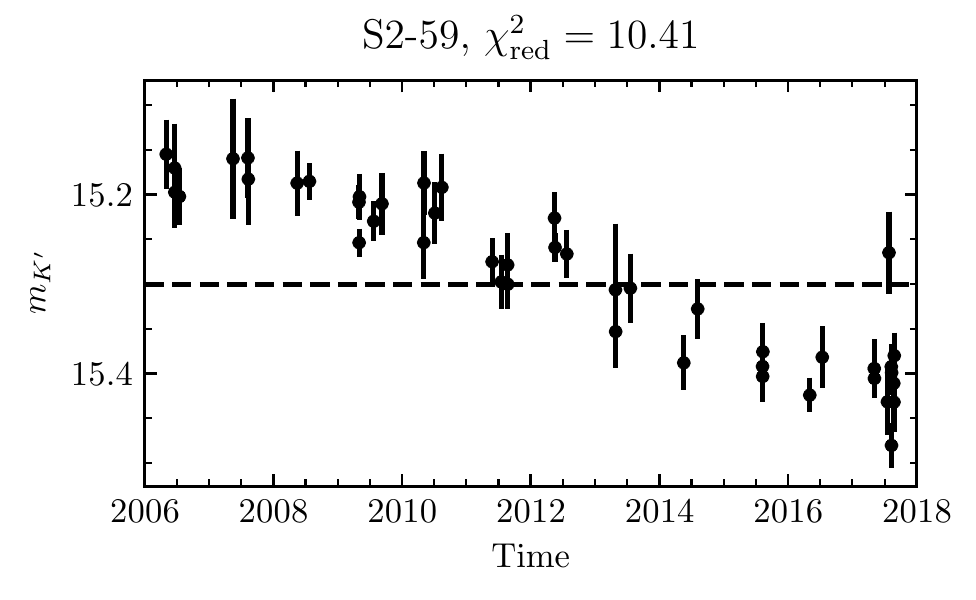}
\end{figure}

\begin{figure}[H]
    \epsscale{1.05}
    \plottwo{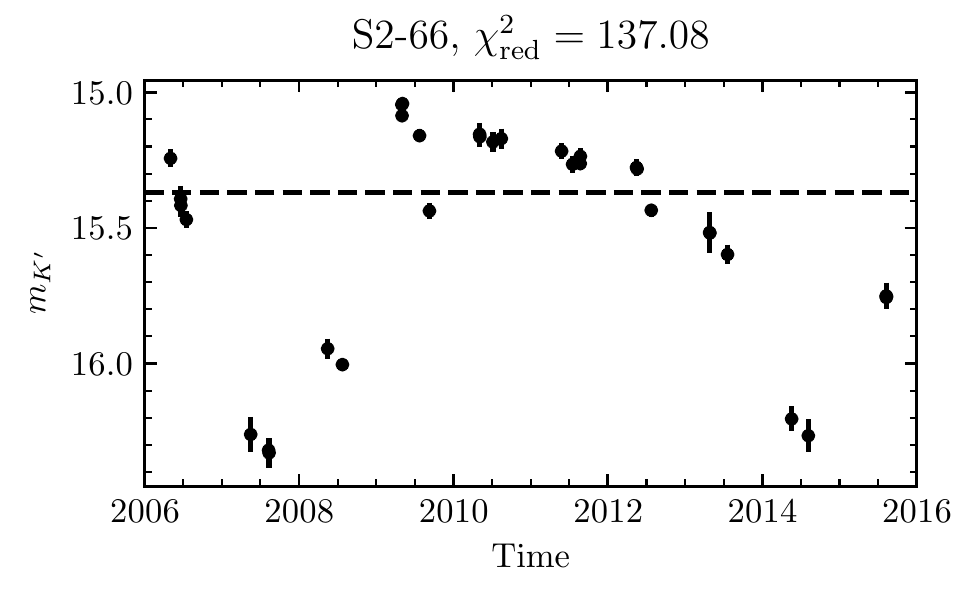}{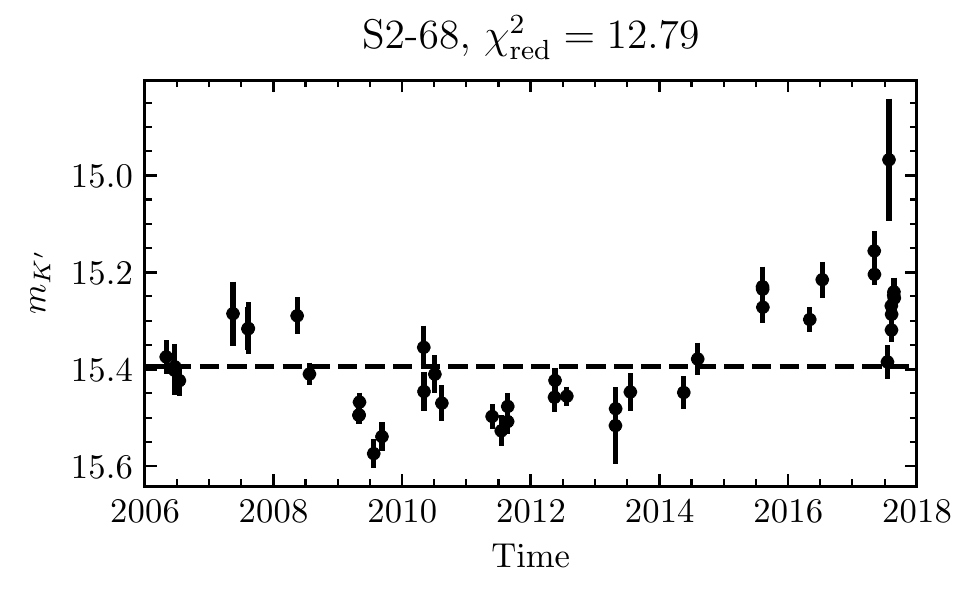}
\end{figure}

\begin{figure}[H]
    \epsscale{1.05}
    \plottwo{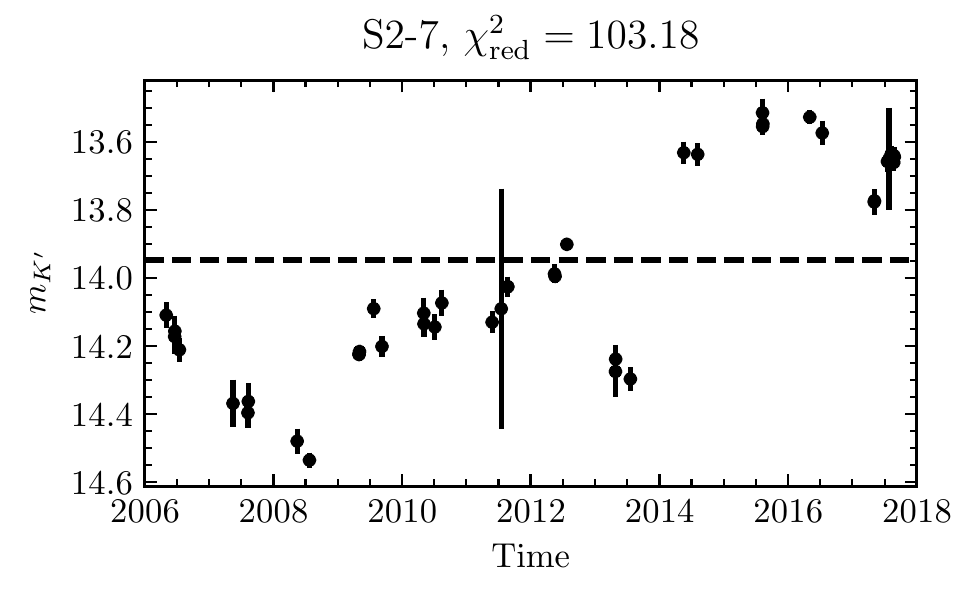}{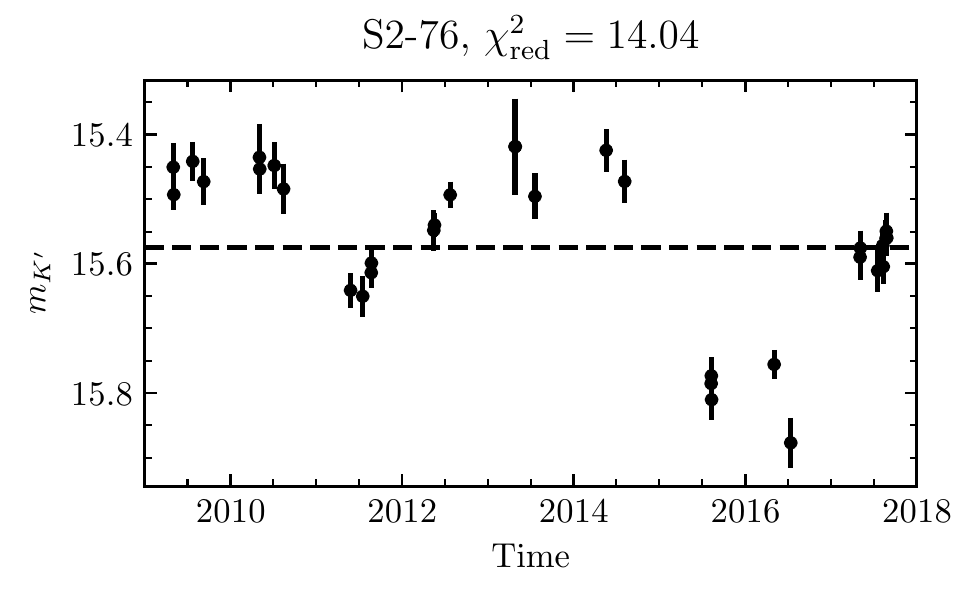}
\end{figure}

\begin{figure}[H]
    \epsscale{1.05}
    \plottwo{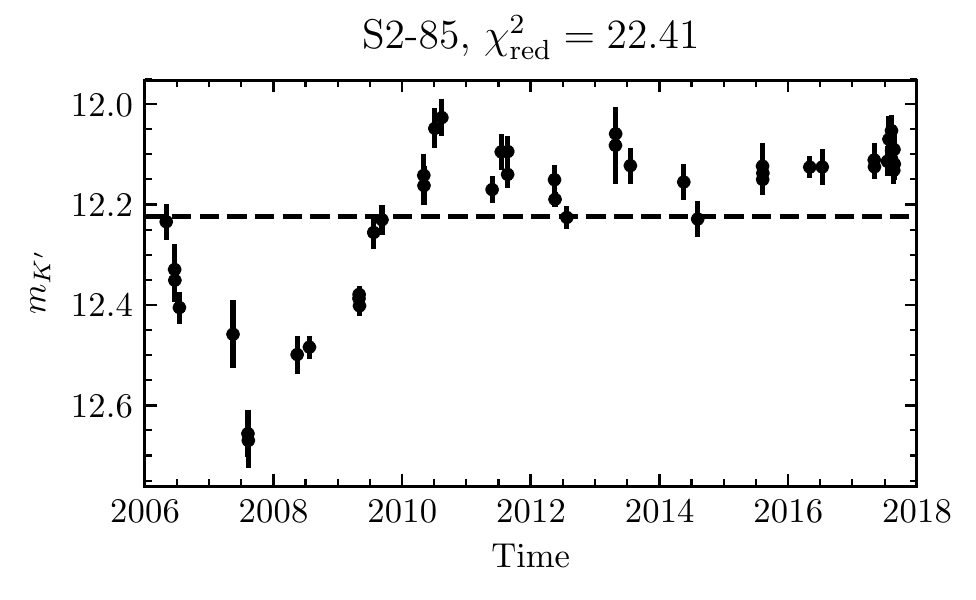}{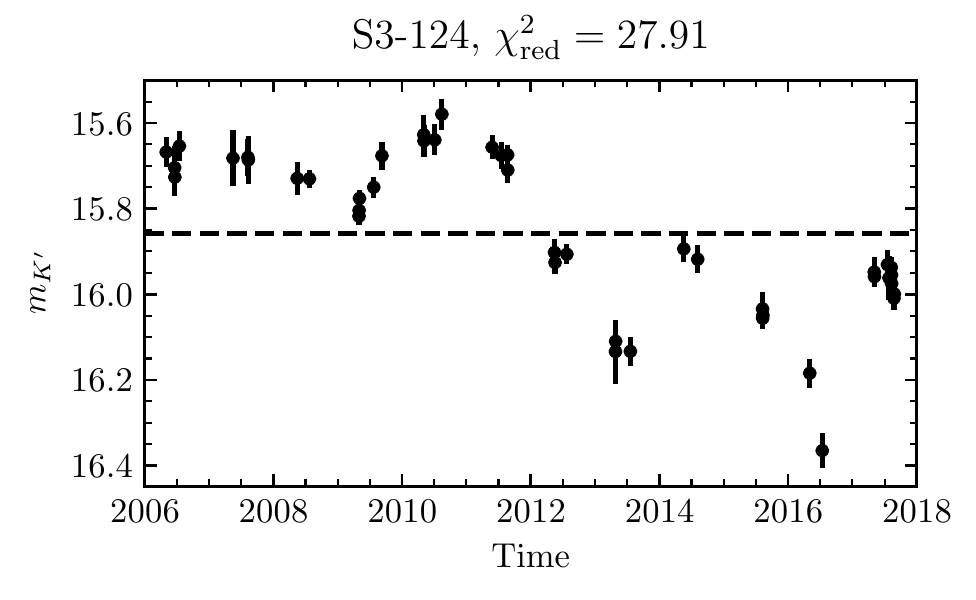}
\end{figure}

\begin{figure}[H]
    \epsscale{1.05}
    \plottwo{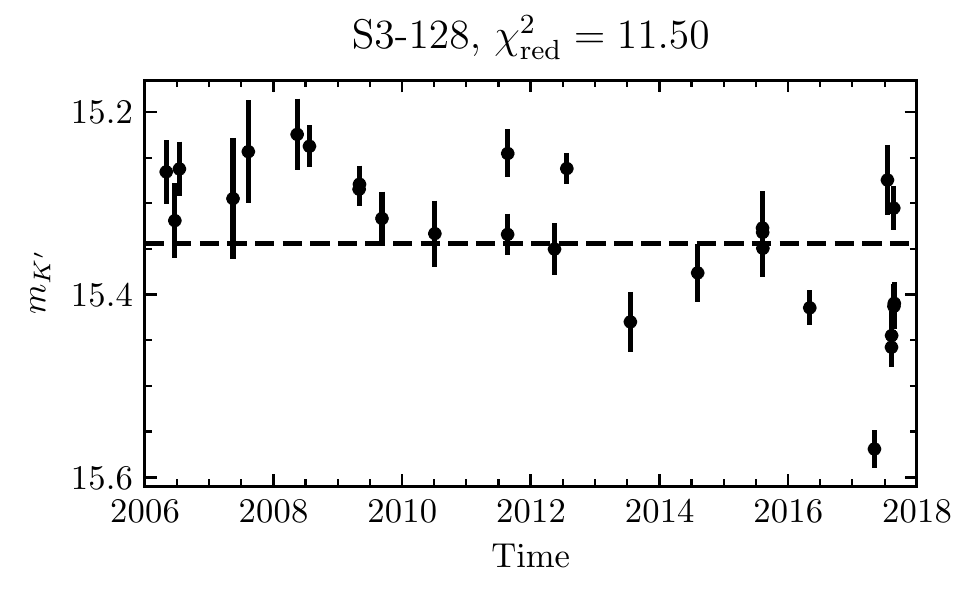}{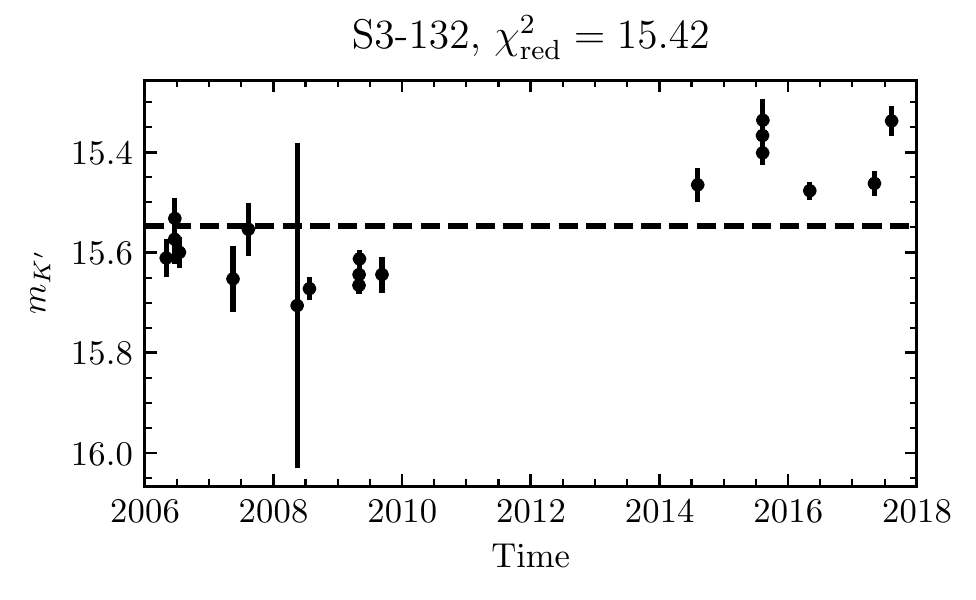}
\end{figure}

\begin{figure}[H]
    \epsscale{1.05}
    \plottwo{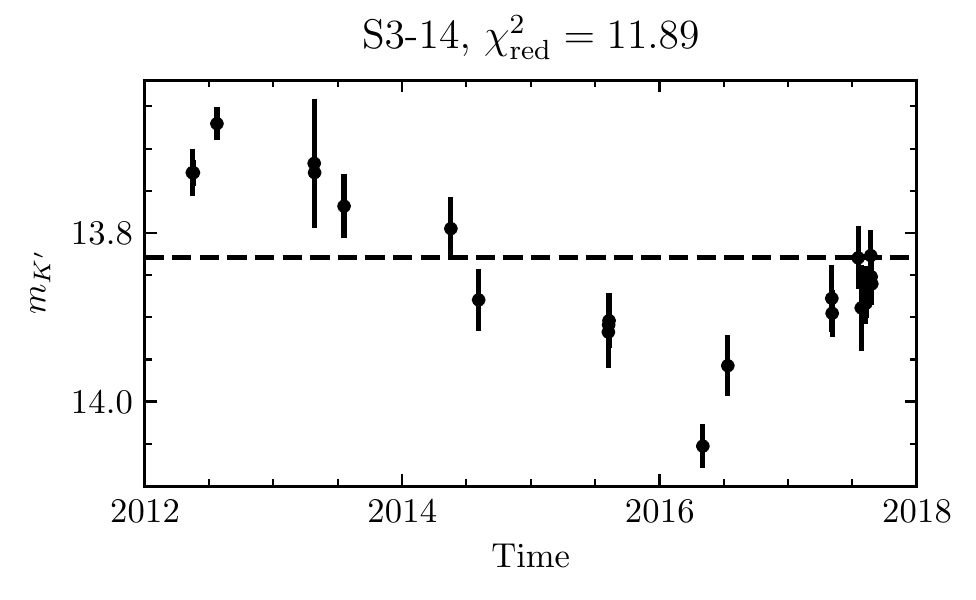}{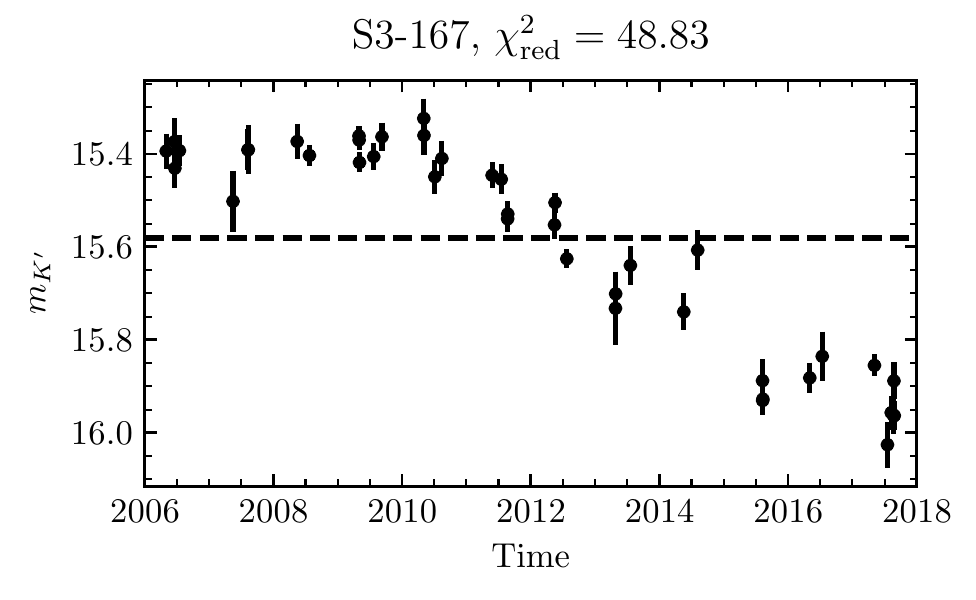}
\end{figure}

\begin{figure}[H]
    \epsscale{1.05}
    \plottwo{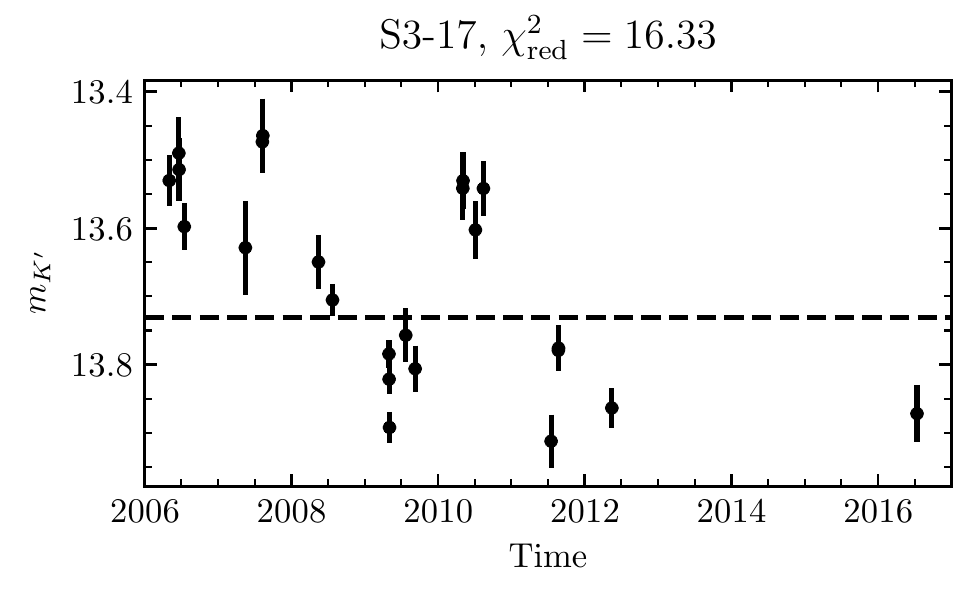}{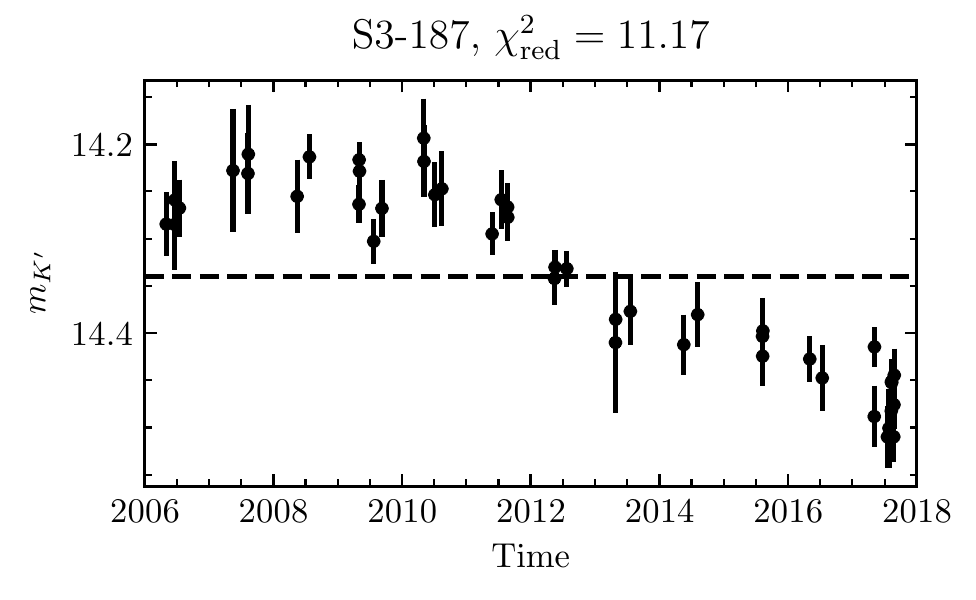}
\end{figure}

\begin{figure}[H]
    \epsscale{1.05}
    \plottwo{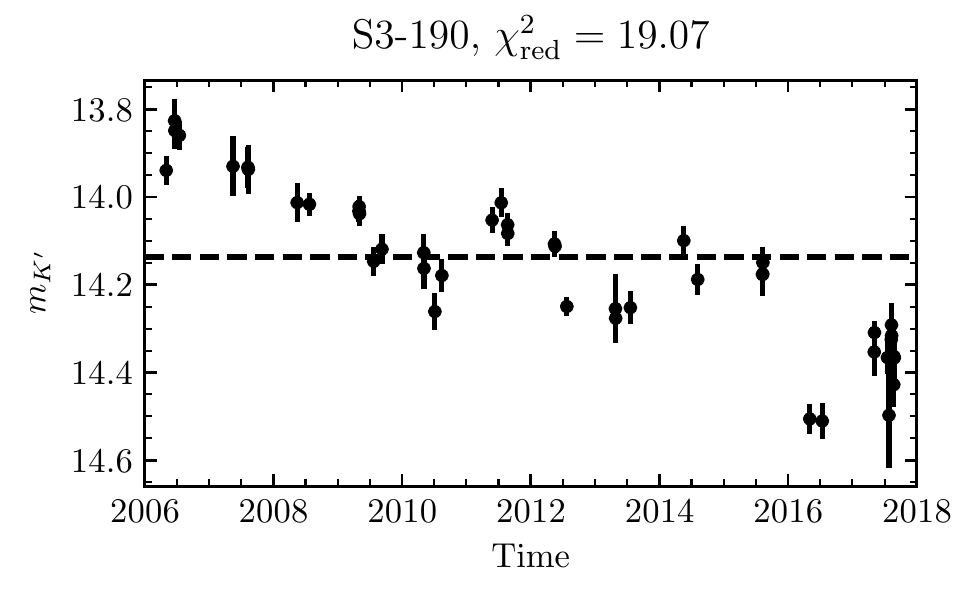}{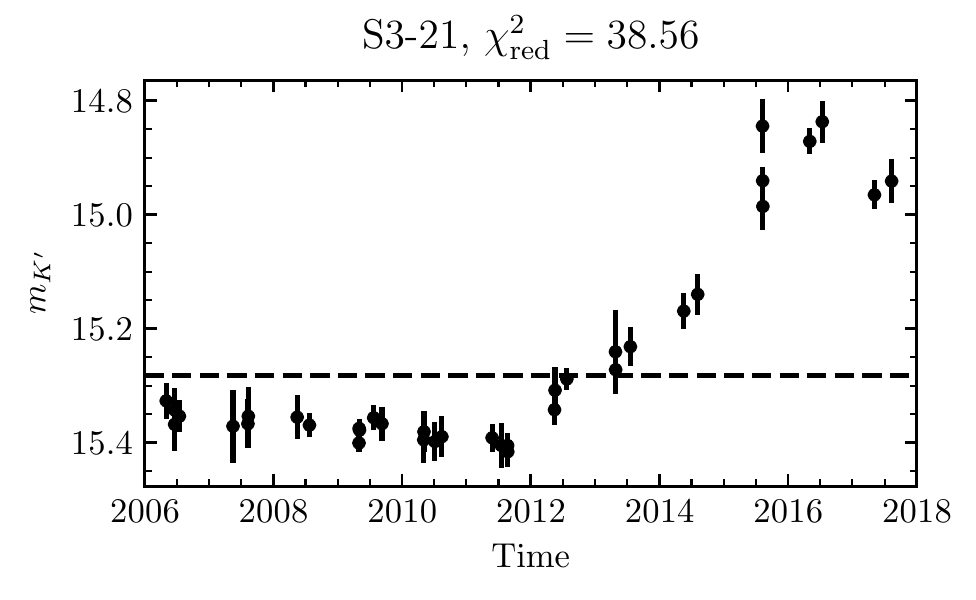}
\end{figure}

\begin{figure}[H]
    \epsscale{1.05}
    \plottwo{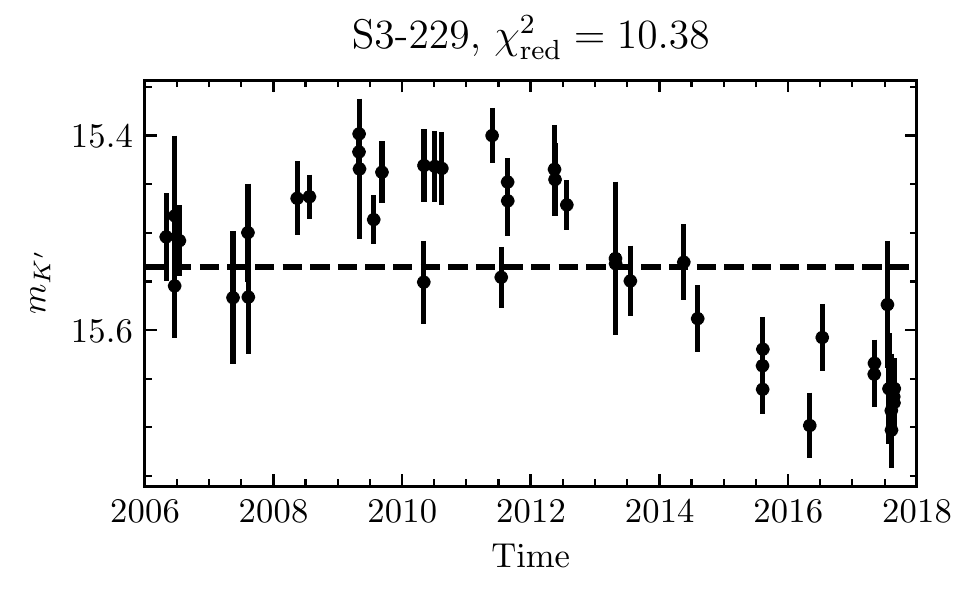}{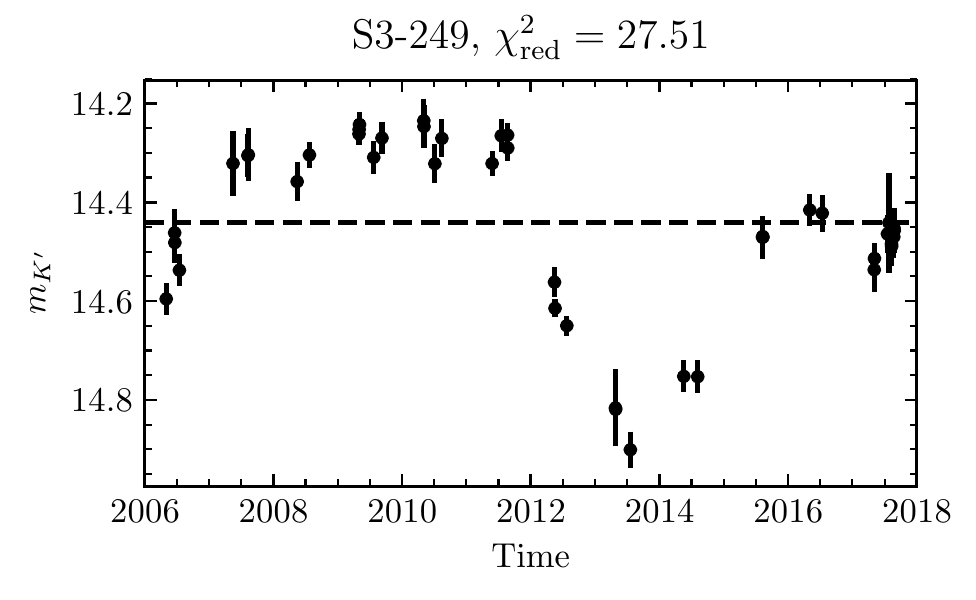}
\end{figure}

\begin{figure}[H]
    \epsscale{1.05}
    \plottwo{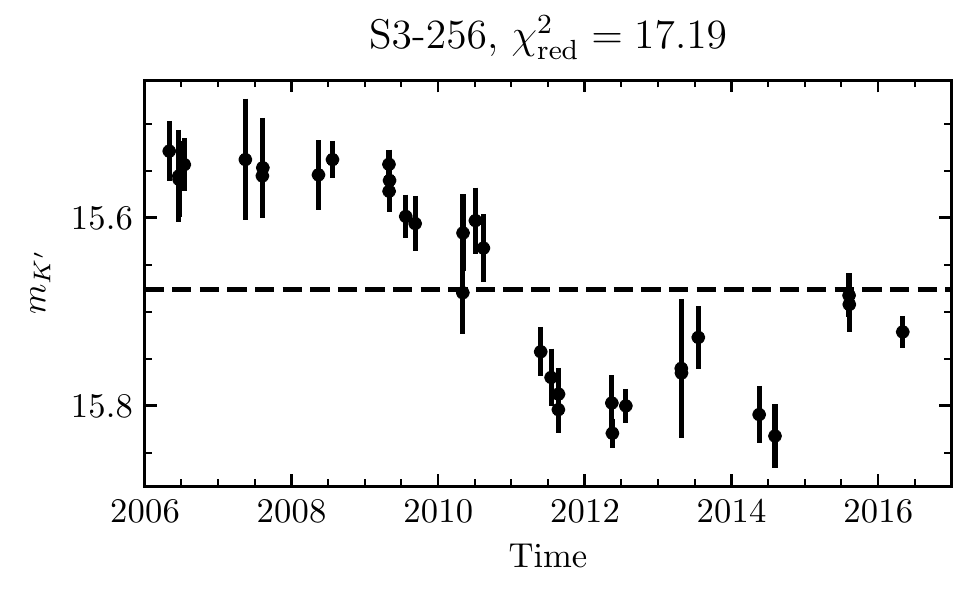}{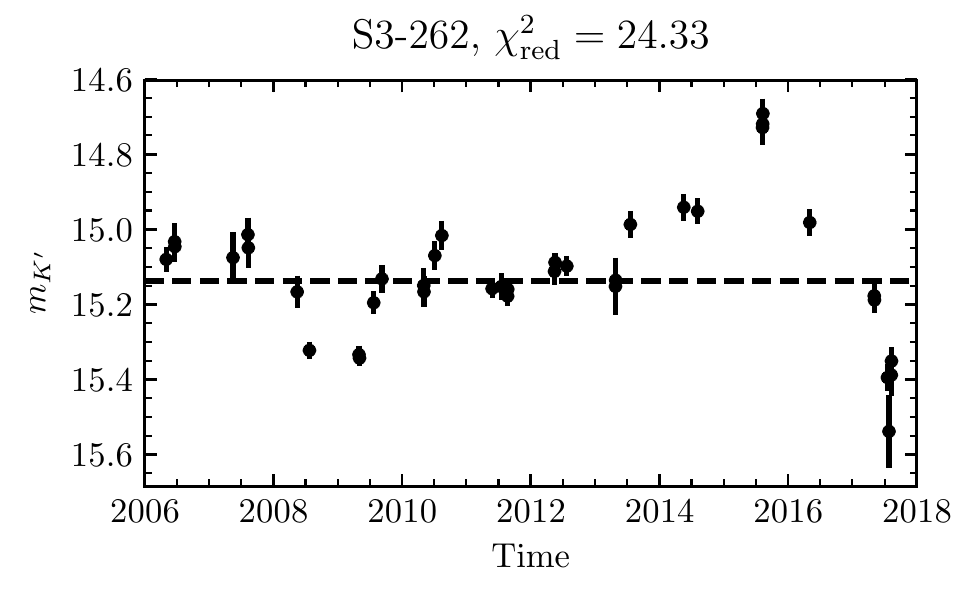}
\end{figure}

\begin{figure}[H]
    \epsscale{1.05}
    \plottwo{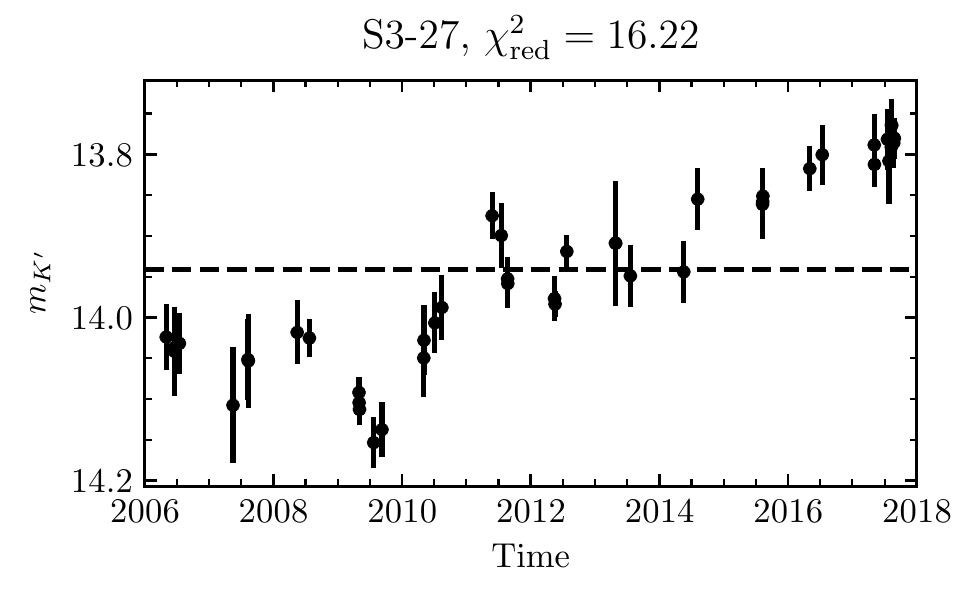}{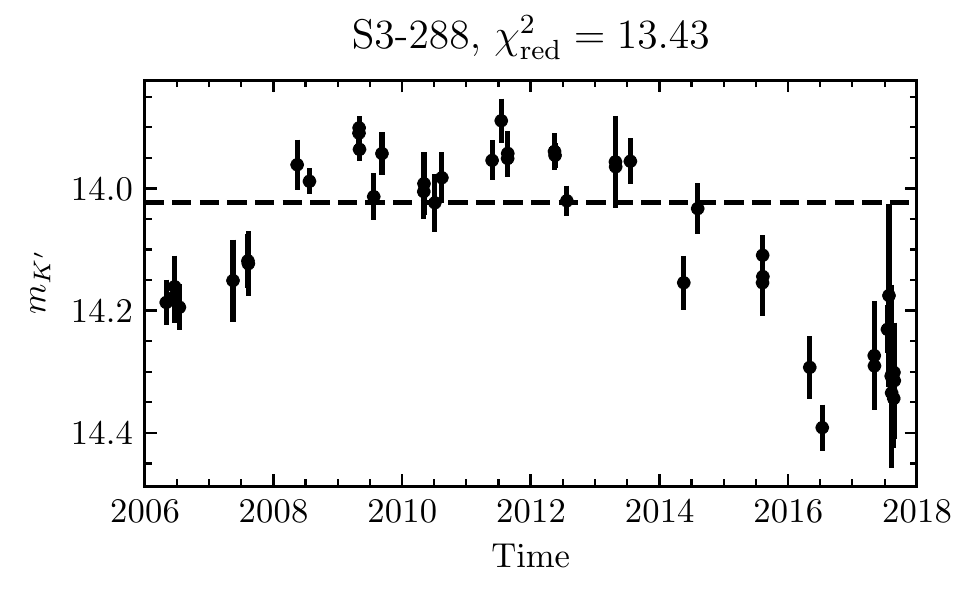}
\end{figure}

\begin{figure}[H]
    \epsscale{1.05}
    \plottwo{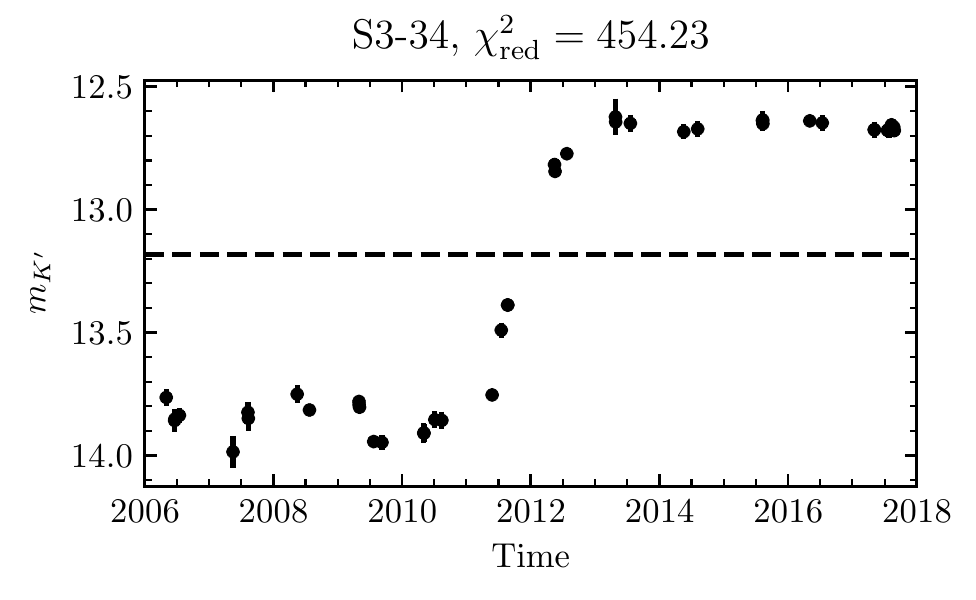}{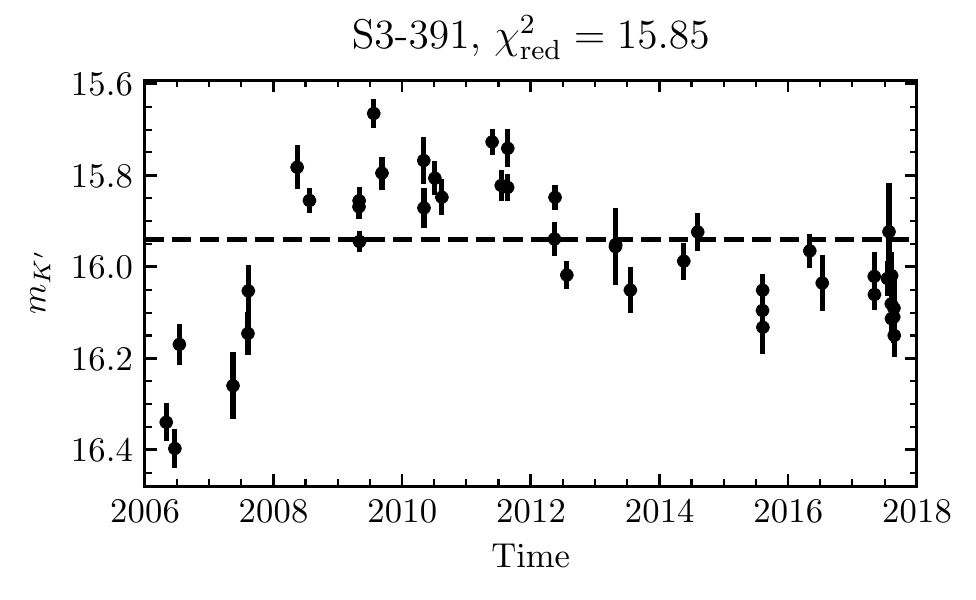}
\end{figure}

\begin{figure}[H]
    \epsscale{1.05}
    \plottwo{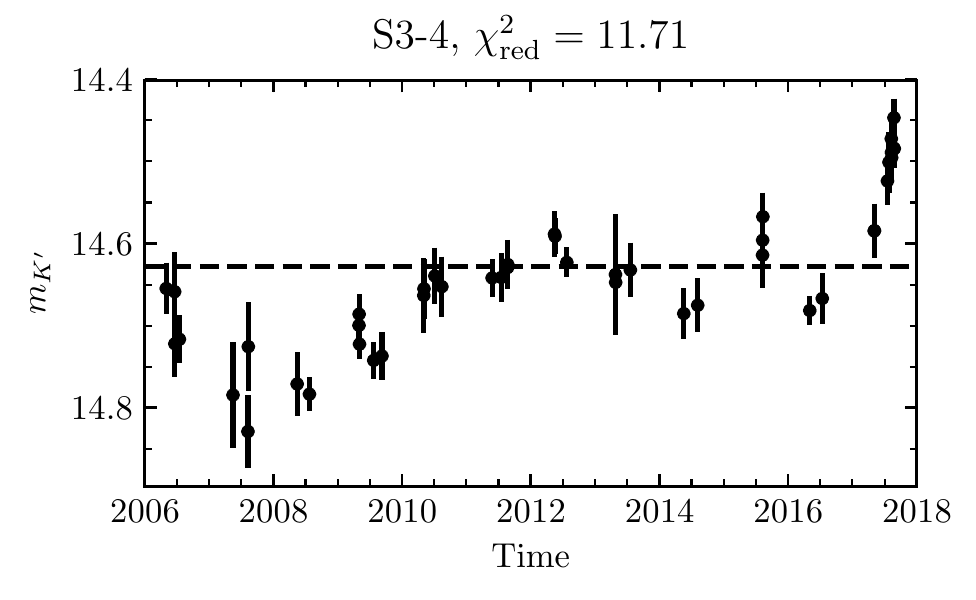}{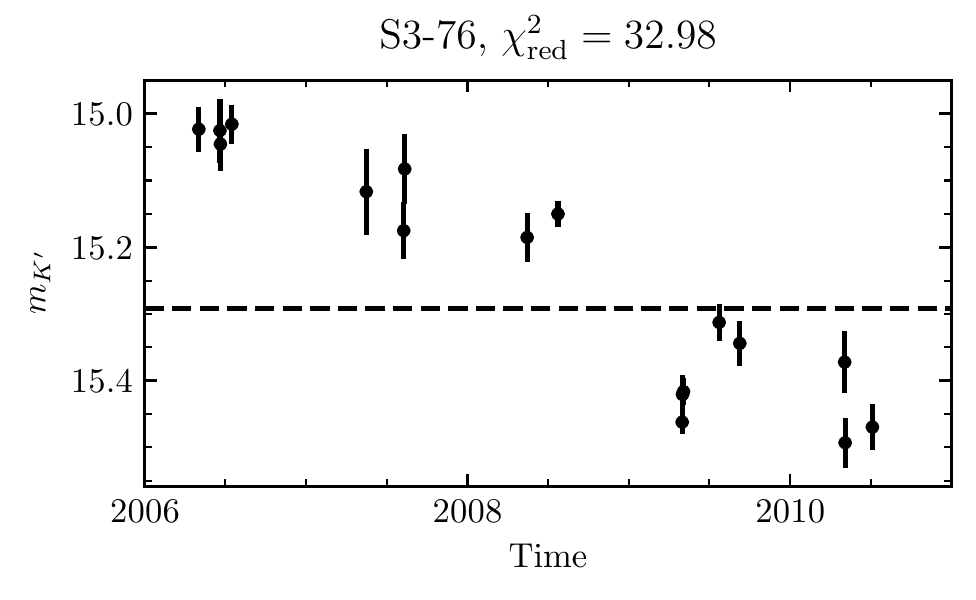}
\end{figure}

\begin{figure}[H]
    \epsscale{1.05}
    \plottwo{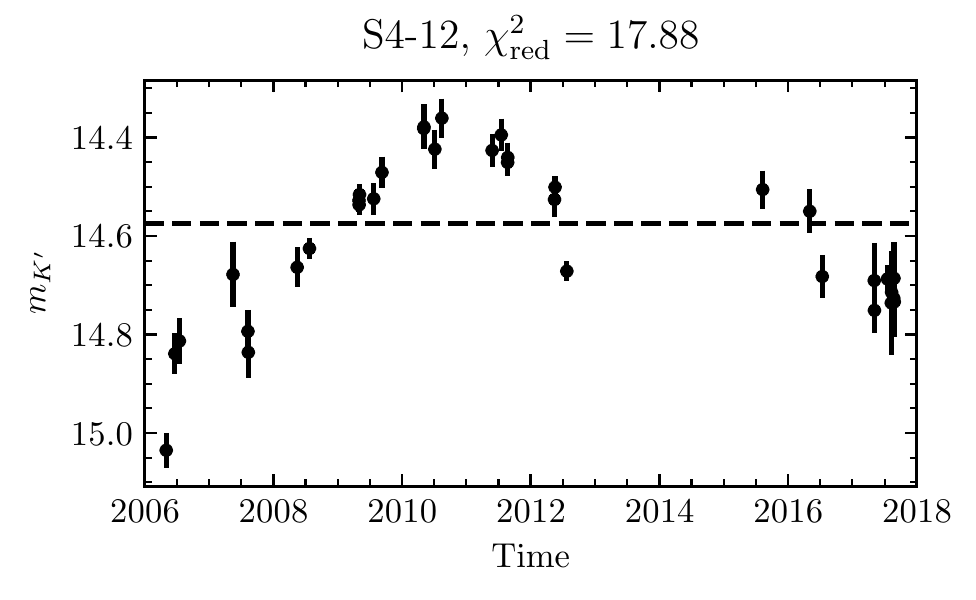}{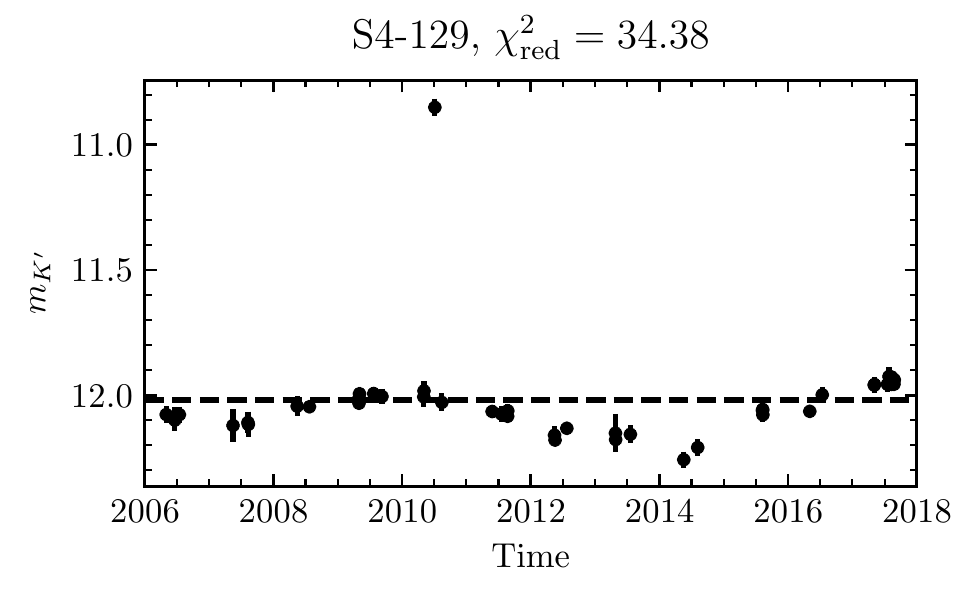}
\end{figure}

\begin{figure}[H]
    \epsscale{1.05}
    \plottwo{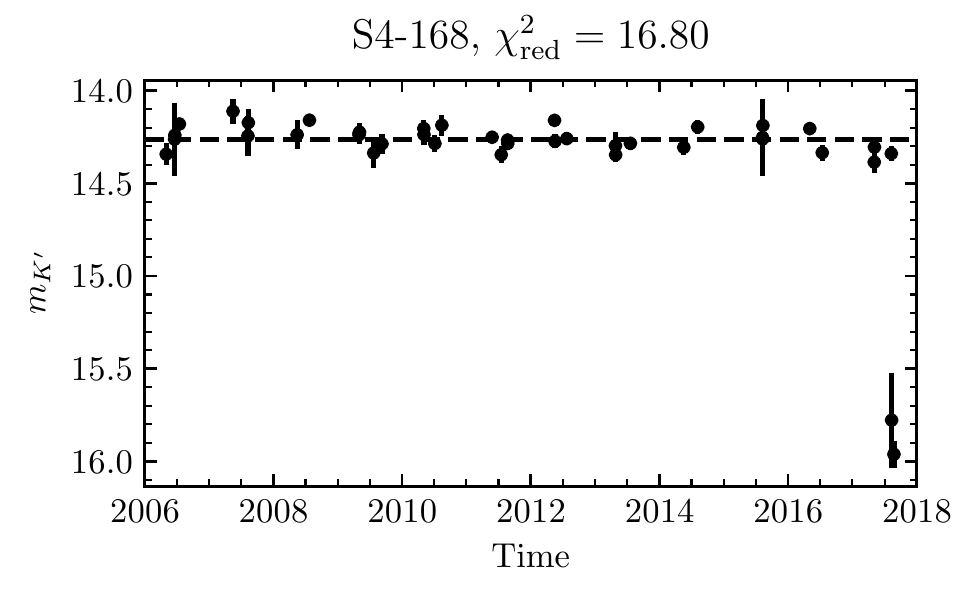}{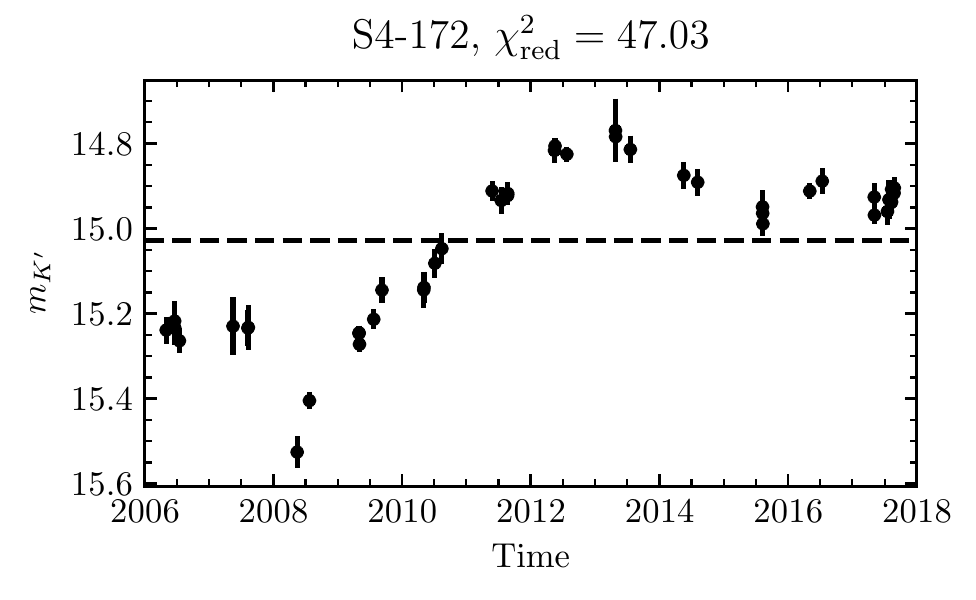}
\end{figure}

\begin{figure}[H]
    \epsscale{1.05}
    \plottwo{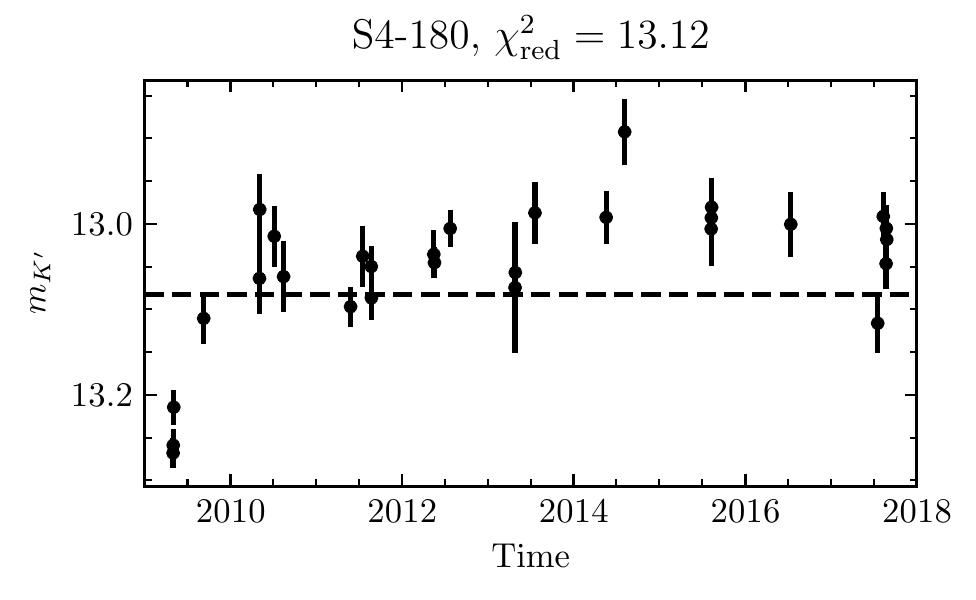}{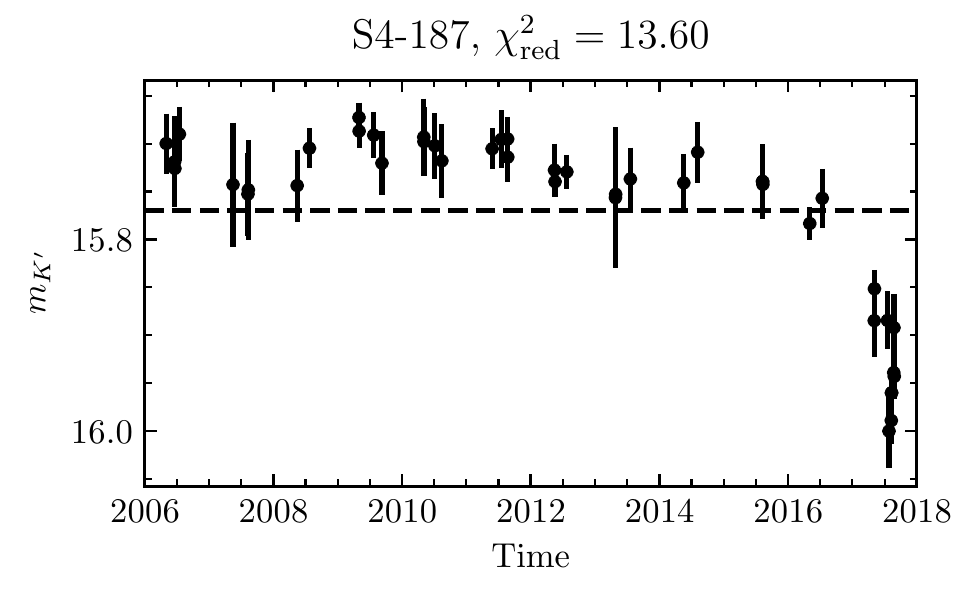}
\end{figure}

\begin{figure}[H]
    \epsscale{1.05}
    \plottwo{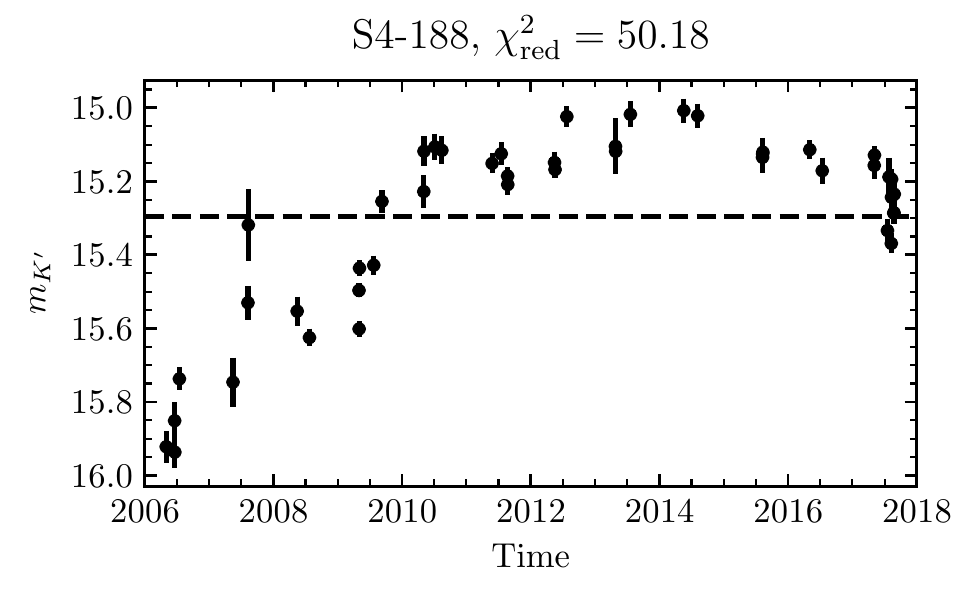}{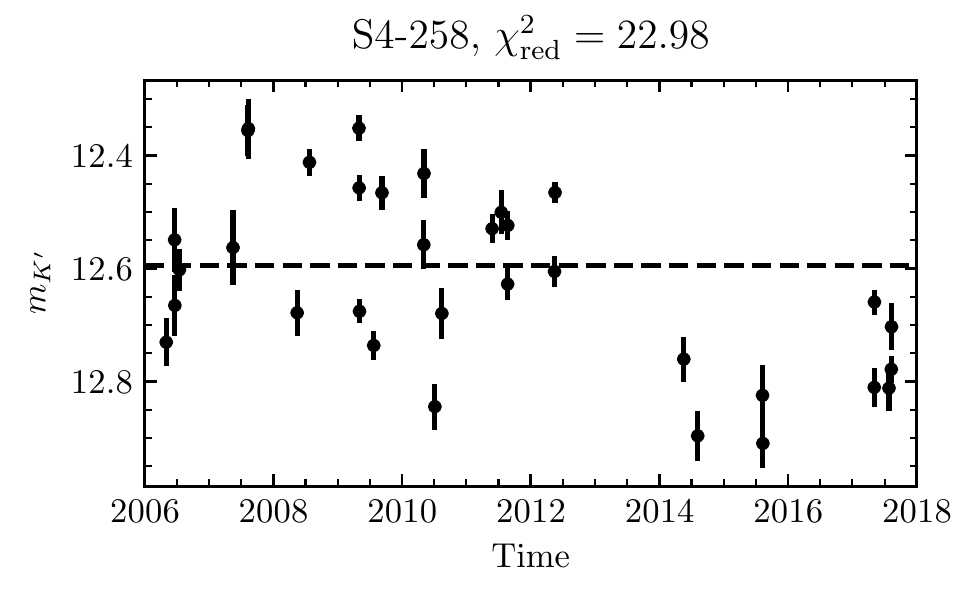}
\end{figure}

\begin{figure}[H]
    \epsscale{1.05}
    \plottwo{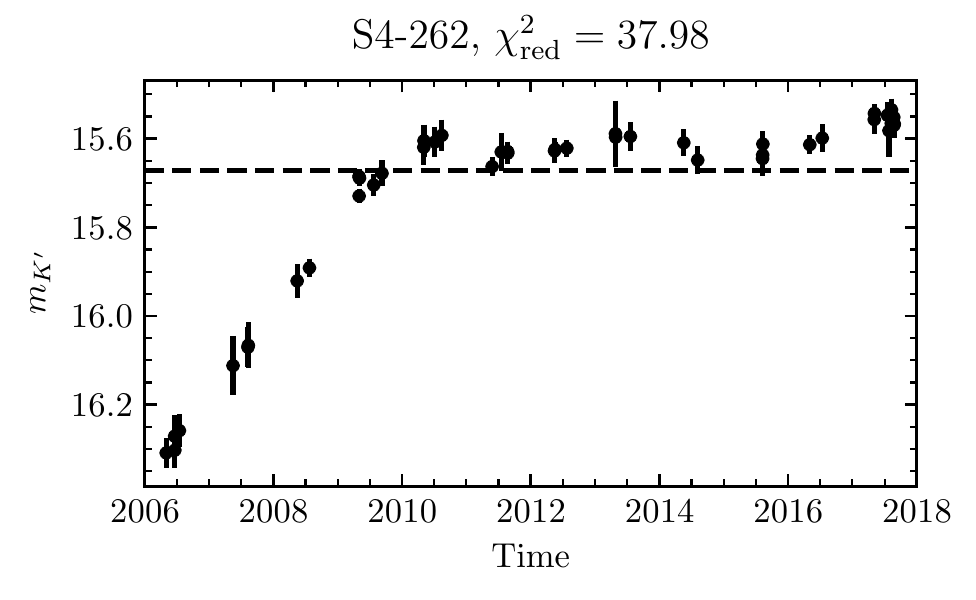}{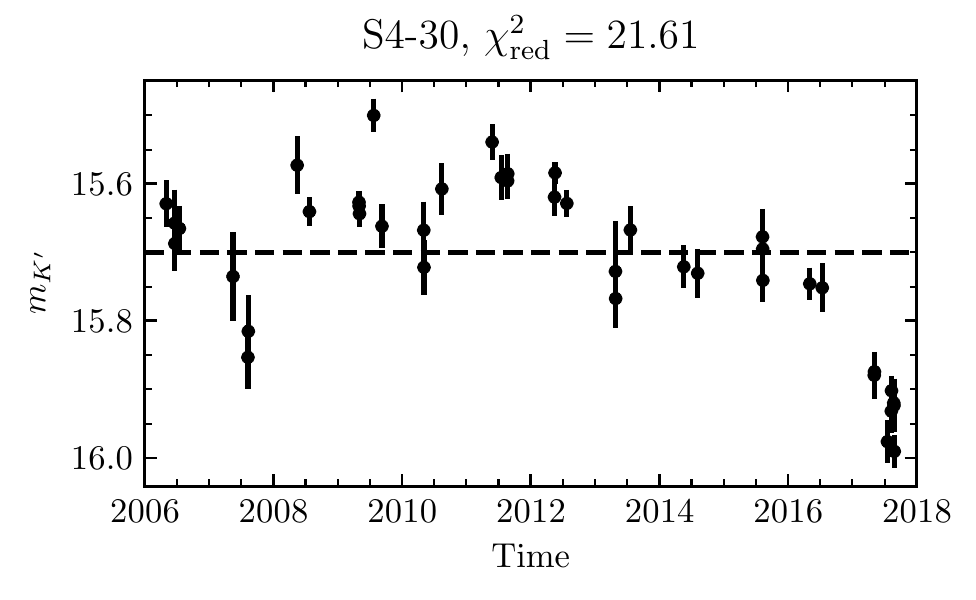}
\end{figure}

\begin{figure}[H]
    \epsscale{1.05}
    \plottwo{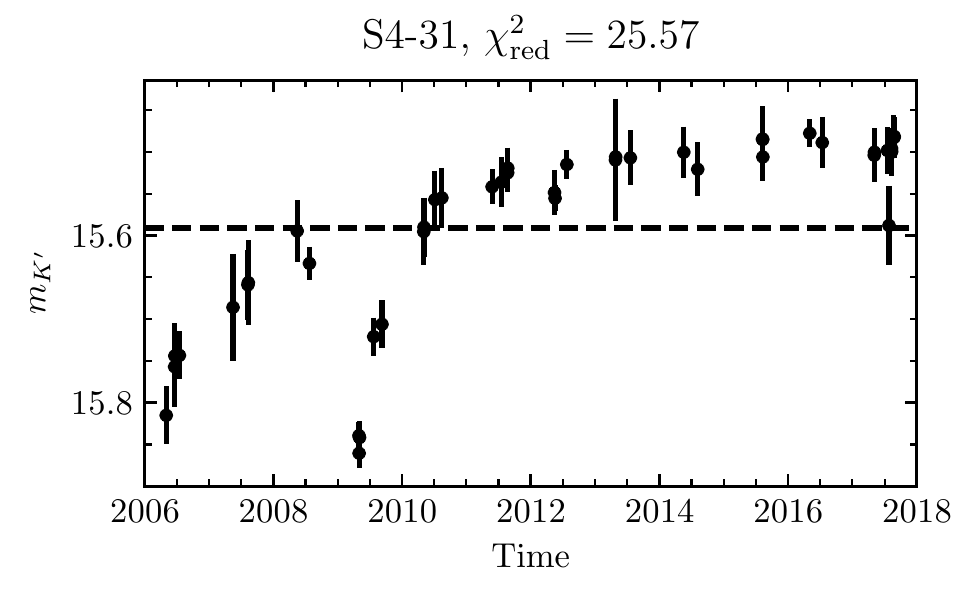}{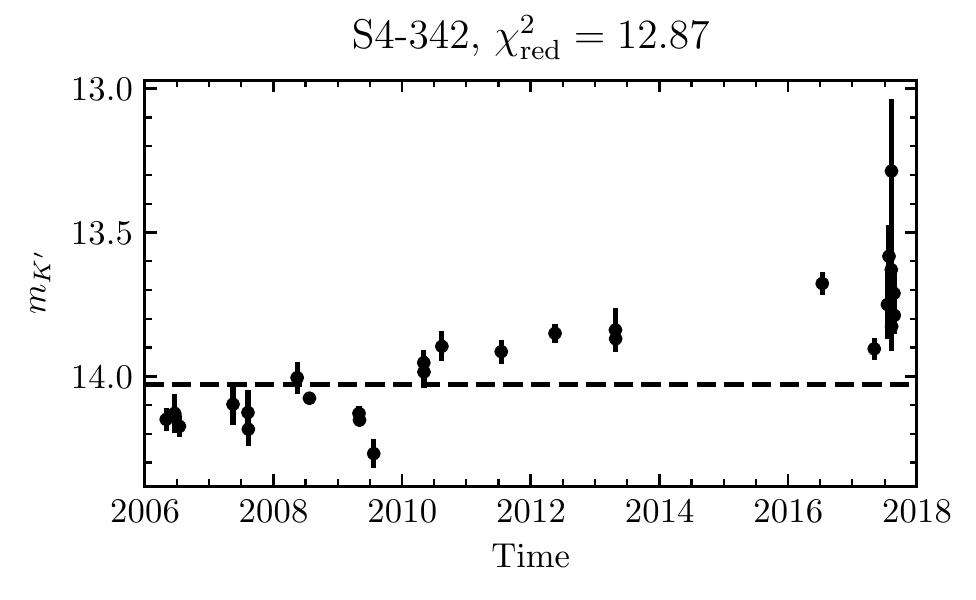}
\end{figure}

\begin{figure}[H]
    \epsscale{1.05}
    \plottwo{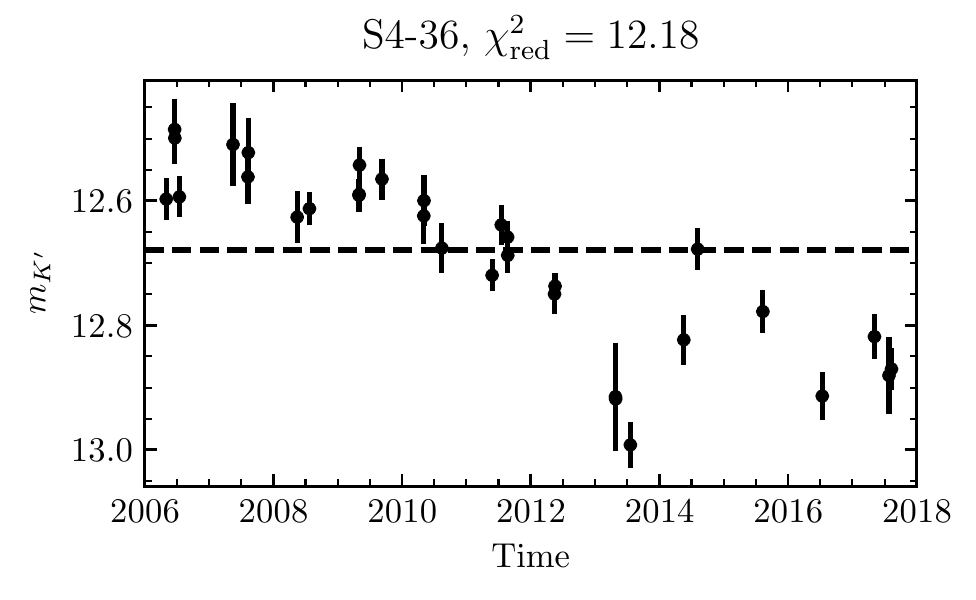}{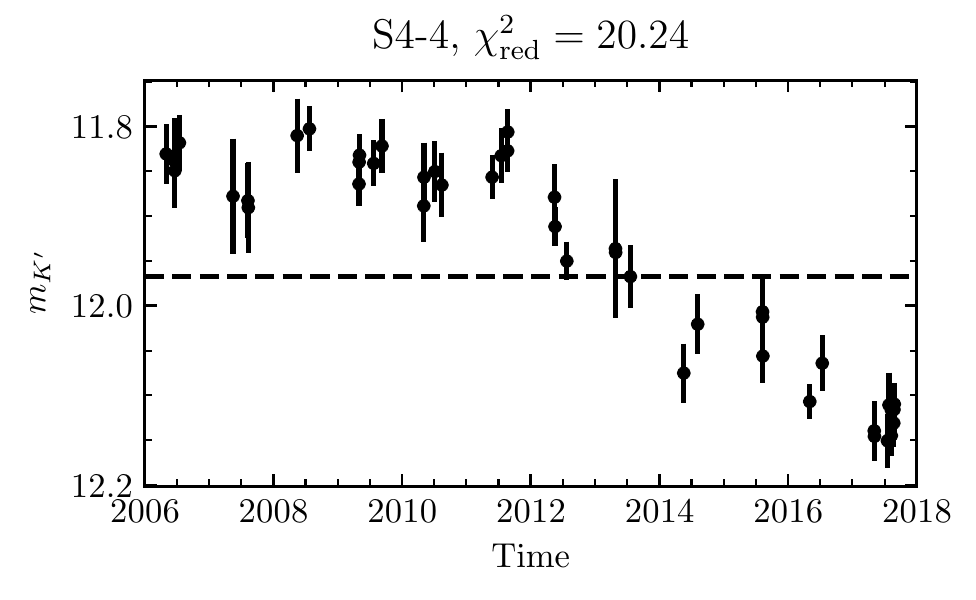}
\end{figure}

\begin{figure}[H]
    \epsscale{1.05}
    \plottwo{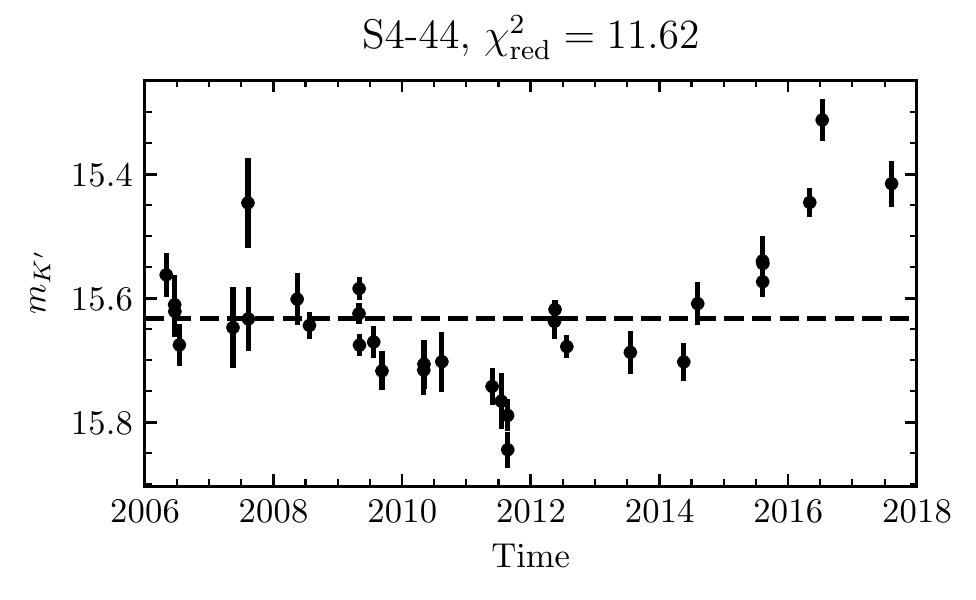}{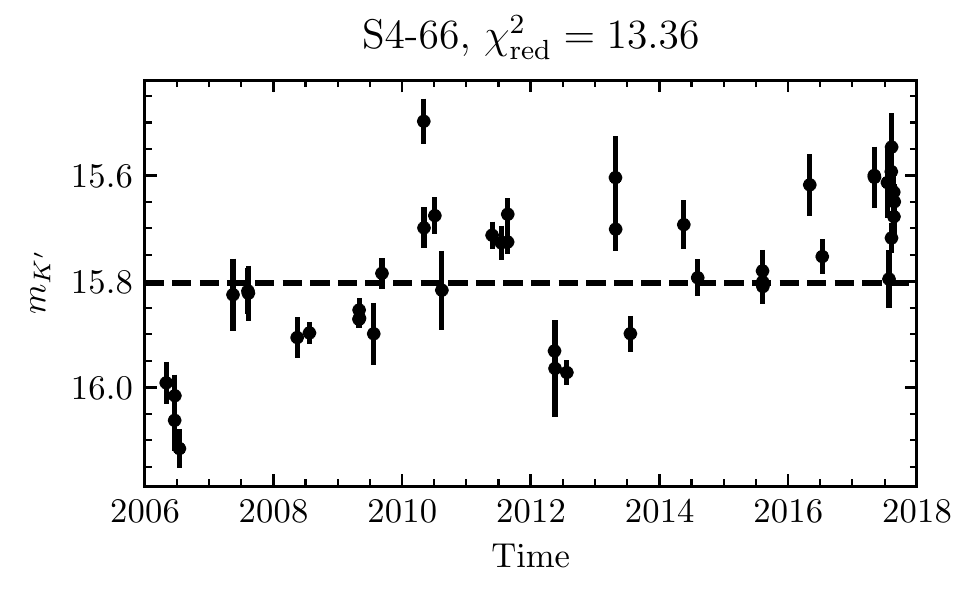}
\end{figure}

\begin{figure}[H]
    \epsscale{1.05}
    \plottwo{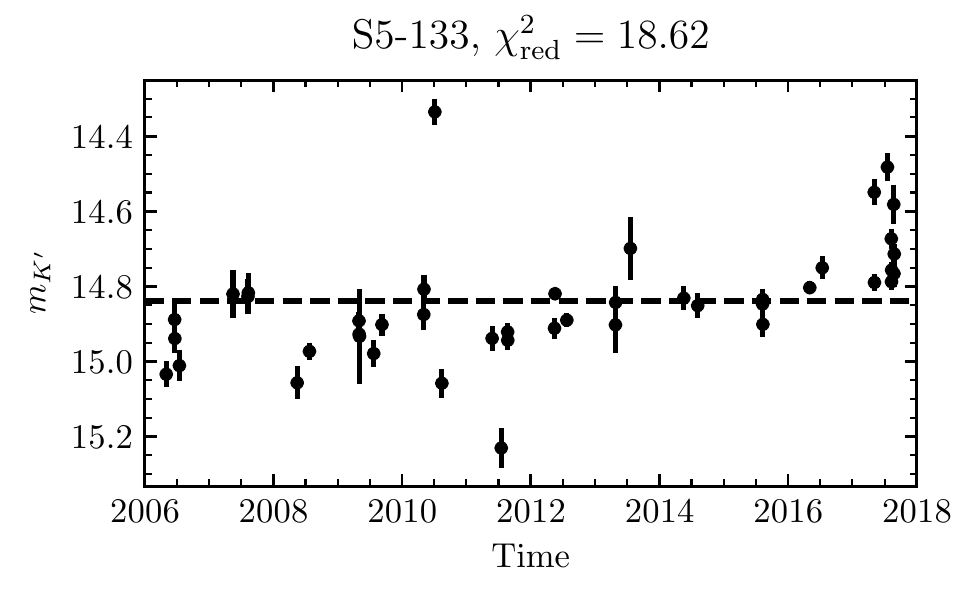}{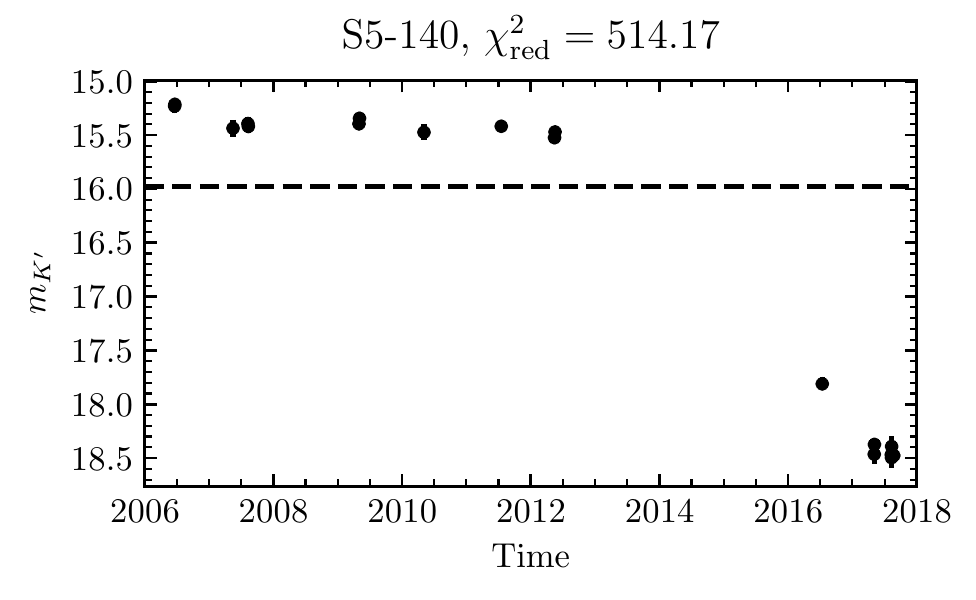}
\end{figure}

\begin{figure}[H]
    \epsscale{1.05}
    \plottwo{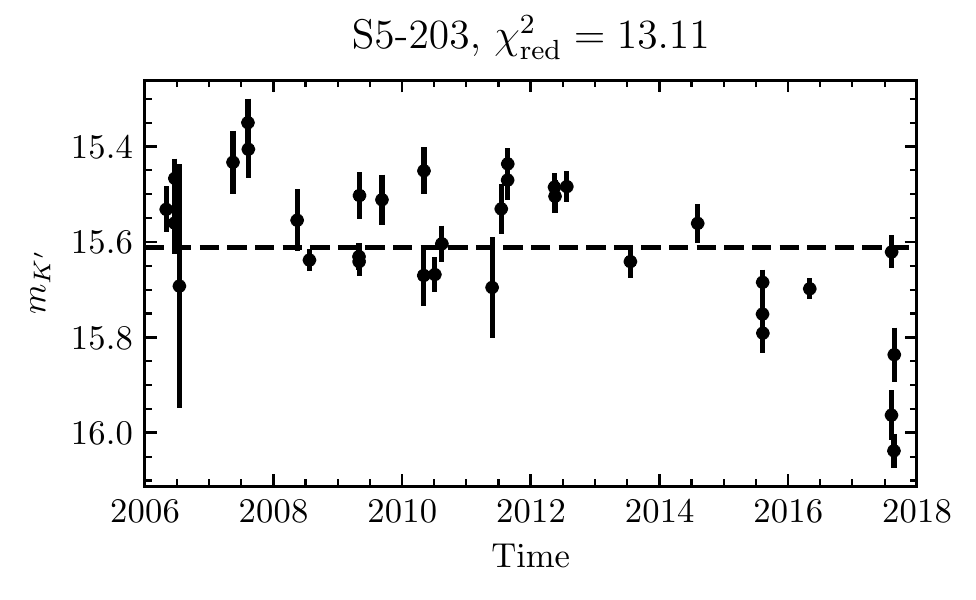}{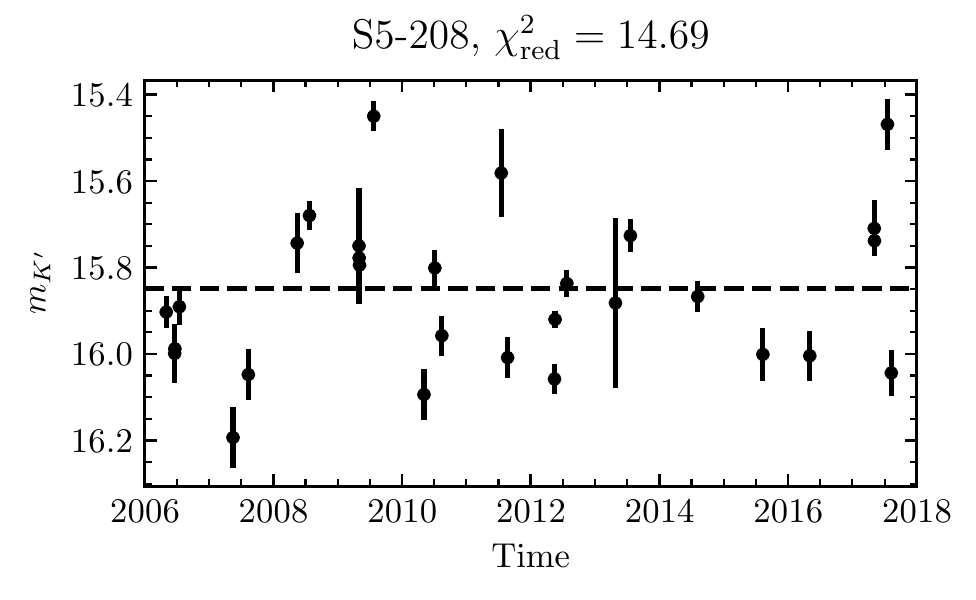}
\end{figure}

\begin{figure}[H]
    \epsscale{1.05}
    \plottwo{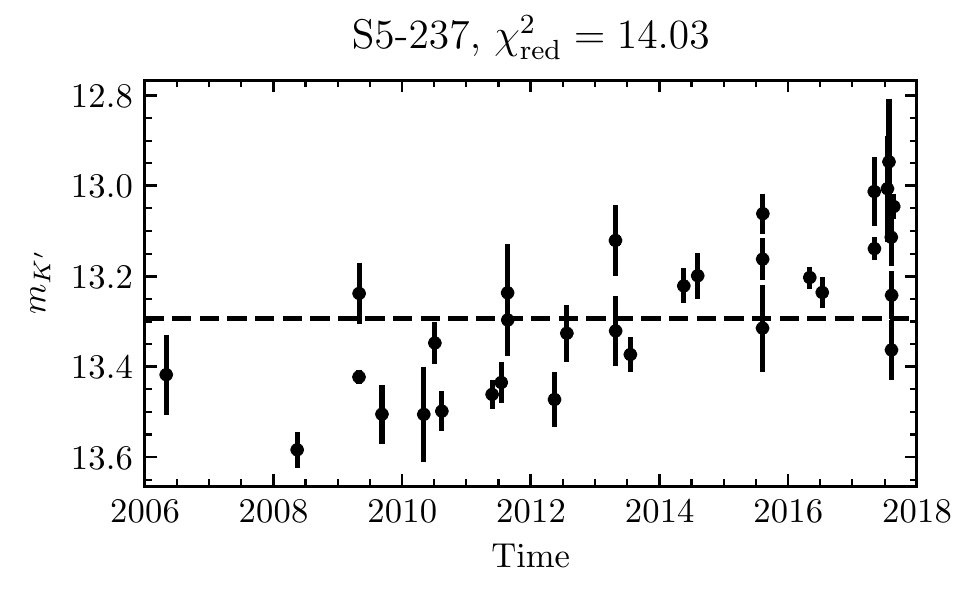}{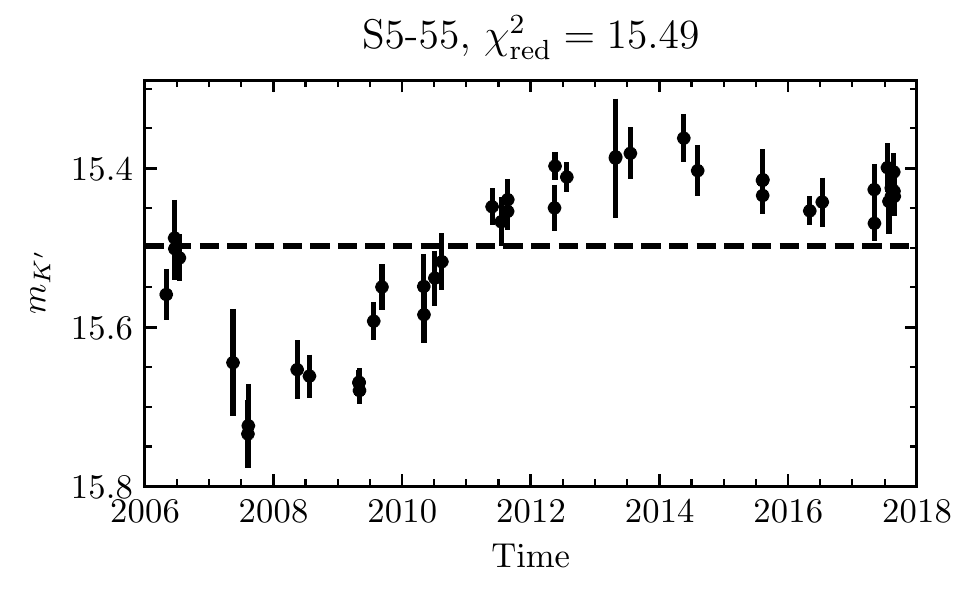}
\end{figure}

\begin{figure}[H]
    \epsscale{1.05}
    \plottwo{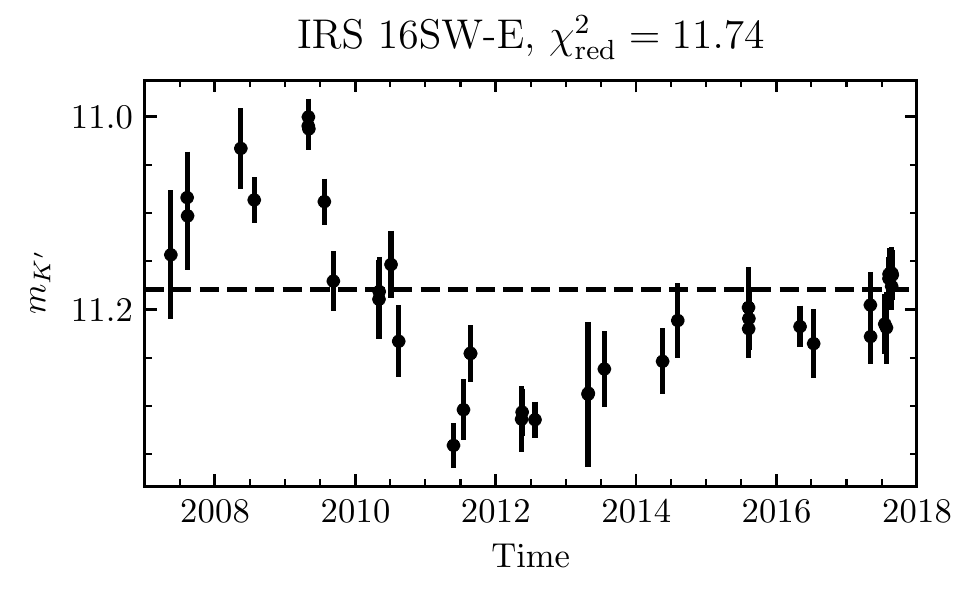}{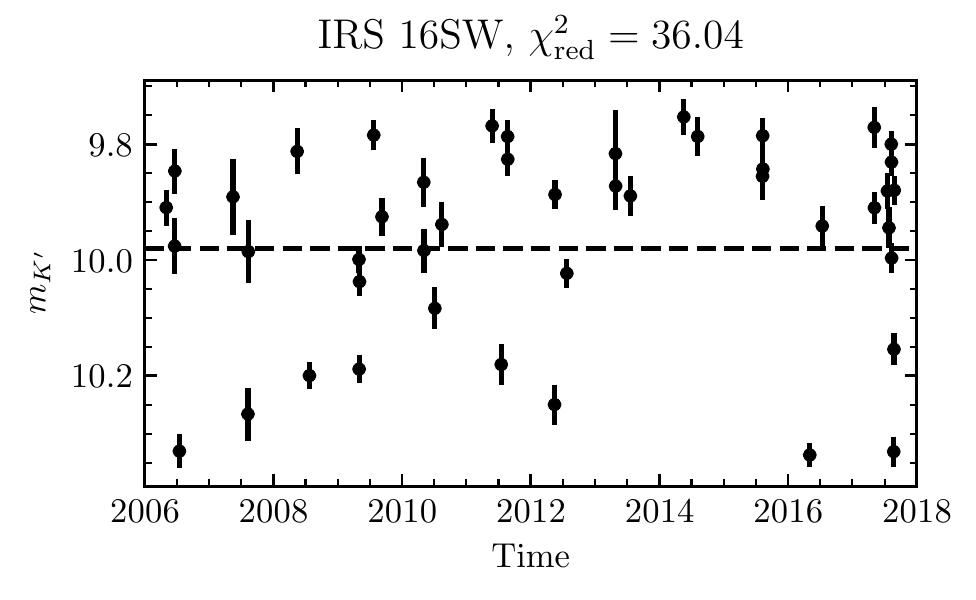}
\end{figure}

\begin{figure}[H]
    \epsscale{1.05}
    \plottwo{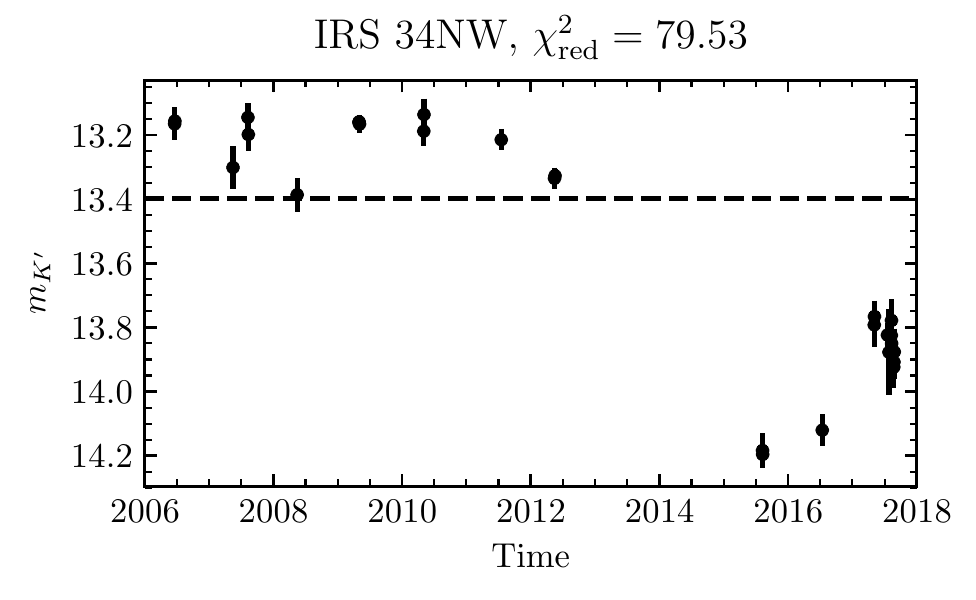}{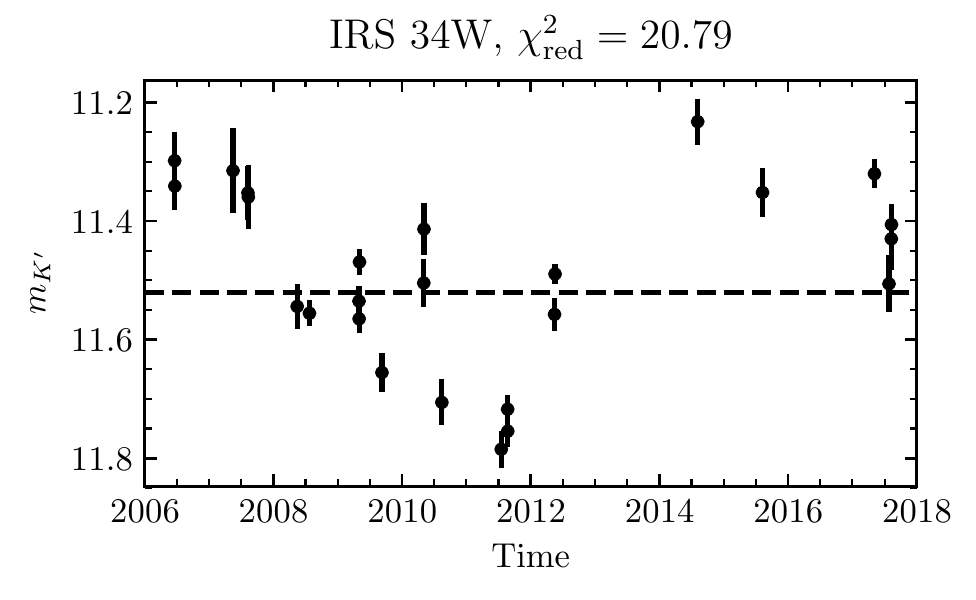}
\end{figure}

\begin{figure}[H]
    \epsscale{0.525}
    \plotone{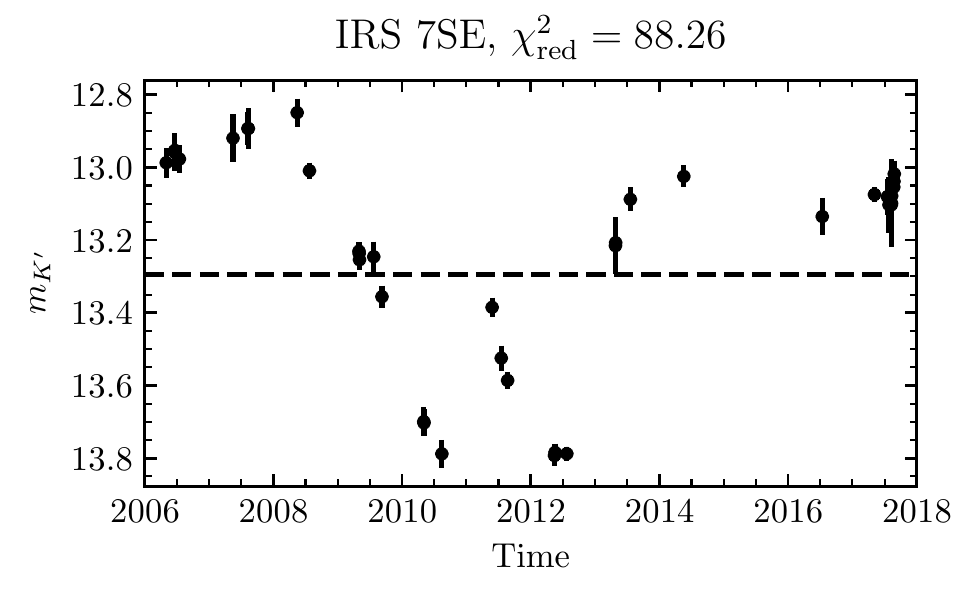}
\end{figure}



\section{Periodicity Methodology Details} 
\label{sec:periodicity_methodology_details}

\subsection{Period Search Range} 
\label{sub:period_search_range}

\begin{figure*}[h]
    \epsscale{1.17}
    \plottwo{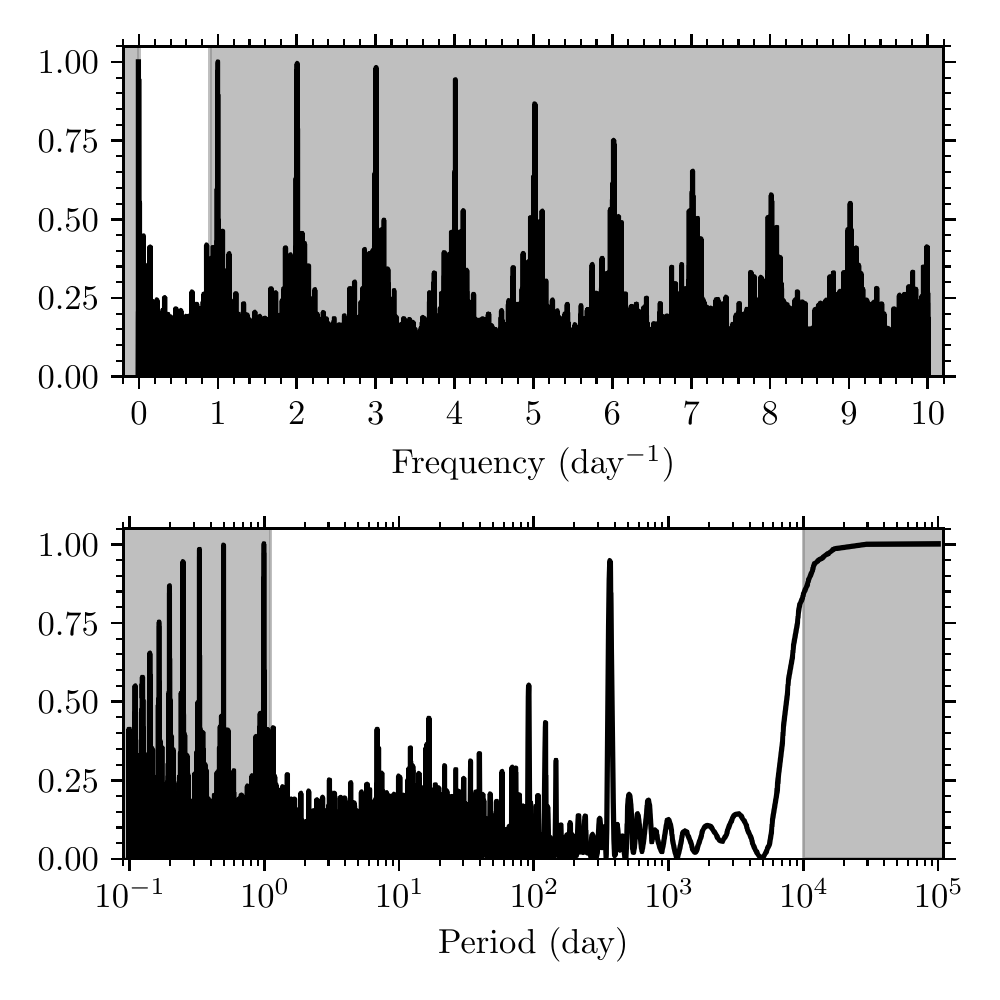}{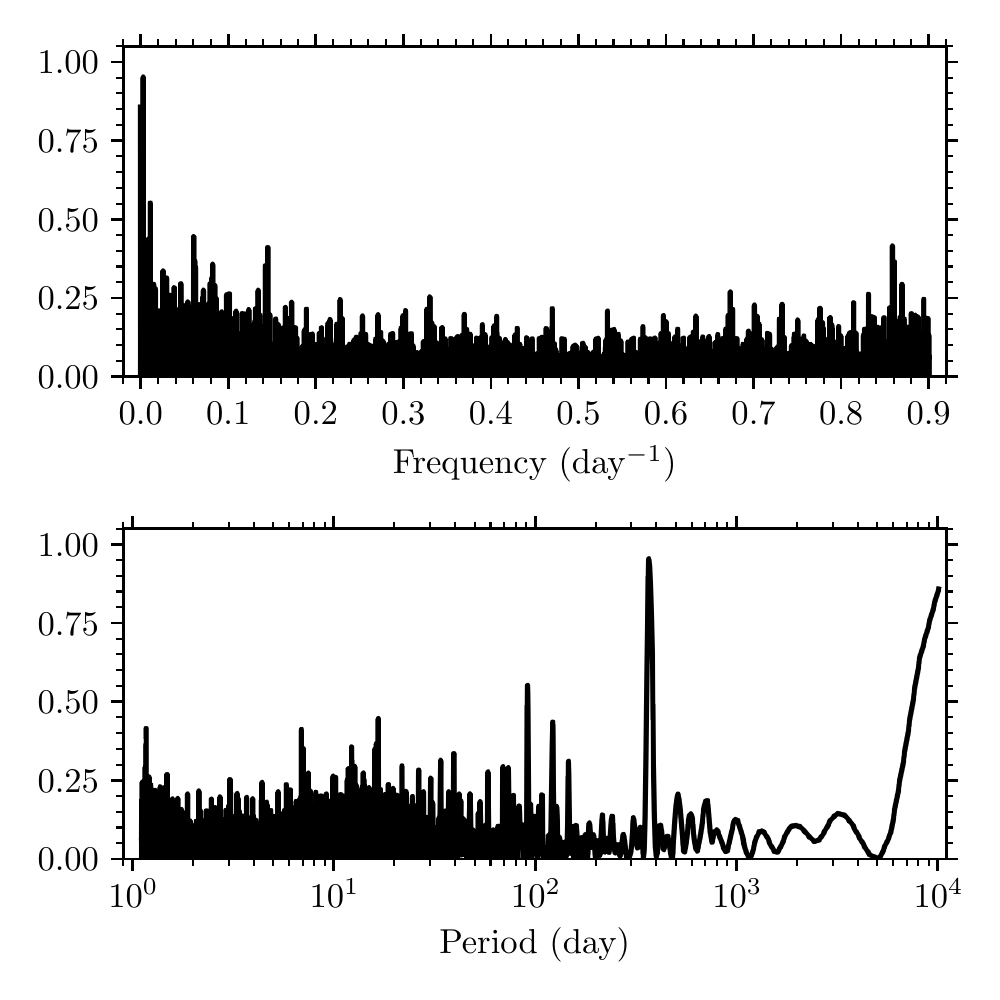}
    \caption{The Lomb-Scargle window power spectrum computed for our experiment's observation times at our periodicity search's frequency spacing is shown in the plots on the left. Note that frequencies larger than 1 day$^{-1}$ are prone to aliasing and at periods larger than $10^4$ days most signals will appear periodic due to this range being much larger than our overall observing time baseline. Our periodicity search range of 1.11 -- 10,000 days is highlighted in white on the left plots and is zoomed in for more detail in the plots on the right.}
    \label{fig:LS_Window}
\end{figure*}

We defined our periodicity search region by computing a window power spectrum for our observations (Figure~\ref{fig:LS_Window}). Notably, in this window power spectrum, periods shorter than about 1.11 days (frequencies $\gtrsim 0.9 \text{ day}^{-1}$ strongly suffer from aliasing due to the spacing of our observations being spaced apart by multiples of $\approx 1$ day. Above 10,000 days, the normalized Lomb-Scargle power extends to 1.0, at periods extending much beyond our observation's time span. With these considerations in mind, we defined our periodicity search region between the frequencies of 0.9 day$^{-1}$ to $10^{-4}$ day$^{-1}$, corresponding to periods between 1.11 days and 10,000 days. The remaining peaks in Figure~\ref{fig:LS_Window} in our periodicity search range originate from our nightly observation cadence and the length of a sidereal day: at a period of $\sim 1$ day (frequency $\sim 1 \text{ day}^{-1}$), and its harmonic rising up at a period of $\sim 10,000$ days (corresponding to a frequency $\sim 0 \text{ day}^{-1}$)). The $\sim 350$ days peak corresponds to our roughly yearly observation cadence, when the GC is visible in the night sky.

\subsection{Removal of long-term linear trends} 
\label{sub:removal_of_long_term_linear_trends}

\begin{figure*}[h]
    \epsscale{1.17}
    \plottwo{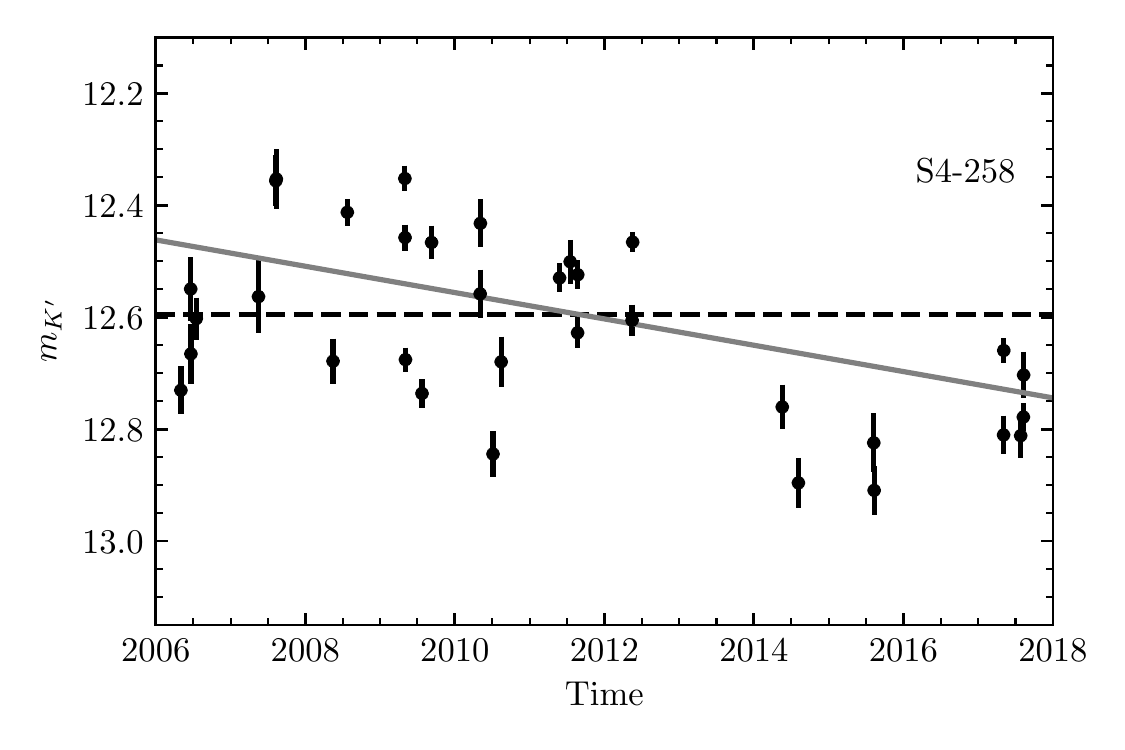}{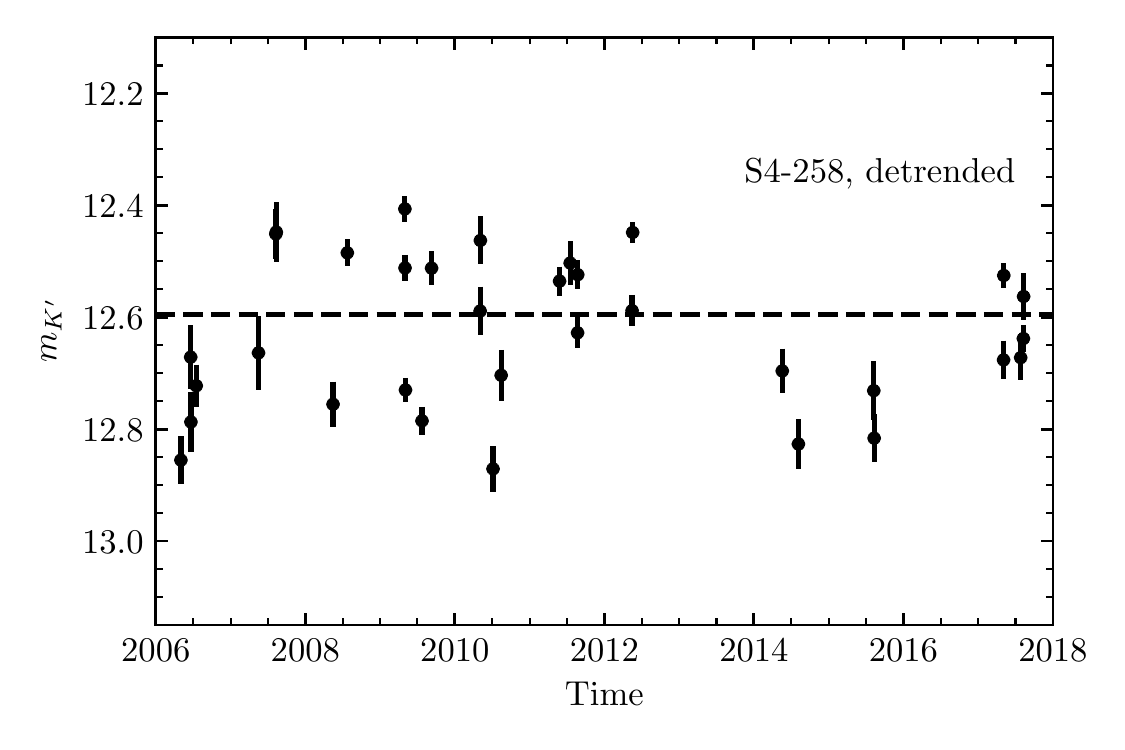}
    \caption{S4-258 \citep[E60:][]{2014ApJ...782..101P} is an eclipsing binary system at the Galactic Center. The left plot shows the light curve from S4-258 across our observations, where the dashed black line indicates the weighted mean magnitude, $\bar{m}_{K'}$, and solid gray line indicates the best-fit linear model to the data. This linear model can indicate a long-term dimming of the binary system. The right plot shows the same data, with the long-term linear dimming trend removed.}
    \label{fig:S4-258_light_curve}
\end{figure*}

\begin{figure*}[h]
    \epsscale{1.17}
    \plottwo{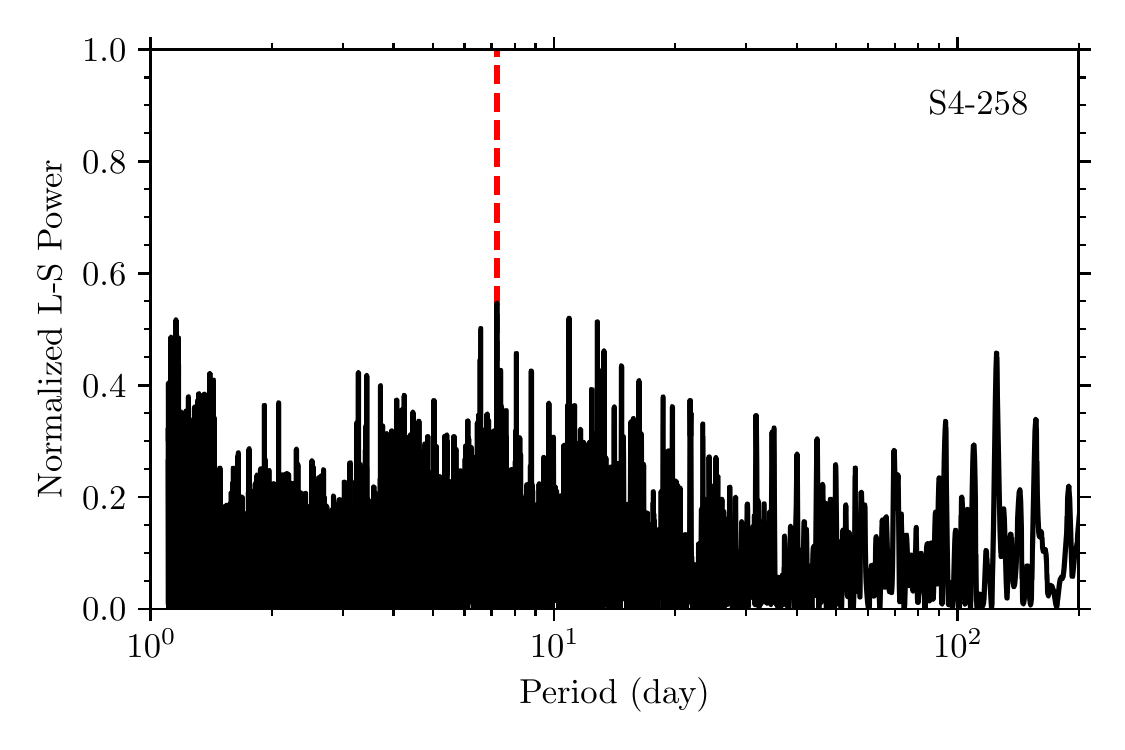}{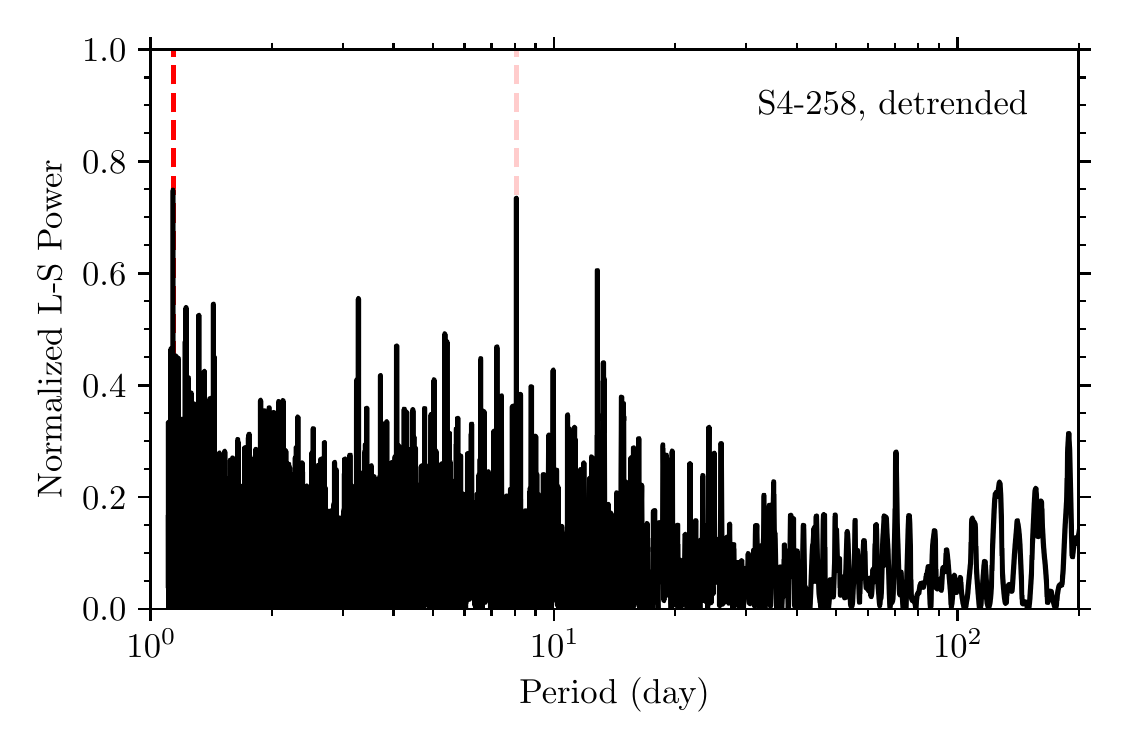}
    \caption{Lomb-Scargle periodogram of S4-258's light curve. The left plot shows the periodogram computed from our observations, while the right plot shows the periodogram once the long-term linear dimming trend is removed. The 1.1380 day peak in the periodogram constructed from the detrended light curve corresponds to the 2.2760 day binary period of the system. The 8.0637 day peak corresponds to an alias of the binary period. Removing the long-term linear dimming trend allows the binary period of the system and its alias to be detected.}
    \label{fig:S4-258_periodogram}
\end{figure*}

\begin{figure*}[h]
    \epsscale{1.17}
    \plottwo{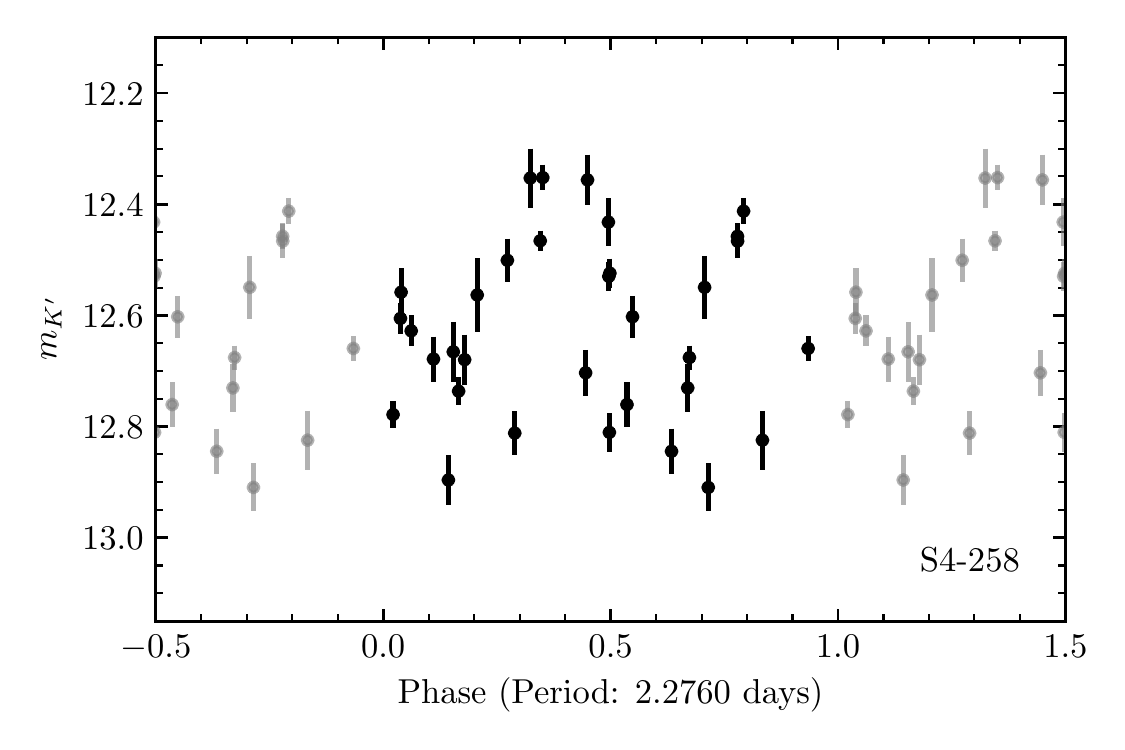}{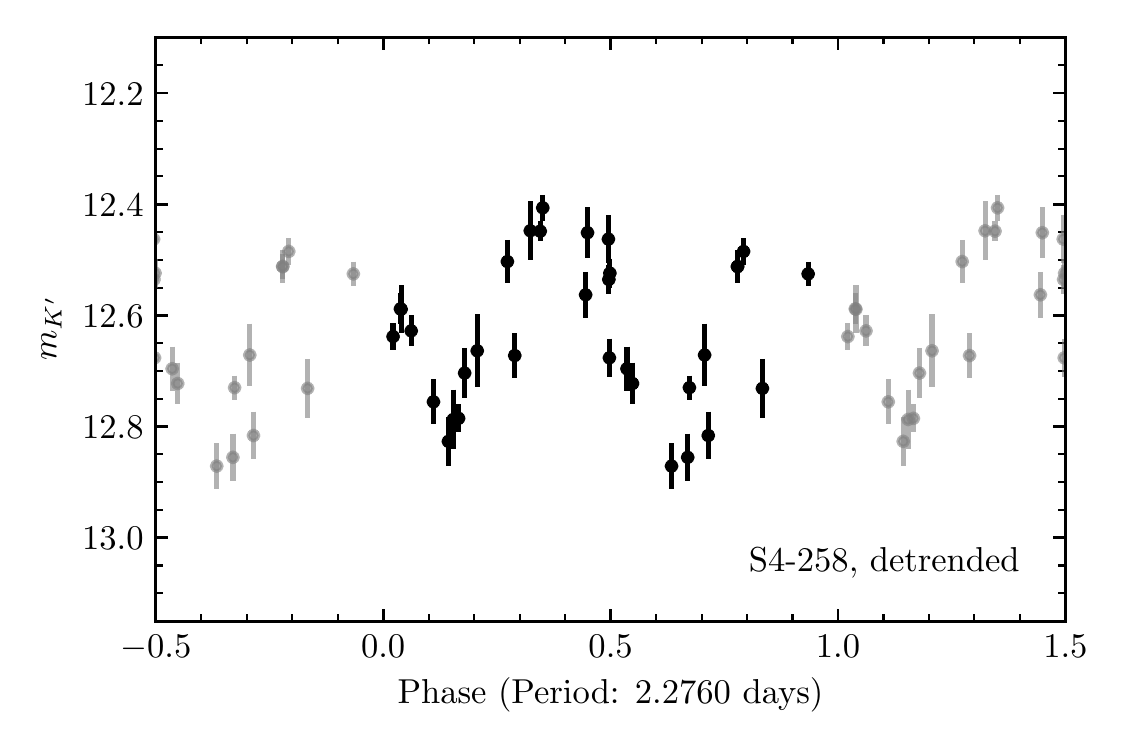}
    \caption{Phased light curve of S4-258 at its binary period of 2.2761 days. The left plot shows the phased light curve from our observations before detrending, while the right plot shows the same light curve with the long-term linear dimming trend removed (shown in Figure~\ref{fig:S4-258_light_curve}).}
    \label{fig:S4-258_phased_light_curve}
\end{figure*}

In our periodicity search, we removed long-term linear trends from the light curves of stars before computing a periodogram. This removal resulted in stronger detections of periodic signals. This can be demonstrated particularly well for the known eclipsing binary system S4-258 \citep[E60:][]{2014ApJ...782..101P}. S4-258 exhibits a long-term linear dimming trend in our dataset, possibly caused by extinction, over our observation baseline (Figure~\ref{fig:S4-258_light_curve}). After removing the long-term linear trend, we find that the periodic signal is detected more strongly in the periodogram (Figure~\ref{fig:S4-258_periodogram}) and that the phased light curve demonstrates is much smoother (Figure~\ref{fig:S4-258_phased_light_curve}).

Several stars in our sample display similar brightening or dimming trends to S4-258 (see Section~\ref{sec:_chi_2__text_red_geq_10_variables}). Any periodic trends that may exist for our sample stars in addition to these low order variations can be detected more strongly once the linear variation is removed.

\begin{figure*}[h]
    \epsscale{1.2}
    \plotone{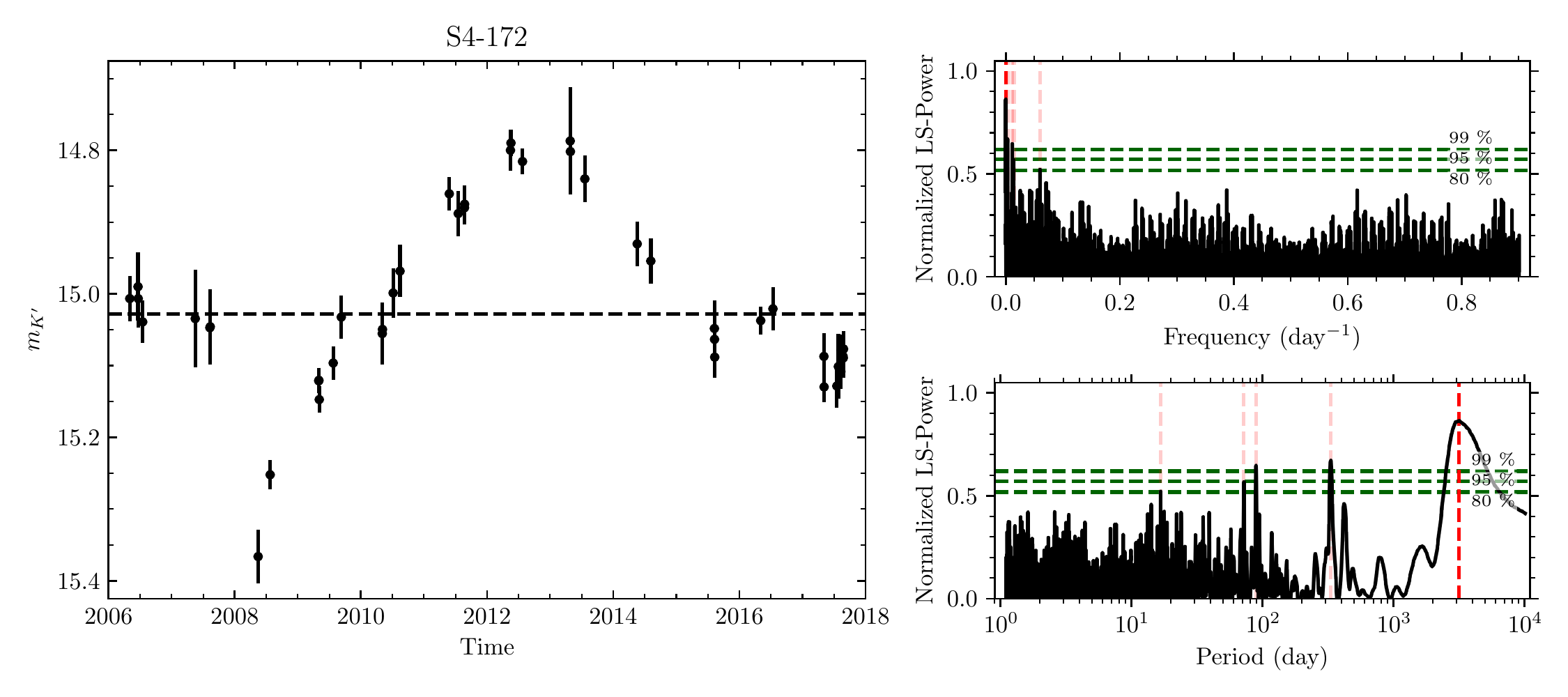}
    \caption{Light curve (left) and periodograms (right) for the star S4-172. The horizontal dashed line in the light curve indicates the weighted mean magnitude. The horizontal dashed green lines in the periodograms indicate the bootstrap test significance levels, while the vertical dashed red lines indicate periodogram peaks above 80\% bootstrap significance.\\
    S4-172 is an example where the long-term variability (corresponding to a peak $\sim 3000$ days) is aliased as powerful peaks in the periodogram at shorter periods.}
    \label{fig:S4-172_lc_per}
\end{figure*}


\section{Periodic Detections} 
\label{sec:_periodic_detections}

\subsection{Likely Periodic Variables} 
\label{sub:_likely_periodic_variables}
\begin{figure}[H]
    \epsscale{1.05}
    \plottwo{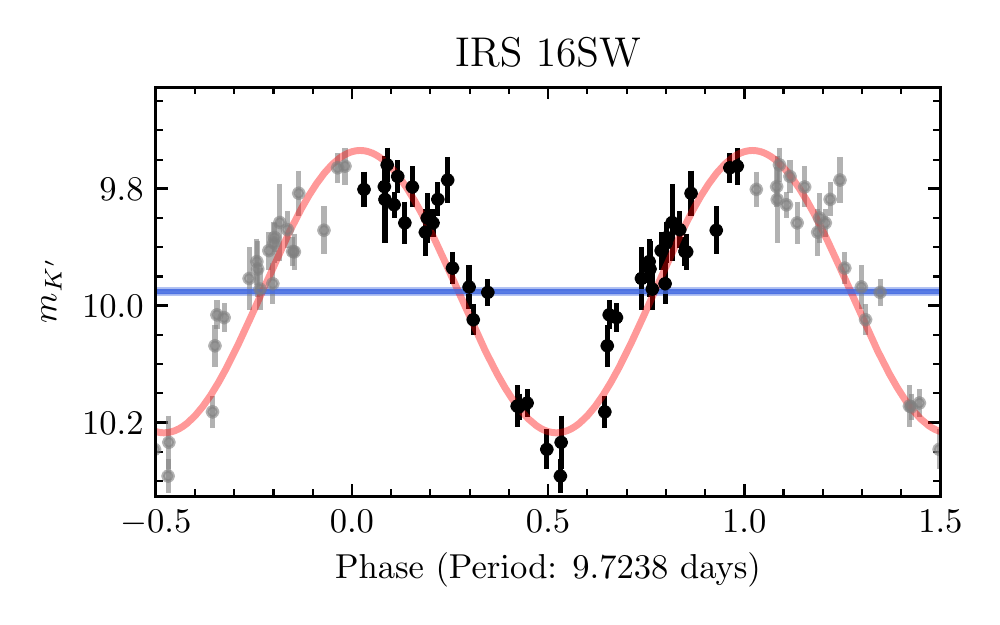}{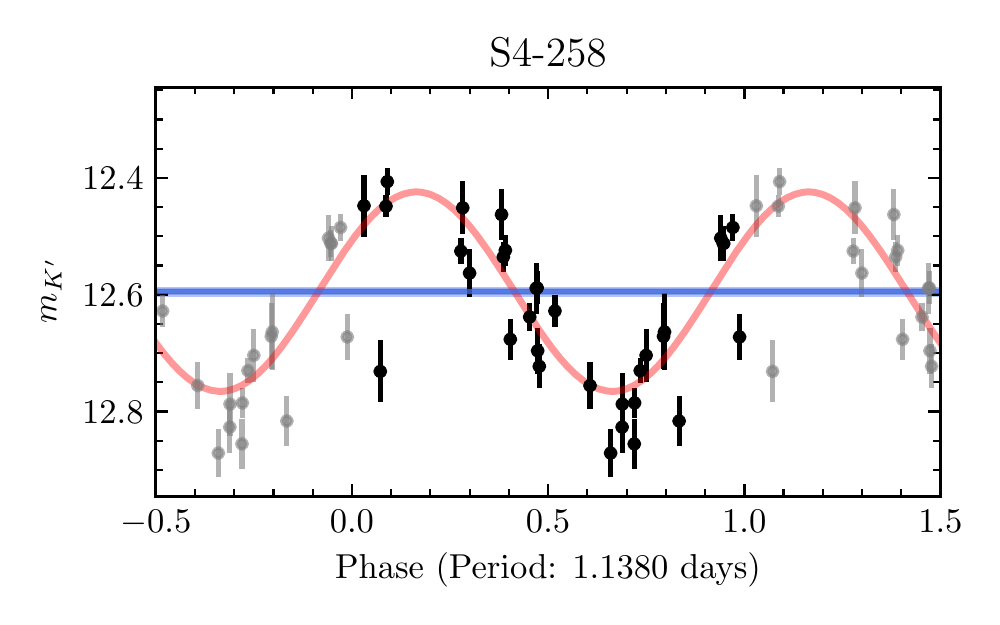}
\end{figure}


\subsection{Possible Periodic Signals} 
\label{sub:_possible_periodic_detections}

\begin{figure}[H]
    \epsscale{1.05}
    \plottwo{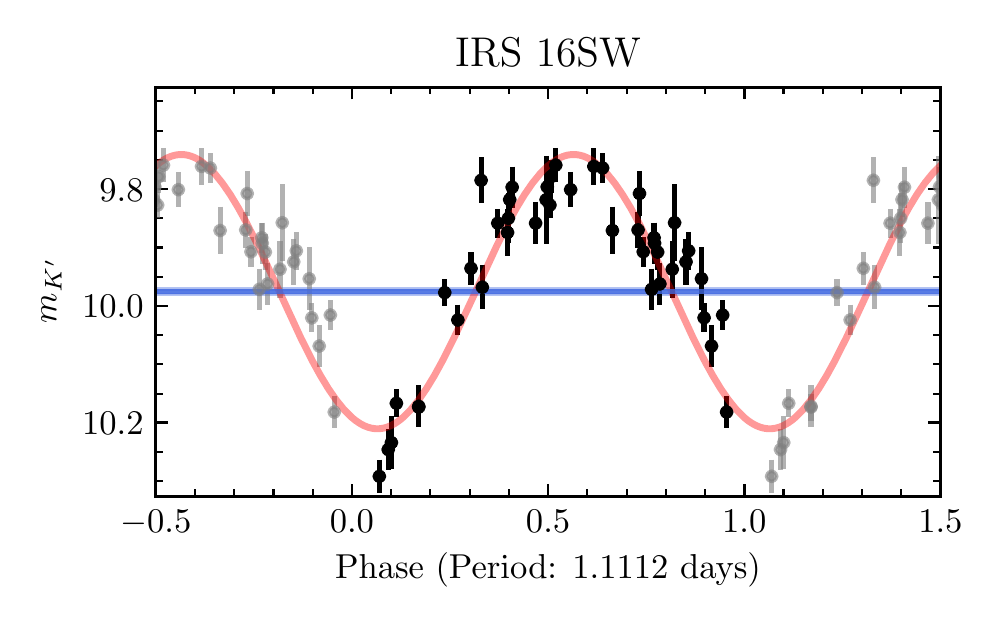}{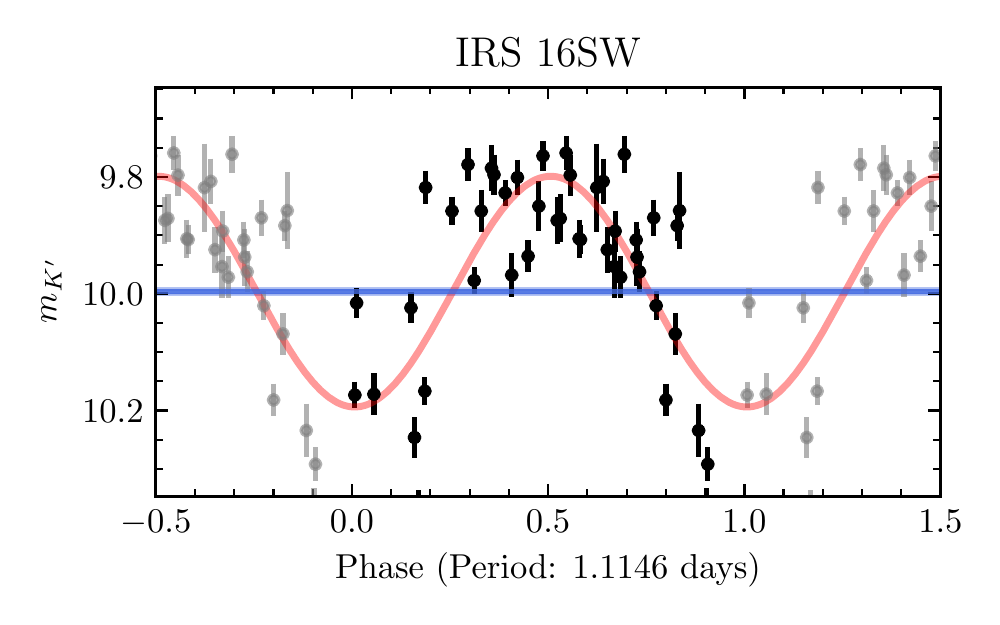}
\end{figure}

\begin{figure}[H]
    \epsscale{1.05}
    \plottwo{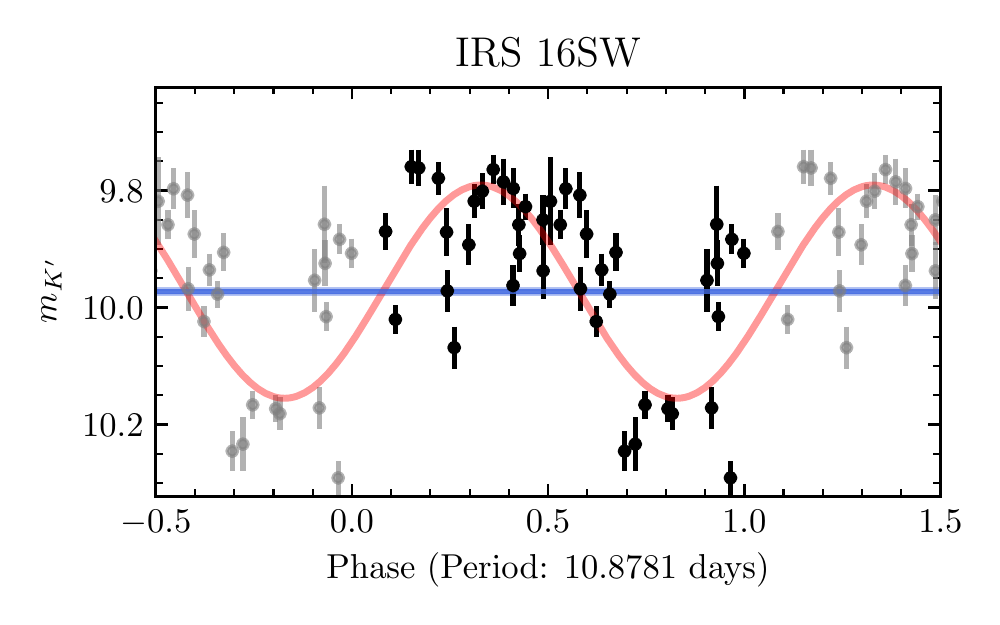}{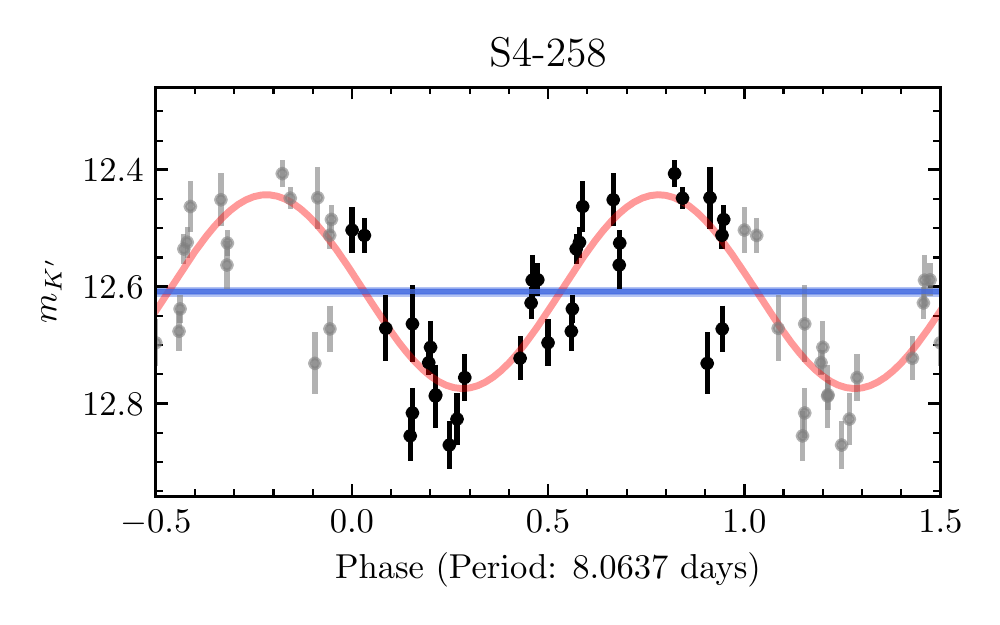}
\end{figure}

\begin{figure}[H]
    \epsscale{1.05}
    \plottwo{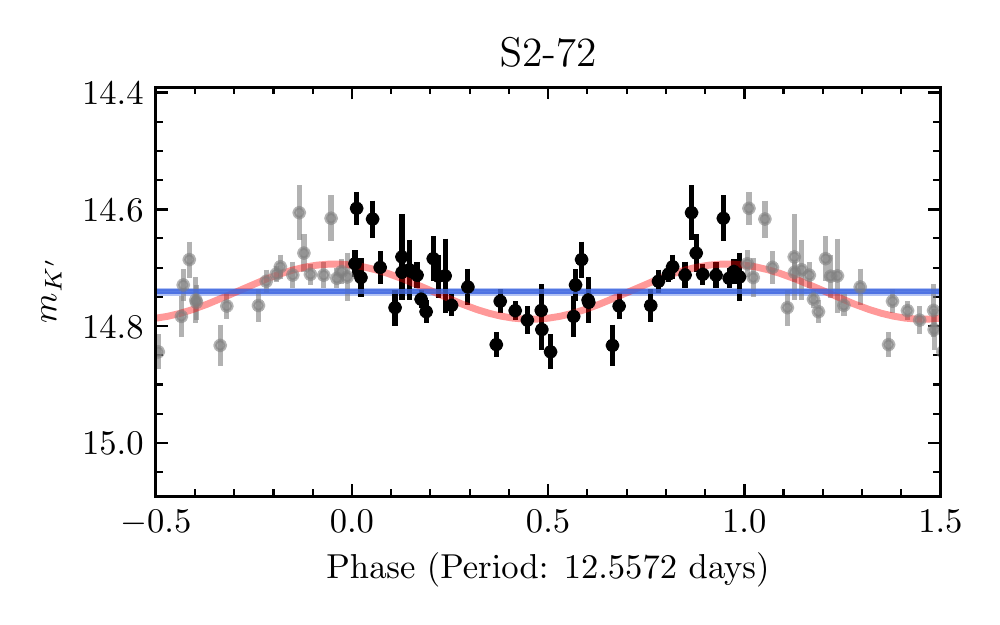}{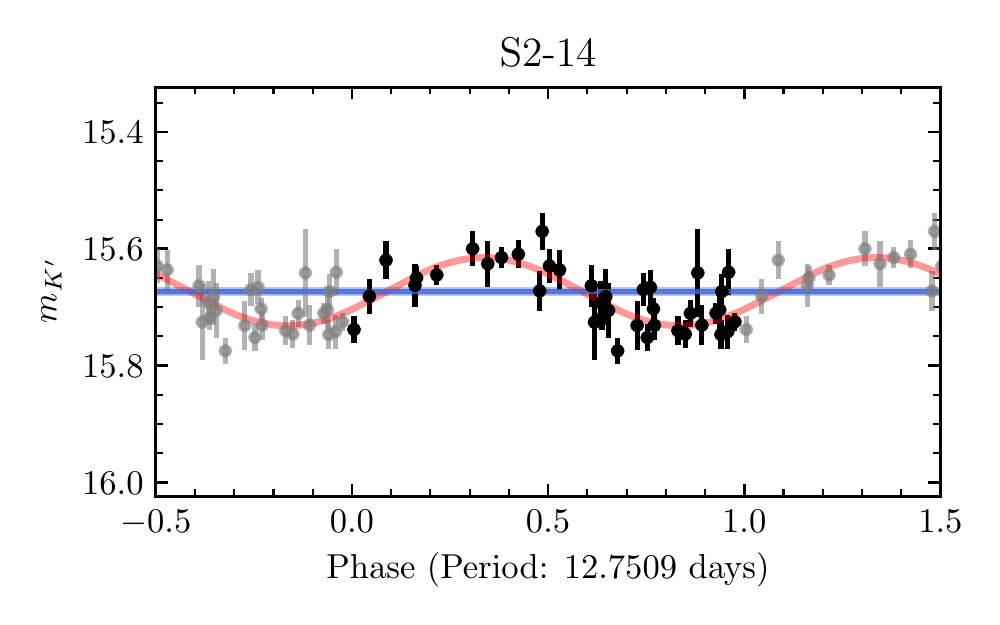}
\end{figure}

\begin{figure}[H]
    \epsscale{1.05}
    \plottwo{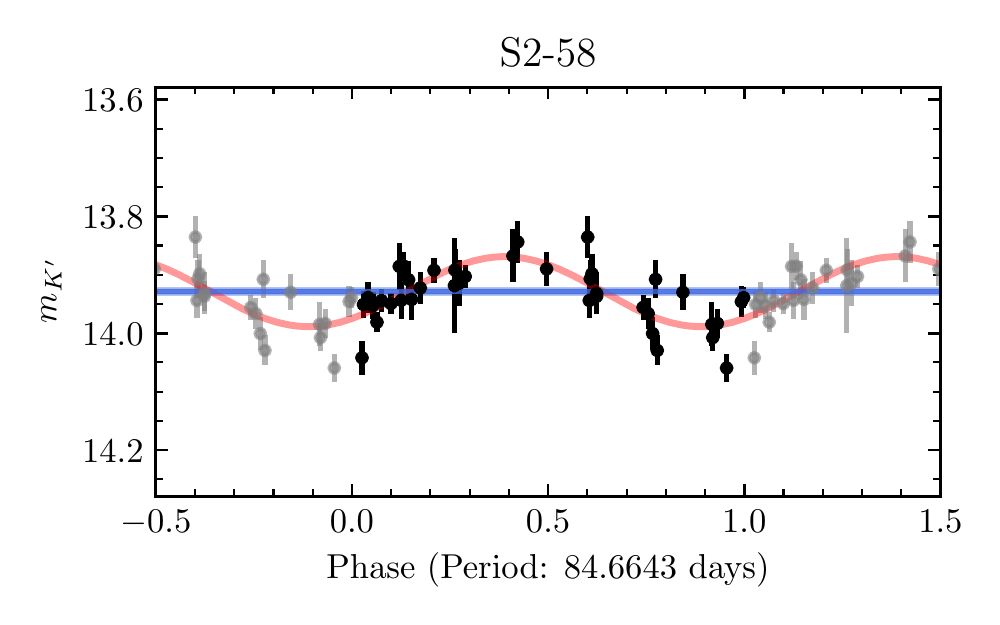}{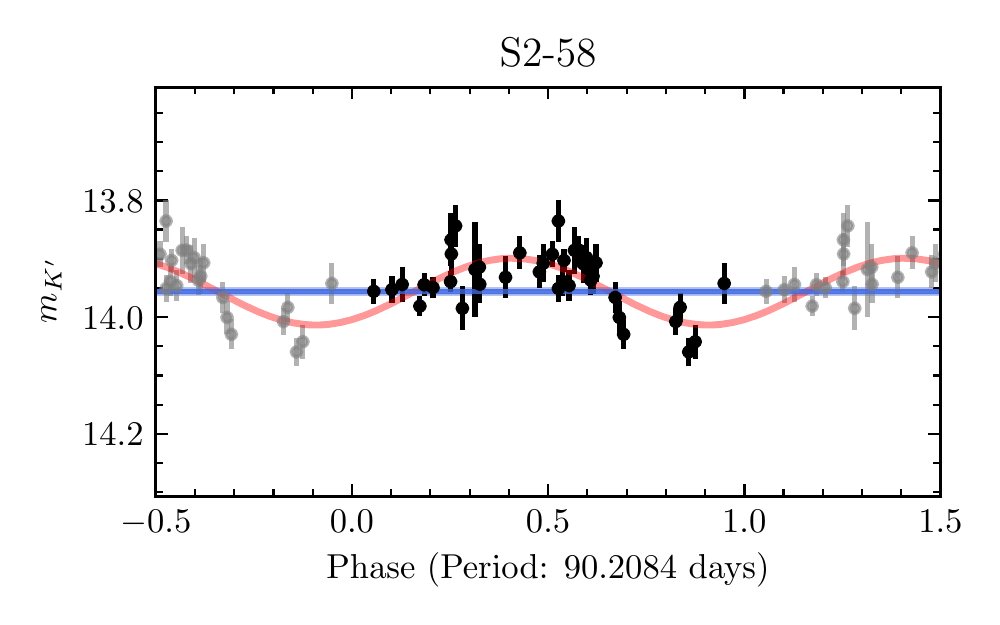}
\end{figure}

\begin{figure}[H]
    \epsscale{1.05}
    \plottwo{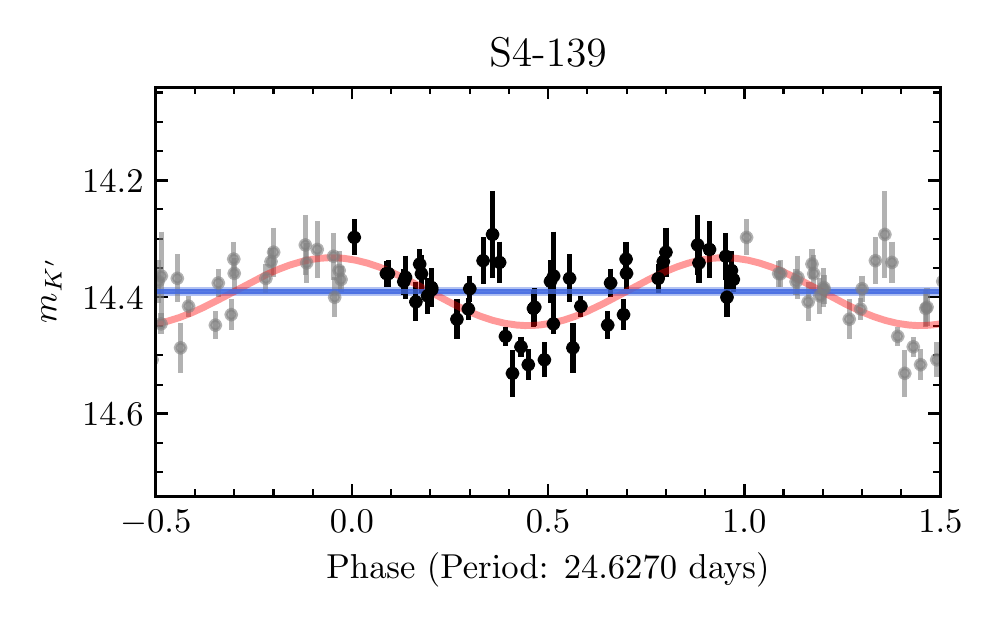}{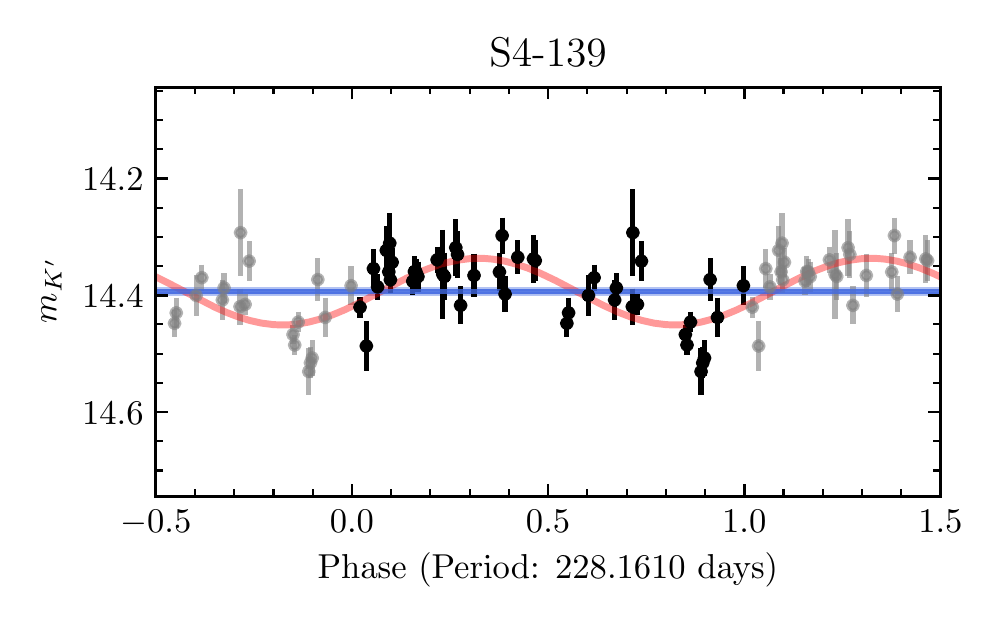}
\end{figure}

\begin{figure}[H]
    \epsscale{1.05}
    \plottwo{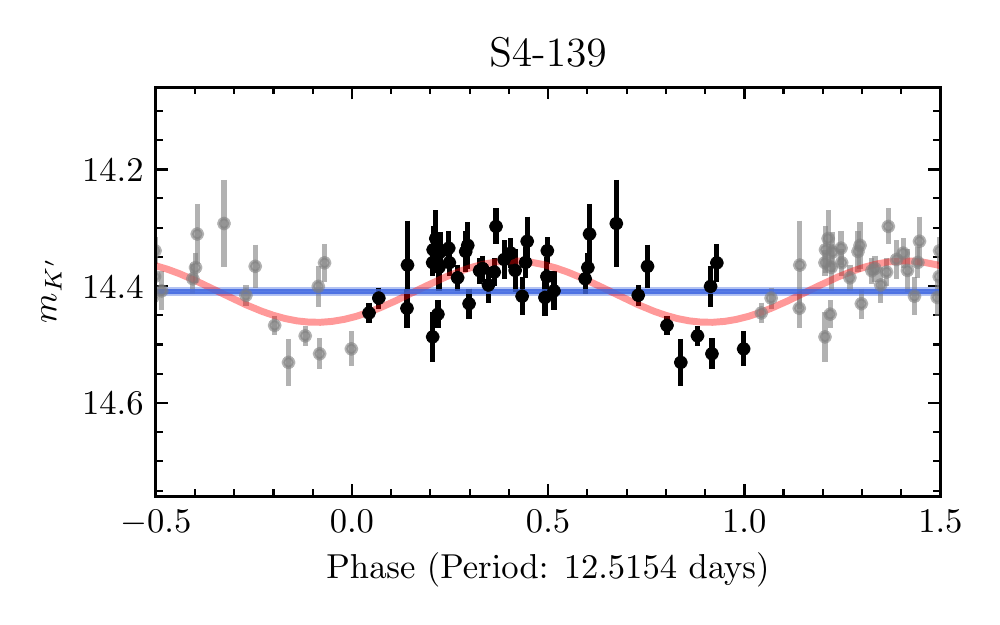}{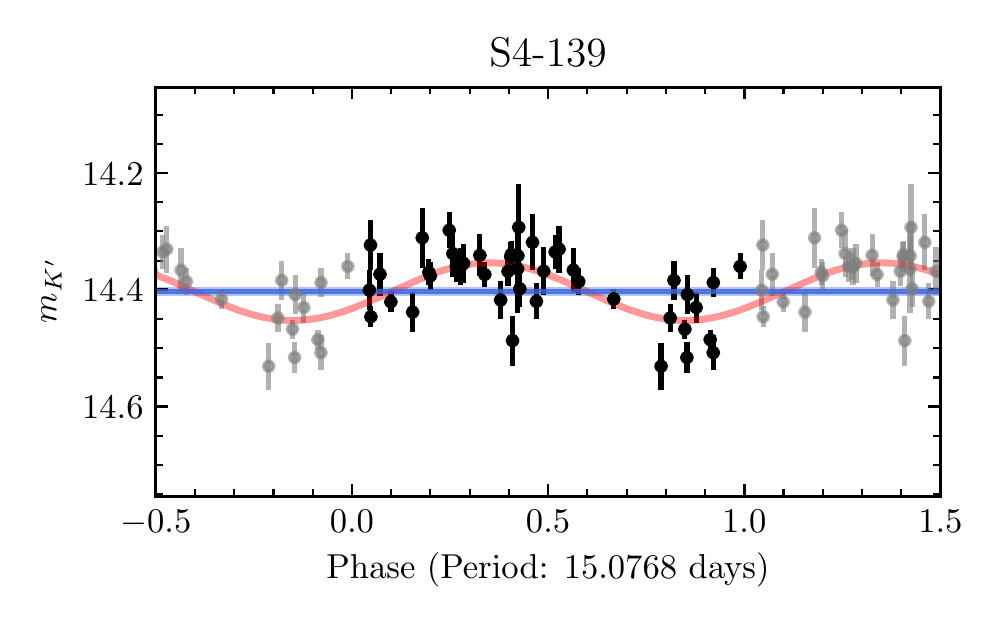}
\end{figure}

\begin{figure}[H]
    \epsscale{0.525}
    \plotone{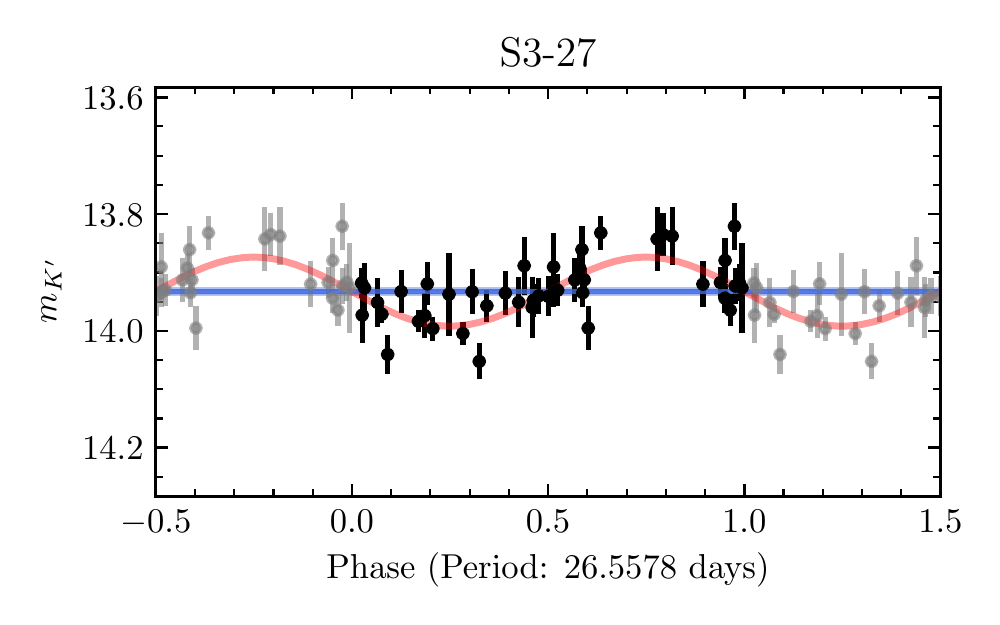}
\end{figure}

\begin{figure}[H]
    \epsscale{1.05}
    \plottwo{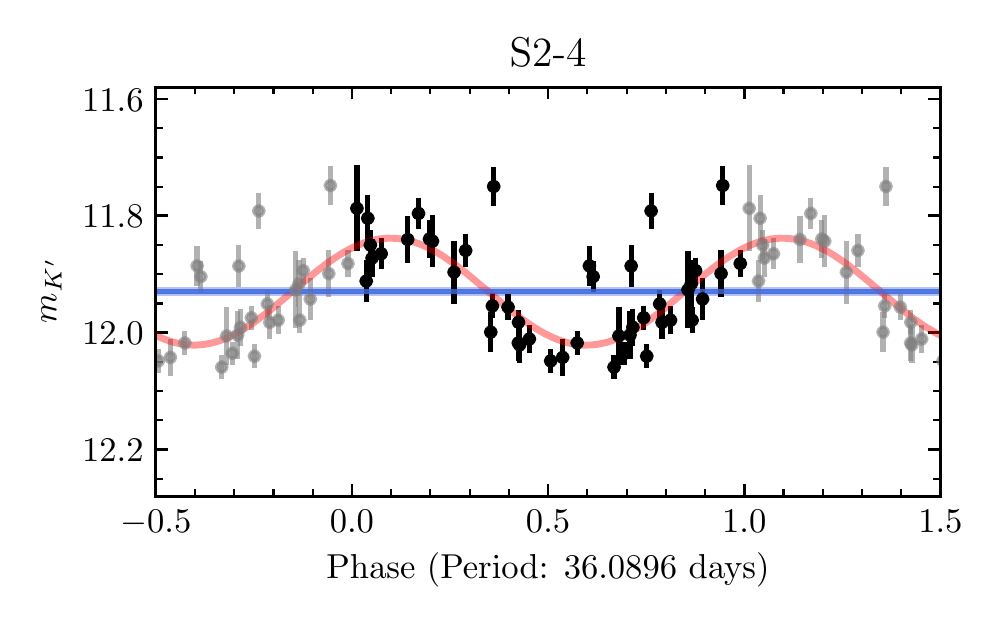}{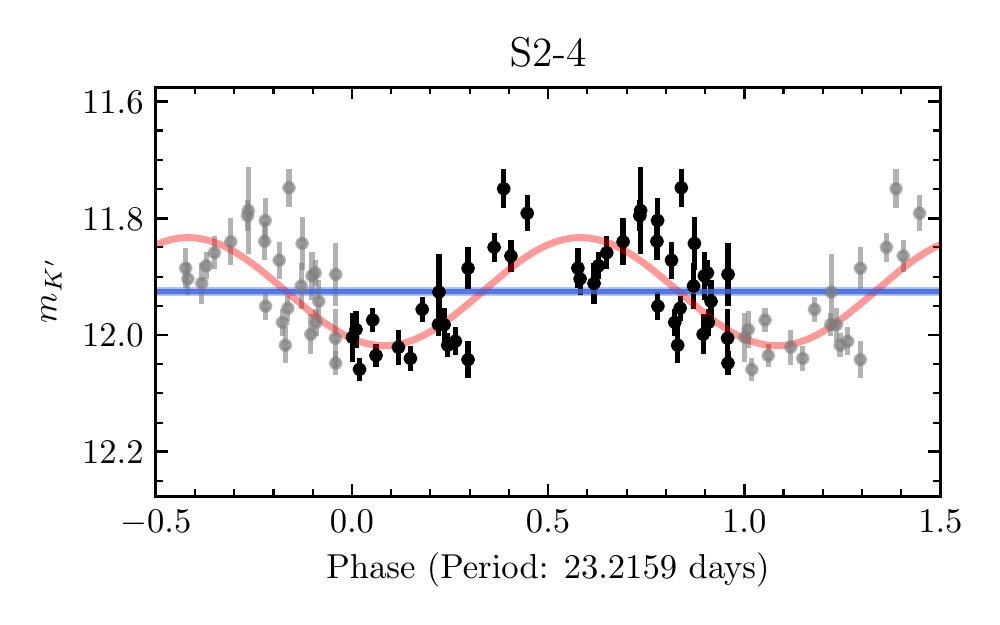}
\end{figure}

\begin{figure}[H]
    \epsscale{1.05}
    \plottwo{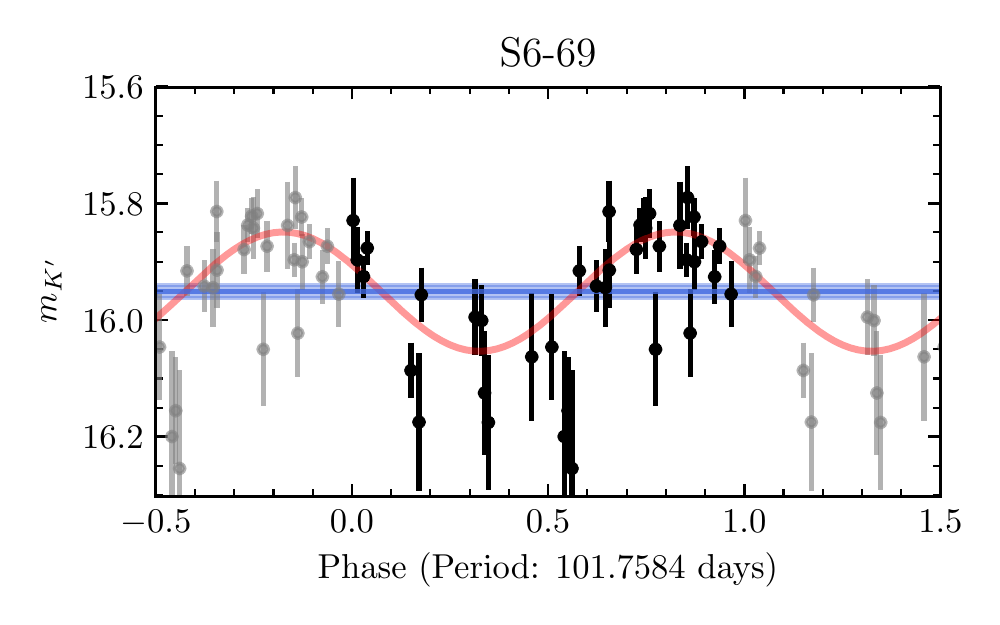}{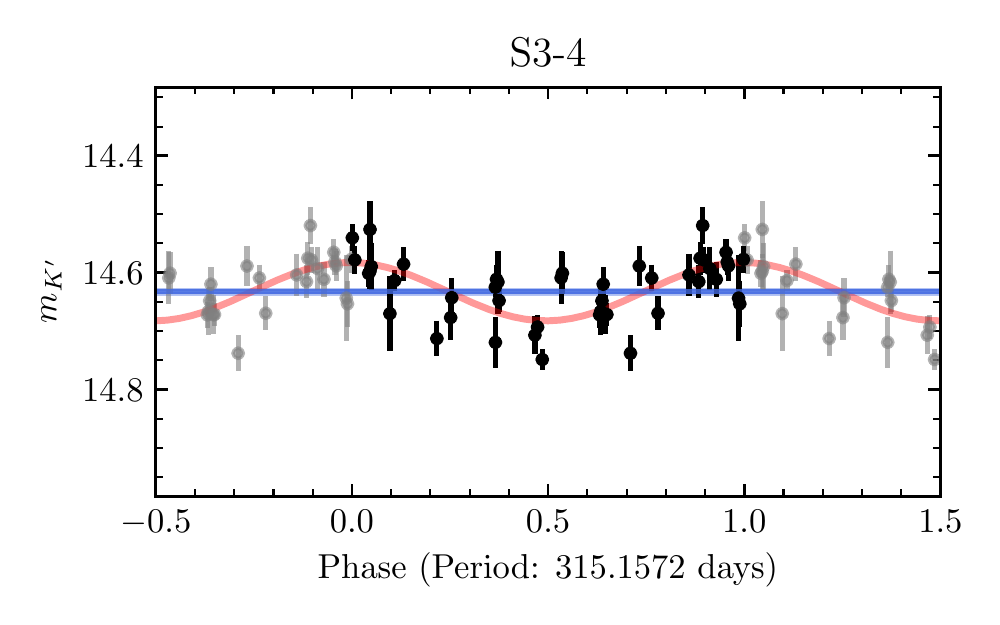}
\end{figure}

\begin{figure}[H]
    \epsscale{1.05}
    \plottwo{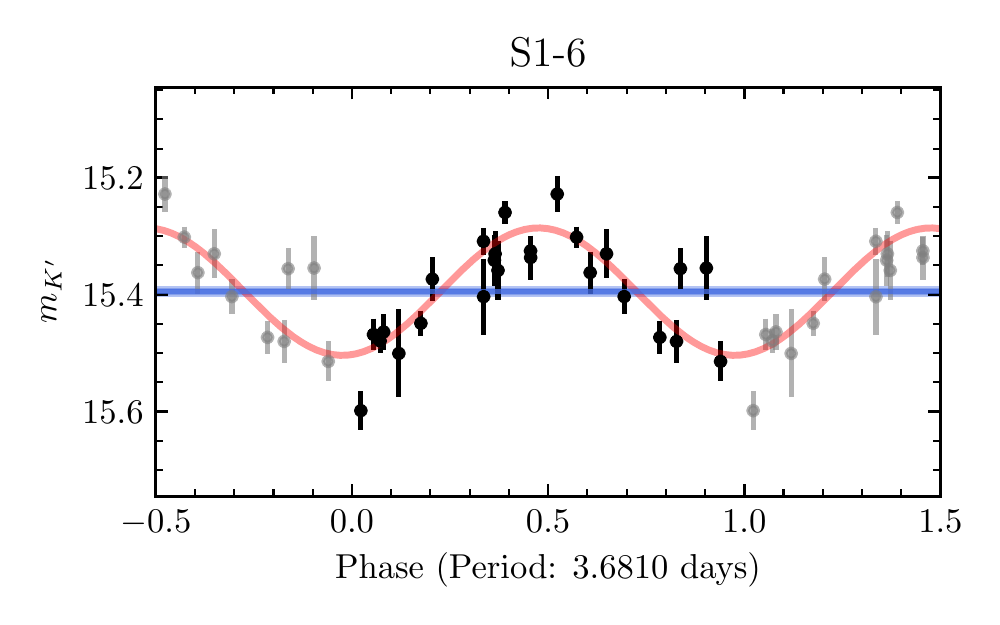}{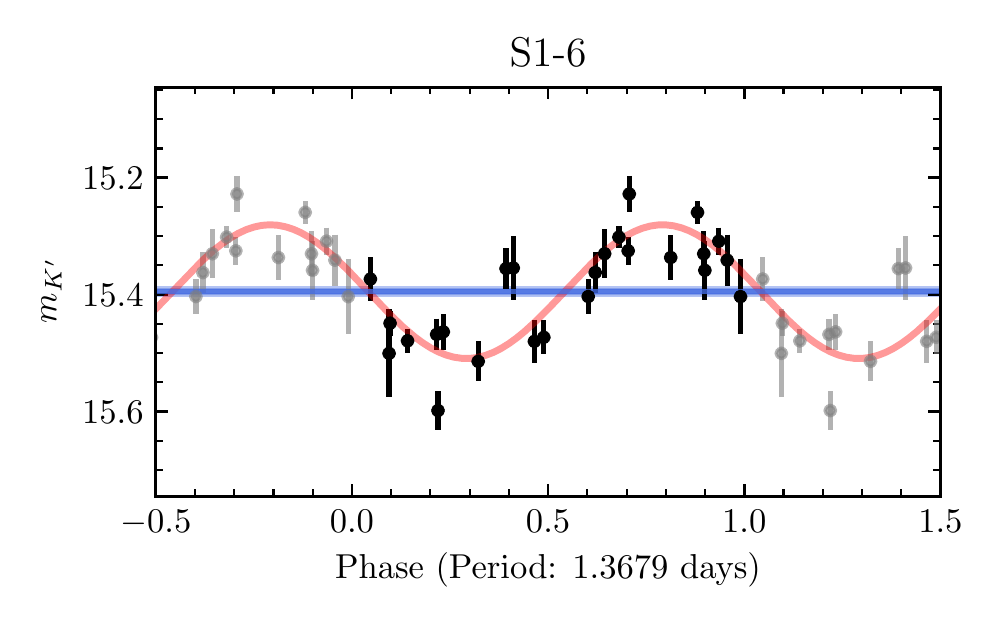}
\end{figure}



\begin{longrotatetable}
\begin{deluxetable}{lrrrrrrrrrrrr}
    \tablecolumns{13}
    \tablecaption{Catalog of stars in sample\label{tab:Sample_Catalog}}
    \tablehead{
        \colhead{Star}                  &
        \colhead{Mean Mag}              &
        \colhead{Spectral Type}         &
        \colhead{Nights Detected}       &
        \colhead{$\chi^2_{\text{red}}$} &
        \colhead{Var.?}                 &
        \colhead{Var.}                  &
        \colhead{$x_0$}                 &
        \colhead{$y_0$}                 &
        \colhead{$t_0$}                 &
        \colhead{RMS}                   &
        \colhead{IQR}                   &
        \colhead{$1/\eta$}              \\
        \colhead{}                      &
        \colhead{($K'$)}                &
        \colhead{}                      &
        \colhead{}                      &
        \colhead{}                      &
        \colhead{}                      &
        \colhead{Type}                  &
        \colhead{($''$ E of Sgr A*)}    &
        \colhead{($''$ N of Sgr A*)}    &
        \colhead{}                      &
        \colhead{}                      &
        \colhead{}                      &
        \colhead{}
    }
\startdata
IRS 16C & 9.91 & Early & 45 & 3.39 & Yes &  & 1.05 & 0.55 & 2009.989 & 0.05 & 0.07 & 1.20 \\
IRS 16SW & 9.98 & Early & 45 & 36.04 & Yes &  & 1.11 & -0.95 & 2009.820 & 0.17 & 0.20 & 0.49 \\
IRS 16NW & 10.16 & Early & 45 & 1.02 & No &  & 0.08 & 1.22 & 2010.047 & 0.04 & 0.05 & 0.99 \\
IRS 33E & 10.18 & Early & 45 & 2.22 & No &  & 0.71 & -3.14 & 2010.182 & 0.04 & 0.06 & 1.05 \\
S2-17 & 10.61 & Early & 45 & 0.90 & No &  & 1.34 & -1.88 & 2010.154 & 0.03 & 0.05 & 1.38 \\
IRS 16CC & 10.67 & Early & 45 & 5.05 & Yes &  & 1.98 & 0.60 & 2010.135 & 0.06 & 0.11 & 3.58 \\
S5-89 & 10.83 & Unknown & 45 & 0.73 & No &  & -0.79 & -5.25 & 2010.162 & 0.03 & 0.04 & 1.08 \\
S3-22 & 11.03 & Late & 45 & 0.68 & No &  & -0.34 & -3.21 & 2010.201 & 0.02 & 0.03 & 1.04 \\
\enddata
\tablecomments{Variability type is determined for highly variable stars ($\chi^2_{\text{red}} \geq 10.0$) by visual inspection of the star's light curve and local field of the star in the images. `C' indicates variability likely caused by confusion and `L' indicates variability on timescales $\gtrsim 1$ year.}
\end{deluxetable}
\end{longrotatetable}

\end{document}